\documentclass[structabstract]{aa}
\usepackage{amsmath}
\usepackage{amssymb}
\usepackage[section]{algorithm}
\usepackage{algorithmic}
\usepackage{aalongtable}
\usepackage{multirow}

\usepackage{graphicx}
\usepackage{txfonts}
\usepackage{natbib}

\bibpunct{(}{)}{;}{a}{}{,}

\newcommand{\quat}[1]{\mathbf{#1}}

\newcommand{\spdot}[1]{\dot{\raisebox{0pt}[7pt][0pt]{${#1}$}}}

\begin{document}
\title{The astrometric core solution for the Gaia mission}
\subtitle{Overview of models, algorithms and software implementation}

\author{L. Lindegren\inst{1}\and
U. Lammers\inst{2}\and
D. Hobbs\inst{1}\and
W. O'Mullane\inst{2}\and
U. Bastian\inst{3}\and
J. Hern{\'a}ndez\inst{2}}

\institute{Lund Observatory, Lund University, Box 43, SE-22100 Lund, Sweden\\
\email{Lennart.Lindegren, David.Hobbs@astro.lu.se}
\and
European Space Agency (ESA), European Space Astronomy Centre (ESAC),
P.O. Box (Apdo.~de Correos) 78, ES-28691 Villanueva de la Ca{\~n}ada, Madrid, Spain\\
\email{Uwe.Lammers, William.OMullane, Jose.Hernandez@sciops.esa.int}
\and
Astronomisches Rechen-Institut, Zentrum f{\"u}r Astronomie der Universit{\"a}t
Heidelberg, M{\"o}nchhofstr. 12--14, DE-69120 Heidelberg, Germany \\
\email{bastian@ari.uni-heidelberg.de}
}

\date{Received 17 August 2011 / Accepted 25 November 2011}

\abstract
{The Gaia satellite will observe about one billion stars and other point-like
sources. The astrometric core solution will determine the astrometric 
parameters (position, parallax, and proper motion) for a subset of these
sources, using a global solution approach which must also include a
large number of parameters for the satellite attitude and optical instrument. 
The accurate and efficient implementation of this solution is an extremely 
demanding task, but crucial for the outcome of the mission.}
{We aim to provide a comprehensive overview of the mathematical and physical
models applicable to this solution, as well as its numerical and algorithmic 
framework.
}
{The astrometric core solution is a simultaneous least-squares estimation
of about half a billion parameters, including the astrometric parameters
for some 100 million well-behaved so-called primary sources. The global nature of the 
solution requires an iterative approach, which can be broken down into a 
small number of distinct processing blocks (source, attitude, calibration 
and global updating) and auxiliary processes (including the frame rotator and selection
of primary sources). We describe each of these processes in some detail, 
formulate the underlying models, from which the observation
equations are derived, and outline the adopted numerical solution methods with due 
consideration of robustness and the structure of the resulting system of
equations. Appendices provide brief introductions to some 
important mathematical tools (quaternions and B-splines for the attitude 
representation, and a modified Cholesky algorithm for positive semidefinite problems)
and discuss some complications expected in the real mission data.}
{A complete software system called AGIS (Astrometric Global Iterative
Solution) is being built according to the methods described in the paper.
Based on simulated data for 2 million primary sources we present some 
initial results, demonstrating the 
basic mathematical and numerical validity of the approach and, 
by a reasonable extrapolation, its practical feasibility in terms of data 
management and computations for the real mission.}
{}

\keywords{Astrometry --  Methods: data analysis --  Methods: numerical -- Space vehicles: instruments}

\maketitle

\section{Introduction}\label{sec:intro}

The space astrometry mission Gaia, planned to be launched by the
European Space Agency (ESA) in 2013, will determine accurate astrometric
data for about one billion objects in the magnitude range from 6 to 20.
Accuracies of 8--25~micro-arcsec ($\mu$as) are typically expected for the trigonometric
parallaxes, positions at mean epoch, and annual proper motions of
simple (i.e., apparently single) stars down to 15th magnitude. The
astrometric data are complemented by photometric and spectroscopic
information collected with dedicated instruments on board the Gaia
satellite. The mission will result in an astronomical database of
unprecedented scope, accuracy and completeness becoming available
to the scientific community around 2021.

The original interferometric concept for a successor mission to
Hipparcos, called GAIA (Global Astrometric Interferometer for
Astrophysics), was described by \citet{gaia1996} but has since evolved
considerably by the incorporation of novel ideas \citep{iaus248:EH} and as a 
result of industrial studies conducted under ESA contracts \citep{gaia2001}.
The mission, now in the final integration phase with
EADS Astrium as prime contractor, is no longer an interferometer but
has retained the name Gaia, which is thus no acronym. For some brief
but up-to-date overviews of the mission, see \citet{iaus248:LL}
and \citet{iaus261:LL}. The scientific case is most comprehensively described 
in the proceedings of the conference The Three-Dimensional Universe with 
Gaia \citep{2005ESASP.576.....T}.

In parallel with the industrial development of
the satellite, the Gaia Data Processing and Analysis Consortium
\citep[DPAC;][]{iaus248:FM} is charged with the task of developing and
running a complete data processing system for analysing the satellite
data and constructing the resulting database (`Gaia Catalogue').
This task is extremely difficult due to the large quantities of
data involved, the complex relationships between different kinds
of data (astrometric, photometric, spectroscopic) as well as between
data collected at different epochs, the need for complex yet efficient
software systems, and the interaction and sustained support of many
individuals and groups over an extended period of time.

A fundamental part of the data processing task is the astrometric core
solution, currently under development in DPAC's Coordination Unit 3 
(CU3), `Core Processing'. Mathematically, the astrometric core solution 
is a simultaneous determination of a very large number of unknowns 
representing three kinds of information: (i) the astrometric parameters for 
a subset of the observed stars, representing the astrometric reference frame;
(ii) the instrument attitude, representing the accurate celestial
pointing of the instrument axes in that reference frame as a function
of time; and (iii) the geometric instrument calibration, representing
the mapping from pixels on the CCD detectors to angular directions
relative to the instrument axes. Although the astrometric core solution
is only made for a subset of the stars, the resulting celestial reference
frame, attitude and instrument calibration are fundamental inputs for
the processing of all observations. Optionally, a fourth kind of
unknowns, the global parameters, may be introduced to describe for
example a hypothetical deviation from General Relativity.

We use the term `source' to denote any astronomical object that Gaia detects
and observes as a separate entity. The vast majority of the Gaia sources are
ordinary stars, many of them close binaries or the components of wide systems,
but some are non-stellar (for example asteroids and quasars). Nearly everywhere 
in this paper, one can substitute `star' for `source' without distortion; however, 
for consistency with established practice in the Gaia community we use `source' 
throughout.

While the total number of distinct sources that will be observed by Gaia is 
estimated to slightly more than one billion, only a subset of them shall 
be used in the astrometric core solution. This subset, known as the 
`primary sources', is selected to be astrometrically well-behaved
(see Sect.~\ref{sec:primsel}) and consists of (effectively) 
single stars and extragalactic sources (quasars and AGNs) that are 
sufficiently stable and point-like. We assume here that the number of
primary sources is about $10^8$, i.e., roughly one tenth of the total
number of objects, although in the end it is possible that an even larger 
number will be used.
 
In comparison with many other parts of the Gaia data processing, the
astrometric core solution is in principle simple, mainly because it
only uses a subset of the observations (namely, those of the primary
sources), which can be accurately modelled in a
relatively straightforward way. In practice, the problem is however
formidable: the total number of unknowns is of the order of $5\times 10^8$,
the solution uses some $10^{11}$ individual observations extracted from
some 70~Terabyte (TB) of raw satellite data, and the entangled nature of
the data excludes a direct solution. A feasible approach has nevertheless
been found, including a precise mathematical formulation, practical
solution method, and efficient software implementation. It is the aim
of this paper to provide a comprehensive overview of this approach.

\begin{figure*}
\begin{center}
\includegraphics[scale=1.0]{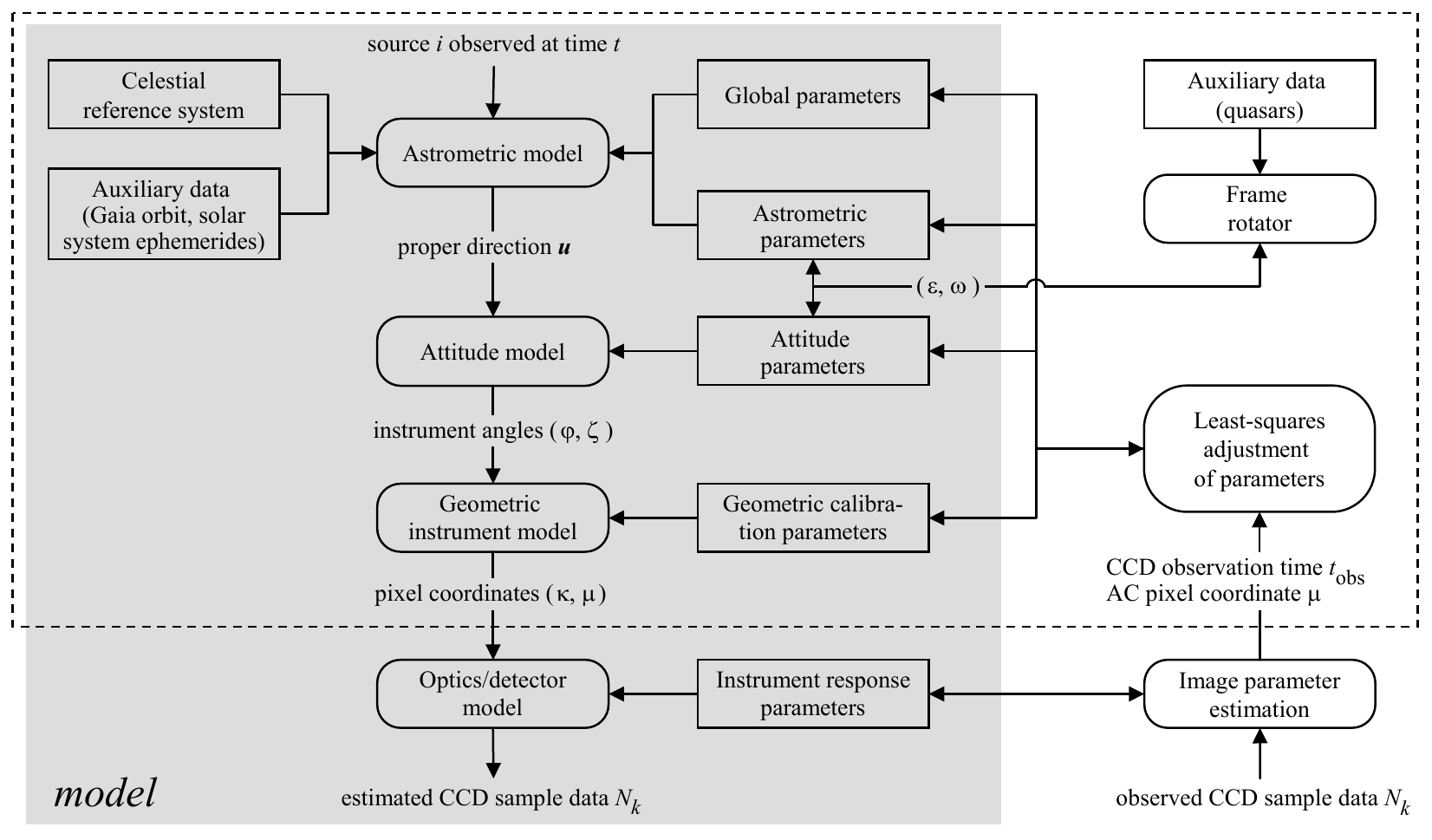}
\caption{ \label{fig:blockdiagram} Schematic representation of the main
elements of the astrometric data analysis. In the shaded area is the
mathematical model that allows to calculate the position of the
stellar image in pixel coordinates, and hence the expected CCD data, for
any given set of model parameters. To the right are the processes that
fit this model to the observed CCD data by adjusting the parameters
in the rectangular boxes along the middle. This paper is primarily
concerned with the geometrical part of the analysis contained in the
dashed box. However, a brief outline of the CCD data modelling and
processing (bottom part of the diagram) is given in Sect.~\ref{sec:signalmodel}
and Appendix~\ref{sec:CTI}.}
\end{center}
\end{figure*}

Concerning notations we have followed the usual convention to denote
all non-scalar entities (vectors, tensors, matrices, quaternions) by
boldfaced characters.
Lower-case bold italics ($\vec{a}$) are used for vectors and one-dimensional
column matrices; upper-case bold italics ($\vec{A}$) usually denote
two-dimensional matrices. Following \citet{murray1983} the prime
($\,'$) signifies both matrix transpose ($\vec{a}'$, $\vec{A}'$) and
scalar multiplication of vectors; thus $\lVert a \rVert=(\vec{a}'\!\vec{a})^{1/2}$
defines the magnitude of the vector $\vec{a}$ as well as the Euclidean
norm of the column matrix $\vec{a}$. 
Angular brackets denote normalization to unit length, as in
$\langle\vec{x}\rangle=\vec{x}\lVert\vec{x}\rVert^{-1}$.
In this notation, no special distinction is thus made between vectors 
as physical entities (also known as geometric, spatial or Euclidean vectors) 
on one hand, and their numerical representations in some coordinate system 
as column matrices (also known as list vectors) on the other hand. Moreover,
list vectors can of course have any dimension: $\vec{a}\in\mathbb{R}^n$.
In the coordinate system whose axes are aligned with unit vectors 
$\vec{x}$, $\vec{y}$, and $\vec{z}$, the components of the arbitrary 
vector $\vec{a}$ are given by $a_x=\vec{x}'\vec{a}$, $a_y=\vec{y}'\vec{a}$, 
and $a_z=\vec{z}'\vec{a}$; if the coordinate system is represented by the
vector triad $\tens{S}=\left[\vec{x}~\vec{y}~\vec{z}\right]$, these
components are given by the column matrix
$\tens{S}'\vec{a}$ \citep[cf.\ Appendix~A in][]{murray1983}. 
This notation, although perhaps unfamiliar to many readers, provides
a convenient and unambiguous framework for representing and
transforming spatial vectors in different coordinate systems.
For a vector-valued function
$\vec{f}(\vec{x})$, $\partial\vec{f}/\partial\vec{x}'$ denotes the
$\dim(\vec{f})\times\dim(\vec{x})$ matrix whose $(i,j)$th element
is $\partial f_i/\partial x_j$. 
Quaternions follow their own algebra
(see Appendix~\ref{sec:quaternions} for a brief introduction) and must
not be confused with vectors/matrices; quaternions are therefore denoted
by bold Roman characters ($\quat{a}$). When taking a
derivative with respect to the quaternion $\quat{a}$,
$\partial x/\partial\quat{a}$ denotes the $4\times 1$ matrix of
derivatives with respect to the quaternion components;
$\partial x/\partial\quat{a}'$ is the transposed matrix.
Tables of acronyms and variables are provided in Appendix~\ref{sec:vars}.

\section{Outline of the approach}\label{sec:outline}

The astrometric principles for Gaia were outlined already in the Hipparcos
Catalogue \cite[Vol.~3, Ch.~23]{hip:catalogue} where, based on the
accumulated experience of the Hipparcos mission, a view was cast to the
future. The general principle of a global astrometric data analysis was
succinctly formulated as the minimization problem:
\begin{equation}
\min_{\vec{s},\,\vec{n}}~\lVert \vec{f}^\text{obs} - \vec{f}^\text{calc}
(\vec{s},\vec{n})\rVert_{\,\mathcal{M}} \, ,
    \label{eq:generalobs}
\end{equation}
with a slight change of notation to adapt to the present paper.
Here $\vec{s}$ is the vector of unknowns (parameters) describing the
barycentric motions of the ensemble of sources used in the astrometric solution,
and $\vec{n}$ is a vector of `nuisance parameters' describing the instrument
and other incidental factors which are not of
direct interest for the astronomical problem but are nevertheless required
for realistic modelling of the data.
The observations are represented by the vector $\vec{f}^\text{obs}$ which
could for example contain the measured detector coordinates 
of all the stellar
images at specific times. $\vec{f}^\text{calc}(\vec{s},\vec{n})$ is the observation
model, e.g., the expected detector coordinates calculated as functions of
the astrometric and nuisance parameters. The norm is calculated in a metric
$\mathcal{M}$ defined by the statistics of the data; in practice the minimization
will correspond to a weighted least-squares solution with due consideration
of robustness issues (see Sect.~\ref{sec:synthesismodel}).

While Eq.~(\ref{eq:generalobs}) encapsulates the global approach to the
data analysis problem, its actual implementation will of course depend
strongly on the specific characteristics of the Gaia instrument and
mission as well as on the practical constraints on the data processing
task.

At the heart of the problem is the modelling of data represented
by the function $\vec{f}^\text{calc}(\vec{s},\vec{n})$ in
Eq.~(\ref{eq:generalobs}). This is schematically represented in the
shaded part of the diagram in Fig.~\ref{fig:blockdiagram}. It shows
the main steps for calculating the expected CCD output in terms of the
various parameters. The data processing, effecting the minimization
in Eq.~(\ref{eq:generalobs}), is represented in the right part of
the diagram. In subsequent sections we describe in some detail the
main elements depicted in Fig.~\ref{fig:blockdiagram}. The astrometric
(source) parameters are represented by the vector $\vec{s}$, while
the nuisance parameters are of three kinds: the attitude parameters
$\vec{a}$, the geometric calibration parameters $\vec{c}$, and the
global parameters $\vec{g}$. The observables $\vec{f}$ are represented
by the pixel coordinates $(\kappa,\mu)$ of point source images on
Gaia's CCDs. (In the actual implementation of the approach, the
minimization problem is formulated in terms of the field angles 
$\eta$, $\zeta$ rather than in the directly measured pixel coordinates 
$\kappa$, $\mu$; see Sects.~\ref{sec:attmodel} and 
\ref{sec:instrumentmodel}.)

The various elements of the astrometric solution are described in some
detail in the subsequent sections. 
Section~\ref{sec:math} provides the mathematical framework needed to model 
the Gaia observations and setting up the least-squares equations for the 
astrometric solution. In the interest of clarity and overview we omit in this 
description certain complications that need to be considered in the final 
data processing system; these are instead briefly discussed in 
Appendix~\ref{sec:complexities}. In Sect.~\ref{sec:numerical} we describe the 
iterative solution method in general terms, and then provide, in 
Sects.~\ref{sec:blocks} and \ref{sec:aux}, the mathematical details of the 
most important procedures. Section~\ref{sec:impl} outlines the existing 
implementation of the solution and presents the results of a 
demonstration run based on simulated observations of about 
2~million primary sources. Appendices~\ref{sec:quaternions} to 
\ref{sec:cholesky} provide brief introductions to three mathematical 
tools that are particularly important for the subsequent development,
namely the use of quaternions for representing the instantaneous satellite 
attitude, the B-spline formalism used to model the attitude as a function of
time, and a modified Cholesky algorithm for the decomposition of positive 
semidefinite normal matrices.

\section{Mathematical formulation of the basic observation model}\label{sec:math}

\subsection{Reference systems}\label{sec:refsystems}

The high astrometric accuracy aimed for with Gaia makes
it necessary to use general relativity for modelling the data. This
implies a precise and consistent formulation of the different reference
systems used to describe the motion of the observer (Gaia), the motion
of the observed object (source), the propagation of light from the source
to the observer, and the transformations between these systems. The
formulation adopted for Gaia \citep{klioner2003,klioner2004} is based
on the parametrized post-Newtonian (PPN) version of the relativistic
framework adopted in 2000 by the International Astronomical Union (IAU);
see \citet{soffel+2003}. In this section only some key concepts from this
formulation are introduced.

The orbit of Gaia and the light propagation from the source to Gaia
are entirely modelled in the Barycentric Celestial Reference System (BCRS)
whose spatial axes are aligned with the International Celestial Reference
System \citep[ICRS,][]{icrs1998}. The associated time coordinate is the
barycentric coordinate time (TCB). Throughout this paper, all time variables
denoted $t$ (with various subscripts) must be interpreted as TCB. The
ephemerides of solar-system bodies (including the Sun and the Earth) are
also expressed in this reference system. Even the motions of the stars
and extragalactic objects are
formally modelled in this system, although for practical reasons certain
effects of the light-propagation time are conventionally ignored in this
model (Sect.~\ref{sec:astromodel}).

In order to model the rotation (attitude) of Gaia and the celestial direction
of the light rays as observed by Gaia, it is expedient to introduce also a
co-moving celestial reference system having its origin at the centre of mass
of the satellite and a coordinate time equal to the proper time at Gaia.
This is known as the Centre-of-Mass Reference System (CoMRS) and the associated
time coordinate is Gaia Time ($T_\text{G}$). \citet{klioner2004} demonstrates
how the CoMRS can be constructed in the IAU 2000 framework in exact analogy
with the Geocentric Celestial Reference System (GCRS), only for a massless
particle (Gaia) instead of the Earth. Like the BCRS and GCRS, the CoMRS is
kinematically non-rotating, and its spatial axes are aligned with the ICRS.
The celestial reference system (either the CoMRS or the ICRS depending on whether 
the origin is at Gaia or the solar-system barycentre) will in the following be 
represented by the vector triad $\tens{C}=[\vec{X}~\vec{Y}~\vec{Z}]$, where
$\vec{X}$, $\vec{Y}$, and $\vec{Z}$ are orthogonal unit vectors aligned with
the axes of the celestial reference system. That is, $\vec{X}$ points towards
the origin ($\alpha=\delta=0$), $\vec{Z}$ towards the north celestial pole
($\delta=+90^\circ$), and $\vec{Y}=\vec{Z}\times\vec{X}$ points to
($\alpha=90^\circ, \delta=0$). 

\begin{figure}[t]
\begin{center}
\includegraphics[width=0.85\columnwidth]{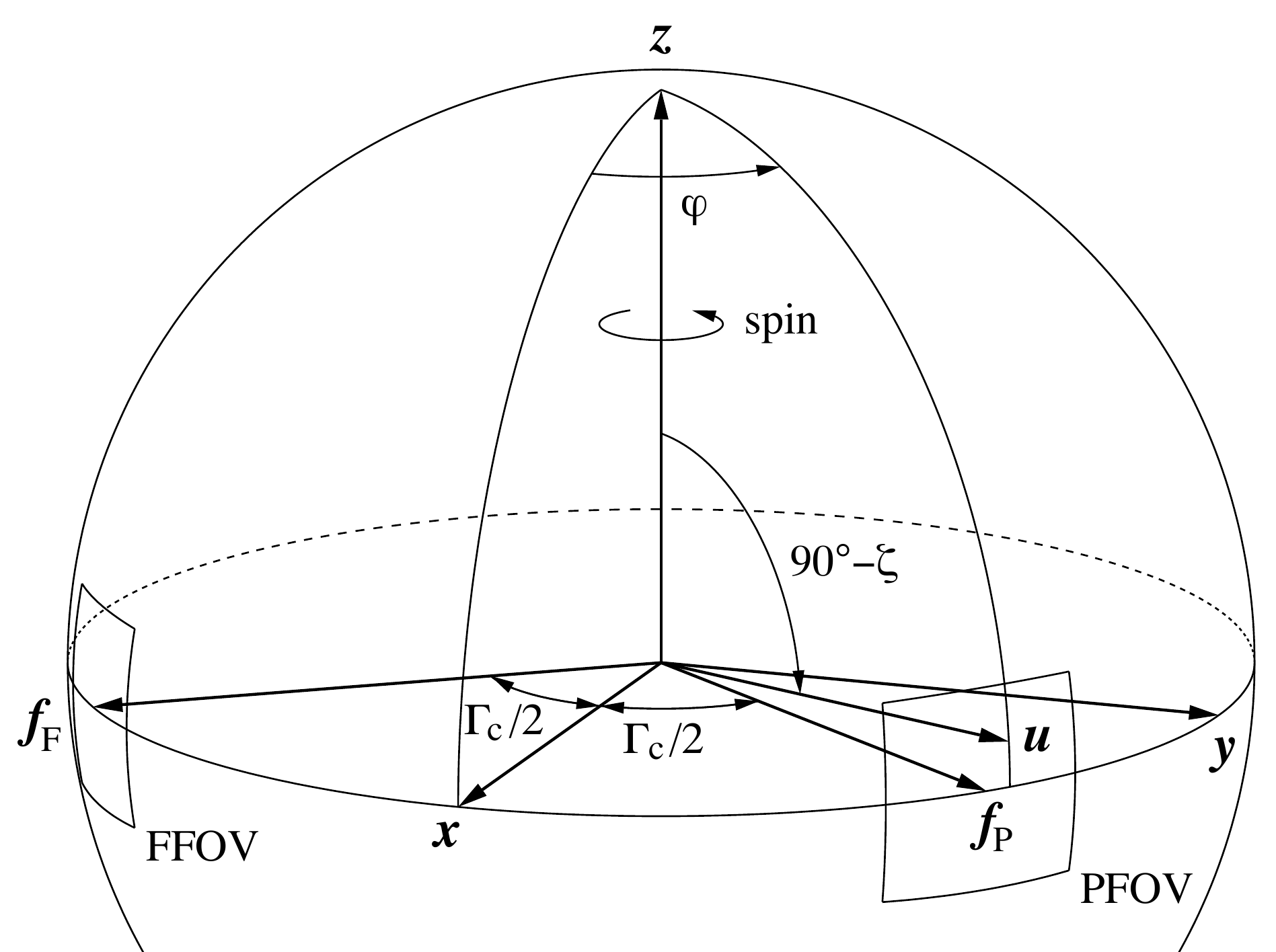}
\caption{ \label{fig:srs} The Scanning Reference System (SRS)
$[\vec{x}~\vec{y}~\vec{z}]$ is defined by the viewing directions $\vec{f}_\text{P}$
and $\vec{f}_\text{F}$ according to Eq.~(\ref{eq:xyz}). The instrument angles
$(\varphi,\zeta)$ are the spherical coordinates in the SRS of the observed (proper)
direction $\vec{u}$ towards the object. See Fig.~\ref{fig:instrcal} for further
definition of the viewing directions. The field sizes are greatly
exaggerated in this drawing; in reality the astrometrically useful field is
$0.72^\circ\times0.69^\circ$ (along$\,\times\,$across scan) for each viewing
direction. The basic angle is
$\Gamma_{\rm c}=\arccos({\vec{f}_\text{P}}'\vec{f}_\text{F})$,
nominally equal to $106.5^\circ$.}
\end{center}
\end{figure}

In the CoMRS the attitude of the satellite is a spatial rotation in three
dimensions, and can therefore be described purely classically, for example
using quaternions (Sect.~\ref{sec:attmodel} and Appendix~\ref{sec:quaternions}). 
The rotated reference system,
aligned with the instrument axes, is known as the Scanning Reference System
(SRS). Its spatial $\vec{x}$, $\vec{y}$, $\vec{z}$ axes (Fig.~\ref{fig:srs})
are defined in terms of the two viewing directions of Gaia $\vec{f}_\text{P}$ (in the
centre of the preceding field of view, PFoV) and $\vec{f}_\text{F}$ (in the
centre of the following field of view, FFoV) as
\begin{equation}\label{eq:xyz}
\vec{x}=\langle\vec{f}_\text{F}+\vec{f}_\text{P}\rangle\, , \quad
\vec{z}=\langle\vec{f}_\text{F}\times\vec{f}_\text{P}\rangle\, , \quad
\vec{y}=\vec{z}\times\vec{x} \, .
\end{equation}
(The precise definition of $\vec{f}_\text{P}$ and $\vec{f}_\text{F}$ is
implicit in the geometric instrument model; see 
Sect.~\ref{sec:instrumentmodel}.) The SRS is represented by the vector
triad $\tens{S}=[\vec{x}~\vec{y}~\vec{z}]$.

For determining the orbit of Gaia and calibrating the on-board clock, it
is also necessary to model the radio ranging and other ground-based observations
of the Gaia spacecraft in the same relativistic framework. For this, we also need the GCRS. These
aspects of the Gaia data processing are, however, not discussed in this paper.

\subsection{Astrometric model }\label{sec:astromodel}

The astrometric model is a recipe for calculating the proper direction
$\vec{u}_i(t)$ to a source ($i$) at an arbitrary time of observation ($t$) in
terms of its astrometric parameters $\vec{s}_i$ and various auxiliary data,
assumed to be known with sufficient accuracy. The auxiliary data include an
accurate barycentric ephemeris of the Gaia satellite, $\vec{b}_\text{G}(t)$,
including its coordinate velocity $\text{d}\vec{b}_\text{G}/\text{d}t$,
and ephemerides of all other relevant solar-system bodies.

For the astrometric core solution every source is assumed to move with uniform
space velocity relative to the solar-system barycentre.
In the BCRS its spatial coordinates at time $t_*$ are therefore given by
\begin{equation}\label{eq:uniform}
\vec{b}_i(t_*) = \vec{b}_i(t_\text{ep}) + (t_*-t_\text{ep})\vec{v}_i
\end{equation}
where $t_\text{ep}$ is an arbitrary reference epoch and 
$\vec{b}_i(t_\text{ep})$, $\vec{v}_i$
define six kinematic parameters for the motion of the source. The six astrometric
parameters in $\vec{s}_i$ are merely a transformation of the kinematic parameters
into an equivalent set better suited for the analysis of the observations.
The six astrometric parameters are:
\begin{description}
\item[$\alpha_i$] the barycentric right ascension at the adopted 
reference epoch $t_\text{ep}$
\item[$\delta_i$] the barycentric declination at epoch $t_\text{ep}$
\item[$\varpi_{i}$] the annual parallax at epoch $t_\text{ep}$
\item[$\mu_{\alpha*i}=(\mbox{d}\alpha_i/\mbox{d}t)\cos\delta_i$] the proper motion in
right ascension at epoch $t_\text{ep}$
\item[$\mu_{\delta i}=\mbox{d}\delta_i/\mbox{d}t$] the proper motion in 
declination at epoch $t_\text{ep}$
\item[$\mu_{ri}=v_{ri}\varpi_i/A_\text{u}$] the `radial proper motion' at epoch 
$t_\text{ep}$, i.e., the radial velocity
of the source ($v_{ri}$) expressed in the same unit as the transverse components of
proper motion ($A_\text{u}=$~astronomical unit).
\end{description}
As explained in Sect.~\ref{sec:refsystems}, the astrometric parameters refer to the
ICRS and the time coordinate used is TCB. The reference epoch $t_\text{ep}$ is
preferably chosen to be near the mid-time of the mission in order to minimize
statistical correlations between the position and proper motion parameters.

The transformation between the kinematic and the astrometric parameters is
non-trivial \citep{klioner2003}, mainly as a consequence of the practical need to
neglect most of the light-propagation time $t-t_*$ between the emission of the light
at the source ($t_*$) and its reception at Gaia ($t$). This interval is typically many
years and its value, and rate of change (which depends on the radial velocity of the source),
will in general not be known with sufficient accuracy to allow modelling of the motion of
the source directly in terms of its kinematic parameters according to Eq.~(\ref{eq:uniform}).
The proper motion components
$\mu_{\alpha*i}$, $\mu_{\delta i}$ and radial velocity $v_{ri}$ correspond to the
`apparent' quantities discussed by \citet[Sect.~8]{klioner2003}.

The coordinate direction to the source at time $t$ is calculated with the same
`standard model' as was used for the reduction of the Hipparcos observations
\citep[][Vol.~1, ~Sect.~1.2.8]{hip:catalogue}, namely
\begin{equation}\label{eq:astro-standard}
\vec{\bar{u}}_i(t) = \bigl\langle \vec{r}_i+(t_\text{B}-t_\text{ep})(\vec{p}_i\mu_{\alpha*i}
+\vec{q}_i\mu_{\delta i}+\vec{r}_i\mu_{ri})-\varpi_i\vec{b}_\text{G}(t)/A_\text{u} \bigr\rangle
\end{equation}
where the angular brackets signify vector length normalization, and
$[\vec{p}_i~\vec{q}_i~\vec{r}_i]$ is the `normal triad' of the source
with respect to the ICRS \citep{murray1983}. In this triad, 
$\vec{r}_i$ is the barycentric coordinate 
direction to the source at time $t_\text{ep}$, 
$\vec{p}_i=\langle\vec{Z}\times\vec{r}_i\rangle$, and 
$\vec{q}_i=\vec{r}_i\times\vec{p}_i$. The components of these unit vectors 
in the ICRS are given by the columns of the matrix
\begin{equation}\label{eq:normal-triad}
\tens{C}'[\vec{p}_i~\vec{q}_i~\vec{r}_i] = \begin{bmatrix} 
-\sin\alpha_i &~  -\sin\delta_i\cos\alpha_i  &~ \cos\delta_i\cos\alpha_i \\ 
\phantom{-}\cos\alpha_i &~  -\sin\delta_i\sin\alpha_i &~  \cos\delta_i\sin\alpha_i \\ 
0 &~  \cos\delta_i &~ \sin\delta_i \end{bmatrix}.
\end{equation}
$\vec{b}_\text{G}(t)$ is the barycentric position of Gaia at the time of
observation, and $A_\text{u}$ the astronomical unit. $t_\text{B}$ is the barycentric
time obtained by correcting the time of observation for the R{\"o}mer delay;
to sufficient accuracy it is given by
\begin{equation}\label{eq:roemer}
t_\text{B} = t + \vec{r}_i'\vec{b}_\text{G}(t)/c \, ,
\end{equation}
where $c$ is the speed of light.

In Eq.~(\ref{eq:astro-standard}) the term containing $\mu_{ri}$ accounts 
for the relative rate of change of the
barycentric distance to the source. This term may produce secular variations of
the proper motions and parallaxes of some nearby stars, which in principle allow
their radial velocities to be determined astrometrically \citep{drav+99}.
However, for the vast majority of these stars, $\mu_{ri}$ can be more accurately
calculated by combining the spectroscopically measured radial velocities with
the astrometric parallaxes. Thus, although all six astrometric parameters are
taken into account when computing the coordinate direction, usually only five 
of them are considered as unknowns in the astrometric solution.

The standard model can be derived by considering the uniform motion of the source
in a purely classical framework, using Euclidean coordinates and neglecting the
light propagation time from the source to the observer (except for the R{\"o}mer
delay). To the accuracy of Gaia, relativistic and light-propagation effects are
by no means negligible, but it can be shown that this model is nevertheless
accurate enough to \emph{model} the observations to sub-microarcsec accuracy.
It is adopted for this purpose because it closely corresponds to the
conventional interpretation of the astrometric parameters. However, when the
astrometric parameters are to be \emph{interpreted} in terms of the barycentric
space velocity of the source, some of these effects may come into play
\citep{lindegren+2003}.

The transformation from $\vec{\bar{u}}_i(t)$ to the observable (proper) direction
$\vec{u}_i(t)$ involves taking into account gravitational light deflection by
solar-system bodies and the Lorentz transformation to the co-moving frame of
the observer (stellar aberration); the relevant formulae are given by
\citet{klioner2003}. This transformation therefore depends also on the global
parameters $\vec{g}$, for example the PPN parameter $\gamma$ which measures
the strength of the gravitational light deflection. The calculation uses some auxiliary
data, not subject to improvement by the solution, and which are here denoted
$\vec{h}$; normally they include for example the barycentric ephemerides of the
Gaia satellite and of solar-system bodies, along with their masses. The complete transformation can therefore be
written symbolically as:
\begin{equation}
    \vec{u}_{i}(t) = \vec{u}(\vec{s}_{i},\vec{g}\,|\,t,\vec{h}) \, ,
\label{eq:astro}
\end{equation}
where the vertical bar formally separates the (updatable) parameters
$\vec{s}_i$ and $\vec{g}$ from the (fixed) given data $t$ and $\vec{h}$.
Strictly speaking, the function $\vec{u}$ returns the coordinates of the 
proper direction in the CoMRS, that is the column matrix $\tens{C}'\vec{u}_{i}(t)$.
The source parameter vector $\vec{s}$ is the concatenation of the 
sub-vectors $\vec{s}_i$ for all the primary sources. 

\subsection{Attitude model }\label{sec:attmodel}

The attitude specifies the instantaneous orientation of the Gaia
instrument in the celestial reference frame, that is the transformation 
from $\tens{C}=[\vec{X}~\vec{Y}~\vec{Z}]$ (more precisely, the
CoMRS) to $\tens{S}=[\vec{x}~\vec{y}~\vec{z}]$ (Sect.~\ref{sec:refsystems})
as a function of time. The spacecraft is controlled to follow a specific
attitude as a function of time, for example the so-called Nominal Scanning Law 
\citep[NSL;][]{2010IAUS..261..331D}, designed to provide good coverage of the
whole celestial sphere while satisfying a number of other requirements. The
NSL is analytically defined for arbitrarily long time intervals by just a few 
free parameters.

However, the actual attitude will deviate from the intended (nominal) scanning
law by up to $\sim\,$1~arcmin in all three axes, and these deviations vary on
time scales from seconds to minutes depending on the level of physical
perturbations and the characteristics of the real-time attitude measurements
and control law (cf.\ Appendix~\ref{sec:attirr}). The CCD integration time of
(usually) 4.42~s means that the true (physical) attitude cannot be observed,
but only a smoothed version of it, corresponding to the convolution of the
physical attitude with the CCD exposure function (Appendix~\ref{sec:TDI}).
This `effective' attitude must however
be a~posteriori estimable, at any instant, 
to an accuracy compatible with the astrometric goals, i.e., at sub-mas level.
For this purpose the effective attitude is modelled in
terms of a finite number of attitude parameters, for which it is necessary to
choose a suitable mathematical representations of the instantaneous
transformation, as well as of its time dependence. For the Gaia data
processing, we have chosen to use quaternions 
(Appendix~\ref{sec:quaternions}) for the former, and B-splines 
(Appendix~\ref{sec:splines}) for the latter representation.

At any time the orientation of the SRS ($\tens{S}$) with respect to the 
CoMRS ($\tens{C}$) may be represented by the attitude matrix
\begin{equation}\label{eq:attmat}
\vec{A} \equiv
\begin{bmatrix}
A_{xx} & A_{xy} & A_{xz} \\
A_{yx} & A_{yy} & A_{yz} \\
A_{zx} & A_{zy} & A_{zz}
\end{bmatrix}
= \tens{S}'\tens{C} =
\begin{bmatrix}
\vec{x}'\vec{X} & \vec{x}'\vec{Y} & \vec{x}'\vec{Z} \\
\vec{y}'\vec{X} & \vec{y}'\vec{Y} & \vec{y}'\vec{Z} \\
\vec{z}'\vec{X} & \vec{z}'\vec{Y} & \vec{z}'\vec{Z}
\end{bmatrix} \, ,
\end{equation}
which is a proper orthonormal matrix, $\vec{A}\vec{A}'=\vec{I}$,
$\det(\vec{A})=+1$. This is also called the direction cosine matrix,
since the rows are the direction cosines of $\vec{x}$, $\vec{y}$ and
$\vec{z}$ in the CoMRS, and the columns are the direction cosines of
$\vec{X}$, $\vec{Y}$ and $\vec{Z}$ in the SRS.

The orthonormality of $\vec{A}$ implies that the matrix elements
satisfy six constraints, leaving three degrees of freedom for the
attitude representation. Although one could parametrize each of
the nine matrix elements as a continuous function of time, for example
using splines, it would in practice not be possible to guarantee
that the orthonormality constraints hold at any time. The adopted solution 
is to represent the instantaneous attitude by a unit quaternion $\quat{q}$,
which has only four components, $\bigl\{q_x, q_y, q_z, q_w \bigr\}$, satisfying
the single constraint $q_x^2+q_y^2+q_z^2+q_w^2=1$. This is
easily enforced by normalization. 

The attitude quaternion $\quat{q}(t)$ gives the rotation from the CoMRS
($\tens{C}$) to the SRS ($\tens{S}$). Using quaternions, their relation 
is defined by the transformation of the coordinates of the arbitrary vector 
$\vec{v}$ in the two reference systems, 
\begin{equation}\label{eq:att-q}
\left\{ \tens{S}'\vec{v}, 0 \right\} = \quat{q}^{-1}
\left\{ \tens{C}'\vec{v}, 0 \right\} \quat{q} \, .
\end{equation} 
In the terminology of Appendix~\ref{sec:vecframe} this is a frame
rotation of the vector from $\tens{C}$ to $\tens{S}$. 
The inverse operation is $\left\{ \tens{C}'\vec{v}, 0 \right\} = \quat{q}
\left\{ \tens{S}'\vec{v}, 0 \right\} \quat{q}^{-1}$.

Using the B-spline representation in Appendix~B, we have
\begin{equation}\label{eq:qspline}
\quat{q}(t) = \left\langle\, \textstyle\sum_{n=\ell-M+1}^{\ell}
\quat{a}_n B_n(t) \,\right\rangle \, ,
\end{equation}
where $\quat{a}_n$ ($n=0\dots N-1$) are the coefficients of the
B-splines $B_n(t)$ of order $M$ defined on the knot sequence
$\left\{\tau_k\right\}_{k=0}^{N+M-1}$. The function $B_n(t)$ is
non-zero only for $\tau_n < t < \tau_{n+M}$, which is why the
sum in Eq.~(\ref{eq:qspline}) only extends over the $M$ terms
ending with $n=\ell$. Here, $\ell$ is the so-called \emph{left index}
of $t$ satisfying $\tau_\ell \le t < \tau_{\ell+1}$. The normalization
operator $\langle\,\rangle$ guarantees that $\quat{q}(t)$ is a unit
quaternion for any $t$ in the interval $[\tau_{M-1},\tau_N]$ over which
the spline is completely defined. Although the coefficients
$\quat{a}_n$ are formally quaternions, they are not of unit length.
The attitude parameter vector $\vec{a}$ consists of the
components of the whole set of quaternions $\quat{a}_n$.

Cubic splines ($M=4$) are currently used in this attitude model.
Each component of the quaternion (before the normalization in
Eq.~\ref{eq:qspline}) is then a piecewise cubic polynomial
with continuous value, first, and second derivative for any $t$;
the third derivative is discontinuous at the knots.
When initializing the solution, it is possible
to select any desired order of the spline. Using a 
higher order, such as $M=5$ (quartic) or $6$ (quintic), provides
improved smoothness but also makes the spline fitting less local,
and therefore more prone to undesirable oscillatory behaviour.
The flexibility of the spline is in principle 
only governed by the number of degrees of freedom (that is, in 
practice, the knot frequency), and is therefore independent of the
order. One should therefore not choose a higher order than is
warranted by the smoothness of the actual effective attitude,
which is difficult to model a~priori (cf.\ Appendix~\ref{sec:attirr}). 
Determining the optimal order 
and knot frequency may in the end only be possible through
analysis of the real mission data. 

Equation~(\ref{eq:astro}) returns the observed direction to the source
relative to the celestial reference system, or 
$\tens{C}'\vec{u}=[u_X,\,u_Y,\,u_Z]'$. In order 
to compute the position of the image in the field of view,   
we need to express this direction in the Scanning Reference System,
SRS (Sect.~\ref{sec:refsystems}), or $\tens{S}'\vec{u}=[u_x,\,u_y,\,u_z]'$. 
The required transformation is obtained by a frame rotation according 
to Eq.~(\ref{eq:att-q}),
\begin{equation}\label{eq:srsprop}
\left\{\tens{S}'\vec{u},\,0\right\}
=\quat{q}^{-1}\left\{\tens{C}'\vec{u},\,0\right\}\,\quat{q} \, ,
\end{equation}
whereupon the instrument angles $(\varphi,\zeta)$ are obtained from
\begin{equation}\label{eq:srsprop1}
   \tens{S}'\vec{u}
\equiv
    \begin{bmatrix}
        u_x \\  u_y \\   u_z
    \end{bmatrix}
=
    \begin{bmatrix}
        \cos\zeta\cos\varphi \\
        \cos\zeta\sin\varphi \\
        \sin\zeta
    \end{bmatrix}
\quad\Leftrightarrow\quad
\left\{\begin{array}{l} \varphi = \text{atan2}\bigl(u_y,u_x\bigr)\\[3pt]
\zeta = \text{atan2}\Bigl(u_z,\,\sqrt{u_x^2+u_y^2}\,\Bigr) \end{array}\right.
\end{equation}
(Fig.~\ref{fig:srs}), where we adopt the convention that $-\pi \le \varphi < \pi$.
The field index $f$ and the along-scan field angle $\eta$ are
finally obtained as
\begin{equation}\label{eq:srspropeta}
f = \text{sign}(\varphi)\, , \quad \eta = \varphi - f\Gamma_{\rm c}/2 \, ,
\end{equation}
where the basic angle, $\Gamma_{\rm c}$, is here a purely conventional value
(as suggested by the subscript).
The field-of-view limitations imply that
$|\eta| \lesssim 0.5^\circ$ and $|\zeta| \lesssim 0.5^\circ$
for any observation.

Given the time of observation $t$ and the set of source
parameters $\vec{s}$, attitude parameters $\vec{a}$, and global
parameters $\vec{g}$, the field index $f$ and the field angles
$(\eta,\zeta)$ can thus be computed by application of
Eqs.~(\ref{eq:astro}), (\ref{eq:qspline})--(\ref{eq:srspropeta}).
The resulting \emph{computed} field angles
are concisely written $\eta(t\,|\,\vec{s},\vec{a},\vec{g})$,
$\zeta(t\,|\,\vec{s},\vec{a},\vec{g})$.

\begin{figure}
\begin{center}
\includegraphics[width=\columnwidth]{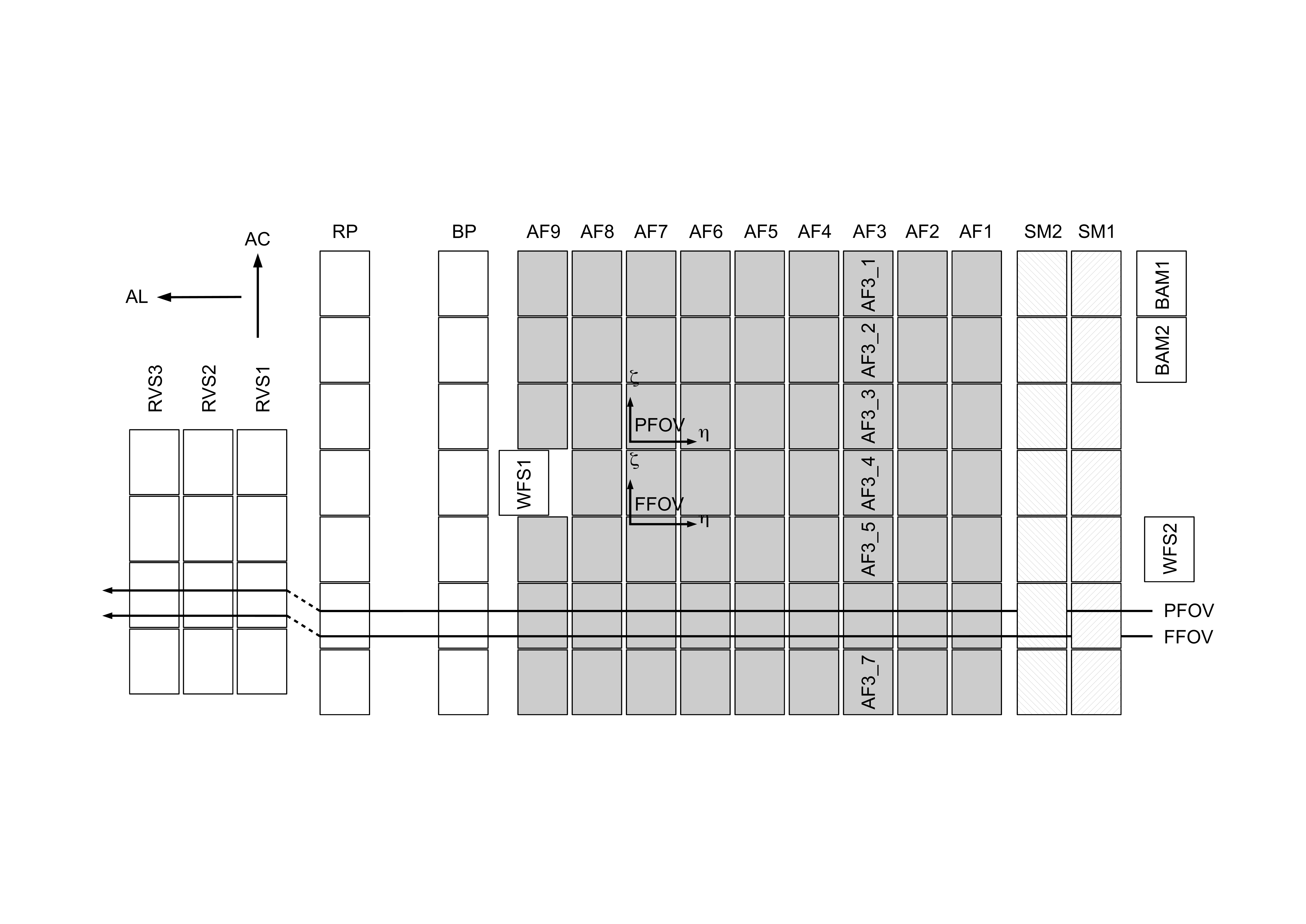}
\caption{\label{fig:fpa} Layout of the CCDs in Gaia's focal plane. 
The star images move from right to left in the diagram. The 
along-scan (AL) and across-scan (AC) directions are indicated 
in the top left corner. The skymappers (SM1, SM2) 
provide source image detection, two-dimensional position
estimation and field-of-view discrimination. The astrometric field
(AF1--AF9) provides accurate AL measurements and (sometimes) AC positions.
Additional CCDs used in the blue and red photometers (BP, RP), the
radial-velocity spectrometer (RVS), for wavefront sensing (WFS) and
basic-angle monitoring (BAM) are not further described in this paper.
One of the rows (AF3) illustrates the system for labelling individual CCDs
(AF3\_1, etc). The nominal paths of two star images, one in the preceding
field of view (PFoV) and one in the following field of view (FFoV) are
indicated. For purely technical reasons the origin of the field 
angles $(\eta,\zeta)$ corresponds to
different physical locations on the CCDs in the two fields.
The physical size of each CCD is $45\times 59$~mm$^2$.
}
\end{center}
\end{figure}

\subsection{Geometric instrument model }\label{sec:instrumentmodel}

The geometric instrument model defines the precise layout of the
CCDs in terms of the field angles $(\eta,\zeta)$. This layout
depends on several different factors, including: the physical geometry
of each CCD; its position and alignment in the focal-plane assembly;
the optical system including its scale value, distortion and other
aberrations; and the adopted (conventional) value of the basic angle
$\Gamma_{\rm c}$. Some of these (notably the optical distortion) are different
in the two fields of view and may evolve slightly over the mission;
on the other hand these variations are expected to be very smooth
as a function of the field angles. Other factors, such as the detailed
physical geometry of the pixel columns, may be extremely stable but
possibly quite irregular on a small scale. As a result, the geometric
calibration model must be able to accommodate both smooth and irregular
patterns evolving on different time scales in the two fields of view.

In the following it should be kept in mind that the astrometric instrument 
of Gaia is optimised for one-dimensional measurements in the along-scan
direction, i.e., of the field angle $\eta$, while the requirements in the 
perpendicular direction ($\zeta$) are much relaxed. This feature of Gaia
(and Hipparcos) is a consequence of fundamental considerations derived 
from the requirements to determine a global reference frame and absolute 
parallaxes \citep{2011EAS....45..109L}. In principle the measurement accuracy 
in the $\zeta$ direction, as well as the corresponding instrument modelling 
and calibration accuracies, may be relaxed by up to a factor given by the inverse angular 
size of the field of view \citep{2005ESASP.576...29L}, or almost two orders of 
magnitude for a field of $0.7^\circ$. In practice a ratio of the order of 10
may be achieved, in which case the across-scan measurement and calibration
errors have a very marginal effect on the overall astrometric accuracy.
 
The focal plane of Gaia contains a total of 106 CCDs \citep{2007SPIE.6690E...8L},
of which only the 14 CCDs in the skymappers (SM1 and SM2) and the 62 
in the astrometric field (AF1 through AF9) are used for the astrometric 
solution (Fig.~\ref{fig:fpa}). The CCDs have 4500 pixel lines in the
along-scan (AL, constant $\zeta$, decreasing $\eta$) direction 
and 1966 pixel columns in the
across-scan (AC, constant $\eta$, increasing $\zeta$) direction.
They are operated in the Time, Delay and Integrate (TDI)
mode (also known as drift-scanning), an imaging technique well-known
in astronomy from ground-based programmes such as the Sloan Digital
Sky Survey \citep{gunn+98}. Effectively, the charges are clocked along
the CCD columns at the same (average) speed as the motion of the optical
images, i.e., 60~arcsec~s$^{-1}$ for Gaia. The exposure (integration) time is 
thus set by the time it takes the image to move across the CCD, or nominally
$T\simeq 4.42$~s, if no gate is activated. 
At this exposure time the central pixels will be saturated for
sources brighter than magnitude $G\simeq 12$.%
\footnote{$G$ is the broad-band magnitude measured on Gaia's 
unfiltered CCDs. $G\simeq V$ for unreddened early-type stars, while
$G\simeq V-2$ for M stars. See \citet{2010A&A...523A..48J} for a 
precise characterization of $G$.}
Gate activation, or `gating' for short, is the adopted method to
obtain valid measurements of brighter sources. Gating temporarily 
inhibits charge transfer across a certain TDI line (row of pixels AC), 
thus effectively zeroing the charge image and reducing the exposure 
time in proportion to the number of TDI lines following 
the gate. A range of discrete exposure times is thus available,
the shortest one, according to current planning, using only 16 TDI lines 
(15.7~ms). Gate selection is made by the on-board software based 
on the measured brightness of the source in relation to the calibrated
full-well capacity of the relevant section of the CCD. Measurement errors 
and the spread in full-well capacities make the gate-activation thresholds 
somewhat fuzzy, and a given bright source is not necessarily always observed 
using the same gate. Moreover, a quasi-random selection of the observations 
of fainter sources will also be gated, viz., if they happen to be read out 
at about the same time, and on the same CCD, as a gated bright star.
The skymappers
are operated in a permanently gated mode, so that in practice only
the last 2900 TDI lines are used in SM1 and SM2.

The skymappers are crucial for the real-time operation of the instrument,
since they detect the sources as they enter the field of view, and allow
determination of an approximate two-dimensional position of the images and
(together with data from AF1) their speed across the CCDs. This information
is used by the on-board attitude computer, in order to determine which 
CCD pixels should be read out and transmitted to ground (Sect.~\ref{sec:signalmodel}). 
The skymappers 
also allow to discriminate between the two viewing directions, since sources
in the preceding field of view (PFoV) are only seen by SM1 and sources
in the following field of view (FFoV) are only seen by SM2. In the
astrometric field (as well as in BP, RP and RVS) the two fields of view are 
superposed.

Most observations acquired in the astrometric field (AF) are purely
one-dimensional through the on-chip AC binning of data
(Sect.~\ref{sec:signalmodel}). However, sources brighter than 
magnitude $G\simeq 13$ are always observed as two-dimensional images, providing
accurate information about the AC pixel coordinate ($\mu$) as well. Some
randomly selected fainter images are also observed two-dimensionally
(`Calibration Faint Stars', CFS). The bright sources, CFS and SM data
provide the AC information necessary for the 
on-ground three-axis attitude determination.

\begin{figure}
\begin{center}
\includegraphics[width=\columnwidth]{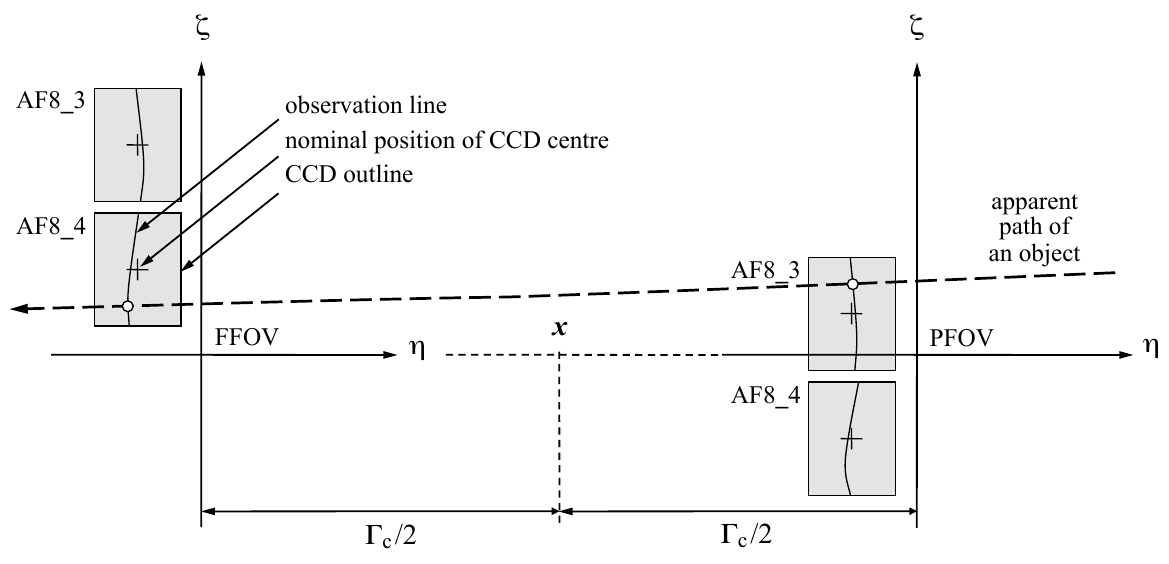}
\caption{\label{fig:instrcal} Schematic illustration of how the field angles
$(\eta,\zeta)$ are defined in terms of the CCD layout in Fig.~\ref{fig:fpa}.
For simplicity only the projections of two CCDs, AF8\_3 and AF8\_4, into the
Scanning Reference System (SRS) are shown (not to scale). The field angles
have their origins at the respective viewing direction $\vec{f}_\text{P}$ or
$\vec{f}_\text{F}$ (Fig.~\ref{fig:srs}), which are defined in relation to
the nominal centres of the CCDs (crosses); the actual configuration of the
detectors is described by fiducial observation lines according to 
Eq.~(\ref{eq:etazeta}).
The dashed curve shows the apparent path of a stellar image across the two
fields of view. Its intersection with the observation lines define the instants
of observations.}
\end{center}
\end{figure}

Because of the TDI mode of observation, AL irregularities of the pixel
geometry are smeared out and need not be calibrated, and any measurement
of the AL or AC position must effectively be referred back to an
`observation time' half-way through the integration. Correspondingly,
the pixel geometry can be represented by a fiducial `observation line'
$\left[\eta(\mu),\,\zeta(\mu)\right]$ traced out in the field angles
$\eta$, $\zeta$ as functions of the AC pixel coordinate $\mu$
(Fig.~\ref{fig:instrcal}). The AC pixel coordinate is defined as a
continuous variable with $\mu=14$ when the image is centred on the
first pixel column, $\mu=15$ at the second pixel column, and so on
until $\mu=1979$ at the last (1966th) pixel column. The offset by
13~pixels allows for the presence of pre-scan pixel data in some
observations.

Nominally, the observation line corresponds to the $(K/2)$th pixel line
projected backwards through the optical instrument onto the Scanning
Reference System on the sky, where $K$ is the number of active AL pixel
lines in the observation (normally $K=4500$).%
\footnote{In reality the definition of the fiducial observation line is
a bit more complex, as some of the pixel lines are blocked out by an
aluminium mask.} For a gated observation
$K$ is much smaller and the observation line is therefore correspondingly
displaced towards the CCD readout register. A separate set of geometrical
calibration parameters is therefore needed for each gate being used.
In the calibration updating (Sect.~\ref{sec:calupdate}) all the calibration parameters 
are however solved together, with the
overlap due to the fuzzy gate-activation thresholds providing the necessary 
connection between the different gates.

Let $n$ be an index identifying the different CCDs used for astrometry,
i.e., for each of the 76 CCDs in the SM and AF part of the focal plane.
Furthermore, let $g$ be a gate index such that $g=0$ is used for non-gated
observations and $g=1,\,\dots,\,12$ for gated observations of progressively
brighter sources. In each field of view (index $f$) the nominal observation
lines $\left[\eta^0_{fng}(\mu),\,\zeta^0_{fng}(\mu)\right]$
could in principle be calculated from the nominal characteristics of the
focal plane assembly (FPA) and ray tracing through the nominal optical
system. However, since the nominal observation lines are only used as
a reference for the calibration of the actual observation lines, a very
simplistic transformation from linear coordinates to angles can be used
without introducing approximation errors in the resulting calibration model.
The nominal observation lines are therefore defined by the transformation
\begin{equation}\label{eq:etazeta0}
\left.
\begin{aligned}
\eta^0_{fng}(\mu) &\equiv \eta^0_{ng} = -Y_\text{FPA}[n,g]/F  \\
\zeta^0_{fng}(\mu) &= -\left(X_\text{FPA}[n]-(\mu-\mu_\text{c})p_\text{AC}
-X_\text{FPA}^\text{centre}[f]\right)/F
\end{aligned}
\right\}
\end{equation}
where $X_\text{FPA}$, $Y_\text{FPA}$ are physical coordinates (in mm) in
the focal plane,%
\footnote{For consistency with notations adopted by the ESA project team
and the industrial contractor, and extensively used, e.g., for on-ground
calibrations, $+X_\text{FPA}$ is oriented along $-\zeta$ and $+Y_\text{FPA}$
along $-\eta$.}
$X_\text{FPA}[n]$ is the physical AC coordinate of the nominal centre of
the $n$th CCD, $Y_\text{FPA}[n,g]$ the physical AL coordinate of the
nominal observation line for gate $g$ on the $n$th CCD, and
$X_\text{FPA}^\text{centre}[f]$ the physical AC coordinate of the
nominal field centre for field index $f$. $\mu_\text{c}=996.5$ is the AC pixel
coordinate of the CCD centre, $p_\text{AC}=30~\mu$m the physical AC pixel
size, and $F=35$~m the nominal equivalent focal length of the telescope.
While $\eta^0_{ng}$ is independent of $f$, in the AC direction the origins
are offset by about $\pm 221$~arcsec between the two fields of view,
corresponding to the physical coordinates
$X_\text{FPA}^\text{centre}[f]=(-37.5~\text{mm})f.$

It is emphasized that the nominal observation lines are purely conventional
reference quantities, and need not be recomputed, e.g., once a more accurate
estimate of $F$ becomes available.

Because of possible changes in the instrument during the mission, the actual
observation lines will be functions of time. The time dependence is quantified
by introducing independent sets of calibration parameters for successive,
non-overlapping time intervals. Different groups of parameters may develop
on different time scales, and the resulting formulation can be quite complex.
For the sake of illustration, let us distinguish between three categories of
geometric calibration parameters:
\begin{enumerate}
\item
Large-scale AL calibration, $\Delta\eta$. This may 
(minutely but importantly) change due to
thermal variations in the optics, the detectors, and their supporting mechanical
structure. These variations could occur on short time scales (of the order of a
day), and would in general be different in the two fields of view. They are
modelled as low-order polynomials in the across-scan pixel coordinate $\mu$.
\item
Small-scale AL calibration, $\delta\eta$. This is mainly related to physical
defects or irregularities in the CCDs themselves, for example `stitching errors'
introduced by the photolithography process used to manufacture the CCDs. These
irregularities are expected to be stable over very long time scales, possibly
throughout the mission, and should be identical in both fields of view. They
require a spatially detailed modelling, perhaps with a resolution of just one
or a few AC pixels.
\item Large-scale AC calibration, $\Delta\zeta$. Although the physical origin
is the same as for $\Delta\eta$, the AC components can be assumed to be constant
on long time scales, since the calibration requirement in the AC direction
is much more relaxed than in the AL direction. They are modelled as low-order
polynomials in the field angles, separately in each field of view.
\end{enumerate}
Let index $j$ identify the `short' time intervals needed for the large-scale
AL calibration, and index $k$ identify the `long' time intervals applicable
to the small-scale AL calibration and large-scale AC calibration. That is,
an observation at time $t$ belongs to some short time interval $j$ and some
long time interval $k$, where $j$ and $k$ are readily computed from the known
$t$ and the corresponding sequences of division times.%
\footnote{Technically, the use of independent parameter values in successive
time intervals represents a spline of order 1 (i.e., degree 0), the separation times constitute
the knot sequence, and $j$ or $k$ correspond to the `left index'
(Appendix~\ref{sec:calcBsplines}).}
Assuming that quadratic polynomials in $\mu$ are sufficient for the large-scale
calibrations, and that full AC pixel resolution is required for the small-scale
AL calibration, the observation lines at time $t$ are modelled as
\begin{equation}\label{eq:etazeta}
\left.
\begin{aligned}
\eta_{fng}(\mu,t) &= \eta^0_{ng}
+ \sum_{r=0}^2 \Delta\eta_{rfngj} L_r^\ast\left(\frac{\mu-13.5}{1966}\right)
+ \delta\eta_{ngkm}  \\
\zeta_{fng}(\mu,t) &= \zeta^0_{fng}(\mu)
+ \sum_{r=0}^2 \Delta\zeta_{rfngk} L_r^\ast\left(\frac{\mu-13.5}{1966}\right)
\end{aligned}
\right\}
\end{equation}
where $L_r^\ast(x)$ is the shifted Legendre polynomial of degree $r$ (orthogonal
on $[0,1]$), i.e., $L_0^\ast(x)=1$, $L_1^\ast(x)=2x-1$, $L_2^\ast(x)=6x^2-6x+1$, etc;
$\Delta\eta_{rfngj}$ are the large-scale AL parameters, $\delta\eta_{ngkm}$ the
small-scale AL parameters (with $m=\text{round}(\mu)$ the index of the nearest
pixel column), and $\Delta\zeta_{rfngk}$ the large-scale AC parameters.

In order to render all the geometric calibration parameters uniquely determinable,
a number of constraints are rigorously enforced by the astrometric solution.
Effectively, they define the origins of the field angles, i.e., the viewing
directions $\vec{f}_\text{P}$ and $\vec{f}_\text{F}$, to coincide with the
average nominal field angles of the CCD centres. The necessary constraints are:
\begin{align}
\sum_f \sum_{n\in\text{AF}} \Delta\eta_{0fn0j} &= 0
     &&\text{for each $j$,}\label{eq:cc1}\\
\sum_m \delta\eta_{n0km}L_r^\ast\left(\frac{m-13.5}{1966}\right) &= 0
     &&\text{for each combination $rnk$,}\label{eq:cc2}\\
\sum_{n\in\text{AF}} \Delta\zeta_{0fn0k} &= 0
     &&\text{for each combination $fk$,}\label{eq:cc3}
\end{align}
Note that $g=0$ throughout in Eq.~(\ref{eq:cc1})--(\ref{eq:cc3}). 
That the constraints are only defined in terms of the
non-gated observations ($g=0$) implies that the observation lines for the
gated observations must be calibrated relative to the non-gated observations.
This is possible since any given bright source will not always be observed
with the same gate.

Constraint (\ref{eq:cc1}) effectively defines the zero point of the AL
field angle $\eta$ by requiring that $\overline{\Delta\eta}=0$ when
averaging over the CCDs and between the two fields of view.
Constraint (\ref{eq:cc2}) implies that the small-scale AL corrections
$\delta\eta$ do not have any components that could be described by
$\Delta\eta$ instead; it therefore ensures a unique division between
the large-scale and small-scale components. Constraint (\ref{eq:cc3})
effectively defines the zero point of the AC field angle $\zeta$
by requiring that $\overline{\Delta\zeta}=0$, separately in each
field of view, when averaged over the CCDs. The sums over $n$ in
Eqs.~(\ref{eq:cc1}) and (\ref{eq:cc3}) are restricted to the CCDs in
the astrometric field (AF), since the skymapper (SM) measurements are
less accurate.

The basic angle $\Gamma_{\rm c}$ introduced in Sect.~\ref{sec:attmodel} is
a fixed reference value, and any real variations of the angle between
the two lines of sight will therefore show up as a variation in
$\Delta\eta$ with opposite signs in the two fields of view.
The offset of the actual basic angle with respect to the conventional
value $\Gamma_{\rm c}$ may be defined as the average difference of 
$\Delta\eta$
between the two fields of view, where the average is computed over the
astrometric CCDs, for gate $g=0$, and over $\mu$; the result for
time period $j$ is
\begin{equation}\label{eq:DeltaGamma}
\Delta\Gamma_j = \frac{1}{62}\sum_{n\in\text{AF}}\sum_f
f\Delta\eta_{0fn0j} \, .
\end{equation}
Although this is obviously a useful quantity to monitor, it does not
appear as a parameter in the geometric calibration model. A number of
additional quantities representing the mean scale offset, the mean
field orientation, etc., can similarly be computed from the large-scale
calibration parameters for the purpose of monitoring.

Equation~(\ref{eq:etazeta}) encapsulates a specific formulation of the
geometric instrument model, with certain assumptions about the shape,
dependencies, and time scales of possible variations. While this particular
model is currently believed to be sufficient to describe the behaviour of
the actual instrument to the required accuracy, it is very likely that
modifications will be needed after a first analysis of the flight data.
Moreover, in the course of the data analysis one may want to try out
alternative models, or examine possible systematics resulting from the
pre-processing (location estimator). In order to facilitate this, a
much more flexible \emph{generic calibration model} has been
implemented. In the generic model, the `observed' field angles
(representing the true observation lines) for any observation $l$ are
written
\begin{equation}\label{eq:generic1}
\left.
\begin{aligned}
\eta_l^\text{obs} &= \eta^0_{ng}
+ \sum_{r=0}^{N_\text{AL}-1} E_r^\text{AL}(l) \\
\zeta_l^\text{obs} &= \zeta^0_{fng}(\mu)
+ \sum_{r=0}^{N_\text{AC}-1} E_r^\text{AC}(l)
\end{aligned}
\right\}
\end{equation}
where each of the $E_r(l)$ (for brevity dropping superscript AL/AC)
represents a basic calibration effect, being a linear combination of
elemental calibration functions $\Phi_{rs}$:
\begin{equation}\label{eq:generic2}
E_r(l) = \sum_{s=0}^{K_r-1} c_{rs}\Phi_{rs}(l) \, .
\end{equation}
$N_\text{AL}$ and $N_\text{AC}$ are the number of effects considered
along and across scan.
The whole set of coefficients $c_{rs}$ forms the calibration vector $\vec{c}$.

In the generic formulation, the multiple indices $f$, $n$, $g$ and the
variables $\mu$ and $t$ are replaced by the single observation index $l$.
This allows maximum flexibility in terms of how the calibration model is
implemented in the software. The functions $\Phi_{rs}$ receive the
observation index $l$, and it is assumed
that this index suffices to derive from it all quantities needed to evaluate
the calibration functions for this observation. Examples of quantities that
can be derived from the observation index are: the FoV index $f$, the CCD and
gate indices $n$ and $g$, the AC pixel coordinate $\mu$, time $t$, and any
relevant astrometric, photometric or spectroscopic parameter of the source
(magnitude, colour index, etc.). Intrinsically real-valued quantities such
as $t$ can be subdivided into non-overlapping intervals with different sets
of calibration parameters applicable to each interval. The basic calibration
model (\ref{eq:etazeta}) can therefore be implemented as a particular instance
of the generic model, with for example $L_r^\ast(x)$ ($r=0,\,1,\,2$)
constituting three of the calibration functions (with $\mu$ derived from $l$).
Once the calibration functions have been coded, the entire calibration model
can be conveniently specified (and changed) through an external configuration
file alone using, e.g., XML structures.

Some elemental calibration functions may be introduced for diagnostic 
purposes rather than actual calibration. An example of this is any function
depending on the colour or magnitude of the source. The origin of such 
effects is briefly explained in Appendix~\ref{sec:chrom} and \ref{sec:CTI}.
Magnitude- and 
colour-dependent variations of the instrument response are expected to be
fully taken into account by the signal modelling on the CCD data level,
as outlined in Sect.~\ref{sec:signalmodel} (notably by the LSF and CDM
calibrations), and should not show up in the astrometric solution. Thus,
non-zero results for such `non-geometric' diagnostic calibration parameters 
indicate that the signal modelling should be improved. In the final solution
the diagnostic calibration parameters should ideally be zero.

The parameters of the generic calibration model must satisfy a number of
constraints similar to Eqs.~(\ref{eq:cc1})--(\ref{eq:cc3}). These can be cast
in the general form
\begin{equation}\label{eq:genericConstraint}
\vec{C}'\vec{c} = \vec{0} \, ,
\end{equation}
where the matrix $\vec{C}$ contains one column, with known coefficients, 
for each constraint. The columns are, by design, linearly independent.

\subsection{Signal (CCD data) model} \label{sec:signalmodel}

As suggested in Fig.~\ref{fig:blockdiagram}, the modelling of CCD data
at the level of individual pixels (i.e., the photon counts) is not part of the 
geometrical model of the observations with which we are concerned in 
this paper. However, the processing of the photon counts, effectively
by fitting the CCD data model, produces the `observations'
that are the input to the astrometric core solution. In order
to clarify the exact meaning of these observations we include here 
a brief overview of the signal model.
  
The pixel size, $10~\mu\mbox{m}\simeq 59$~mas in the along-scan
(AL) direction and $30~\mu\mbox{m}\simeq 177$~mas in the across-can
(AC) direction, roughly matches the theoretical diffraction image
for the $1.45\times 0.50$~m$^2$ telescope pupil of Gaia (effective
wavelength $\sim$~650~nm). Around each
detected object, only a small rectangular window (typically 6--18 pixels
long in the AL direction and 12 pixels wide in the AC direction) is
actually read out and transmitted to the ground. Moreover, for most of
the observations in the astrometric field (AF), on-chip binning in the
serial register is used to sum the charges over the window in the
AC direction. This effectively results in a one-dimensional image of
6--18 AL `samples', where the signal ($N_k$) in each sample $k$ is the 
sum of 12 AC pixels.
The exact time $t_k$ when a sample was transferred to the serial register,
expressed on the TCB scale, is in principle known from the time
correlation of the on-board clock with ground-based time signals.
Because of the known one-to-one relation between the TDI period counter
$k$ and $t_k$, we may use $k$ as a proxy for $t_k$ in subsequent
calculations.

For single stars, and in the absence of the effects discussed in 
Appendix~\ref{sec:CTI}, the sample values in the window are modelled as a
stochastic variable (Poisson photon noise plus electronic readout noise) 
with expected value
\begin{equation}
    \lambda_k \equiv \mbox{E}\left( N_k \right) = \beta + \alpha L(k - \kappa)
    \label{eq:signal01}
\end{equation}
where $\beta$, $\alpha$ and $\kappa$ are the so-called image parameters
representing the (uniform) background level, the amplitude (or flux) of
the source, and the AL location (pixel coordinate) of the image centroid.
The continuous, non-negative function $L(x)$ is the Line Spread Function
(LSF) appropriate for the observation. $L(x)$  depends, for example, on the
spectrum of the source and on the position of the image in the focal plane.
The (in general non-integer) pixel coordinate $\kappa$ is expressed on
the same scale as the (integer) TDI index $k$, and may be translated to
the equivalent TCB $t(\kappa)$ by means of the known relation between
$k$ and $t_k$. The \emph{CCD observation time} is defined as
$t(\kappa - K/2)$, where $K$ is the number of TDI periods used for
integrating the image (see Appendix~\ref{sec:TDI}). Formally, the CCD observation 
time is the instant of time at which the centre of the source image
passed across the CCD fiducial line halfway between the first and the
last TDI line used in the integration (this will depend on the gating).

Fitting the CCD signal model to the one-dimensional sample values $N_k$
thus gives, as the end result of observation $l$, an estimate of the AL pixel
coordinate $\kappa$ of the image in the pixel stream, which is then transformed
to the observation time $t_l$. The fitting procedure also provides
an estimate of the uncertainty in $\kappa$, which can be expressed in
angular measure as a formal standard deviation of the AL measurement,
$\sigma_l^\text{AL}$. It is derived by error propagation through the
fitted signal model, taking into account the dominant noise sources,
photon noise and readout noise.

For some observations, AC information is also provided through
two-dimensional sampling of the pixel window around the detected object.
This applies to all SM observations, AF observations of bright
($G\lesssim 13$) sources, and AF observations of Calibration Faint Stars
(Sect.~\ref{sec:instrumentmodel}). The modelling of the two-dimensional
images follows the same principles as outlined above for the one-dimensional
(AL only) case, only that the LSF is replaced by a two-dimensional
Point Spread Function (PSF) and that there is one more location parameter
to estimate, namely the AC pixel coordinate $\mu$. The astrometric result
in this case consists of the observation time $t_l$, the
observed AC coordinate $\mu_l$, and their formal standard uncertainties
$\sigma_l^\text{AL}$ and $\sigma_l^\text{AC}$ (both expressed as angles).

The estimation errors for different images are, for the subsequent analysis, 
assumed to be statistically independent (and therefore uncorrelated). This is a 
very good approximation to the extent that they only depend on the photon 
and readout noises. 
However, modelling errors at the various stages of the processing
(in particular CTI effects in the signal modelling [Appendix~\ref{sec:TDI}]
and attitude irregularities [Appendix~\ref{sec:attirr}]) are likely to introduce errors
that are correlated at least over the nine AF observations in a field transit.
The resulting correlations as such are not taken into account in the astrometric
solution (i.e., the weight matrix of the least-squares equations is taken to
be diagonal), but the sizes of the modelling errors are estimated, and
are employed to reduce the statistical weights of the observations as 
described in Sect.~\ref{sec:synthesismodel}.  
The AL and AC estimates of a given (two-dimensional) 
image are roughly independent at least in the limit of small optical
aberrations.

\subsection{Synthesis model}\label{sec:synthesismodel}

By synthesis of the models described in the preceding sections,
we are now in a position to formulate very precisely the core
astrometric data analysis problem as outlined in
Sect.~\ref{sec:outline}. The unknowns are represented by the
vectors $\vec{s}$, $\vec{a}$, $\vec{c}$, and $\vec{g}$ of
respectively the source, attitude, calibration, and global
parameters. For any observation $l$ the observed quantities are the
observation time $t_l$ and, where applicable, the observed AC pixel
coordinate $\mu_l$, with their formal uncertainties $\sigma_l^\text{AL}$,
$\sigma_l^\text{AC}$. We then have the global minimization problem
\begin{equation}\label{eq:synth}
\min_{\vec{s},\vec{a},\vec{c},\vec{g}}\;
Q = \sum_{l\,\in\,\text{AL}} \frac{(R_l^\text{AL})^2\,w_l^\text{AL}}%
{(\sigma_l^\text{AL})^2+(\epsilon_l^\text{AL})^2}
+ \sum_{l\,\in\,\text{AC}} \frac{(R_l^\text{AC})^2\,w_l^\text{AC}}%
{(\sigma_l^\text{AC})^2+(\epsilon_l^\text{AC})^2}\, ,
\end{equation}
where
\begin{align}
R_l^\text{AL}(\vec{s},\vec{a},\vec{c},\vec{g}) &= \eta_{fng}(\mu_l,t_l\,|\,\vec{c})-
\eta(t_l\,|\,\vec{s},\vec{a},\vec{g})\label{eq:resAL}\, ,\\[3pt]
R_l^\text{AC}(\vec{s},\vec{a},\vec{c},\vec{g}) &= \zeta_{fng}(\mu_l,t_l\,|\,\vec{c})-
\zeta(t_l\,|\,\vec{s},\vec{a},\vec{g})\label{eq:resAC}
\end{align}
are the residuals in the field angles, taken as functions of the 
unknowns\footnote{Note that in Eq.~(\ref{eq:resAL}) the quantity $\mu_l$ is just a 
given value (observed or computed); in the case of one-dimensional images 
the observed $\mu_l$ is replaced by an approximate value derived from 
current knowledge on the source and attitude.},
and $l\in\text{AL}$ refers to observations with a valid AL component, etc.
The applicable indices $f$, $n$, $g$ are of course known for every
observation $l$. In Eq.~(\ref{eq:synth}), $\epsilon_l^\text{AL}$ and
$\epsilon_l^\text{AC}$ represent all AL and AC error sources extraneous
to the formal uncertainties; they include in particular modelling errors in the
source behaviour (e.g., for unrecognized binaries), attitude and calibration, 
which have to be estimated as functions
of time and source in the course of the data analysis process.
$w_l^\text{AL}$ and $w_l^\text{AC}$ are weight factors, always in the range 0 to 1;
for most observations they are equal to 1, but `bad' data (outliers) are assigned
smaller weight factors. The determination of these factors is discussed in
Sect.~\ref{sec:source-inner}.

For the sake of conciseness we shall hereafter consider the AL and AC components
of an observation to have separate observation indices $l$, so that for example
$R_l$ stands for either $R_l^\text{AL}$ or $R_l^\text{AC}$, as the case may be. 
This allows the two sums in Eq.~(\ref{eq:synth}) to be
contracted and written concisely as
\begin{equation}\label{eq:Jfunc}
Q(\vec{s},\vec{a},\vec{c},\vec{g}) =
\sum_{l} \frac{R_l^2\,w_l}{\sigma_l^2+\epsilon_l^2} =
\sum_{l} R_l^2\,W_l\, ,
\end{equation}
where $W_l=w_l/(\sigma_l^2+\epsilon_l^2)$ is the statistical weight of the 
observation.

The excess noise $\epsilon_l$ represents modelling errors and should ideally
be zero. However, it is unavoidable that some sources do not behave exactly 
according to the adopted astrometric model (Sect.~\ref{sec:astromodel}), or
that the attitude sometimes cannot be represented by the spline model in 
Sect.~\ref{sec:attmodel} to sufficient accuracy. The excess noise term
in Eq.~(\ref{eq:Jfunc}) is introduced to allow these cases to be handled in a 
reasonable way, i.e., by effectively reducing the statistical weight of the
observations affected. It should be noted that these modelling errors are
assumed to affect \emph{all} the observations of a particular star, or
\emph{all} the observations in a given time interval. (By contrast, the 
downweighting factor $w_l$ is intended to take care of isolated outliers,
for example due to a cosmic-ray hit in one of the CCD samples.) This
is reflected in the way the excess noise is modelled as the sum of two
components,
\begin{equation}\label{eq:excess}
\epsilon_l^2 = \epsilon_i^2 + \epsilon^2_a(t_l) \, ,
\end{equation}
where $\epsilon_i$ is the excess noise associated with source $i$ (if
$l\in i$, that is, $l$ is an observation of source $i$), and $\epsilon_a(t)$ 
is the excess 
attitude noise, being a function of time. For a `good' primary source, we 
should have $\epsilon_i=0$, and for a `good' stretch of attitude data 
we may have $\epsilon_a(t)=0$.  Calibration modelling errors are not 
explicitly introduced in Eq.~(\ref{eq:excess}), but could be regarded as 
a more or less constant 
part of the excess attitude noise. The estimation of $\epsilon_i$
is described in Sect.~\ref{sec:source-inner}, and the estimation of 
$\epsilon_a(t)$ in Sect.~\ref{sec:attNoise}.

Three separate functions are needed to describe the excess attitude 
noise, corresponding to AL observations, AC observations in the
preceding field of view (ACP), and AC observations in the following field
of view (ACF).
We distinguish between these functions by letting the subscript $a$ in
$\epsilon_a(t)$ stand for either AL, ACP or ACF.

\section{Solving the global minimization problem} \label{sec:numerical}

Assuming $10^8$ primary sources, 
the number of unknowns in the global minimization problem, Eq.~(\ref{eq:synth}),
is about $5\times 10^8$ for the sources ($\vec{s}$), $4\times 10^7$ for the
attitude ($\vec{a}$, assuming a knot interval of 15~s for the 5~yr mission;
cf.\ Sect.~\ref{sec:att-normals}), $10^6$ for the calibration $\vec{c}$, and less than
100 global parameters ($\vec{g}$). The number of elementary observations
($l$) considered is about $8\times 10^{10}$. However, the size of the data
set, and the large number of parameters, would not by themselves be a
problem if the observations, or sources, could be processed sequentially.
The difficulty is caused by the strong connectivity among the observations:
each source is effectively observed relative to a large number of other sources
simultaneously in the field of view, or in the complementary field of view
some $106.5^\circ$ away on the sky, linked together by the attitude and
calibration models. The complexity of the astrometric solution in terms
of the connectivity between the sources provided by the attitude modelling
was analysed by \citet{bombrun+09}, who concluded that a direct solution
is infeasible, by many orders of magnitude, with today's computational
capabilities. The study neglected the additional connectivity due to
the calibration model, which makes the problem even more unrealistic to
attack by a direct method. Note that this impossibility is not a defect, 
but a virtue of the mathematical system under consideration: 
it guarantees that a unique, coherent  and completely independent 
global solution for the whole sky can be derived from the system.

The natural alternative to a direct solution is to use some iterative
method. This is in fact the standard way to deal with large, sparse systems
of equations. The literature in the field is vast and a plethora of methods
exist for various kinds of applications. However, the iterative method
adopted for Gaia did not spring from a box of ready-made algorithms.
Rather, it was developed over several years in parallel with the software
system in which it could be implemented and tested. Originally based on
an intuitive and somewhat simplistic approach, the algorithm has developed
through a series of experiments, insights and improvements into a rigorous,
efficient and well-understood procedure, completely adapted to its
particular application. In this section we first describe the approach
in broad outline, then provide the mathematical background for its better
understanding and further development.

\subsection{Outline of the iterative solution} \label{sec:agis}

The numerical approach to the Gaia astrometry is a block-iterative least-squares
solution, referred to as the Astrometric Global Iterative Solution (AGIS).
In its simplest form, four blocks are evaluated in a cyclic sequence until
convergence. The blocks map to the four different kinds of unknowns outlined
in Sect.~\ref{sec:math}, namely:
\begin{itemize}
\item[S:] the source (star) update, in which the astrometric parameters $\vec{s}$
of the primary sources are improved;
\item[A:] the attitude update, in which the attitude parameters $\vec{a}$ are
improved;
\item[C:] the calibration update, in which the calibration parameters $\vec{c}$
are improved;
\item[G:] the global update, in which the global parameters $\vec{g}$ are improved.
\end{itemize}
The G block is optional, and will perhaps only be used in some of the final solutions,
since the global parameters can normally be assumed to be known a~priori to high
accuracy.

The blocks must be iterated because each one of them needs data from the three other
processes. For example, when computing the astrometric parameters in the S block,
the attitude, calibration and global parameters are taken from the previous iteration.
The resulting (updated) astrometric parameters are used the next
time the A block is run, and so on. This iterative approach to the astrometric 
solution was proposed early on in the Hipparcos project as an
alternative to the `three-step method' subsequently adopted for the original Hipparcos
reductions; see \citet[][Vol.~3, p.~488]{hip:catalogue} and references therein.
Indeed, the later re-reduction of the Hipparcos raw data by \citet{book:newhip}
used a very similar iterative method, and yielded significantly improved results 
mainly by virtue of the much more elaborate attitude modelling implemented as part
of the approach.

While the block-iterative solution as outlined above is intuitive and appealing in
its simplicity, it is from a mathematical standpoint not obvious that it must converge;
and if it does indeed converge, it is not obvious how
many iterations are required, whether the order of the blocks in each iteration
matter, and whether the converged results do in fact constitute a solution to the
global minimization problem. These are fundamental questions for the validity of
the adopted iterative approach, and we therefore take some care in the following
subsections to establish its theoretical foundations 
(Sects.~\ref{sec:lsqr}--\ref{sec:iter}). Section~\ref{sec:blocks} then describes
each of the S, A, C and G blocks in some detail.
In addition to these blocks, separate processes are required for the alignment 
of the astrometric solution with the ICRS, the selection of primary sources,
and the calculation of standard uncertainties; these auxiliary processes are discussed 
in Sect.~\ref{sec:aux}.

\subsection{Least-squares approach} \label{sec:lsqr}

Strictly speaking, Eq.~(\ref{eq:synth}) is not a least-squares problem, because
of the weight factors $w_l^\text{AL}$, $w_l^\text{AC}$ (as well as the excess
noises $\epsilon_l^\text{AL}$, $\epsilon_l^\text{AC}$), which depend on the AL and
AC residuals and hence on the unknowns $(\vec{s},\vec{a},\vec{c},\vec{g})$.
In Eq.~(\ref{eq:generalobs}) this dependence is formally included in the unspecified
metric $\mathcal{M}$, which therefore is not simply a (weighted) Euclidean norm.

In principle, the minimization problem (\ref{eq:synth}) can be solved by finding
a point where the partial derivatives of the objective function $Q$ with respect to
all the unknowns are simultaneously zero.
In practice, however, the partial derivatives are not computed completely rigorously,
and the problem solved is therefore a slightly different one from what is outlined
above. In order to understand precisely the approximations involved, it is necessary
to consider how different kinds of non-linearities enter the problem.

The functions $\eta_{fng}(\mu,t\,|\,\vec{c})$ and $\zeta_{fng}(\mu,t\,|\,\vec{c})$
appearing in Eqs.~(\ref{eq:resAL})--(\ref{eq:resAC}) are strictly linear in the
calibration parameters $\vec{c}$, by virtue of the basic geometric calibration model
in Eq.~(\ref{eq:etazeta}) or the generic model in
Eqs.~(\ref{eq:generic1})--(\ref{eq:generic2}). On the other hand, the functions
$\eta(t\,|\,\vec{s},\vec{a},\vec{g})$ and $\zeta(t\,|\,\vec{s},\vec{a},\vec{g})$
are non-linear in $\vec{s}$, $\vec{a}$, and $\vec{g}$ due to the complex
transformations involved (Sects.~\ref{sec:astromodel}--\ref{sec:attmodel}).
However, thanks to the data processing prior to the astrometric solution, the
initial errors in these parameters are already so small that the corresponding
errors in $\eta$ and $\zeta$ are only some 0.1~arcsec ($\sim 10^{-6}$~rad).
Second-order terms are therefore typically less than
$10^{-12}~\text{rad}\simeq 0.2~\mu$as, that is negligible in comparison with
the noise of a single AL observation (some $100~\mu$as). This means that the
partial derivatives of the residuals $R_l^\text{AL}$ and $R_l^\text{AC}$ with
respect to all the unknowns do not change in the course of the solution.
(In practice they are in fact recomputed in each iteration, although that is mainly
done because it is more convenient than to store and retrieve the values;
nevertheless, this takes care of any remaining non-linearity, however small.)
The non-linearities of the underlying astrometric, attitude, calibration
and global models are therefore not an issue for the minimization problem as such.

The weight factors $w_l$ 
represent a different kind of non-linearity, potentially much more important for
the final solution. These factors are introduced to make the solution robust against
outliers, by reducing their influence on $Q$ and hence on the estimated parameter
values (Sect.~\ref{sec:source-inner}). Ideally, outlying observations should not
contribute at all to the solution (by having $w_l=0$), while `normal' observations
should receive full weight ($w_l=1$). In reality there will however be a transition
region where the weight factors are between 0 and 1. Since the weight factors are
in practice determined by the normalized residuals,
$\hat{R}_l\equiv R_l(\sigma_l^2+\epsilon_l^2)^{-1/2}$, which in turn depend on
the parameter values $(\vec{s},\vec{a},\vec{c},\vec{g})$, it follows that the
partial derivatives contain extra terms of the form
$\hat{R}_l^2\partial w_l/\partial\vec{s}$, $\hat{R}_l^2\partial w_l/\partial\vec{a}$,
etc., that are non-zero for some observations. Analogous considerations apply
to the excess noise terms $\epsilon_l$: they too are estimated by means of
the residuals (Sects.~\ref{sec:source-inner} and \ref{sec:attNoise}), 
and therefore in principle introduce additional terms in the partial derivatives.

We take the somewhat pragmatic approach of neglecting the terms depending on
the partial derivatives of $w_l$ and $\epsilon_l$ with respect to the unknowns
when seeking the solution to the global minimization problem. The consequences
of this approximation can be appreciated by observing that the down-weighting
($w_l<1$) only kicks in when the absolute value of the residual exceeds a few
times the total standard uncertainty, or some 0.5--1~mas for typical observations of
bright sources. Similarly the estimated $\epsilon_l$ are only sensitive to changes of the 
residuals of a similar size. Experience with AGIS runs on simulated data show
that the typical changes of the residuals fall below this level already in the first few iterations. 
Thereafter
the weight factors and the estimated excess noises do not change significantly.
Neglecting the derivatives of $w_l$ and $\epsilon_l$ in the global minimization
problem is therefore equivalent to solving the weighted least-squares problem
with $w_l$ and $\epsilon_l$ fixed at whatever values they settle to after the
initial iterations. This is a reasonable assumption given that the statistical
weight of any observation, and the size of the modelling errors, are not
a~priori expected to depend on the actual values of the parameters. The
precise solution does of course depend on how $w_l$ and $\epsilon_l$ are
estimated, but that is an unavoidable consequence of any practical data
analysis approach.

\subsection{Normal equations} \label{sec:norm}

The minimization of $Q$ in Eq.~(\ref{eq:Jfunc}) is thus solved by the
weighted least-squares method, assuming fixed weights $W_l$ that are
however determined as
part of the solution. The normal equations for the sources are given by
$\frac{1}{2}\partial Q/\partial\vec{s}=\vec{0}$, and similarly for the
other unknowns. Linearising around any reference point
$(\vec{s}^\text{ref},\vec{a}^\text{ref},\vec{c}^\text{ref},\vec{g}^\text{ref})$
within the linear regime of parameter space, i.e., setting
$\vec{s}=\vec{s}^\text{ref}+\vec{x}_s$, and similarly for the other unknowns,
and expanding to first order in the differential vector
$\vec{x}=(\vec{x}_s',\vec{x}_a',\vec{x}_c',\vec{x}_g')'$, we find the normal
equations as
\begin{equation}\label{eq:norm}
\left[ \sum_l
\frac{\partial R_l}{\partial\vec{x}}\frac{\partial R_l}{\partial\vec{x}'}
W_l\right]\vec{x} = -\sum_l \frac{\partial R_l}{\partial\vec{x}}
R_l(\vec{s}^\text{ref},\vec{a}^\text{ref},\vec{c}^\text{ref},\vec{g}^\text{ref})W_l
\, .
\end{equation}
This can be written in matrix form as
\begin{equation}\label{eq:norm1}
\vec{N}\,\vec{x} = \vec{b} \, ,
\end{equation}
where $\vec{N}$ is a symmetric matrix. We now proceed to analyse the structure
of this linear system of equations in terms of the previously mentioned block
updates S, A, C, G.

The matrix $\vec{N}$ and the vectors (column matrices) $\vec{x}$, $\vec{b}$
can be partitioned into sub-matrices and sub-vectors corresponding to the
different parameter vectors $\vec{s}$, $\vec{a}$, $\vec{c}$, and $\vec{g}$:
\begin{equation}\label{eq:norm2}
    \begin{bmatrix}
        \vec{N}_{ss}  & \vec{N}_{sa}  & \vec{N}_{sc}  & \vec{N}_{sg}  \\
        \vec{N}_{as}  & \vec{N}_{aa}  & \vec{N}_{ac}  & \vec{N}_{ag}  \\
        \vec{N}_{cs}  & \vec{N}_{ca}  & \vec{N}_{cc}  & \vec{N}_{cg}  \\
        \vec{N}_{gs}  & \vec{N}_{ga}  & \vec{N}_{gc}  & \vec{N}_{gg}
    \end{bmatrix} \,
    \begin{bmatrix}
        \vec{x}_s \\
        \vec{x}_a \\
        \vec{x}_c \\
        \vec{x}_g
    \end{bmatrix} =
    \begin{bmatrix}
        \vec{b}_{s} \\
        \vec{b}_{a} \\
        \vec{b}_{c} \\
        \vec{b}_{g}
    \end{bmatrix} \, ,
\end{equation}
where $\vec{N}_{as}=\vec{N}_{sa}'$, etc. Of importance here is that
$\vec{N}_{ss}$ and $\vec{N}_{aa}$ have a particularly simple structure.
Since $\vec{s}$ is sub-divided into vectors $\vec{s}_i$ of length 5
for the individual primary sources ($i$), it is natural to sub-divide
$\vec{N}_{ss}$ into blocks of $5\times 5$ elements.
From Eq.~(\ref{eq:norm}) it follows that the $(i,j)$th such block is
given by
\begin{equation}\label{eq:Nss}
\left[\vec{N}_{ss}\right]_{ij} = \sum_l
\frac{\partial R_l}{\partial\vec{s}_i}\frac{\partial R_l}{\partial\vec{s}'_{\!j}}
W_l =
\begin{cases}\displaystyle
\sum_{l\,\in\,i} \frac{\partial R_l}{\partial\vec{s}_i}\frac{\partial R_l}{\partial\vec{s}'_{\!i}}
W_l & \text{if $i=j$,}\\[12pt]
\vec{0} & \text{if $i\ne j$,}
\end{cases}
\end{equation}
where $l\in i$ signifies that the sum is taken over the observations of
source $i$. The result for $i\ne j$ follows because no observation $l$ is
associated with more than one primary source. $\vec{N}_{ss}$ is consequently
block-diagonal, and $\vec{N}_{ss}^{-1}\vec{b}_s$ can trivially be computed
for arbitrary vector $\vec{b}_s$ by looping through the sources and
solving the corresponding $5\times 5$ system for each source.%
\footnote{Here, and elsewhere in this paper, an expression like
$\vec{A}^{-1}\vec{b}$ is shorthand notation for solving the system
$\vec{A}\vec{y}=\vec{b}$. The inverse matrix $\vec{A}^{-1}$ is (usually) not computed, but only the
solution vector $\vec{y}=\vec{A}^{-1}\vec{b}$ itself.}
This is exactly what is done in the source update block (S).

The vector of attitude unknowns is naturally sub-divided into vectors
$\vec{a}_n$ of length 4, containing the elements of the quaternions
$\quat{a}_n$ that serve as coefficients in the B-spline representation,
Eq.~(\ref{eq:qspline}). If $\vec{N}_{aa}$ is correspondingly sub-divided
into blocks of $4\times 4$ elements, it follows from Eq.~(\ref{eq:norm})
that the $(n,m)$th such block is given by
\begin{equation}\label{eq:Naa}
\left[\vec{N}_{aa}\right]_{nm} = \sum_l
\frac{\partial R_l}{\partial\vec{a}_n}\frac{\partial R_l}{\partial\vec{a}'_{m}}
W_l \, .
\end{equation}
With $\ell$ denoting the left index of $t_l$ (Sect.~\ref{sec:attmodel}), we
have $\partial R_l/\partial\vec{a}_n=\vec{0}$ whenever $n<\ell-M+1$ or $n>\ell$,
where $M$ is the order of the spline. It follows that
$\left[\vec{N}_{aa}\right]_{nm}=\vec{0}$ if $|n-m|>M-1$
(cf.\ Appendix~\ref{sec:bsplines}). The non-zero blocks in $\vec{N}_{aa}$
are therefore confined to the diagonal and $M-1$ blocks above and below the
diagonal (Fig.~\ref{fig:Naa}). 
Thus, $\vec{N}_{aa}^{-1}\vec{b}_a$ can be efficiently computed for
arbitrary $\vec{b}_a$ since the Cholesky decomposition of the matrix
does not produce any additional fill-in (Appendix~\ref{sec:cholesky}).
This system is solved in the attitude update block (A).

In the geometric instrument model (Sect.~\ref{sec:instrumentmodel}), each
CCD/gate and time interval combination (index $ngk$ for example in Eq.~\ref{eq:etazeta}) has
its own set of unknowns. Moreover, a given observation $l$ can only refer
to one CCD/gate combination. By the same reasoning as above, the sub-matrix
$\vec{N}_{cc}$ is therefore block-diagonal, and $\vec{N}_{cc}^{-1}\vec{b}_c$
can be computed for arbitrary $\vec{b}_c$ by looping over the CCD/gates
combinations. This is done in the calibration update block (C).
Although the number of calibration parameters per CCD/gate combination can
be fairly large ($\sim\!10^4$), the resulting systems are well within the
bounds that can readily be handled by direct matrix methods, even without
taking into account their sparseness.

By definition all the global parameters affect each observation, and
the sub-matrix $\vec{N}_{gg}$ is therefore full. However, since the number
of global parameters is never large, $\vec{N}_{gg}^{-1}\vec{b}_g$ can
easily be computed, which is done in the global update block (G).

In the description above we have implicitly assumed that each of the
diagonal blocks $\vec{N}_{ss}$, $\vec{N}_{aa}$, $\vec{N}_{cc}$, and 
$\vec{N}_{gg}$ is non-singular, and even well-conditioned in order to
avoid numerical instability. This is equivalent to the statement that
each of the blocks S, A, C and G is a well-posed problem: for example,
that the determination of the source parameters is `easy' if we assume
that we know the attitude, calibration and global parameters.
This will in practice be guaranteed by the choice of primary sources
(which will make $[\vec{N}_{ss}]_{ii}$ well-conditioned 
for every $i$) for the S block, and 
by the adopted attitude, calibration and global parametrizations, 
including the constraints necessary to render the updates unique --
in particular the quaternion length normalization in Eq.~(\ref{eq:qspline})
for the attitude model, and Eq.~(\ref{eq:genericConstraint}) for the 
calibration model.

Turning now to the off-diagonal sub-matrices of $\vec{N}$, it is 
natural to sub-divide for example the sub-matrix $\vec{N}_{as}$
into blocks of $4\times 5$
elements corresponding to the 4 components of the quaternion and the
5 astrometric parameters. The $(n,i)$th block,
\begin{equation}\label{eq:Nas}
\left[\vec{N}_{as}\right]_{ni} = \sum_l
\frac{\partial R_l}{\partial\vec{a}_n}\frac{\partial R_l}{\partial\vec{s}'_{\!i}}
W_l \, ,
\end{equation}
is non-zero only if source $i$ was observed in the time interval
$[\tau_n,\tau_{n+M}]$, which is the support of the B-spline $B_n(t)$.
$\vec{N}_{as}$ is therefore very sparse, but it also has no simple
structure because the distribution of the non-zero blocks is linked to
the scanning law. Fortunately, as will be explained in Sect.~\ref{sec:iter}, there
is no need to explicitly compute, much less store, this sub-matrix, nor
any of the other off-diagonal sub-matrices in Eq.~(\ref{eq:norm2}).

\subsection{Rank of the normal equations}\label{sec:rank}

From the nature of the astrometric observations, which are in effect
differential within the (combined) field of view, and the modelling of
the primary sources, which does not assume that any of their positions
or proper motions are known a~priori, it is clear that there is no
unique astrometric solution to the problem as outlined above. The
fundamental reason for this is that any (small) change in the orientation
of the celestial reference system, as well as the introduction of a
(small) inertial spin of the system, would leave all observations
invariant. In principle the non-uniqueness of the solution is not a
problem as such, since the resulting system of positions and proper
motions are afterwards aligned with the ICRS by a special process
(Sect.~\ref{sec:framerotator}). However, it does imply that the
normal matrix $\vec{N}$ is in principle singular,%
\footnote{More precisely, $\vec{N}$ is properly semidefinite,
so that $\vec{x}'\vec{N}\vec{x}\ge 0$ for all $\vec{x}\ne\vec{0}$,
with equality for some $\vec{x}\ne\vec{0}$.\label{footnote1}}
which may have consequences for the numerical solution of the
normal equations. We say singular `in principle' because arithmetic
rounding errors will in practice prevent it from becoming truly
singular, although it remains extremely ill-conditioned.

More precisely, we expect $\vec{N}$ to have rank $n-6$, if
$n=\dim(\vec{N})$ is the total number of unknowns. The null space
of the matrix,
\begin{equation}\label{eq:null}
\mathcal{N}(\vec{N}) = \{ \vec{v}\in\mathbb{R}^n\,|\,\vec{N}\vec{v}=\vec{0}\}
\end{equation}
therefore has rank six. Indeed, it is easy to find six linearly
independent vectors $\vec{v}_0,\,\dots,\,\vec{v}_5$ that span the null space:
the first three are found by introducing small changes in the orientation
of the celestial reference system around each of its principal axes,
and deriving the corresponding changes in $\vec{s}$ and $\vec{a}$
($\vec{c}$ and $\vec{g}$ being independent of the reference system);
the last three are correspondingly found by introducing a small
inertial spin of the reference system around each axis (see
Sect.~\ref{sec:nullsp} for details). Introducing
the $n\times 6$ matrix $\vec{V}=(\vec{v}_0,\,\dots,\,\vec{v}_5)$ we have
\begin{equation}\label{eq:nullV}
\vec{N}\vec{V} = \vec{0}\, .
\end{equation}
The singularity can in principle be removed by adding the six constraints
$\vec{V}'\vec{x}=\vec{0}$, but in practice that is not necessary. It suffices
to derive \emph{one} particular solution $\vec{\tilde{x}}$ to the normal
equations, then the whole solution space can be written
$\vec{\tilde{x}}+\vec{V}\vec{z}$ for arbitrary $\vec{z}\in\mathbb{R}^6$.
The vector $\vec{z}$ is effectively determined by the frame rotator
(Sect.~\ref{sec:framerotator}), yielding the unique solution
that best agrees with the adopted definition of the ICRS.

Quite apart from numerical rounding errors, it is not completely
true that $\vec{N}$ is mathematically singular. Stellar aberration and
parallax introduce some absolute knowledge about the reference system
via the barycentric ephemeris of Gaia, which is expressed in ICRS and
is not part of the adjustment process. However, since stellar
aberration is at most $10^{-4}$~rad, the orientation error would have
to be of the order of 1~mas for the aberration effect to change by
$0.1~\mu$as (say). So, although absent in principle, the indeterminacy
of the reference frame orientation and spin exists in practice.
Since the orientation can be determined to much
higher accuracy than 1~mas (by means of the optical counterparts of radio sources),
the contribution of stellar aberration to the absolute frame knowledge
can be neglected in practice. The same holds, a~fortiori, for the
much smaller parallax effect.

\subsection{The simple iteration step}\label{sec:iter}

Consider the system
\begin{equation}\label{eq:precond}
\vec{K}\vec{d}=\vec{b}\, , \qquad \vec{K} =
    \begin{bmatrix}
        \vec{N}_{ss} & \vec{\varnothing} & \vec{\varnothing} & \vec{\varnothing} \\
        \vec{N}_{as} & \vec{N}_{aa}      & \vec{\varnothing} & \vec{\varnothing} \\
        \vec{N}_{cs} & \vec{N}_{ca}      & \vec{N}_{cc}      & \vec{\varnothing} \\
        \vec{N}_{gs} & \vec{N}_{ga}      & \vec{N}_{gc}      & \vec{N}_{gg}
    \end{bmatrix} \, .
\end{equation}
where $\vec{b}$ is the same right-hand side as in Eq.~(\ref{eq:norm1}), 
and each $\vec{\varnothing}$ stands for a zero-filled sub-matrix of the
appropriate dimensions. Although this system is of the same size as
Eq.~(\ref{eq:norm1}), is it fundamentally different in that it can be directly solved
through a sequence of smaller systems,
\begin{equation}\label{eq:precond1}
\left.
\begin{aligned}
\vec{N}_{ss}\vec{d}_s &= \vec{b}_s \\
\vec{N}_{aa}\vec{d}_a &= \vec{b}_a-\vec{N}_{as}\vec{d}_s \\
\vec{N}_{cc}\vec{d}_c &= \vec{b}_c-\vec{N}_{cs}\vec{d}_s-\vec{N}_{ca}\vec{d}_a \\
\vec{N}_{gg}\vec{d}_g &= 
\vec{b}_g-\vec{N}_{gs}\vec{d}_s-\vec{N}_{ga}\vec{d}_a-\vec{N}_{gc}\vec{d}_c
\quad
\end{aligned}
\right\} 
\end{equation}
where each sub-system is of the form S, A, C, G described above, allowing it
to be solved with relative ease. (Naturally, the resulting solution $\vec{d}$
is also quite different from the $\vec{x}$ in Eq.~\ref{eq:norm1}!) 
By means of Eqs.~(\ref{eq:norm})
and (\ref{eq:Nas}) the right-hand side in the second sub-system becomes
\begin{align}
\vec{b}_a-\vec{N}_{as}\vec{d}_s &= -\sum_l \frac{\partial R_l}{\partial\vec{a}}
\left[ R_l(\vec{s}^\text{ref},\vec{a}^\text{ref},\vec{c}^\text{ref},\vec{g}^\text{ref})
+\frac{\partial R_l}{\partial\vec{s}'}\vec{d}_s\right]W_l \nonumber \\
&= -\sum_l \frac{\partial R_l}{\partial\vec{a}}
R_l(\vec{s}^\text{ref}+\vec{d}_s,\vec{a}^\text{ref},\vec{c}^\text{ref},\vec{g}^\text{ref})W_l
\, ,\label{eq:precond2}
\end{align}
where the linearity of $R_l(\vec{s},\vec{a},\vec{c},\vec{g})$ has been used in a 
Taylor expansion around the reference values. This shows that the off-diagonal 
matrix $\vec{N}_{as}$ is not needed in order to compute the right-hand side
of the second sub-system in Eq.~(\ref{eq:precond1}), if only the residuals
are computed \emph{after} the source parameters have been updated by the 
solution of the first sub-system. Similarly, we find that
the right-hand side in the third sub-system can be obtained from the residuals
after updating both the source and attitude parameters, and so on. The off-diagonal
sub-matrices in $\vec{K}$ are therefore not needed, provided that the
sub-vectors of unknowns are successively updated before the new right-hand sides
are computed.

From the above it is clear that a single AGIS iteration, consisting of the
successive application of the four update blocks S, A, C and G, is mathematically
equivalent to an updating of the unknowns by $\vec{d}=\vec{K}^{-1}\vec{b}$.
In the context of iterative solution algorithms, the matrix $\vec{K}$ is referred
to as the preconditioner of the normal equations system \citep{axelsson96}.

As previously noted, we assume that the block-diagonal matrices 
$\vec{N}_{ss}$, $\vec{N}_{aa}$, $\vec{N}_{cc}$, $\vec{N}_{gg}$ are all 
non-singular, and in fact positive definite, for a proper formulation 
of the S, A, C, and G blocks. This ensures that the preconditioner 
$\vec{K}$ is non-singular, even though $\vec{N}$ is not.

In the course of the iterations, new right-hand sides of the normal
equations will be computed, while the matrix remains essentially unchanged.
From now on, let us express the residuals $R_l$ not as functions of the
parameter values $\vec{s}$, $\vec{a}$, $\vec{c}$, $\vec{g}$ but in terms 
of the differential parameter vector $\vec{x}$ relative to the reference values.
The original right-hand side, denoted $\vec{b}^{(0)}$, corresponds to
the initial differential parameter vector $\vec{x}^{(0)}=\vec{0}$. 
If $\vec{x}$ (without a superscript) denotes the exact solution of 
Eq.~(\ref{eq:norm1}), the initial error vector is 
$\vec{e}^{(0)}=\vec{x}^{(0)}-\vec{x}=-\vec{x}$. Although this
vector is of course not known, we do know
$-\vec{N}\vec{e}^{(0)}=\vec{N}\vec{x}=\vec{b}^{(0)}$.
Solving the preconditioner system (\ref{eq:precond}) gives
$\vec{d}^{(0)}=\vec{K}^{-1}\vec{b}^{(0)}$ and the updated parameter vector
$\vec{x}^{(1)}=\vec{x}^{(0)}+\vec{d}^{(0)}$. The new error vector 
$\vec{e}^{(1)}=\vec{x}^{(1)}-\vec{x}$ is again not known, but
$-\vec{N}\vec{e}^{(1)}=\vec{b}^{(0)}-\vec{N}\vec{d}^{(0)}=\vec{b}^{(1)}$
is obtained by inserting the updated parameters in the right-hand side
of Eq.~(\ref{eq:norm}), by the same argument as in Eq.~(\ref{eq:precond2}).
Generalizing, we have the so-called \emph{simple iteration scheme}
\begin{equation}\label{eq:iter}
\left.
\begin{aligned}
\vec{b}^{(k)} &= -\sum_l \frac{\partial R_l}{\partial\vec{x}}
R_l(\vec{x}^{(k)})W_l\quad \\
\vec{d}^{(k)} &= \vec{K}^{-1}\vec{b}^{(k)}\\
\vec{x}^{(k+1)} &= \vec{x}^{(k)}+\vec{d}^{(k)}\\
\end{aligned}
\right\} \quad k=0,\,1,\,\dots
\end{equation}
For the successive right-hand sides we find by recursion
\begin{equation}\label{eq:iterb}
\vec{b}^{(k+1)} = \vec{b}^{(k)} - \vec{N}\vec{d}^{(k)}
= (\vec{I}-\vec{N}\vec{K}^{-1})\vec{b}^{(k)}
= \vec{\tilde{B}}^{k+1}\vec{b}^{(0)} \, ,
\end{equation}
where
\begin{equation}\label{eq:iterbmat}
\vec{\tilde{B}} = \vec{I}-\vec{N}\vec{K}^{-1} \, .
\end{equation}
For the successive updates and errors we find
\begin{align}
\vec{d}^{(k+1)} &= \vec{K}^{-1}\vec{b}^{(k+1)}
= (\vec{I}-\vec{K}^{-1}\vec{N})\vec{d}^{(k)}
= \vec{B}^{k+1}\vec{d}^{(0)} \, ,\label{eq:iterx} \\
\vec{e}^{(k+1)} &= \vec{e}^{(k)}+\vec{d}^{(k)}
= (\vec{I}-\vec{K}^{-1}\vec{N})\vec{e}^{(k)}
= \vec{B}^{k+1}\vec{e}^{(0)} \, . \label{eq:itere}
\end{align}
where
\begin{equation}\label{eq:itermat}
\vec{B}=\vec{I}-\vec{K}^{-1}\vec{N}
\end{equation}
is called the iteration matrix \citep{axelsson96}.
Equations~(\ref{eq:iterb})--(\ref{eq:itermat}) are of great theoretical
interest, as explained below, although none of them is used
in the actual computations.

The convergence behaviour of the simple iteration scheme can
largely be understood by means of these relations and especially in
terms of the properties of the iteration matrix $\vec{B}$ governing
the sequence of updates $\vec{d}$ and errors $\vec{e}$, and the 
adjunct matrix $\vec{\tilde{B}}$ governing
the sequence of right-hand sides. It is well known
\citep[e.g.,][]{axelsson96} that the simple iteration scheme in
Eq.~(\ref{eq:iter}) converges (that is $\vec{x}^{(k)}\rightarrow\vec{x}$)
for arbitrary starting approximation if and only if $\rho(\vec{B})<1$.
Here, $\rho(\vec{B})$ is the spectral radius of $\vec{B}$, i.e., the
largest absolute value of its eigenvalues. Under this condition we
have $\vec{d}^{(k)}\rightarrow\vec{0}$ and
$\vec{e}^{(k)}\rightarrow\vec{0}$ for $k\rightarrow\infty$.
Also the right-hand side
$\vec{b}^{(k)}=\vec{K}\vec{d}^{(k)}\rightarrow\vec{0}$
under the same condition.%
\footnote{$\vec{B}$ and $\vec{\tilde{B}}=\vec{K}\vec{B}\vec{K}^{-1}$
have the same spectral radius; indeed, their eigenvalues are the same,
as can be seen from the characteristic polynomial
$\det(z\vec{I}-\vec{B})=\det(\vec{K})(z\vec{I}-\vec{B})\det(\vec{K}^{-1})
=\det(z\vec{I}-\vec{K}\vec{B}\vec{K}^{-1})$ being the same for the
two matrices (A.~Bombrun, private communication).}

As discussed in Sect.~\ref{sec:rank}, the normal matrix $\vec{N}$ is 
singular and its null space spanned by the $n\times 6$ matrix $\vec{V}$. 
Therefore,
\begin{equation}\label{eq:eigenV}
\vec{B}\vec{V}=\vec{V}-\vec{K}^{-1}\vec{N}\vec{V}=\vec{V}
\end{equation}
which shows that $\vec{B}$ has a six-fold eigenvalue equal to 1, with
the corresponding eigenvectors spanning $\mathcal{N}(\vec{N})$. The
spectral radius of $\vec{B}$ is therefore not less than 1, and the simple
iteration scheme does not converge for arbitrary initial errors.

This is not a real problem, for the following reason. First, let us note
that $\vec{B}$ is not a full-rank matrix. Indeed, a direct calculation of
Eq.~(\ref{eq:itermat}) using the expression for $\vec{K}$ in
Eq.~(\ref{eq:precond}) shows that the first $n_s$ columns of $\vec{B}$
are zero.%
\footnote{That the first $n_s$ columns in the iteration matrix are zero
means that the results of the next iteration are independent of the current
source parameters $\vec{x}_s^{(k)}$. This may seem surprising at first, but is
a simple consequence of the fact that each iteration starts with the source
update block (S). In this block, the updated source parameters depend on the
previous attitude, calibration and global parameters, but not on the
previous source parameters.\label{footnote10}}
Thus (at least) $n_s$ of its eigenvalues are identically zero.
A corresponding set of orthogonal unit vectors is given by the columns
of the $n\times n_s$ matrix
\begin{equation}\label{eq:eigenZ}
\vec{Z}= \left[\begin{array}{c} \vec{I} \\[3pt] \vec{\varnothing} \end{array}\right]
\quad\begin{array}{cl} \} & n_s~\text{rows}\\[3pt] \} & n-n_s~\text{rows} \end{array} \, ,
\end{equation}
so that $\vec{B}\vec{Z}=\vec{0}$. We assume that the remaining $n-n_s-6$
eigenvalues satisfy $0<|\lambda|<1$. Thus, if $\vec{\Lambda}$ is the diagonal
matrix containing these eigenvalues and $\vec{U}$ a matrix of size
$n\times(n-n_s-6)$ whose columns are made up of the corresponding eigenvectors,
we have $\vec{B}\vec{U}=\vec{U}\vec{\Lambda}$. The columns of $\vec{Z}$,
$\vec{U}$ and $\vec{V}$ together span $\mathbb{R}^n$, and it is therefore
possible to decompose the solution vector as
\begin{equation}\label{eq:xdecomp}
\vec{x} = \vec{Z}\vec{x}_s+\vec{U}\vec{y}+\vec{V}\vec{z} \,
\end{equation}
where $\vec{x}_s\in\mathbb{R}^{n_s}$ is the `source' part of $\vec{x}$,
$\vec{y}\in\mathbb{R}^{n-n_s-6}$ and $\vec{z}\in\mathbb{R}^{6}$.
Since $-\vec{e}^{(0)}=\vec{x}$ we find
\begin{equation}\label{eq:xdecomp1}
-\vec{e}^{(1)}
= -\vec{B}\vec{e}^{(0)}
= \vec{B}\vec{Z}\vec{x}_s+\vec{B}\vec{U}\vec{y}+\vec{B}\vec{V}\vec{z}
= \vec{U}\vec{\Lambda}\vec{y}+\vec{V}\vec{z} \, ,
\end{equation}
and by recursion
\begin{equation}\label{eq:ekdecomp}
-\vec{e}^{(k)} = \vec{U}\vec{\Lambda}^k\vec{y}+\vec{V}\vec{z} \, .
\end{equation}
The first term clearly vanishes for $k\rightarrow\infty$ if
$\rho(\vec{\Lambda})<1$, as we have assumed. The second term,
which is the projection of $\vec{x}$ on the null space of $\vec{N}$,
remains unchanged by the iterations. The update vector can be written
\begin{equation}\label{eq:deltakdecomp}
\vec{d}^{(k)} = \vec{e}^{(k+1)} - \vec{e}^{(k)} =
\vec{U}\vec{\Lambda}^k(\vec{I}-\vec{\Lambda})\vec{y} \, ,
\end{equation}
which vanishes under the same condition. The same is then true
for the right-hand side $\vec{b}^{(k)}=\vec{K}\vec{d}^{(k)}$.
Effectively, this means that we can ignore the singularity of $\vec{N}$
in the simple iteration scheme; it will converge to some valid solution
of the (singular) normal equations, and the convergence process can be
monitored by means of the vectors $\vec{d}^{(k)}$ and $\vec{b}^{(k)}$.
After convergence, the required null-space components can be found
and added by means of the frame rotator.

To summarize, we have shown that the simple iteration scheme converges
in the desired way, provided that the spectral radius of $\vec{B}$, not
counting the eigenvalues related to the null space of $\vec{N}$, is less
than 1. It is well known \citep[e.g.,][]{golu+96} that this condition is
satisfied for any symmetric positive definite $\vec{N}$, using the 
Gauss--Seidel preconditioner. However, the spectral radius may be very 
close to 1, implying very slow convergence. In the case of AGIS the 
situation is more complex, and convergence is in practice demonstrated 
through simulations (Sect.~\ref{sec:impl}), but the theoretical background 
outlined above is of great help when interpreting the results.

\subsection{Order of the block processes}\label{sec:order}

In the preceding sections we have assumed that the four blocks are
always executed in the sequence SACG, and that the updated parameters
resulting from each block is used in the subsequent blocks. The
particular preconditioner $\vec{K}$ in Eq.~(\ref{eq:precond})
incorporates these assumptions. We will now discuss variants of this
scheme and show that they can be mathematically represented by
different preconditioners.

Each AGIS iteration always starts with the source update block (S).
The main reason for starting with S rather than A, for example, is
that the observations can be inspected and analysed one source at a
time in order to estimate the quality of the data, identify possible
outliers, and check whether the source is suitable as a primary source.
The identification of outliers, in particular, requires several
`inner' iterations of the source observation equations
(Sect.~\ref{sec:source-inner}), which can
be done with relative ease because of the small number of data
points involved.

The order of the remaining blocks ACG adopted in the preceding
sections is more or less arbitrary. It is easy to see through an
analogy with Eqs.~(\ref{eq:precond})--(\ref{eq:precond1}) that
the sequence SCAG (for example) is mathematically represented by
the alternative preconditioner
\begin{equation}\label{eq:precondSCAG}
\vec{K} =
    \begin{bmatrix}
        \vec{N}_{ss} & \vec{\varnothing} & \vec{\varnothing} & \vec{\varnothing} \\
        \vec{N}_{as} & \vec{N}_{aa}      & \vec{N}_{ac}      & \vec{\varnothing} \\
        \vec{N}_{cs} & \vec{\varnothing} & \vec{N}_{cc}      & \vec{\varnothing} \\
        \vec{N}_{gs} & \vec{N}_{ga}      & \vec{N}_{gc}      & \vec{N}_{gg}
    \end{bmatrix} \, ,
\end{equation}
and that other permutations of the sequence are similarly obtained by
transposing the corresponding off-diagonal blocks. Changing the
preconditioner means changing the iteration matrix (\ref{eq:itermat})
and therefore possibly also its eigenvalues, which in turn govern the rate of
convergence.

From a mathematical viewpoint, the choice of the starting block
(S in our case) determines the initial update $\vec{d}^{(0)}$
(cf.\ footnote~\ref{footnote10}) and therefore influences all subsequent updates.
However, this particular choice is not expected to have much influence
on the convergence rate after a number of iterations, since the S block
has no special status in the periodically repeated sequence
$\dots$SACGSACGSACGSA$\dots$. Thus we may surmise that the different
iteration matrices for the cyclically permuted sequences SACG, ACGS,
CGSA and GSAC do in fact have the same set of eigenvalues. For symmetry
reasons it can also be surmised that the reversed sequences GCAS, SGCA,
ASGC and CASG have the same eigenvalues as the original sequences. Thus,
there are in fact only three sets of sequences with possibly distinct
convergence behaviour, represented by SACG (or SGCA), SAGC (or SCGA),
and SCAG (or SGAC) if we take S to be the starting block. There is no
obvious way of knowing a~priori which of the three possibilities is
to be preferred, or even if they are significantly different.

Apart from the various permutations discussed above, there is another
way to modify the AGIS iteration scheme, which can also be described
in terms of the preconditioner. This modification is related to the
practical organization of the flow of data to the different block
processes. The current implementation of the simple iteration scheme
differs somewhat from the description in Sect.~\ref{sec:iter}, and the
block updates are in fact practically organized as follows:
\begin{enumerate}
\item
Initialize the iteration counter, $k=0$
\item
Choose starting values for all unknowns: $\vec{s}^{(0)}$, $\vec{a}^{(0)}$,
$\vec{c}^{(0)}$, $\vec{g}^{(0)}$.
\item
Estimate $\vec{s}^{(k+1)}$ using $\vec{a}^{(k)}$, $\vec{c}^{(k)}$, $\vec{g}^{(k)}$.
\item
Estimate $\vec{a}^{(k+1)}$ using $\vec{s}^{(k+1)}$, $\vec{c}^{(k)}$, $\vec{g}^{(k)}$.
\item
Estimate $\vec{c}^{(k+1)}$ using $\vec{s}^{(k+1)}$, $\vec{a}^{(k)}$, $\vec{g}^{(k)}$.
\item
Estimate $\vec{g}^{(k+1)}$ using $\vec{s}^{(k+1)}$, $\vec{a}^{(k)}$, $\vec{c}^{(k)}$.
\item
Increment $k$ and go to Step~3.
\end{enumerate}
The crucial difference compared with Eq.~(\ref{eq:precond1}) is that the C block
in Step~5 does not use the updated attitude but the old one, and that the G block
in Step~6 similarly uses the `old' attitude and calibration parameters. This has
the practical advantage that the A, C and G blocks can be carried out in parallel,
with big savings in terms of the amounts of data that have to be exchanged between
the processes (see Sect.~\ref{sec:impl}). Schematically, this can be represented
as S[ACG], where the blocks in brackets are (or can be) executed in parallel
(and so the order of the bracketed blocks does not matter). The corresponding
preconditioner is
\begin{equation}\label{eq:precondS(CAG)}
\vec{K} =
    \begin{bmatrix}
        \vec{N}_{ss} & \vec{\varnothing} & \vec{\varnothing} & \vec{\varnothing} \\
        \vec{N}_{as} & \vec{N}_{aa}      & \vec{\varnothing} & \vec{\varnothing} \\
        \vec{N}_{cs} & \vec{\varnothing} & \vec{N}_{cc}      & \vec{\varnothing} \\
        \vec{N}_{gs} & \vec{\varnothing} & \vec{\varnothing} & \vec{N}_{gg}
    \end{bmatrix} \, .
\end{equation}
Yet other variants may be considered, for example S[AC]G, where Step~6 uses the
updated attitude and calibration parameters, with preconditioner
\begin{equation}\label{eq:precondS(CA)G}
\vec{K} =
    \begin{bmatrix}
        \vec{N}_{ss} & \vec{\varnothing} & \vec{\varnothing} & \vec{\varnothing} \\
        \vec{N}_{as} & \vec{N}_{aa}      & \vec{\varnothing} & \vec{\varnothing} \\
        \vec{N}_{cs} & \vec{\varnothing} & \vec{N}_{cc}      & \vec{\varnothing} \\
        \vec{N}_{gs} & \vec{N}_{ga}      & \vec{N}_{gc}      & \vec{N}_{gg}
    \end{bmatrix} \, ,
\end{equation}
and [SACG], for which
\begin{equation}\label{eq:precond(SCAG)}
\vec{K} =
    \begin{bmatrix}
        \vec{N}_{ss}      & \vec{\varnothing} & \vec{\varnothing} & \vec{\varnothing} \\
        \vec{\varnothing} & \vec{N}_{aa}      & \vec{\varnothing} & \vec{\varnothing} \\
        \vec{\varnothing} & \vec{\varnothing} & \vec{N}_{cc}      & \vec{\varnothing} \\
        \vec{\varnothing} & \vec{\varnothing} & \vec{\varnothing} & \vec{N}_{gg}
    \end{bmatrix} \, .
\end{equation}
This last case is known as the (block) Jacobi method, while the use of a full triangular
preconditioner as in Eq.~(\ref{eq:precond}) is known as the (block) Gauss--Seidel method
\citep{axelsson96}. As we have seen, the currently implemented simple iteration scheme
is intermediate between the Jacobi and Gauss--Seidel methods.

Intuitively, we expect the Gauss--Seidel method to converge more quickly than the Jacobi
or any intermediate method, simply because each block then uses the most recent (and,
presumably, best) estimates of the parameters. However, our practical experience with
AGIS shows that the `difficult' part of the problem is to disentangle the source and
attitude parameters. For example, the calibration parameters are generally found to
converge much faster than the source and attitude parameters. Thus it does not seem to
matter much if Step~5 above uses $\vec{a}^{(k+1)}$ or $\vec{a}^{(k)}$, i.e., whether
the sub-matrix $\vec{N}_{ca}$ is included or not in the preconditioner. A similar
argument can be made concerning the G block, provided some measures are taken to
decorrelate the global parameters from the source parameters 
(Sect.~\ref{sec:globalupdate}). Thus, the various intermediate methods are 
probably nearly as good as the Gauss--Seidel
method, in terms of the convergence rate, and the precise scheme may then rather be
determined by practical considerations. With the present data processing architecture,
the favoured scheme is S[ACG] as described above.

\subsection{Accelerated iteration, conjugate gradients and the hybrid iteration scheme}\label{sec:CG}

The `simple iteration' (SI) scheme described above was the starting point for a long
development towards a fully functional scheme with much improved convergence
properties. The main stages in this development were the `accelerated simple
iteration' (ASI), the conjugate gradients (CG), and finally the fully flexible `hybrid 
scheme' (A/SI-CG) to be used in the final implementation of AGIS. As much of this
development has at most historical interest, only a brief outline is given here.

Already in the very early implementation of the simple iteration scheme it was observed that
convergence was slower than (naively) expected, and that after some iterations,
the updates always seemed to go in the same direction, forming a geometrically
(exponentially) decreasing series. With the hindsight of the analysis in  
Sect.~\ref{sec:iter} this behaviour is very easily understood: the persistent pattern 
of updates is roughly proportional to the eigenvector of the largest eigenvalue
of the iteration matrix, and the (nearly constant) ratio of the sizes of successive
updates is the corresponding eigenvalue. From this realization it was natural to
test an acceleration method based on a Richardson-type extrapolation of the
updates. The idea is simply that if the updates in two successive iterations are
roughly proportional to each other, $\vec{d}^{(k+1)}\simeq\lambda\vec{d}^{(k)}$,
with $|\lambda| < 1$, then we can infer that the next update is again a factor
$\lambda$ smaller than $\vec{d}^{(k+1)}$, and so on. The sum of all the
updates after iteration $k$ can therefore be estimated as $\vec{d}^{(k+1)}+
\lambda\vec{d}^{(k+1)}+\lambda^2\vec{d}^{(k+1)}+\dots=
(1-\lambda)^{-1}\vec{d}^{(k+1)}$. Thus, in iteration $k+1$ we apply an
acceleration factor $1/(1-\lambda)$ based on the current estimate of the ratio
$\lambda$. This accelerated simple iteration (ASI) scheme is seen 
to be a variant of the well-known successive overrelaxation method \citep{axelsson96}.
The factor $\lambda$ is estimated by statistical analysis of the parallax updates 
for a small fraction of the sources; the parallax updates are used for this analysis, 
since they are unaffected by a possible change in the frame orientation between 
successive iterations. With this simple device, the number of iterations for full 
convergence was reduced roughly by a factor 2.

Both the simple iteration and the accelerated simple iteration belongs to a much
more general class of solution methods known as Krylov subspace approximations.
The sequence of updates $\vec{d}^{(k)}$, $k=0\dots K-1$ generated by the first
$K$ simple iterations constitute the basis for the $K$-dimensional subspace of the
solution space, known as the Krylov subspace for the given matrix and
right-hand side \citep[e.g.,][]{book:greenbaum-1997,vanVorst:03}. Krylov methods
compute approximations that, in the $k$th iteration, belongs to the $k$-dimensional
Krylov subspace. But whereas the simple and accelerated iteration schemes, in the
$k$th iteration, use updates that are just proportional to the $k$th basis vector, 
more efficient algorithms generate approximations that are (in some sense) optimal
linear combinations of all $k$ basis vectors. Conjugate gradients (CG) is one of the 
best-known such methods, and possibly the most efficient one for general
symmetric positive-definite matrices.
\citep[e.g.,][]{axelsson96,book:bjork-1996,vanVorst:03}. Its
implementation within the AGIS framework is more complicated, but 
has been considered in detail by \citet{bombrun+10}. As it 
provides significant advantages over the SI and ASI schemes in terms of 
convergence speed, this algorithm has been 
chosen as the baseline method for the astrometric core solution of Gaia
(see below however). From practical experience, we have found that CG
is roughly a factor 2 faster than ASI, or a factor
4 faster than the SI scheme. Like SI, the CG algorithm uses a preconditioner 
and can be formulated in terms of the S, A, C and G blocks, so the subsequent 
description of these blocks remains valid. In the terminology of \citet{bombrun+10}
the process of solving the preconditioner system $\vec{K}\vec{d}=\vec{b}$
is the \emph{kernel} operation common to all these solution methods, which 
only differ in how the updates are applied according to the various 
\emph{iteration schemes}. 

The CG algorithm assumes that the normal matrix
is constant in the course of the iterations. This is not strictly true if
the observation weights are allowed to change as functions of the
residuals, as will be required for efficient outlier elimination 
(Sect.~\ref{sec:source-inner}). Using the CG algorithm together with
the weight-adjustment scheme described below could therefore lead
to instabilities, i.e., a reduced convergence rate or even non-convergence.
On the other hand, the SI scheme is extremely stable with respect to
all such modifications in the course of the iterations, as can be expected 
from the interpretation of the SI scheme as the successive and independent
application of the different solution blocks. The finally adopted algorithm
is therefore a \emph{hybrid scheme} combining SI (or ASI) and CG, where 
SI is used initially, until the weights have settled, after which CG is turned on. 
A temporary switch back to SI, with an optional re-adjustment of the 
weights, may be employed after a certain number of CG iterations; this
could avoid some problems due to the accumulation of numerical rounding 
errors in CG.

\section{Updating processes} \label{sec:blocks}

In this section we describe in some detail each of the updating blocks
S, A, C and G that form the basis (or kernel process) for the AGIS iteration loop.

\subsection{Source updating (S)} \label{sec:sourceupdate}

\subsubsection{The normal equations}\label{sec:source-normals}

The astrometric model for the sources is given in Sect.~\ref{sec:astromodel}.
In the source update block (S) the source parameters $\vec{s}$ are improved
by solving the first line in Eq.~(\ref{eq:precond1}). According to
Eqs.~(\ref{eq:norm}) and (\ref{eq:Nss}) this can be done for one source ($i$)
at a time by solving the following system of equations for the update
$\vec{d}_i$ of the five astrometric parameters in $\vec{s}_i$:
\begin{equation} \label{eq:normS}
\left[ \sum_{l\,\in\,i} \frac{\partial R_l}{\partial\vec{s}_i}
\frac{\partial R_l}{\partial\vec{s}'_{\!i}}W_l \right]\,\vec{d}_i
= -\sum_{l\,\in\,i} \frac{\partial R_l}{\partial\vec{s}_i}
R_l(\vec{s}_i)W_l \, .
\end{equation}
Here $W_l=w_l/(\sigma_l^2+\epsilon_l^2)$, with $\sigma_l$ denoting the given
formal standard uncertainty of observation $l$, expressed as an angle. $w_l$ and
$\epsilon_l$ are, respectively, the downweighting factor and excess noise
introduced in Sect.~\ref{sec:synthesismodel}. In Eq.~(\ref{eq:normS}) the 
dependence of $R_l$ on $\vec{a}$, $\vec{c}$ and $\vec{g}$ has been suppressed, 
since the system is solved with these parameters fixed.

In matrix notation, the normal equations (\ref{eq:normS}) can be written
\begin{equation} \label{eq:normSmat}
\vec{A}_i' \vec{W}_i \vec{A}_i\,\vec{d}_i = \vec{A}_i' \vec{W}_i \vec{h}_i \, ,
\end{equation}
where
\begin{equation} \label{eq:designS}
\vec{A}_i=\left[ -\frac{\partial R_l}{\partial\vec{s}'_{\!i}} \right]_{l\in i}
\end{equation}
is an $n_i\times 5$ matrix with $n_i=\sum_{l\in i}1$ the number
of observations of source $i$ (typically $n_i\sim 800$),
\begin{equation} \label{eq:designSrhs}
\vec{h}_i=\left[ R_l(\vec{s}_i) \right]_{l\in i}
\end{equation}
is a column matrix of length $n_i$, and $\vec{W}_i$ is a diagonal matrix
containing the weight factors $W_l$. Although the residuals $R_l$
are in principle non-linear functions of $\vec{s}_i$, this non-linearity can
be neglected if the parameters are close enough to the final solution, 
i.e., if the resulting update $\vec{d}_i$ is small enough. The
partial derivatives in $\vec{A}_i$ can then be regarded as fixed throughout
the updating process, and $\vec{A}_i$ and $\vec{h}_i$ can immediately 
be computed when entering the source updating. The weight matrix 
$\vec{W}_i$, on the other hand, depends on $w_l$ and $\epsilon_l$, 
which are modified as part of the process as explained below. 

For given $\vec{W}_i$ the system of equations~(\ref{eq:normSmat}) is solved 
using the Cholesky algorithm (Appendix~\ref{sec:cholesky}). At the end of
the source updating, the full ($5\times 5$) inverse matrix is computed, 
providing a first-order estimate of the covariance of the astrometric parameters 
in $\vec{s}_i$ (see Sect.~\ref{sec:cov}).

We note that Eq.~(\ref{eq:normSmat}) corresponds to the overdetermined 
system of observation equations
\begin{equation} \label{eq:obseqS}
\vec{A}_i\,\vec{d}_i \simeq \vec{h}_i \quad\text{(weight~matrix~$\vec{W}_i$)} \, .
\end{equation}
After solution, the updated residuals are contained in the vector
$\vec{R}_i=\vec{h}_i-\vec{A}_i\vec{d}_i$, and the contribution of the source
to the objective function $Q$ is given by
\begin{equation} \label{eq:Qi}
Q_i = \vec{R}_i' \vec{W}_i \vec{R}_i \, .
\end{equation}

\subsubsection{The inner iteration: identifying outliers and estimating the 
excess source noise}\label{sec:source-inner}

Since the weight matrix $\vec{W}_i$ depends on the downweighting factors $w_l$ 
and the excess noises $\epsilon_l$, which in turn depend on the updated residuals
$\vec{R}_i$, the normal equations (\ref{eq:normSmat}) are non-linear and must be
solved by iteration. We refer to this as the \emph{inner iteration} of the source update, 
to distinguish it from the AGIS iteration discussed in Sect.~\ref{sec:iter}. 

Using Eq.~(\ref{eq:excess}) we can write
\begin{equation} \label{eq:W}
W_l = \frac{w_l}{\sigma_l^2 + \epsilon^2_a(t_l) + \epsilon^2_i} =
\frac{w_l}{\tilde{\sigma}_l^2 + \epsilon^2_i} \, ,
\end{equation}
where $\tilde{\sigma}_l=\left[\sigma_l^2+\epsilon_a^2(t_l)\right]^{1/2}$
is the formal standard uncertainty of the observation adjusted for the excess
attitude noise, which, when entering the source update,  is assumed to 
be known from a previous attitude update (Sect.~\ref{sec:attNoise}). In the very first source update the excess attitude noise must be set
to the mean errors of the pre-AGIS attitude parameter estimates which are
used for starting up the iterations. 
The downweighting factors depend on the normalized residuals, i.e.,
\begin{equation} \label{eq:wl}
w_l = w\left(\frac{R_l}{\sqrt{\tilde{\sigma}_l^2 + \epsilon^2_i}}\right) \, ,
\end{equation}
where the function $w(z)$ is such that $w(z)=1$ for $|z|\lesssim 3$ and 
gradually decreasing to 0 for larger $|z|$ (see Eq.~\ref{eq:dw}). 

The inner iteration actually consists of two nested procedures: an outer 
one which determines the downweighting factors $w_l$, and an inner one 
which determines the excess source noise $\epsilon_i$ for a fixed set of 
$w_l$. We begin by considering the inner procedure.

For fixed downweighting
factors, the source update aims to minimize $Q_i$ in Eq.~(\ref{eq:Qi})
with respect to the unknown $\vec{d}_i$ and $\epsilon_i$. But it is 
immediately seen that $Q_i$ can be made arbitrarily small just by making 
$\epsilon_i$ large enough. Consequently, we cannot use unconstrained
minimization to solve this problem. The necessary constraint is provided 
by the condition that $Q_i$, under the assumption that the excess noises 
have been correctly estimated and the outliers properly removed, should 
follow the chi-square distribution with $\nu = n_i - n_\text{out} - 5$ 
degrees of freedom. Here $n_i$ is the number of observations of source
$i$, $n_\text{out}$ the number of outliers and 5
the number of astrometric parameters estimated. In particular, the expected 
value is $\mbox{E}(Q_i)=\nu$. The number of outliers $n_\text{out}$ is estimated
by counting the number of observations with $w_l < 0.2$. This limit was
empirically found to give a reasonable estimate of the actual number of
outliers in a variety of simulated cases. 

For given $\epsilon_i$ the weight matrix is now known, Eq.~(\ref{eq:normSmat}) 
can be trivially solved and the residuals $\vec{R}_i$ computed. We thus define 
the function $Q(\epsilon_i^2)=\vec{R}_i' \vec{W}_i \vec{R}_i$. The excess source 
variance is then taken to be
\begin{equation} \label{eq:epsi}
\epsilon_i^2 = \begin{cases} ~0 & \text{if $Q(0)\le\nu$}, \\
~\text{solution of $Q(\epsilon_i^2)=\nu$} & \text{otherwise}. \end{cases}
\end{equation}
In the second case, the non-linear equation $Q(y)=\nu$ is iteratively solved 
by a series of improvements $\Delta y=(1-Q(y)/\nu)Q(y)/Q'(y)$, starting from 
$y=0$. Typically, 2--3 iterations are sufficient.%
\footnote{This iteration formula can be derived by matching the rational 
approximation $Q(y)\simeq a/(b+y)$ to the value and derivative of $Q(y)$
at the current point $y$. Compared with the standard Newton--Raphson 
method, which uses a linear approximation around the current $y$, the
present formula converges much quicker due to the more reasonable 
behaviour of the rational approximation especially for large $y$.}

This procedure returns a positive $\epsilon_i$ as soon as $Q(0)>\nu$.
If the resulting $\epsilon_i$ is much smaller than the typical $\sigma_l$
of the source, it is probably not significant. The significance of the 
$\epsilon_i$ can more easily be judged from an auxiliary statistic that
can be computed almost for free. Under the null hypothesis ($\epsilon_i=0$)
we know that $Q(0) \sim \chi^2_\nu$, so the expected value is $\nu$
and the variance $2\nu$. Thus we may take
\begin{equation} \label{eq:D}
D = \frac{Q(0)-\nu}{\sqrt{2\nu}}
\end{equation}
as a measure of the significance of the estimated $\epsilon_i$, with 
$D > 2$ indicating a probably significant value.

Having determined $\epsilon_i$, and a
corresponding set of residuals $\vec{R}_i$, the downweighting factors
can immediately be computed from Eq.~(\ref{eq:wl}). However, having
changed the downweighting factors (and possibly $\nu$) it is now
necessary to repeat the estimation of $\epsilon_i$ and $\vec{d}_i$
with the new set $w_l$. Typically, four such iterations are found to
be sufficient. The only remaining problem is how to start the iterations,
that is the initial selection of the downweights $w_l$. It is not possible to
start by assuming $w_l=1$ for all the observations, since we must take
into account that some small fraction of the data could be utterly wrong.
Such gross outliers, if not removed already from the start, would severely
slow down or even prevent the convergence of the inner iterations. The
adopted solution is to make an extremely robust estimation of the
standard deviation of the initial residuals (contained in $\vec{h}_i$),
from which the initial downweightings are obtained. This robust standard
deviation $\varsigma_i$ is calculated as half the intersextile range of
the elements in $\vec{h}_i$, whereupon the initial $w_l=w(h_l/\varsigma_i)$.

After convergence of the inner iteration, the statistical weight of the 
source $W_i$ is computed according to Eq.~(\ref{eq:Wi}). This quantity,
together with $\epsilon_i$ and the magnitude, are the most important 
indicators for the selection of primary sources (Sect.~\ref{sec:relegate}).

The weight function $w(z)$ currently used is the following:
\begin{equation}\label{eq:dw}
  w(z) =
  \begin{cases}
     1 & \text{if $|z|\le 2$} \\
     1 - 1.773735t^2 + 1.141615t^3 & \text{if $2\leq |z|<3$} \\
     \exp(-|z|/3) & \text{otherwise,} \\
  \end{cases}
\end{equation}
where $t=|z|-2$ and the numerical constants have been chosen to make a smooth
transition at $|z|=3$. The exponential decay for $|z|>3$ provides a dramatic
weight reduction for large residuals; e.g., at 10~sigmas we have
$w(10)\simeq 0.036$, while at 100~sigmas we have $w(100)\simeq 3\times 10^{-15}$.

Among the many different weight functions proposed in the literature, the
so-called Huber estimator \citep{huber:1981} using
\begin{equation}\label{eq:dwH}
  w_\text{H}(z) =
  \begin{cases}
     1 & \text{if $|z|\le c$} \\
     c(2|z|-c)z^{-2} & \text{otherwise} \\
  \end{cases}
\end{equation}
has been considered as an alternative to Eq.~(\ref{eq:dw}), e.g., with $c=2$.
This gives a much slower weight decay for large residuals, e.g.,
$w_\text{H}(10)=0.36$ and $w_\text{H}(100)=0.0396$. Future experiments
may decide which weight function will finally be used for AGIS.

\subsubsection{Calculation of partial derivatives}\label{sec:source-pd}

The calculation of the partial derivatives in Eq.~(\ref{eq:designS}) is
done as follows. From Eqs.~(\ref{eq:resAL}) and (\ref{eq:resAC}) we
have, by means of (\ref{eq:srspropeta}),
\begin{equation}\label{eq:dfads}
-\frac{\partial R_l^\text{AL}}{\partial\vec{s}'_{\!i}}
= \frac{\partial\varphi_l}{\partial\vec{s}'_{\!i}}\, , \quad\quad
-\frac{\partial R_l^\text{AC}}{\partial\vec{s}'_{\!i}}
= \frac{\partial\zeta_l}{\partial\vec{s}'_{\!i}}\, .
\end{equation}
In analogy with Eq.~(\ref{eq:normal-triad}) we introduce the auxiliary 
vectors $\vec{m}_l=\langle\vec{z}\times\vec{u}_l\rangle$ and
$\vec{n}_l=\vec{u}_l\times\vec{m}_l$, which together with 
$\vec{u}_l$ form the normal triad $[\vec{m}_l~\vec{n}_l~\vec{u}_l]$
with respect to the SRS; its components in the SRS are given by the 
columns of the matrix
\begin{equation}\label{eq:srspqr}
\tens{S}'[\vec{m}_l~\vec{n}_l~\vec{u}_l] = \begin{bmatrix} 
-\sin\varphi_l &~  -\sin\zeta_l\cos\varphi_l  &~ \cos\zeta_l\cos\varphi_l \\ 
\phantom{-}\cos\varphi_l &~  -\sin\zeta_l\sin\varphi_l &~  \cos\zeta_l\sin\varphi_l \\ 
0 &~  \cos\zeta_l &~ \sin\zeta_l \end{bmatrix}.
\end{equation}
By differentiation of the last column we obtain
\begin{equation}\label{eq:srsprop1diff}
\frac{\partial\vec{u}_l}{\partial\vec{s}'_{\!i}} =
\vec{m}_l\,\frac{\partial\varphi_l}{\partial\vec{s}'_{\!i}}\cos\zeta_l +
\vec{n}_l\,\frac{\partial\zeta_l}{\partial\vec{s}'_{\!i}} \, ,
\end{equation}
and thus
\begin{equation}\label{eq:srsprop1diffsol}
-\frac{\partial R_l^\text{AL}}{\partial\vec{s}'_{\!i}} =
\vec{m}_l'\,\frac{\partial\vec{u}_l}{\partial\vec{s}'_{\!i}}\sec\zeta_l \, ,
\quad\quad
-\,\frac{\partial R_l^\text{AC}}{\partial\vec{s}'_{\!i}} =
\vec{n}_l'\,\frac{\partial\vec{u}_l}{\partial\vec{s}'_{\!i}} \, .
\end{equation}
These expressions can be evaluated in any coordinate system, but perhaps 
most conveniently in the SRS using $\tens{S}'\vec{m}_l$ and $\tens{S}'\vec{n}_l$
from Eq.~(\ref{eq:srspqr}), and 
\begin{equation}\label{eq:srspropdiff}
\left\{\frac{\partial(\tens{S}'\vec{u}_l)}{\partial\vec{s}'_{\!i}},\,0\right\}=
\quat{q}_l^{-1}\left\{\frac{\partial(\tens{C}'\vec{u}_l)}{\partial\vec{s}'_{\!i}},\,0\right\}\quat{q}_l
\end{equation}
from Eq.~(\ref{eq:srsprop}).
To compute the partial derivatives of $\tens{C}'\vec{u}_i$, we first obtain 
from Eq.~(\ref{eq:astro-standard}) the derivatives%
\footnote{As indicated by the asterisk in the first derivative in 
Eq.~(\ref{eq:ubardiff}), the differential in right ascension is a true arc, 
thus: $\partial\alpha_{*i}\equiv (\partial\alpha_i)\cos\delta_i$.
The corresponding update in right ascension, i.e., the first element of 
the $\vec{d}_i$ obtained by solving (\ref{eq:normSmat}), is therefore
$\Delta\alpha_i\cos\delta_i$.\label{footn:ast}}
of the coordinate direction as:
\begin{multline}\label{eq:ubardiff}
\frac{\partial\vec{\bar{u}}_l}{\partial\alpha_{*i}} = \vec{p}_i \, , \quad
\frac{\partial\vec{\bar{u}}_l}{\partial\delta_{i}} = \vec{q}_i \, , \quad
\frac{\partial\vec{\bar{u}}_l}{\partial\varpi_{i}} =
-(\vec{I}-\vec{r}_i\vec{r}'_i)\vec{b}_\text{G}(t_l)/A_\text{u} \, , \\
\frac{\partial\vec{\bar{u}}_l}{\partial\mu_{\alpha *i}} = 
\vec{p}_i\tau_l \, , \quad
\frac{\partial\vec{\bar{u}}_l}{\partial\mu_{\delta i}} = 
\vec{q}_i\tau_l \, , \\
\frac{\partial\vec{\bar{u}}_l}{\partial\mu_{r i}} = 
(\vec{I}-\vec{r}_i\vec{r}'_i)\vec{b}_\text{G}(t_l)\varpi_i\tau_l/A_\text{u}
- (\vec{p}_i\mu_{\alpha *i}+\vec{q}_i\mu_{\delta i})\tau_l^2 \, ,
\end{multline}
where $\tau_l=t_{\text{B}l}-t_\text{ep}$ for brevity, and we have used 
the normal triad in the celestial reference system, Eq.~(\ref{eq:normal-triad}).
Although usually only the first five derivatives are needed (see however Sect.~\ref{sec:cov}), 
the last equation gives, for completeness, the derivative with respect to the sixth 
astrometric parameter $\mu_{ri}$: the first term corresponds to the secular 
change in parallax and the second to the perspective acceleration.
 
The rigorous transformation from $\vec{\bar{u}}_l$ to $\vec{u}_l$ is quite
complex, but by far the largest difference ($\sim\! 10^{-4}$~rad) is due to
stellar aberration. By comparison, gravitational light deflection by the Sun
is $\sim\! 2\times 10^{-8}$~rad. While the rigorous transformation is
required to compute the vector $\vec{u}_l$ itself, some simplifications can
be accepted when computing the partial derivatives. Indeed, for this purpose
it is sufficient to consider the classical stellar aberration formula,
\begin{equation}\label{eq:classAberr}
\vec{u}_l \simeq \langle \vec{\bar{u}}_l + \vec{v}_\text{G}(t_l)c^{-1} \rangle \, ,
\end{equation}
accurate to first order in $v_\text{G}/c$, 
where $\vec{v}_\text{G}=\text{d}\vec{b}_\text{G}/\text{d}t$ is the barycentric 
coordinate velocity of Gaia and $c$ the speed of light. To a relative precision 
better than $10^{-6}$ we then have
\begin{equation}\label{eq:duds}
\frac{\partial\vec{u}_l}{\partial\vec{s}'_{\!i}} \simeq 
\left[\left(1-\frac{\vec{\bar{u}}'_l\vec{v}_\text{G}}{c}\right)\vec{I}
-\frac{\vec{\bar{u}}_l\vec{v}_\text{G}'}{c}\right]
\frac{\partial\vec{\bar{u}_l}}{\partial\vec{s}'_{\!i}} \, ,
\end{equation}
where $\vec{I}$ is the $3\times 3$ identity matrix.

\subsection{Attitude updating (A)} \label{sec:attitudeupdate}

\subsubsection{The normal equations}\label{sec:att-normals}

The attitude model using B-splines to represent the components of
the attitude quaternion as functions of time is described in
Sect.~\ref{sec:attmodel}. (For reference purposes, conventions for
notation and some important properties of splines and B-splines are
explained in Appendix \ref{sec:bsplines}.) In the attitude update
process (A) the attitude parameters $\vec{a}$ are improved by solving
the second sub-system in Eq.~(\ref{eq:precond1}). Recalling that $\vec{a}$
is divided into sub-vectors of length 4, representing the quaternions
$\quat{a}_n$ in Eq.~(\ref{eq:qspline}), the $n$th set of four equations
can be written, using Eqs.~(\ref{eq:norm}) and (\ref{eq:Naa}),
\begin{equation} \label{eq:normA}
\sum_{m=n-M+1}^{n+M-1}
\left[ \sum_{l\,\in\, L_n\cap L_m} \frac{\partial R_l}{\partial\quat{a}_n}
\frac{\partial R_l}{\partial\quat{a}'_{\!m}}W_l \right]\,\vec{d}_m
= -\sum_{l\,\in\, L_n} \frac{\partial R_l}{\partial\quat{a}_n}
R_l(\vec{s},\vec{a},\vec{c},\vec{g})W_l \, .
\end{equation}
Here $L_n$ stands for the set of observations occurring within the
support of $B_n(t)$, i.e., $L_n=\{l \, | \, \tau_n \le t_l < \tau_{n+M}\}$,
where $M$ is the order of the spline. On the right-hand side, it is
understood that the residuals $R_l$ are calculated for the most recent
source parameters $\vec{s}$ (i.e., from the preceding source updating),
while the attitude, calibration and global parameters are the
not-yet-updated ones, as explained in Sect.~\ref{sec:order}. The
weights $W_l$ are the ones calculated in the source updating.

\begin{figure}[t!]
\begin{center}
\includegraphics[width=0.95\columnwidth]{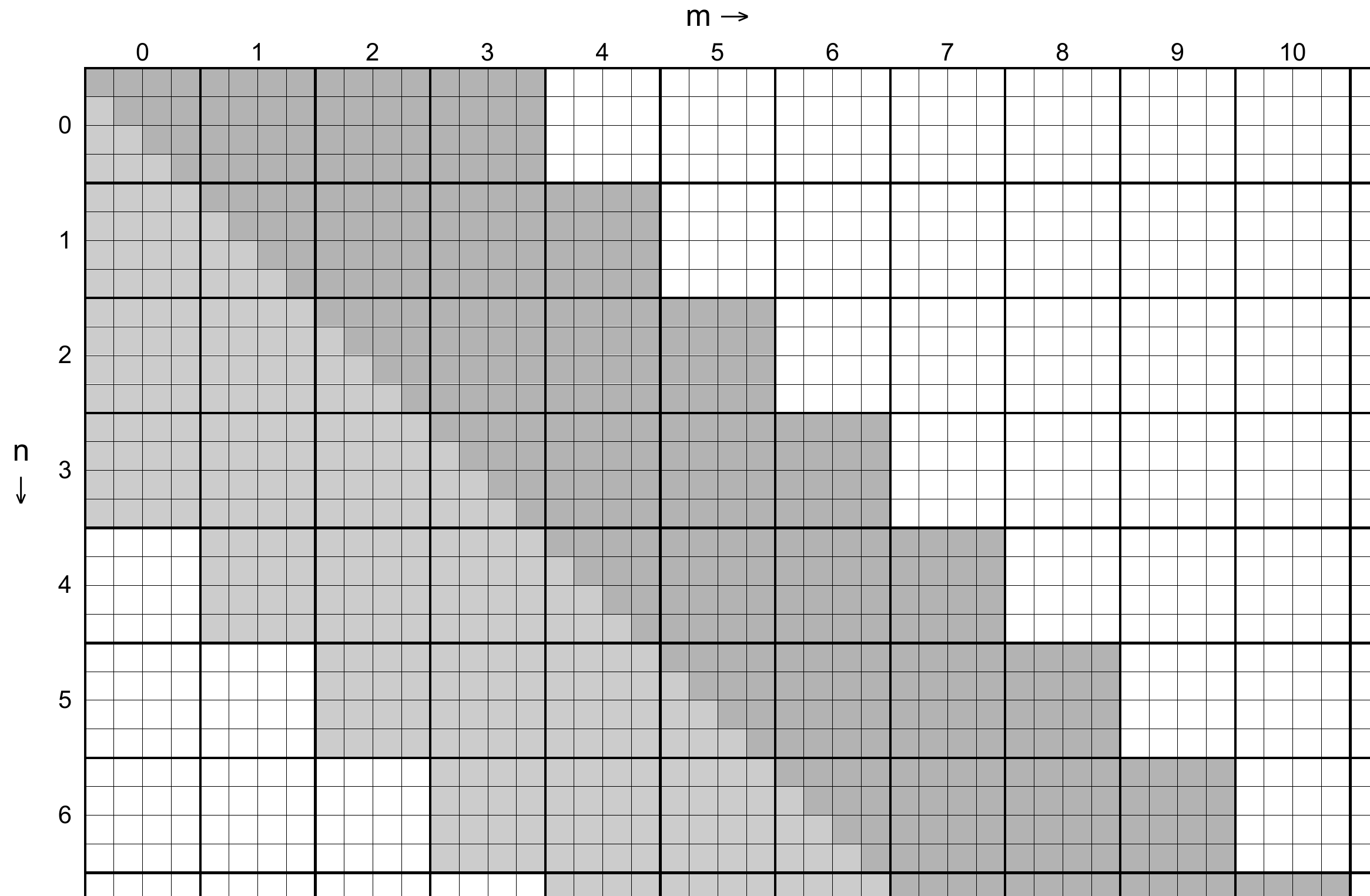}
\null\vspace*{3pt}
\caption{Structure of the attitude normal equations matrix $\vec{N}_{aa}$
for a cubic spline (order $M=4$). The blocks of size $4\times 4$ are indexed
by $(n,m)$ as in Eq.~(\ref{eq:Naa}). The grey cells represent non-zero
elements. Since the matrix is symmetric, the elements below the diagonal
(in lighter grey) need not be stored.}
\label{fig:Naa}
\end{center}
\end{figure}

The structure of the symmetric matrix $\vec{N}_{aa}$ is shown in
Fig.~\ref{fig:Naa}. If each quaternion component is represented by a
spline of order $M$ with $N$ degrees of freedom (so $n=0\dots N-1$),
the total number of unknowns is $4N$ and the average bandwidth of
$\vec{N}_{aa}$ (counting non-zero elements from the diagonal up)
is $4(M-1)+2.5$. Including the right-hand side, the total number of
reals that need to be stored for the normal equations is therefore
$\simeq (16M-2)N$ or $62N$ for cubic splines. With a knot interval
of about 15~s, about 3~MB is required to store the attitude normal equations
for one day of observations. Thus it is completely realistic to
store the attitude normal equations for the entire mission in
primary memory. Cholesky factorization of the normal equations
does not produce any more non-zero elements in the matrix; the
factorization and solution can therefore use the same storage
as the normal equations. Moreover, since the number of arithmetic
operations grows only linearly with $N$, it is computationally
feasible to solve the normal equations for any stretch of data.

\subsubsection{Segmentation of the data}\label{sec:att-segment}

Even though it is feasible to treat the complete set of normal
equations for the attitude updating as a single system, it is
desirable for several reasons to divide up the data temporally.
This allows one to set up a very straightforward and efficient distributed
attitude updating, simply by handing out the processing of different
time segments to different computing nodes. Also the inspection of residuals
in order to detect stretches of bad fit (caused, for example, by
micrometeoroid impacts), and the subsequent reprocessing of these
stretches, is greatly facilitated if it can be done on shorter
data segments.

\begin{figure}[t!]
\begin{center}
\includegraphics[width=75mm,]{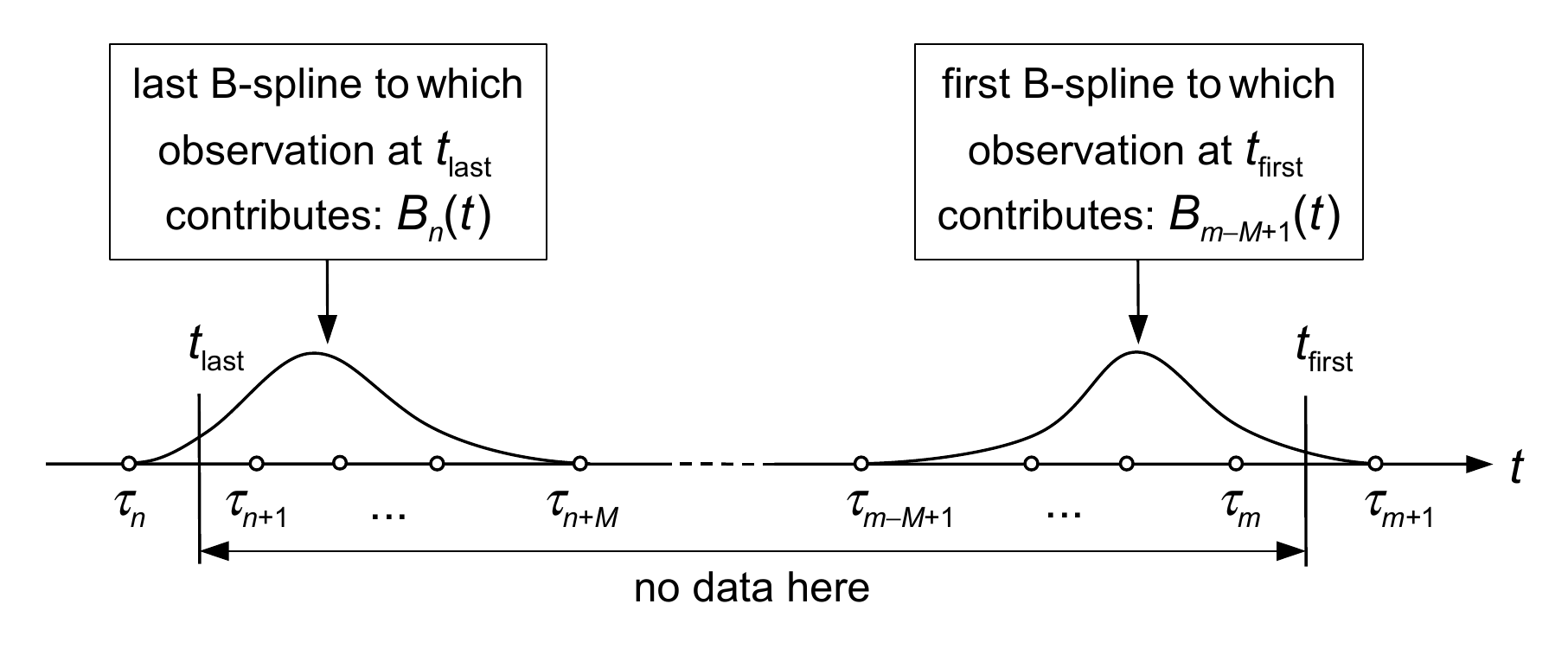}
\caption{A natural break in the definition of the attitude spline
occurs if there is a gap in the observations containing at least
$M$ knots, where $M$ is the order of the spline. $t_\text{last}$
is the time of the last observation before the gap, $t_\text{first}$
the time of the first observation after the gap.}
\label{natural-gap}
\end{center}
\end{figure}

The spline model is capable of interpolating sensibly (if not
accurately) over short data gaps. However, if the data gap contains
at least $M$ knots (with $M=4$ for cubic splines), the two splines
on each side of the gap become completely disconnected. This is
illustrated in Fig.~\ref{natural-gap}, where $n$ and $m$ are the
left indices of, respectively, the last observation before the gap
($t_\text{last}$) and the first observation after the gap
($t_\text{first}$). $B_n(t)$ and $B_{m-M+1}(t)$ are the last and first
B-splines whose coefficients depend on the observations before and
after the gap. Clearly the two segments of the attitude spline are
disconnected if $n<m-M+1$ or $m-n\ge M$. We call this a natural
attitude break.

\begin{figure}[t!]
\begin{center}
\includegraphics[width=90mm,]{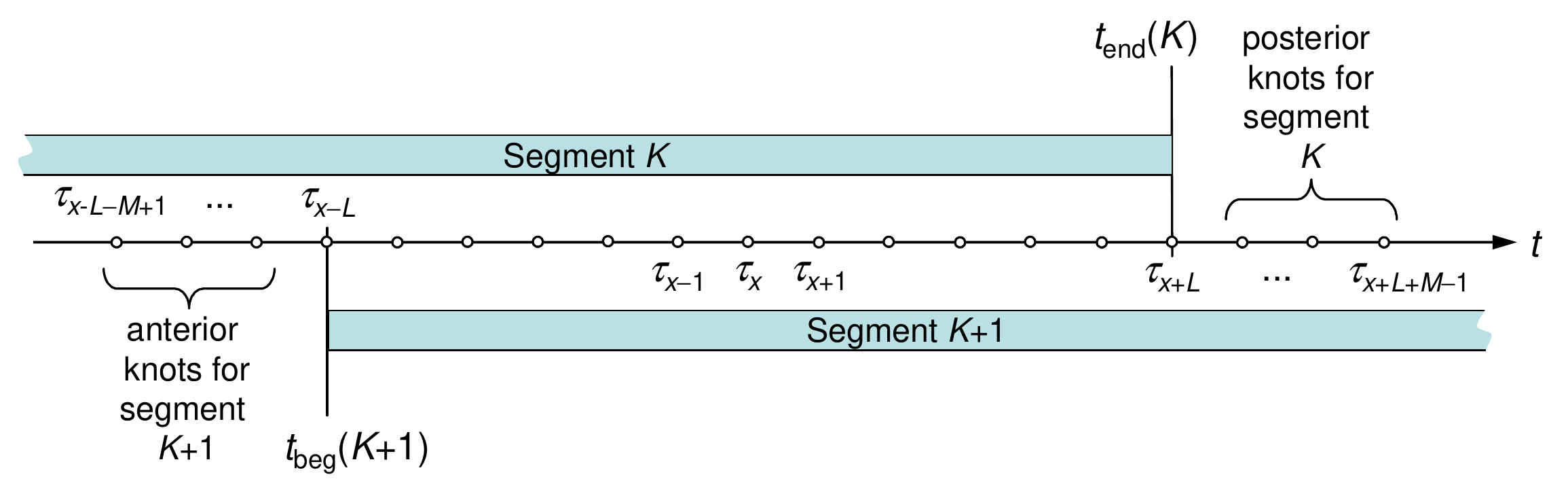}
\caption{Illustrating the assignment of knots for the attitude update
solutions in two consecutive segments $K$ and $K+1$, with a breakpoint
at knot index $x$ and using $2L$ overlapping knot intervals.}
\label{GAIA-Model2}
\end{center}
\end{figure}

In the absence of natural breaks, artificial ones can be introduced
at suitable intervals by a simple method and without any significant
loss of accuracy. The idea is to make separate solutions for overlapping
segments, as illustrated in Fig.~\ref{GAIA-Model2}. The segments use
a common knot sequence $\{\tau_k\}$ that may extend over the whole
length of the mission. Each segment $K$ defines an attitude spline in
the interval $[t_\text{beg}(K),t_\text{end}(K)]$, based on data
with observation times in that same interval. The endpoints coincide
with certain knots in such a way that $t_\text{end}(K)=\tau_{x+L}$
and $t_\text{beg}(K+1)=\tau_{x-L}$, where $\tau_x$ is the cross-over
knot between segments $K$ and $K+1$ and $2L$ the number of
overlapping knot intervals. The anterior and posterior knots for
each segment are also taken from the common knot sequence.
The local character of the splines means that the resulting fit
around $\tau_x$ is practically the same for the two segments,
provided $L$ is large enough. For cubic splines ($M=4$) it is
found that $L=12$ is sufficient.

Each segment gives a system of normal equations (\ref{eq:normA})
for the updates $\vec{d}_n$ to the attitude parameters
$\quat{a}_n$ for a certain range of index $n$. For example, with
reference to Fig.~\ref{GAIA-Model2}, in segment $K$ updates are
computed up to and including $n=x+L-1$, while in segment $K+1$
updates are computed starting with $n=x-L-M+1$. At least in
the middle part of the overlap region, the updates for a given
$n$ should be essentially the same in the two segments. It therefore
does not matter much which of the results is chosen. The mid-point
is at index $n=x-M/2$ if $M$ is even, or half-way between $x-(M+1)/2$
and $x-(M-1)/2$ if $M$ is odd. We may therefore agree to use the
solution from segment $K$ to update $\quat{a}_n$ up to and including
index $n=x-\lfloor M/2\rfloor$, while the solution from segment
$K+1$ is used starting with $n=x-\lfloor M/2\rfloor+1$. The important
thing is that no $n$ is missed by the updating, nor updated twice.

Once all the coefficients $\quat{a}_n$ have been updated from the
different segmented solutions, the segmentation loses its meaning and
can in principle be forgotten. For example, when evaluating
the attitude at a specific time $t$, it does not matter to which segment
that instant belonged. In the next iteration of the attitude update, a
different segmentation can in principle be used.

The overlapping segments mean that a fraction of the observations
need to be processed twice in the attitude updating. The fraction equals
the ratio of the overlap length to the mean length of the segment, and
increases the shorter the segments are made. For example, with a segment
length of one day (it would not seem reasonable to have shorter segments)
and a mean knot interval of 15~s, the fractional overlap for $L=12$ is
only 0.4\%.

\subsubsection{Calculation of partial derivatives}\label{sec:att-pd}

For the partial derivatives we obtain in analogy with
Eq.~(\ref{eq:srsprop1diffsol})
\begin{equation}\label{eq:dRdq}
-\frac{\partial R_l^\text{AL}}{\partial\quat{q}_l} =
\vec{m}_l'\,\frac{\partial\vec{u}_l}{\partial\quat{q}_l}\sec\zeta_l \, ,
\quad\quad
-\,\frac{\partial R_l^\text{AC}}{\partial\quat{q}_l} =
\vec{n}_l'\,\frac{\partial\vec{u}_l}{\partial\quat{q}_l} \, ,
\end{equation}
where the derivatives with respect to the attitude quaternion should
be interpreted componentwise. Differentiating Eq.~(\ref{eq:srsprop})
and using that $\mbox{d}(\tens{C}'\vec{u}_l)=\vec{0}$, we have
\begin{equation}\label{eq:dodq}
\bigl\{\mbox{d}(\tens{S}'\vec{u}_l),\,0\bigr\}
= 2\bigl\{\tens{S}'\vec{u}_l,\,0\bigr\}\quat{q}_l^{-1}\mbox{d}\quat{q}_l\, ,
\end{equation}
which after some manipulation gives
\begin{equation}\label{eq:dfadq}
-\frac{\partial R_l^\text{AL}}{\partial\quat{q}_l}
= -2\sec\zeta_l\,\quat{q}_l\,\Bigl\{\tens{S}'\vec{n}_l,0\Bigr\}\, ,
\quad
-\,\frac{\partial R_l^\text{AC}}{\partial\quat{q}_l}
= 2\quat{q}_l\,\Bigl\{\tens{S}'\vec{m}_l,0\Bigr\}\, .
\end{equation}
The derivatives with respect to the spline coefficients $\quat{a}_n$ are
obtained after multiplying the above expressions by $B_n(t_l)$, assuming
that the normalization factor in Eq.~(\ref{eq:qspline}) is close to unity
(Sect.~\ref{sec:att-constr}).

\subsubsection{Constraining the attitude updating}\label{sec:att-constr}

Since the attitude is represented by a unit quaternion, its components
should at all times satisfy $q_x^2+q_y^2+q_z^2+q_w^2=1$. All four components
are nevertheless needed, for while the magnitude of any of them can be inferred
from the other three components, its sign cannot. The redundancy of the
representation manifests itself in that the length of the quaternion cannot
be determined from the observations. Indeed, as can be seen from
Eqs.~(\ref{eq:srsprop})--(\ref{eq:srsprop1}), applying an arbitrary non-zero
scale factor to the attitude quaternion $\quat{q}$ has no effect on the computed
instrument angles, and is therefore unobservable. The attitude parameters
$\quat{a}_n$ are therefore also undefined with respect to a certain scale
value. As a consequence, the normal matrix $\vec{N}_{aa}$ computed from
Eq.~(\ref{eq:normA}) is singular, and constraints are needed for computing
a unique solution.

The normalization in Eq.~(\ref{eq:qspline}) was introduced to guarantee that
the calculated quaternion is always of unit length, although, as we have
seen, this is not strictly necessary for some of the subsequent calculations.%
\footnote{On the other hand we have implicitly assumed $\|\quat{q}\|=1$, for
example in deriving Eq.~(\ref{eq:dfadq}).}

Naively, one might expect that the coefficients $\quat{a}_n$ could be
scaled independently for each B-spline, i.e., that different scale factors
could apply to each index $n$. This is not the case, however. At any time,
the (non-normalized) attitude quaternion is a linear combination of $M$
adjacent coefficients $\quat{a}_n$, and unless all four coefficients
are scaled by exactly the same factor, the result will not be a simple
scaling of the quaternion.%
\footnote{Note that the normalization in Eq.~(\ref{eq:qspline}) is effected
by applying a normalization factor that is a continuous function of time.
Since any given spline coefficient $\quat{a}_n$ is used in a finite time 
interval (namely, the support of the corresponding B-spline), one cannot 
obtain a correct normalization for all $t$ by scaling the spline coefficients
by some value depending on $n$.}
Applying this argument to every observation time $t_l$, it is readily
seen that the same scaling factor must be used for all the coefficients
in any attitude segment without natural or artificial breaks. In principle,
therefore, the attitude normal matrix for such a segment has a rank defect
of 1 (that is, the rank is one less than the number of attitude parameters),
and would only need a single constraint equation to become non-singular.

Numerical experiments, using Singular Value De\-com\-po\-si\-tion
\citep[SVD; see, e.g.,][]{golu+96} of the
matrix $\vec{N}_{aa}$ computed from simulated observations over successively
longer time intervals, indeed show the expected rank defect of 1 for intervals
up to several hundred knots. That is, there is a clear gap (of several orders
of magnitude) between the smallest singular value and the second smallest one.
For longer time intervals the gap gradually closes and the problem thus
becomes ill-posed \citep{book:hansen1998}. Thus, any reasonably long time
interval will in practice require some form of regularization rather than
the application of just a single constraint.

The adopted solution method is a variant of the well known Tikhonov
regularization \citep{book:hansen1998}. The objective function
in Eq.~(\ref{eq:Jfunc}) is modified to include a term depending on the
deviation of the normalization factor in Eq.~(\ref{eq:qspline}) from unity
for each observation. We write the deviation as
\begin{equation}\label{eq:Dl}
 D_l \equiv 1 - \left\|\, \sum_{n=\ell-M+1}^{\ell} \quat{a}_n B_n(t_l)\,\right\|^2
\end{equation}
and the modified objective function as
\begin{equation}\label{eq:JfuncMod}
Q(\vec{s},\vec{a},\vec{c},\vec{g}) =
\sum_{l} \left( R_l^2 + \lambda^2 D_l^2\right) W_l\, ,
\end{equation}
where $\lambda$ is a small but non-zero regularization parameter. We have found that
$\lambda = 10^{-3}$ to $10^{-2}$ gives a solution that is always numerically stable,
and quite insensitive to the precise value of $\lambda$. As a result,
the normal equations (\ref{eq:normA}) become
\begin{multline} \label{eq:normAmod}
\sum_{m=n-M+1}^{n+M-1}
\left[ \sum_{l\,\in\, L_n\cap L_m} \left(
\frac{\partial R_l}{\partial\quat{a}_n}\frac{\partial R_l}{\partial\quat{a}'_{\!m}}
+ \lambda^2
\frac{\partial D_l}{\partial\quat{a}_n}\frac{\partial D_l}{\partial\quat{a}'_{\!m}}
\right)W_l \right]\,\vec{d}_m \\
= -\sum_{l\,\in\, L_n} \left(
\frac{\partial R_l}{\partial\quat{a}_n}R_l + \lambda^2
\frac{\partial D_l}{\partial\quat{a}_n}D_l \right)W_l \, .
\end{multline}
The required partial derivatives, obtained from Eq.~(\ref{eq:Dl}), are
\begin{equation}\label{eq:dDldan}
-\frac{\partial D_l}{\partial\quat{a}_n} =
2\!\!\!\sum_{m=\ell-M+1}^{\ell}\!\quat{a}_m B_m(t_l) B_n(t_l)
\simeq 2\quat{q}(t_l)B_n(t_l) \, ,
\end{equation}
where the approximation makes use of the fact that the normalization
factor in Eq.~(\ref{eq:qspline}) is close to unity.

\subsubsection{Estimating the excess attitude noise}\label{sec:attNoise}

The excess attitude noise $\epsilon_a(t)$ introduced in Eq.~(\ref{eq:excess}) 
accounts for modelling errors in the attitude representation. Such errors could be 
caused for example by (unmodelled) micrometeoroid impacts, `clanks' due to 
sudden redistributions of satellite inertia, propellant sloshing, thruster noise, 
or mechanical vibrations (Appendix~\ref{sec:attirr}). 
Due to the cubic spline representation, any localized
effect that cannot be fitted by the spline will result in systematic residuals
that span over a few consecutive knot intervals. Indeed, discontinuities in the 
rate (e.g., from micrometeoroid impacts) or angle (e.g., from clanks) produce
characteristic patterns of residuals that can be used to identify such events.
A significant effort will be devoted to the possibly semi-manual and interactive 
process of finding these events. When identified, they can be handled for example 
by modifying the knot sequence (Sect.~\ref{sec:attInit}). But even after this
process, the model will be imperfect due to for example high-frequency
thruster noise.%
\footnote{`High-frequency' here means roughly the range 
$1/2\Delta\tau\simeq 0.03$~Hz to  0.2~Hz, where $\Delta\tau$ is the
typical spline knot interval ($\sim$15~s); lower frequencies are absorbed
by the spline and higher frequencies are smoothed out by the integration
across the CCD.}
Similarly, there will be a large number of impacts that are too small to be 
individually recognized; collectively they add some unmodelled attitude 
errors, which $\epsilon_a(t)$ may account for. However, it should be noted that 
$\epsilon_a(t)$ does \emph{not} include any component of the observation
noise (principally from CCD photon noise), nor is it an estimation of the 
attitude uncertainty (cf.\ Sect.~\ref{sec:cov}).

Three components of the excess noise, designated $\epsilon_\text{AL}(t)$, 
$\epsilon_\text{ACP}(t)$, and $\epsilon_\text{ACF}(t)$, need to be derived
independently of each other, representing modelling errors in the
AL attitude, the AC attitude of the preceding field of view, and the
AC attitude in the following field of view. The three components
are statistically nearly independent thanks to the way the attitude
measurements are made, and the fact that the basic angle is not far
from $90^\circ$.

The algorithm to estimate $\epsilon_a(t)$ (for $a=$~AL, ACP or ACF)
is based on a simple statistical processing of the residuals
$R_l$ derived in the source updating.
The time line $t$ is divided into `buckets' $[t_j,\, t_{j+1})$ such that
each bucket ($j$) will contain a sufficient number of observations,
also in the AC direction. The size (duration) of a bucket should be
several knot intervals for the attitude
spline, but the boundaries $t_j$ need not in any other way be related
to the attitude knot sequence. One set of buckets is needed for
each attitude component (AL, ACP, ACF). Let $l\in ja$ signify that
observation $l$ belongs to bucket $j$ and attitude component $a$.
After having completed the source updating for all primary sources, 
the excess attitude noise in bucket $j$ is estimated as
\begin{equation}\label{e09}
\epsilon_a^2(t_j\le t < t_{j+1}) = \max\left(
0,\, \underset{l\in ja}{F_{0.68}}(
R_l^2-\sigma_l^2-\epsilon_i^2)\right) \, ,
\end{equation}
where $F_{0.68}()$ is the 68\% quantile (68th percentile) of the argument values.
It is important to note that the downweighting factors $w_l$
determined during the source updating are not used here to eliminate
possible outliers; this function is instead taken care of by
using the quantile to compute a robust estimate of the typical excess 
variance in the attitude bucket. This means that if the `outliers' detected 
by the source update were actually caused by a stretch of bad attitude,
then this will be recognized by a large value of the quantile in Eq.~(\ref{e09}),
and consequently by an increased $\epsilon_a^2$.

In the subsequent attitude update, the downweighting factors $w_l$
are re-computed based on the residual from the previous source update
but with a value for the total variance,
$\sigma_\ell^2+\sigma_i^2+\epsilon_a^2$, using the newly estimated
$\epsilon_a^2$. Thus, only the `true' outliers -- that are not due to the 
bad attitude -- are now downweighted. The data may thus contribute to 
the attitude updating even if they had been flagged as outliers in the 
preceding source updating. 

The functions $\epsilon_a(t)$ are obviously an extremely useful
diagnostic for the progress of the AGIS iterations as well as
(after convergence) for the quality of the attitude modelling
and data. They can be
plotted as a function of time, and the quantity of data is such
that human inspection is feasible. They are also
needed for setting the detection threshold for micrometeoroid
impacts.

The accumulation of statistics in the buckets is best done in parallel 
with the source updating, when the residuals are readily at hand.  
One remaining problem is how to compute the quantile in Eq.~(\ref{e09})
in an efficient way, without having to store billions of residuals. Indeed,
exact calculation of quantiles would require to store all the values 
$R_l^2-\sigma_l^2-\epsilon_i^2$ in a bucket before the quantile 
can be computed. However, if we are content
with an approximate estimate of the quantile, there are a number of
sequential estimation algorithms available that only need to store a
much smaller amount of data per bucket, see for example 
\citet{greenwald+01}, \citet{gilbert+02} and references therein.
We have chosen to use the Incremental Quantile estimation algorithm
due to \citet{chambers+06}.

\subsubsection{Initialization of the attitude parameters}\label{sec:attInit}

An approximate estimate of the attitude is already provided by the
initial data treatment (IDT) preceding the astrometric solution.
This may be given as a discrete time series, for example one quaternion
every second of time. The first time the attitude update is executed
for a certain time interval, a regular knot sequence is set up and the
B-spline coefficients $\quat{a}_n$ in Eq.~(\ref{eq:qspline}) are determined
by a least-squares fit. For a given time series of attitude estimates,
this is a linear problem and therefore easily solved. The resulting initial
attitude $\vec{a}^{(0)}$ is used in the first source update (S) and
subsequently improved by the attitude update process (A) as part of the
AGIS iteration scheme.

By default, a regular knot sequence is adopted, i.e., the knot interval
$\Delta\tau=\tau_{n+1}-\tau_n$ is taken to be more or less constant.
Given the endpoints $t_\text{beg}$, $t_\text{end}$ of a data segment,
the knots are set up at regular intervals respecting a given maximum
value of $\Delta\tau$ (of the order of 5 to 20~s). The assignment of knots
must also take into account the need for anterior and posterior knots,
as discussed in Appendix~\ref{sec:splines}, and in the case of
segmented data, the overlapping knots as
discussed in Sect.~\ref{sec:att-segment}.

Occasionally the knot sequence needs to be redefined as a result of
the adjustment process. Possible causes could be:
\begin{itemize}
\item
If the spline is not flexible enough to accurately model the data,
it may be necessary to decrease the maximum allowed $\Delta\tau$.
\item
Conversely, overfitting of the data may require the maximum allowed
$\Delta\tau$ to be increased.
\item
Locally, a scarcity of accurate data or a short gap could make it
necessary to remove some knots or introduce a natural break in
the attitude representation (Sect.~\ref{sec:att-segment}).
\item
Very locally, a bad fit may result from a micrometeoroid hit causing
an almost instantaneous change in the angular velocity of the satellite.
This may be dealt with by introducing multiple knots at the appropriate
instants (Appendix~\ref{sec:attirr} and \ref{sec:multknot}).
\end{itemize}
Having redefined the knot sequence, it is necessary to re-initialize
the spline coefficients $\quat{a}_n$, which must now refer to the
new knot sequence. This is most easily done by evaluating $\quat{q}(t)$
for a regular time series, with a sampling interval much smaller than
$\Delta\tau$ (e.g., 1~s), and fitting the new spline to the time
series.

\subsection{Calibration updating (C)} \label{sec:calupdate}

The geometric instrument model is given in Sect.~\ref{sec:instrumentmodel}.
We assume here the generic calibration model in 
Eqs.~(\ref{eq:generic1})--(\ref{eq:generic2}), in which the parameters are 
indexed by $rs$. In the calibration update block (C) the calibration parameters 
$\vec{c}$ are improved by solving the third sub-system in Eq.~(\ref{eq:precond1}),
i.e., the normal equations
\begin{equation} \label{eq:normC}
\left[\sum_l \frac{\partial R_l}{\partial\vec{c}}
\frac{\partial R_l}{\partial\vec{c}'} W_l \right]\,\vec{d}_c
= -\sum_l \frac{\partial R_l}{\partial\vec{c}}
R_l(\vec{s},\vec{a},\vec{c},\vec{g}) W_l \, .
\end{equation}
The residuals in the right-hand side are computed from the parameters
values in the current or preceding iteration according to the discussion
in Sect.~\ref{sec:order}. Because the calibration model is linear, the partial 
derivatives are uniquely determined by the observation index $l$,
\begin{equation} \label{eq:pdC}
\frac{\partial R_l}{\partial c_{rs}} = \frac{\partial\eta^\text{obs}_l}{\partial c_{rs}} 
= \begin{cases} ~\Phi_{rs}(l) & \text{if $l\in rs$}, \\ ~0 & \text{otherwise}.
\end{cases}
\end{equation}
In the normal matrix, the element with subscripts $(rs)_1$ and $(rs)_2$ is
non-zero only if there is at least one observation $l$ such that
$l\in (rs)_1$ and $l\in (rs)_2$. Depending on how the calibration parameters 
are grouped into sets with no common observations (for example according 
to the CCD/gate combination; cf.\ Sect.~\ref{sec:norm}), the 
normal matrix will therefore be block-diagonal, which the calibration updating
takes advantage of in order to save computations. It also
facilitates distributed processing.
 
Since the weights $W_l$ are fixed from the preceding source and attitude
updating processes, the update $\vec{d}_c$ can be calculated in a 
single direct solution, using the robust Cholesky decomposition 
(Appendix~\ref{sec:cholesky}). However, due to the degeneracy between for 
example the large-scale and small-scale AL calibration parameters, this will
produce an arbitrary feasible solution $\vec{\tilde{d}}_c$, which does not 
necessarily satisfy the constraints in Eq.~(\ref{eq:genericConstraint}). The
constrained update is obtained as
\begin{equation} \label{eq:calCon}
\vec{d}_c = \vec{\tilde{d}}_c - \vec{C}(\vec{C}'\vec{C})^{-1}\vec{C}'\vec{\tilde{d}}_c \, ,
\end{equation}
whereupon the updated $\vec{c}$ can be computed.

The above-mentioned degeneracy among the calibration parameters
means that the normal matrix calculated according to Eq.~(\ref{eq:normC})
is singular, which seems to contradict our assumption (Sect.~\ref{sec:norm})
that $\vec{N}_{cc}$ is positive-definite. However, if $\vec{C}$ spans
the null space of $\vec{N}_{cc}$, as it should for a properly formulated set
of constraints, then it can be seen that Eq.~(\ref{eq:calCon}) gives the same 
result as solving the updates with the modified normal matrix 
$\vec{N}_{cc}+\lambda^2\vec{C}\vec{C}'$, which is positive definite for
any $\lambda\ne 0$. Thus, the procedure outlined above is equivalent to
solving the constrained system with positive-definite matrix.

\subsection{Global updating} \label{sec:globalupdate}

An arbitrary number of global parameters may be solved for in the AGIS system.
Global parameters should be defined in such a way that their default values, 
equal to zero, correspond to the baseline solution. By not solving for the
globals, we implicitly set them to zero, resulting in the baseline solution. For example,
we have a very high confidence in General Relativity, which in the parametrized
post-Newtonian (PPN) formalism implies the parameter $\gamma=1$. 
The global parameter related to the gravitational deflection of light should 
therefore not be $\gamma$ itself, but for example the parameter $g_0$ in
\begin{equation}\label{eq:gammaPPN}
\gamma = 1 + g_0 \, .
\end{equation}
That is, $g_0=0$ corresponds to the baseline case of General Relativity.
The global parameter vector is $\vec{g}=(g_0,~g_1,~\dots)'$.

The normal equations for the update $\vec{d}_g$ to the global parameter
vector are
\begin{equation} \label{eq:normG}
\left[ \sum_l \frac{\partial R_l}{\partial\vec{g}}
\frac{\partial R_l}{\partial\vec{g}'} W_l \right]\,\vec{d}_g
= -\sum_l \frac{\partial R_l}{\partial\vec{g}}
R_l(\vec{s},\vec{a},\vec{c},\vec{g})W_l \, ,
\end{equation}
where the sums are taken over all the observations $l$, and the statistical
weights $W_l$ follow from the preceding source and attitude updates.
The partial derivatives in Eq.~(\ref{eq:normG}) are computed in exact
analogy with Eqs.~(\ref{eq:srsprop1diffsol})--(\ref{eq:srspropdiff}) for
the source updating, only with $\vec{g}$ replacing $\vec{s}_i$. The
calculation of $\partial\vec{u}_l/\partial\vec{g}'$ is not detailed here.

In the simple iteration scheme (Sect.~\ref{sec:iter}), the inclusion of
$g_0$ representing the PPN parameter $\gamma$ considerably slows 
down the convergence of the astrometric solution. As explained by
\citet{hobbs+2010}, this behaviour is caused by the relatively strong 
correlation between the gravitational light deflection by the Sun 
(proportional to $1+\gamma$, and directed away from the Sun) and 
trigonometric parallax (directed towards the solar-system barycentre, 
never far from the Sun). 
\citet{hobbs+2010} found that the convergence rate could be
restored by the introduction of a  pseudo-parameter $g_1$ representing 
a global shift of all parallaxes. (The update to this parameter is solved 
in each iteration but never applied -- its value remains at zero and the
converged values of all the other parameters are unaffected; 
hence the prefix `pseudo'.) It was later found that this artefact is not needed 
when using the conjugate gradients scheme, which gives roughly the same
rate of convergence whether or not $g_1$ is included.

\section{Auxiliary processes}\label{sec:aux}

In this section we describe some auxiliary processes that are not
necessarily part of the astrometric solution as such, but nevertheless
needed in order to construct the astrometric catalogue. They concern
the definition of the reference system for the source positions and
proper motions by means of the frame rotator (Sect.~\ref{sec:framerotator}),
the selection of primary sources (Sect.~\ref{sec:primsel}),
and the computation of the standard uncertainties and correlations of the
astrometric parameters (Sect.~\ref{sec:cov}).

\subsection{Frame rotator} \label{sec:framerotator}

As explained in Sect.~\ref{sec:rank}, the measurement principle of Gaia 
results in a system of positions and proper motions that is essentially
undefined with respect to an arbitrary (small) offset in the orientation and
spin of the reference frame. As a consequence, the normal matrix $\vec{N}$ 
is in principle singular with a rank defect of 6. 

While the solution of rank-deficient problems in general requires special 
attention to the singularities, for example by adding constraints to avoid 
numerical instability, no such complication arises here because of the 
way AGIS works. Basically, a solution is found by iterating between the 
source and attitude updatings (the calibration and global updatings play 
no role here because they are to first order independent of the reference 
frame). When the sources are updated, the reference frame is in reality 
set by the (then assumed) attitude; similarly, when the attitude is updated, 
the frame is set by the (then assumed) source parameters. In terms of the
matrix formulation of Sect.~\ref{sec:iter} this is equivalent to the statement
that the preconditioner $\vec{K}$ is non-singular. The end result 
is that AGIS converges to a solution with both the source and attitude 
parameters expressed in the same, but largely arbitrary, reference frame.

The intention is however that the final source parameters (positions and 
proper motions) shall be expressed in a celestial reference frame that 
represents, as closely as possible, the International Celestial Reference 
System (ICRS). For consistency, it is moreover necessary that the attitude 
parameters are expressed in exactly the same frame as the source parameters. 
It is the task of the frame rotator to accomplish this. A similar process 
was used to align the Hipparcos Catalogue with the extragalactic reference 
frame \citep{ndac1995}. 

In the following we start with the rigorous definition of the rotation correction,
then derive a linear approximation applicable to the small corrections that
we have in practice. Finally, we discuss the determination of the rotation 
parameters and their application in the AGIS iteration scheme.

\subsubsection{Relation between the ICRS and AGIS frames}\label{sec:framerel}

Ideally, the astrometric solution should result in parameters that are expressed
in the BCRS (Sect.~\ref{sec:refsystems}), whose axes are aligned with the ICRS
here represented by $\tens{C}=\left[ \vec{X}~\vec{Y}~\vec{Z} \right]$. 
However, due to the in 
principle undefined reference frame of AGIS, the astrometric solution is 
in effect expressed relative to a slightly different triad, which we denote 
$\tilde{\tens{C}}=\left[ \vec{\tilde{X}}~\vec{\tilde{Y}}~\vec{\tilde{Z}} \right]$. 
The two reference systems, which for simplicity will be referred to as the 
ICRS and AGIS frames, are related by a time-dependent spatial rotation given 
by the quaternion $\quat{f}(t)$; thus the coordinates of the arbitrary (fixed) 
vector $\vec{v}$ are transformed according to the frame rotation formula
\begin{equation}\label{eq:fr}
\left\{\tens{C}'\vec{v},\,0\right\} = \quat{f}(t)^{-1}
\,\big\{\tilde{\tens{C}}'\vec{v},\,0\big\}\, \quat{f}(t) 
\end{equation} 
(cf.\ Eq.~\ref{eq:quatframerot}). Due to the kinematical constraints of 
the AGIS solution, $\quat{f}(t)$ describes a uniform spin motion of 
the two frames with respect to each other.

For consistency with \citet{ndac1995} we parametrize $\quat{f}(t)$
by means of two vectors $\vec{\varepsilon}$ and $\vec{\omega}$ representing
\emph{corrections} to the orientation and spin of the AGIS frame. More precisely,
the parameters of the frame rotator are the six coordinates of the vectors in
the AGIS frame at some adopted frame rotator epoch $t_\text{fr}$ (not 
necessarily the same as the reference epoch $t_\text{ep}$ of the astrometric 
parameters). These coordinates are denoted $\varepsilon_{\tilde{X}}$, 
$\varepsilon_{\tilde{Y}}$, $\varepsilon_{\tilde{Z}}$, $\omega_{\tilde{X}}$, 
$\omega_{\tilde{Y}}$, and $\omega_{\tilde{Z}}$; according to our kinematical
assumption they are strictly constant numbers.
For the arbitrary epoch $t$ the frame rotator quaternion is, therefore, 
\begin{equation}\label{eq:fr1}
\quat{f}(t) = \quat{Q}\big[(t-t_\text{fr})\tilde{\tens{C}}'\vec{\omega}\big]
\quat{Q}\big(\tilde{\tens{C}}'\vec{\varepsilon}\big) \, ,
\end{equation}
where $\quat{Q}$ is the function introduced by Eq.~(\ref{eq:qrot}).
Equations~(\ref{eq:fr}) and (\ref{eq:fr1}) provide the basis for the rigorous
transformation of any data between the two frames, given the rotation
parameters $\tilde{\tens{C}}'\vec{\varepsilon}=\left[
\varepsilon_{\tilde{X}}~\varepsilon_{\tilde{Y}}~\varepsilon_{\tilde{Z}}\right]'$ 
and $\tilde{\tens{C}}'\vec{\omega}=\left[
\omega_{\tilde{X}}~\omega_{\tilde{Y}}~\omega_{\tilde{Z}}\right]'$.

While the above expressions are strictly valid for arbitrarily large rotation 
parameters, we have in practice $\|\vec{\varepsilon}\|$, 
$\|(t-t_\text{fr})\vec{\omega}\|<20~\text{mas}\simeq 10^{-7}$~rad, 
at least in the final iterations of AGIS. This means that second-order terms are 
completely negligible ($<0.002~\mu$as). To first order we have
\begin{equation}\label{eq:fr2}
\quat{f}(t) \simeq \quat{Q}\big[\tilde{\tens{C}}'\vec{\varepsilon} + 
(t-t_\text{fr})\tilde{\tens{C}}'\vec{\omega}\big] \, ,
\end{equation}
and the vector part of Eq.~(\ref{eq:fr}) becomes, to the same approximation,
\begin{equation}\label{eq:fr3}
\begin{bmatrix} v_X \\ v_Y \\ v_Z \end{bmatrix} =
\begin{bmatrix} v_{\tilde{X}} \\ v_{\tilde{Y}} \\ v_{\tilde{Z}} \end{bmatrix} +
\begin{bmatrix} 0 & -v_{\tilde{Z}} & +v_{\tilde{Y}} \\
+v_{\tilde{Z}} & 0 & -v_{\tilde{X}} \\ 
-v_{\tilde{Y}} & +v_{\tilde{X}} & 0 \end{bmatrix}
\begin{bmatrix} \varepsilon_{\tilde{X}}+(t-t_\text{fr})\,\omega_{\tilde{X}} \\ 
\varepsilon_{\tilde{Y}}+(t-t_\text{fr})\,\omega_{\tilde{Y}} \\
\varepsilon_{\tilde{Z}}+(t-t_\text{fr})\,\omega_{\tilde{Z}}
\end{bmatrix}
\, .
\end{equation}

\subsubsection{Transformation of the astrometric parameters}\label{sec:frameast}

Let $\tilde{\alpha}$, $\tilde{\delta}$, $\tilde{\mu}_{\alpha*}$, 
$\tilde{\mu}_{\delta}$ be the position and proper motion parameters 
for a source as derived in AGIS, that is referring to $\tilde{\tens{C}}$. For
brevity we omit here the source index ($i$), and do not consider 
the parallax $\varpi_i$ and radial proper motion $\mu_{ri}$ which 
are independent of the frame orientation. 
In analogy with Eq.~(\ref{eq:normal-triad}) we have the normal 
triad $[\vec{\tilde{p}}~\vec{\tilde{q}}~\vec{r}]$ with respect 
to the AGIS frame, where $\vec{r}$ is the barycentric direction to the
source at time $t_\text{ep}$, 
$\vec{\tilde{p}}=\langle\vec{\tilde{X}}\times\vec{r}\rangle$
and $\vec{\tilde{q}}=\vec{r}\times\vec{\tilde{p}}$; its coordinates
in the AGIS frame are given by the columns of the matrix
\begin{equation}\label{eq:normal-triad-AGIS}
\tilde{\tens{C}}'\left[\vec{\tilde{p}}~~\vec{\tilde{q}}~~\vec{r}\right]
= \begin{bmatrix} -\sin\tilde{\alpha} &~ -\sin\tilde{\delta}\cos\tilde{\alpha}
&~ \cos\tilde{\delta}\cos\tilde{\alpha}~ \\ 
\phantom{-}\cos\tilde{\alpha} &~ -\sin\tilde{\delta}\sin\tilde{\alpha} &~
\cos\tilde{\delta}\sin\tilde{\alpha}~ \\ 
0 &~ \cos\tilde{\delta} &~ \sin\tilde{\delta}~ \end{bmatrix}.
\end{equation}
At the source reference epoch $t_\text{ep}$ the direction cosines
of $\vec{r}$ are related by the frame rotation in Eq.~(\ref{eq:fr1});
thus
\begin{equation}\label{eq:fr-r}
\left\{ \tens{C}'\vec{r},~0 \right\} = \quat{f}(t_\text{ep})^{-1}
\,\big\{ \tilde{\tens{C}}'\vec{r},~0 \big\}\, \quat{f}(t_\text{ep}) \, .
\end{equation} 
From $\tens{C}'\vec{r}=[r_x~r_y~r_z]'$ the position parameters in the 
ICRS frame are obtained as
\begin{equation}\label{eq:fr-ad}
\alpha = \text{atan2}(r_y,~r_x)\, , \quad 
\delta = \text{atan2}\left(r_z,~\sqrt{r_x^2+r_y^2}\right) \, .
\end{equation} 

The transformation of the proper motion components is a bit more
complicated, as they are expressed with respect to axes that are
physically (slightly) different in the two frames, viz., $\vec{\tilde{p}}$, 
$\vec{\tilde{q}}$ in the AGIS frame, and $\vec{p}$, $\vec{q}$ in the 
ICRS frame. However, the time derivative (at epoch $t_\text{ep}$) of 
the barycentric direction to the source is a fixed vector in 
space, known as the proper motion vector. In a kinematically 
non-rotating system it can be written
\begin{equation}\label{eq:mu}
\vec{\mu} = \vec{p}\mu_{\alpha*} + \vec{q}\mu_\delta
= \vec{\tilde{p}}\tilde{\mu}_{\alpha*} +
\vec{\tilde{q}}\tilde{\mu}_\delta - \vec{\omega}\times\vec{r}\, ,
\end{equation}
where the last term is the correction for the spin of the AGIS frame.
The coordinates of the proper motion vector in the two frames,
\begin{equation}\label{eq:mu-icrs}
\tens{C}'\vec{\mu} = \tens{C}'\vec{p}\mu_{\alpha*} + \tens{C}'\vec{q}\mu_\delta
\end{equation}
and
\begin{equation}\label{eq:mu-agis}
\tilde{\tens{C}}'\vec{\mu} = \tilde{\tens{C}}'\vec{\tilde{p}}\tilde{\mu}_{\alpha*} +
\tilde{\tens{C}}'\vec{\tilde{q}}\tilde{\mu}_\delta 
- \left(\tilde{\tens{C}}'\vec{\omega}\right)\times\left(\tilde{\tens{C}}'\vec{r}\right)
\end{equation}
are related by a frame rotation analogous to Eq.~(\ref{eq:fr-r}),
\begin{equation}\label{eq:fr-mu}
\left\{ \tens{C}'\vec{\mu},~0 \right\} = \quat{f}(t_\text{ep})^{-1}
\,\big\{ \tilde{\tens{C}}'\vec{\mu},~0 \big\}\, \quat{f}(t_\text{ep}) \, .
\end{equation} 
From Eq.~(\ref{eq:mu}) the proper motion components in the ICRS 
frame are then
\begin{equation}\label{eq:fr-mu-icrs}
\mu_{\alpha*} = \vec{p}'\vec{\mu} 
= \left(\tens{C}'\vec{p}\right)'\left(\tens{C}'\vec{\mu}\right)\, , \quad 
\mu_\delta = \vec{q}'\vec{\mu}
= \left(\tens{C}'\vec{q}\right)'\left(\tens{C}'\vec{\mu}\right)\, . 
\end{equation} 

For given $(\tilde{\alpha}, \tilde{\delta}, \tilde{\mu}_{\alpha*}, \tilde{\mu}_{\delta})$ 
and $(\tilde{\tens{C}}'\vec{\varepsilon},\tilde{\tens{C}}'\vec{\omega})$, the 
sequence of calculations is therefore: 
\begin{enumerate}
\item Calculate $\quat{f}(t_\text{ep})$ by Eq.~(\ref{eq:fr1});
\item Calculate $\tilde{\tens{C}}'[\vec{\tilde{p}}~\vec{\tilde{q}}~\vec{r}]$
by Eq.~(\ref{eq:normal-triad-AGIS});
\item Calculate $\tens{C}'\vec{r}$ by Eq.~(\ref{eq:fr-r}) and $\tens{C}'\vec{\mu}$ 
by Eq.~(\ref{eq:fr-mu}); 
\item Calculate $\alpha$ and $\delta$ by Eq.~(\ref{eq:fr-ad}) and
$\tens{C}'\vec{p}$ and $\tens{C}'\vec{q}$ by Eq.~(\ref{eq:normal-triad}); 
\item Calculate $\mu_{\alpha*}$ and $\mu_{\delta}$ by Eq.~(\ref{eq:fr-mu-icrs}).
\end{enumerate}
As we have not employed the approximations in Eqs.~(\ref{eq:fr2})--(\ref{eq:fr3}),
these transformations are rigorous.

\subsubsection{Transformation of the attitude parameters}\label{sec:frameatt}

In analogy with Eq.~(\ref{eq:att-q}) the attitude quaternion $\quat{\tilde{q}}(t)$ 
derived in AGIS defines the transformation from the AGIS frame 
$\tilde{\tens{C}}$ to the SRS $\tens{S}$ as a function of time; thus for the
arbitrary vector $\vec{v}$
\begin{equation}\label{eq:fr-q1}
\left\{ \tens{S}'\vec{v}, 0 \right\} = \quat{\tilde{q}}(t)^{-1}
\,\big\{ \tilde{\tens{C}}'\vec{v}, 0 \big\}\, \quat{\tilde{q}}(t) \, .
\end{equation}
Solving $\big\{ \tilde{\tens{C}}'\vec{v}, 0 \big\}$ and inserting into
Eq.~(\ref{eq:fr}) yields
\begin{equation}\label{eq:fr-q2}
\left\{ \tens{C}'\vec{v}, 0 \right\} = \quat{f}(t)^{-1}\quat{\tilde{q}}(t)
\,\left\{ \tens{S}'\vec{v}, 0 \right\}\,\quat{\tilde{q}}(t)^{-1}\quat{f}(t) \, .
\end{equation}
Comparison with Eq.~(\ref{eq:att-q}) shows that the corrected attitude
is given by
\begin{equation}\label{eq:fr-q3}
\quat{q}(t) = \quat{f}(t)^{-1}\quat{\tilde{q}}(t) \, .
\end{equation} 
In practice the AGIS attitude $\quat{\tilde{q}}(t)$ is expressed in terms
of B-splines by means of coefficients $\quat{\tilde{a}}_n$ as in
Eq.~(\ref{eq:qspline}). The result of the time-dependent transformation 
by $\quat{f}(t)^{-1}$ in Eq.~(\ref{eq:fr-q3}) cannot, in general, be exactly 
represented by means of B-splines. However, since the transformation is 
changing extremely slowly in comparison with the duration of the support
of each B-spline ($\sim\,$1~min), 
and also the changes of $\quat{a}$ from knot to knot are very small,
we make a negligible error by 
transforming the coefficients instead of the attitude quaternion. Thus
we use
\begin{equation}\label{eq:fr-q4}
\quat{a}_n = \quat{f}(\bar{t}_n)^{-1}\quat{\tilde{a}}_n \, ,
\end{equation} 
where
\begin{equation}\label{eq:RF4}
\bar{t}_n = \frac{1}{2}(\tau_n + \tau_{n+M})
\end{equation}
is the time half-way through the support of $B_n(t)$.

\subsubsection{Determination of the frame rotator parameters}\label{sec:framedet}

The parameters $\tilde{\tens{C}}'\vec{\varepsilon}=\left[
\varepsilon_{\tilde{X}},~\varepsilon_{\tilde{Y}},~\varepsilon_{\tilde{Z}}\right]'$
and $\tilde{\tens{C}}'\vec{\omega}=\left[
\omega_{\tilde{X}},~\omega_{\tilde{Y}},~\omega_{\tilde{Z}}\right]'$
are determined by a weighted least-squares solution, using as input the 
differences in positions and proper motions, for a subset of the sources,
between the AGIS results and a~priori data. Three kinds of sources may be 
used for this purpose:
\begin{itemize}
\item
A subset $S_\text{NR}$ of the primary sources can be assumed to define a
kinematically non-rotating celestial frame. Typically this subset will contain 
some $10^5$ to $10^6$ quasars and point-like galactic nuclei, mainly 
identified from ground-based surveys and photometric criteria. This subset
effectively determines $\vec{\omega}$. 
\item 
A subset $S_\text{P}$ of $S_\text{NR}$ in addition have positions
$(\hat{\alpha},\hat{\delta})$ accurately determined with respect to the 
ICRS by means that are completely independent of Gaia. Typically it will 
contain the optical counterparts of extragalactic objects with accurate 
positions from radio interferometry (VLBI). Due to the cosmological
acceleration effect described below it is necessary to assign an epoch 
$t_\text{P}$ to each such position. This subset effectively determines 
$\vec{\varepsilon}$.
\item 
The third subset $S_\text{PM}$ consists of primary sources that do not
a~priori belong to the non-rotating subset, but have positions and/or 
proper motions that are accurately determined with respect to the ICRS 
independent of Gaia. This could include radio stars observed by VLBI,
or stars whose absolute proper motions have been determined by some
other means. The astrometric parameters of a source in this subset are denoted 
$\hat{\alpha}$, $\hat{\delta}$, $\hat{\mu}_{\alpha*}$, 
$\hat{\mu}_{\delta}$ and refer to the epoch $t_\text{PM}$ (the parallax is 
irrelevant here, as it is identical in both frames). It is not expected that this 
subset will contribute very significantly to the determination of $\vec{\varepsilon}$
and/or $\vec{\omega}$, but they are included in the discussion below 
since they may provide important consistency checks.
\end{itemize}
In the following we derive the appropriate observation equations for the
different kinds of sources. The derivation
assumes that the frame rotator parameters are numerically small so that
the linear approximation in Eq.~(\ref{eq:fr3}) applies. 
For the (weighted) least-squares estimation of the frame rotator parameters 
it is, furthermore, necessary to assign the appropriate statistical weights to 
the observations and to have procedures for identifying and handling outliers. 
These issues are however not discussed here. 

In the frame rotator solution, the six parameters must be
complemented by three more parameters $a_X$, $a_Y$, $a_Z$
taking into account the acceleration of the solar-system barycentre 
in a cosmological frame \citep{bastian95,gwinn+97,kopeikin+06}.
Such an acceleration, by the vector $\vec{\alpha}$, will cause a systematic
`streaming' (dipole) pattern of the apparent proper motions of extragalactic 
objects, described by
\begin{equation}\label{eq:mu0}
\vec{\mu}_0 = (\vec{I}-\vec{r}\vec{r}')\vec{a} \, .
\end{equation} 
Here $\vec{r}$ is the direction to the source and $\vec{a}=\vec{\alpha}/c$
to first order in $c^{-1}$, where $c$ is the speed of light.
The galactocentric acceleration of the solar-system barycentre by 
$\|\vec{\alpha}\|\simeq 2\times 10^{-10}$~m~s$^{-2}$ is expected to produce 
a proper motion pattern with an amplitude of $\|\vec{a}\|\simeq 4~\mu$as~yr$^{-1}$,
which Gaia should be able to detect given a sufficient number of quasars
among the primary sources.%
\footnote{The expected acceleration due to, for example, the Andromeda 
galaxy or the Shapley Supercluster \citep{proust+06} are at most $\sim\! 10^{-3}$
of the galactocentric acceleration. On the other hand, nearby massive galactic
objects and large-scale deviations of the galactic potential from 
axisymmetry could conceivably produce a larger deviation in the direction
of the total acceleration.}
The additional parameters introduced in the
frame rotator solution are the components of $\vec{a}$ in the ICRS, or
$\left[a_X~a_Y~a_Z\right]=\tens{C}'\vec{a}$; they may be expressed in 
the same unit as the proper motions. 

\paragraph{Observation equations for a source in $S_\text{NR}$.}
A kinematically non-rotating source should only have an apparent
proper motion due to the cosmological acceleration. Equating $\vec{\mu}$ 
in (\ref{eq:mu}) with $\vec{\mu}_0$ from Eq.~(\ref{eq:mu0}) and taking 
the scalar products with $\vec{\tilde{p}}$ and $\vec{\tilde{q}}$ results 
in the two observation equations
\begin{equation}\label{eq:cond-mu1}
\left.
\begin{aligned}
\vec{\tilde{p}}'\vec{a} + \vec{\tilde{q}}'\vec{\omega} &= \tilde{\mu}_{\alpha*}
\\[3pt]
\vec{\tilde{q}}'\vec{a} - \vec{\tilde{p}}'\vec{\omega} &= \tilde{\mu}_{\delta}
\end{aligned}
\quad\right\} \, ,
\end{equation}
where we have used the scalar triple product rule%
\footnote{$\vec{a}'(\vec{b}\times\vec{c})=\vec{b}'(\vec{c}\times\vec{a})=
\vec{c}'(\vec{a}\times\vec{b})$}
for the terms including $\vec{\omega}$. These equations are linear in
the unknown acceleration and spin parameters, and the coefficients
are the known coordinates of $\vec{\tilde{p}}$ and $\vec{\tilde{q}}$ in
either $\tilde{\tens{C}}$ or $\tens{C}$ (to the adopted approximation 
the coefficients are the same in the two frames).

\paragraph{Observation equations for a source in $S_\text{P}$.}
In order to compare positions it is necessary to choose an epoch at which
to make the comparison. At the chosen epoch of comparison, $t$, the
barycentric coordinate direction of the source is, to first order in the
proper motion,
\begin{align}\label{eq:fr-ubar}
\vec{\bar{u}}_\text{B}(t) &= \vec{r}(t_\text{P})+(t-t_\text{P})\vec{\mu}_0
\nonumber \\[3pt]
&=\vec{r}(t_\text{ep})+(t-t_\text{ep})
\left(\vec{\tilde{p}}\tilde{\mu}_{\alpha*} + \vec{\tilde{q}}\tilde{\mu}_\delta 
- \vec{\omega}\times\vec{r}\right) \, .
\end{align}
In the first equality we have made the assumption that the source
has the apparent proper motion $\vec{\mu}_0$ when observed in the 
ICRS frame; the second uses the same proper motion vector derived from 
the AGIS data according to Eq.~(\ref{eq:mu}). If we now compute the
coordinates of $\vec{\bar{u}}_\text{B}(t)$ in $\tens{C}$ (using the first
equality) and $\tilde{\tens{C}}$ (using the second equality), they must be
related according to Eq.~(\ref{eq:fr3}). Resolving the coordinate differences
along $\alpha$ and $\delta$ we obtain the observation equations
\begin{equation}\label{eq:cond-Delta}
\left. \begin{aligned}
(t-t_\text{P})\vec{\tilde{p}}'\vec{a}+
\vec{\tilde{q}}'\big(\vec{\varepsilon}+(t-t_\text{fr})\vec{\omega}\big)
&= \Delta\alpha^* \\[3pt]
(t-t_\text{P})\vec{\tilde{q}}'\vec{a}-
\vec{\tilde{p}}'\big(\vec{\varepsilon}+(t-t_\text{fr})\vec{\omega}\big)
&= \Delta\delta
\end{aligned}
\quad\right\} \, ,
\end{equation}
where 
\begin{equation}\label{eq:fr-Delta}
\begin{bmatrix} \Delta\alpha^*\\ \Delta\delta \end{bmatrix}
= 
\begin{bmatrix} 
\tilde{p}_{\tilde{X}} & \tilde{p}_{\tilde{Y}} & \tilde{p}_{\tilde{Z}}\\ 
\tilde{q}_{\tilde{X}} & \tilde{q}_{\tilde{Y}} & \tilde{q}_{\tilde{Z}}
\end{bmatrix}
\begin{bmatrix} 
\cos\tilde{\delta}\cos\tilde{\alpha} - \cos\hat{\delta}\cos\hat{\alpha}\\ 
\cos\tilde{\delta}\sin\tilde{\alpha} - \cos\hat{\delta}\sin\hat{\alpha}\\ 
\sin\tilde{\delta} - \sin\hat{\delta}
\end{bmatrix}
+ (t-t_\text{ep})
\begin{bmatrix} \tilde{\mu}_{\alpha*}\\ \tilde{\mu}_\delta \end{bmatrix} .
\end{equation}
The observation equations in proper motion are of course the same as in
Eq.~(\ref{eq:cond-mu1}).

Returning to the choice of comparison epoch $t$, it is clear that the result 
in terms of the estimated frame rotator parameters should in principle 
not depend on this choice. However, that will only be the case if the 
statistical correlations among the data are taken into account in the 
least-squares estimation. Otherwise one should choose $t$ to minimize 
these correlations. From Eq.~(\ref{eq:fr-Delta}) it is seen that the right-hand
sides of Eq.~(\ref{eq:cond-mu1}) and (\ref{eq:cond-Delta}) are uncorrelated
if $t=t_\text{ep}$, provided that the position and proper motion errors in
AGIS are also uncorrelated (which is generally the case if $t_\text{ep}$ is
appropriately chosen, i.e., close to mid-mission). Consequently we 
suggest using $t=t_\text{ep}$.

\paragraph{Observation equations for a source in $S_\text{PM}$.}
Here it will be necessary to distinguish two cases depending on how 
the proper motions have been measured.
If $\hat{\mu}_{\alpha*}$, $\hat{\mu}_{\delta}$ are the proper motion
components of a source measured relative to background quasars, then
the observation equations are simply
\begin{equation}\label{eq:cond-mu2}
\left.
\begin{aligned}
\vec{\tilde{p}}'\vec{a} + \vec{\tilde{q}}'\vec{\omega} &= \tilde{\mu}_{\alpha*}
- \hat{\mu}_{\alpha*}\\[3pt]
\vec{\tilde{q}}'\vec{a} - \vec{\tilde{p}}'\vec{\omega} &= \tilde{\mu}_{\delta}
-\hat{\mu}_{\delta}
\end{aligned}
\quad\right\} \, .
\end{equation}
As expected, the equations simplify to Eq.~(\ref{eq:cond-mu1}) in case the 
measured proper motion is zero. The observation equations obtained from
the comparison of positions are the same as in Eq.~(\ref{eq:cond-Delta}),
but with right-hand side
\begin{multline}\label{eq:fr-Delta1}
\begin{bmatrix} \Delta\alpha^*\\ \Delta\delta \end{bmatrix}
= 
\begin{bmatrix} 
\tilde{p}_{\tilde{X}} & \tilde{p}_{\tilde{Y}} & \tilde{p}_{\tilde{Z}}\\ 
\tilde{q}_{\tilde{X}} & \tilde{q}_{\tilde{Y}} & \tilde{q}_{\tilde{Z}}
\end{bmatrix}
\begin{bmatrix} 
\cos\tilde{\delta}\cos\tilde{\alpha} - \cos\hat{\delta}\cos\hat{\alpha}\\ 
\cos\tilde{\delta}\sin\tilde{\alpha} - \cos\hat{\delta}\sin\hat{\alpha}\\ 
\sin\tilde{\delta} - \sin\hat{\delta}
\end{bmatrix} \\
+ (t-t_\text{ep})
\begin{bmatrix} \tilde{\mu}_{\alpha*}\\ \tilde{\mu}_\delta \end{bmatrix}
- (t-t_\text{P})
\begin{bmatrix} \hat{\mu}_{\alpha*}\\ \hat{\mu}_\delta \end{bmatrix} .
\end{multline}

If, on the other hand, the proper motion is not measured relative to the
local extragalactic background, but in a global non-rotating frame, then
it already includes a contribution from $\vec{\mu}_0$ and the appropriate 
observation equations are obtained by deleting the terms depending on
$\vec{a}$:
\begin{equation}\label{eq:cond-mu3}
\left.
\begin{aligned}
\vec{\tilde{q}}'\vec{\omega} &= \tilde{\mu}_{\alpha*}-\hat{\mu}_{\alpha*}\\[3pt]
-\vec{\tilde{p}}'\vec{\omega} &= \tilde{\mu}_{\delta}-\hat{\mu}_{\delta}\\[3pt]
\vec{\tilde{q}}'\big(\vec{\varepsilon}+(t-t_\text{fr})\vec{\omega}\big)
&= \Delta\alpha^* \\[3pt]
-\vec{\tilde{p}}'\big(\vec{\varepsilon}+(t-t_\text{fr})\vec{\omega}\big)
&= \Delta\delta
\end{aligned}
\quad\right\} \, .
\end{equation}

\subsubsection{The null space vectors}\label{sec:nullsp}

In Sect.~\ref{sec:rank} we introduced the $n\times 6$ matrix $\vec{V}$
whose columns span the null space of the normal matrix $\vec{N}$.
For completeness we give here the explicit expressions for one 
such set of null vectors.
Any small change in the unknowns $\vec{x}$, by a linear combination
of the columns in $\vec{V}$, will leave the calculated residuals unchanged.
Applying the frame rotator for arbitrary $\vec{\varepsilon}$ and 
$\vec{\omega}$ obviously leaves the residuals unchanged, and we can
therefore compute the columns of $\vec{V}$ as the partial derivatives
of $\vec{x}$ with respect to the six frame rotator parameters. Since
we are concerned with infinitesimal changes, the distinction between
the AGIS and ICRS frames is no longer necessary. If $\vec{V}$ is
partitioned similarly to $\vec{x}$ and $\vec{b}$ in Eq.~(\ref{eq:norm2}),
or $\vec{V}=\big[\vec{V}'_s,\,\vec{V}'_a,\,\vec{V}'_c,\,\vec{V}'_g\big]'$,
we find by means of Eq.~(\ref{eq:cond-mu3}),
\begin{equation}
\left[\vec{V}_s\right]_i = 
\begin{bmatrix}
\phantom{-}q_X & \phantom{-}q_Y & \phantom{-}q_Z 
   & \phantom{-}\tau q_X & \phantom{-}\tau q_Y & \phantom{-}\tau q_Z \\
-p_X & -p_Y & -p_Z & -\tau p_X & -\tau p_Y & -\tau p_Z \\
0 & 0 & 0 & 0 & 0 & 0 \\
0 & 0 & 0 & \phantom{-}q_X & \phantom{-}q_Y & \phantom{-}q_Z \\
0 & 0 & 0 & -p_X & -p_Y & -p_Z 
\end{bmatrix} , 
\end{equation}
where $\tau = t_\text{ep}-t_\text{fr}$ and we have omitted the source index 
$i$ on the matrix elements. The order of the astrometric parameters is 
$(\alpha*,\,\delta,\,\varpi,\,\mu_{\alpha*},\,\mu_\delta)$. From 
Eqs.~(\ref{eq:fr2}) and (\ref{eq:fr-q4}) we similarly obtain for the attitude
parameters $\quat{a}_n=\big\{a_x,\,a_y,\,a_z,\,a_w\big\}$,
\begin{equation}
\left[\vec{V}_a\right]_n = \frac{1}{2}
\begin{bmatrix}
-a_w & -a_z & \phantom{-}a_y & -\tau_n a_w & -\tau_n a_z 
& \phantom{-}\tau_n a_y \\
\phantom{-}a_z & -a_w & -a_x & \phantom{-}\tau_n a_z & -\tau_n a_w 
& -\tau_n a_x \\
-a_y & \phantom{-}a_x & -a_w & -\tau_n a_y & \phantom{-}\tau_n a_x 
& -\tau_n a_w \\
\phantom{-}a_x & \phantom{-}a_y & \phantom{-}a_z & \phantom{-}\tau_n a_x 
& \phantom{-}\tau_n a_y & \phantom{-}\tau_n a_z \\
\end{bmatrix} , 
\end{equation}
where $\tau_n = \bar{t}_n-t_\text{fr}$. The calibration and global
parameters are not affected by the frame rotator, so $\vec{V}_c=\vec{0}$ 
and $\vec{V}_g=\vec{0}$.

Let $\vec{\tilde{d}}$ be a vector of small changes to the unknowns $\vec{x}$,
as for example the update computed in one of the AGIS iterations. In some
situations it is desirable to remove from $\vec{\tilde{d}}$ its component in
the null space, i.e., to project it on the row space of $\vec{N}$. This will for
example ensure that the orientation and spin of the AGIS frame is not, on
the average, changed by the update. In principle, we could achieve this by a process 
analogous to Eq.~(\ref{eq:calCon}):
\begin{equation} \label{eq:derotate}
\vec{d} = \vec{\tilde{d}} - \vec{V}(\vec{V}'\vec{V})^{-1}\vec{V}'\vec{\tilde{d}} \, ,
\end{equation}
where $\vec{d}$ is the projection of the update on the row space of $\vec{N}$.
This is equivalent to solving the unweighted least-squares problem
$\vec{V}\vec{z}\simeq\vec{\tilde{d}}$, yielding the orientation and spin 
components as $\vec{\hat{z}}=(\vec{V}'\vec{V})^{-1}\vec{V}'\vec{\tilde{d}}$, 
followed by the subtraction of the null space component $\vec{V}\vec{\hat{z}}$.
In practice, this can equivalently be achieved by means of the frame 
rotator, without the need to compute $\vec{V}$ explicitly.

\subsubsection{Role of the frame rotator in AGIS}\label{sec:frole}
The frame rotator process consists of the three steps: (i) determine the
frame rotator parameters according to Sect.~\ref{sec:framedet}; 
(ii) correct the astrometric parameters for all sources according to
Sect.~\ref{sec:frameast}; (iii) correct the attitude parameters
according to Sect.~\ref{sec:frameatt}. In principle, this process only
needs to be run after convergence of the AGIS iterations; nevertheless,
there may be a case for running it after each AGIS iteration, preventing
a progressive deviation from the ICRS in the course of the iterations.

In particular, for the simulation experiments described in Sect.~\ref{sec:impl} 
it was found necessary to run the frame rotator after each iteration, in 
order to be able to compare the results of each iteration with the `true' 
astrometric parameters (used as input to the simulations). Without this
correction, the differences between the `estimated' and `true' parameters 
for the source positions and proper motions would have been grossly 
distorted by frame errors originating from the starting values of the 
attitude parameters. 
In this case all the primary sources were treated as belonging to the 
subset $S_\text{PM}$, for which Eq.~(\ref{eq:cond-mu3}) is appropriate.

\subsection{Selection of primary sources} \label{sec:primsel}

The astrometric core solution does not use all the sources observed by Gaia, 
but only a subset of them known as the primary sources. The selection of
this subset is made iteratively, based on the results of earlier astrometric
solutions and other processes such as double-star and variability analysis
(not discussed in this paper). The main criterion for a primary source is that
its proper direction, at all times when it is observed by Gaia, is adequately
modelled by the standard astrometric model outlined in 
Sect.~\ref{sec:astromodel}. Examples of sources that should be excluded
based on this criterion are solar-system objects, short-period astrometric 
binaries with a significant size of the photocentre orbit, long-period 
astrometric binaries with a significant curvature of the photocentre orbit, 
certain unresolved binaries where one or both of the components are variable, 
and active galactic nuclei (AGNs) with significant variation of their 
photocentre positions. Variable stars, long-period resolved binaries, 
eclipsing and spectroscopic binaries need not be excluded a~priori from
the set of primary sources, although many of them are potentially problematic
from an astrometric modelling viewpoint (e.g., a variable star might be part of an
unresolved double or multiple star, resulting in a variable photocentre position).
On the other hand, partially resolved double/multiple stars and other 
extended sources will be problematic
even if their photocentres strictly adhere to the basic astrometric model, 
Eq.~(\ref{eq:uniform}), because technically the determination of the 
photocentre becomes more difficult and less precise.

For the calibrations there are other requirements on the primary sources, in
particular that there are enough of them at various magnitudes and colours,
while their sky distribution is less important. Securing a sufficient number of
primary sources for the calibrations will tend to include many more 
primary sources in some areas, such as the galactic plane, resulting in a 
very non-uniform distribution across the celestial sphere.

\subsubsection{The number of sources needed for AGIS} \label{sec:numsources}

The number of sources required for the Astrometric Global Iterative Solution is
driven by the calibration needs of having representative numbers of sources of
different magnitudes, and the attitude needs of having a sufficient
number of sources in every knot interval.

Let us consider first the requirements for the geometric small-scale 
calibration of the CCDs (Sect.~\ref{sec:instrumentmodel}). The angular extent
of a single CCD in the across-scan direction is about $0.1^\circ$. It scans the
celestial sphere at a rate of $1^\circ$~min$^{-1}$ and therefore covers about 
2000~deg$^2$ per week, if both fields of view are counted. Since there are
1966~pixel columns across the CCD, we have the convenient rule of thumb
that each pixel column covers 1~deg$^2$ per week. Thus, if the average density
of suitable primary sources is $D$~deg$^{-2}$ and we require a minimum of $N$
observations of such sources per pixel column for its calibration, then the 
minimum time needed is $N/D$~weeks. For example, with $10^8$ primary 
sources we have $D=2400$~deg$^{-2}$, and it is then reasonable that the
small-scale calibration can be made, at a resolution of one or a few pixels,
in a matter of weeks. However, for the gated observations of bright sources
($G\lesssim 12$~mag), the available numbers are much smaller and it will be
necessary to sacrifice resolution in time, number of pixels, or both in order
to have a sufficient number of observations per calibration cell. For example,
the gate used for the brightest sources will perhaps mainly be used for 
$G\simeq 5.7$ to 8.8; the \emph{total} number of such stars is about 
130\,000 \citep[from the Tycho-2 Catalogue;][]{hog+00} or
$D\simeq 3$~deg$^{-2}$, of which only a fraction will be suitable as
primary sources. Thus, even over the whole five-year mission there will
only be a few hundred observations per pixel column. From these considerations
it is clear that as many as possible of the bright stars should be selected
as primary sources. 

Consider next the requirements for the attitude determination. The minimum
reasonable knot interval is of the same order as the transit time over a
CCD, or about 5~s. Since we require both along-scan and across-scan
measurements, and the latter are normally only provided by the SM and
some of the AF1 observations, we assume conservatively one observation in each
coordinate per field-of-view transit. There are seven CCDs across the
width of the field of view; the area scanned is therefore 1000~deg$^2$
per day in each field of view, or 0.06~deg$^2$ per 5~s interval. If we
require, say, 100 transits per knot interval for a reliable attitude
determination, then the minimum density of primary sources is
$D=1700$~deg$^{-2}$, or some 70~million sources in total. To 
achieve this density in the galactic pole regions requires that stars 
as faint as $G\simeq 19$ are included among the primary sources. 
Thus it is clear that the design aim of $\simeq 10^8$ primary sources
is quite reasonable, and that these will have to include both very bright
and very faint stars, as well as many quasars down to $G=20$ for
the extragalactic link.

Apart from these minimum requirements, it must be maintained that 
the quality of the solution will only improve, the more (good) primary 
sources are included. Stated in the negative sense, the solution quality 
cannot improve by removing good-quality primary sources. From this
viewpoint one should aim to include as many primary sources as 
possible in the final solution.

On the other hand, one should not forget that it is possible to run AGIS 
with much fewer primary sources by disabling short-term small-scale 
calibrations and using longer knot intervals for the attitude. This will 
increase modelling errors, which however is acceptable for initial runs 
where the input data have not yet been properly calibrated.

\subsubsection{Selection criteria} \label{sec:relegate}

As outlined above a source has to pass several tests, derived from 
different processes, in order to qualify as a primary source. The most
important test is derived from the AGIS solution itself, and is based 
on how well the standard astrometric model (Sect.~\ref{sec:astromodel}) 
fits the data. However, if initially we want to limit the number of primary
sources, a somewhat more sophisticated selection procedure is needed
to guarantee the minimum requirements.

Each source carries an attribute representing the `relegation factor' $U$,
which is a floating-point number ranging from about 1 to infinity.
$U\simeq 1$ implies the source is perfect for use in the astrometric 
solution, while successively larger values indicate less suitable sources.
The relegation factor may incorporate the results of several different 
tests, and therefore provides a continuous variable for use in the selection 
process. The name derives from the need to `relegate' a primary source
into a secondary one when $U$ exceeds a predefined value, which may
happen for example in the course of the AGIS iterations, or from one
solution to the next. On the other hand, a secondary source may be
promoted to a primary if its $U$ value decreases below the set threshold.
This suggests that all potential primary sources should be processed
through the source update (Sect.~\ref{sec:sourceupdate}), after which its 
status as primary/secondary may be decided.

The excess source noise $\epsilon_i$ estimated during the source
updating (Sect.~\ref{sec:source-inner}) may be a good starting point 
for calculating the relegation factor, e.g.:
\begin{equation}\label{eq:relegFac}
U_i = \sqrt{1 + \left[\epsilon_i / e(G_i)\right]^2} \, ,
\end{equation}
where $e(G)$ is a normalization factor depending on the magnitude.
The choice of the function $e(G)$ determines the balance between
absolute and relative contributions to the modelling error budget.
With the choice $e(G)\simeq\sigma_l^\text{AL}$ (the formal along-scan observational 
standard uncertainty for a source of magnitude $G$), $U_i$ approximates
the RMS normalized residual of the source. Selecting sources based
on this $U_i$ tends to discriminate against bright stars where
modelling errors may dominate over photon-noise errors. 
On the other hand, choosing $e(G)=\text{constant}$ means that
only sources with the smallest $\epsilon_i$ are accepted; this
may remove too many of the faint sources, where $\epsilon_i$
is still small in comparison with $\sigma_l^\text{AL}$. A reasonable compromise
between these two extreme cases should be found.
The value from Eq.~(\ref{eq:relegFac}) can later be combined
with other factors indicating for example photometric variability, or 
some other potentially problematic property, so that in general the 
relegation factor can be determined with some formula using a 
combination of the source attributes.

Apart from the relegation factor, which indicates whether a source is at
all suitable as a primary source, the general principle should be to 
maximize the total weight of the primary sources. The weight of a
source $i$ is defined as
\begin{equation}\label{eq:Wi}
W_i= \frac{1}{n_i}\sum_{l\in i} \frac{w_l}{\sigma_l^2 + \epsilon_i^2} \, ,
\end{equation}
where the average is taken over the $n_i=\sum_{l\in i}1$ accepted 
AL observations of the source.
Note that we do not use the more obvious definition
$W_i=\sum_{l\in i}W_l$ with $W_l$ from Eq.~(\ref{eq:W}), since we
do not want to penalize a source by the excess attitude noise in some
of its observations, nor favour a source because it is observed many 
times due to the scanning law.

Having defined the relegation factor and weight per source, the selection 
of primary sources can be made to maximize the total weight with due 
regard to sky uniformity (for attitude determination) and magnitude 
distribution (for instrument calibration). A possible procedure is
the following.

We start by specifying the minimum density $D_\text{min}$ (deg$^{-2}$) 
required for the attitude determination, the targeted total number 
$N_\text{tot}$ of primary sources, with $N_\text{tot}\ge (129600/\pi)D_\text{min}$,
and the maximum acceptable relegation factor $U_\text{max}$. For the 
geometric calibration of gated observations we may also specify 
minimum numbers $\{N_g\}$ for several intervals in $G$. Then:
\begin{enumerate}
\item 
Using a coarse-grained tessellation of the celestial sphere (see below),
select in each pixel the $N_p$ sources with the largest $W_i$ that satisfy 
$U_i\le U_\text{max}$, where $N_p$ is the minimum number per pixel
that will ensure $D_\text{min}$.
\item
For each magnitude bin with a required minimum number $N_g$, count
the actual number of primary sources already selected; if it is less than 
$N_g$ add sources with $U_i\le U_\text{max}$ based on $W_i$.
\item
If the total number after Step~1 and 2 is less than $N_\text{tot}$, 
add sources with $U_i\le U_\text{max}$ based on $W_i$. If the total 
number exceeds $N_\text{tot}$, reduce $U_\text{max}$ and repeat 
the process.
\end{enumerate}
If the required number cannot be reached in a particular pixel or 
magnitude bin, then it is necessary to increase $U_\text{max}$ 
locally for that pixel or bin.

For the tessellation in Step~1 in principle any reasonable way to divide up
the sphere into cells of approximately the same area could be used, but 
for statistical operations the HEALPix scheme \citep{gor05} has some advantages 
\citep{omu01}. The cell size is determined by the choice of HEALPix parameter
$\text{\it NSIDE}$ and is important as it predicts the level of homogeneity over 
the sphere. A cell area of about $1/3$ of the field of view might be close
to optimal, and is achieved with $\text{\it NSIDE}=128$ yielding 196\,608~cells.

For the first run with real data some selection must be made using the initial 
star catalogue. In this case the relegation factor may be set to 1 for all
sources and the selection based entirely on their spatial distribution 
and magnitudes. This will reduce the input to the first run of the astrometric 
solution, after which the relegation factor will be updated as described above. 
The secondary-source update step runs on 
all sources not included in AGIS; hence this will set a relegation factor 
for all sources observed by Gaia.

\subsection{Computation of standard uncertainties and correlations} \label{sec:cov}

It is mandatory that the catalogue of astrometric parameters resulting from
the astrometric core solution includes complete and reliable information 
about the expected error statistics. The most important quantity is the 
estimated standard uncertainty of each astrometric parameter. However, the
statistical correlation between the different astrometric parameters -- 
both between the different parameters of the same source and between 
the parameters of different sources -- is also important and should
be quantified. Such correlations are produced both by attitude modelling 
errors (Sect.~\ref{sec:attNoise}) and the statistical uncertainty due to the
finite astrometric weight of the sources contributing to the attitude
determination. 
More generally, we need a method to estimate the $5\times 5$
covariance matrix $\text{Cov}(\vec{s}_i,\vec{s}_j)$ of the astrometric
parameters of any two sources~$i$, $j$ (including the case $i=j$). In principle,
these are sub-matrices of the upper-left $n_s\times n_s$ part of 
$\vec{N}^\dagger$, the pseudo-inverse of the complete normal equations 
matrix in Eq.~(\ref{eq:norm1}). (The pseudo-inverse should be used since
the matrix is singular.) Although there are methods to compute 
selected elements in $\vec{N}^\dagger$ that may be feasible even for a system 
as large as this, it is utterly impossible to produce any significant fraction
of the covariances by a direct computation. Instead, it will be necessary to
rely on approximations and statistical estimates. A first approximation
is obtained by ignoring the statistical uncertainty contributed by the 
errors of the attitude, calibration and global parameters; in this 
case we can ignore all parts of $\vec{N}$ in Eq.~(\ref{eq:norm2}) except
$\vec{N}_{ss}$ and find
\begin{equation}\label{eq:cov}
\text{Cov}(\vec{s}_i,\vec{s}_j) \equiv \left[\big(\vec{N}^\dagger\big)_{ss}\right]_{ij}
\simeq 
\begin{cases} 
~\left[\vec{N}_{ss}\right]_{ii}^{-1} & \text{if $i=j$,}\\[3pt]
~\vec{0} & \text{if $i\ne j$,}
\end{cases}
\end{equation}
where the inverse of $\left[\vec{N}_{ss}\right]_{ii}=\vec{A}_i' \vec{W}_i \vec{A}_i$
is regularly computed as part of the source updating using Eq.~(\ref{eq:normSmat}). 
However, in this approximation we clearly cannot 
estimate the covariance between sources ($i\ne j$), which is unacceptable;
moreover, we underestimate the within-source covariance
($i=j$) because of the neglected attitude and calibration errors.
Refining this estimate is a very important problem which however is 
addressed elsewhere \citep{holl+2010,holl+2010a}.

A somewhat related problem is the need to be able to transform the astrometric 
results, without loss of information, to an arbitrary epoch different from the 
$t_\text{ep}$ used in the astrometric solution. The standard model in 
Eq.~(\ref{eq:astro-standard}) allows the astrometric parameters to be transformed
in a completely reversible manner, based on the assumption of uniform space 
motion relative to the solar-system barycentre. In this process it is necessary to 
include the sixth astrometric parameter $\mu_{ri}$, even if it is (partially) derived 
from a spectroscopic radial velocity. Similarly, the transformation of the 
covariance matrix must consider all six parameters. The relevant
formulae are given in Vol.~1, Sect.~1.5.5 of The Hipparcos and Tycho Catalogues 
\citep{hip:catalogue} and are not repeated here.%
\footnote{The sixth parameter $\mu_r$ is denoted $\zeta$ in \citet{hip:catalogue}.}
Indeed, the normal equations for the six parameters contain the full information 
of the Gaia observations of a particular source with respect to the standard 
astrometric model, and for this reason it is desirable to compute and store
these normals even if only a subset of them is used in the actual source update 
(cf.\ Sect.~\ref{sec:source-pd}).

\section{Software implementation and demonstration solutions}\label{sec:impl}

The practical realisation of the AGIS scheme as outlined in the preceding
sections is contained in software that is being developed, since early 2006, jointly
by teams at the European Space Astronomy Centre (ESAC) and Lund Observatory
within the Gaia Data Processing and Analysis Consortium (DPAC). It is a central 
software module embedded into a complex, overall data processing
system \citep{WOM+2007}, whose ultimate goal is the creation of the
final Gaia catalogue with a targeted release date around 2021.

This section gives a concise overview of the architectural design
of the AGIS software (Sect.~\ref{sec:agissw-overview}) on the one hand,
and on the other presents some selected results from a recently (June--November 2011) 
conducted large-scale astrometric solution (Sect.~\ref{sec:agis-results}) using as
input simulated data for more than 2~million sources. 
While this number is still a factor 50 smaller than the number of primary
sources foreseen in the final AGIS runs (around 2018--2020), and the present
simulations are much simplified especially with respect to the attitude modelling,
we believe that this proof-of-concept run demonstrates the practical validity 
and correctness of the key theoretical concepts described in this paper. 
Future tests will involve simulated data sets that are both larger and more
realistic, with a parallel further development of the algorithms and software in 
view of the added new complexities.

\begin{figure}[t]
\begin{center}
\includegraphics[scale=.7]{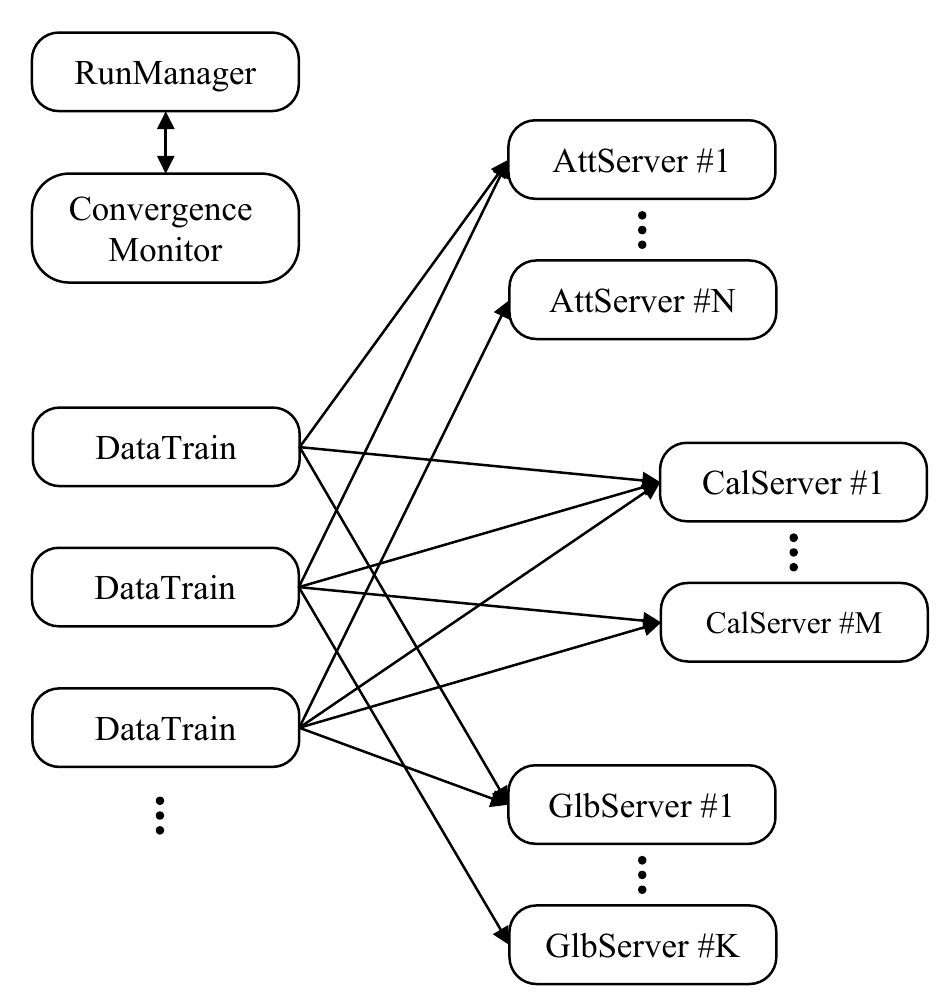}
\caption{Simplified architectural design diagram of AGIS. The rounded boxes 
are independent Java processes running
in parallel on different nodes of a multi-CPU processing cluster.
The arrows indicate main data exchanges between the various processes.
Input/output-related data flow from/to the storage system is not shown,
and likewise some important but conceptually irrelevant interactions between 
some elements (e.g., the \emph{Server}s and the \emph{RunManager}).
\label{fig:agis-design}}
\end{center}
\end{figure}

\subsection{AGIS software overview\label{sec:agissw-overview}}

The term AGIS is subsequently often used to refer to the actual software
implementation of the scheme. Like virtually all Gaia data processing software,
AGIS is entirely written in the object-oriented Java programming language
\citep{2010ASPC..434..135O}.
The implementation has been briefly outlined in \citet{Lammers+09} and 
more comprehensively in \citet{WOM2011}, to which the interested reader is
referred for more details. In this section, some key classes are briefly 
described; following Java naming conventions their names (given in italics) 
are concatenated capitalized nouns, as in \emph{RunManager}. 

Owing to the number of sources and the associated large data volumes that 
have to be handled (see Sect.~\ref{sec:intro}) it is clear that a well-performing
system must be distributable on modern, multi-node, multi-core processing
hardware environments and make optimal use of parallelism as far as permitted
by the AGIS scheme. Another elementary consideration is that inherently
slow disk input-output operations should be minimized and never allowed to 
be a bottleneck for any of the computing processes.

These basic requirements have led to the system schematically depicted
in Fig.~\ref{fig:agis-design} with its main components and data flow.
The central elements are \emph{DataTrain}s, \emph{Server}s, a \emph{RunManager},
and a \emph{ConvergenceMonitor}. When an AGIS run starts the \emph{RunManager}
splits the entire processing task of the first iteration into
separate and independent jobs which are then taken and executed in
parallel by \emph{DataTrain}s that have been started on the different CPUs 
of the processing system.
Each such job involves the processing of all observations for a group
of sources (with typically 100--1000 sources per group), which is done 
by looping over the sources, one at a time.
Before the loop is entered all the data that are needed for the processing
(observations, as well as all source, attitude, and calibration data) have 
been loaded into memory and are
passed on to the core algorithms. This is a key design aspect
which, together with a suitable grouping of the data on the storage system,
ensures that the system is never input/output limited and that it has a 
constantly high utilization of the CPUs (typically at the 90\% level).

Each AGIS iteration starts with the updating of the respective
sources (see Sect.~\ref{sec:sourceupdate}). Computed provisional
updates are then written to the storage system and, finally,
the observations together with the updated source data are sent
to attitude, calibration, and global servers via the CPU-interconnecting
network. The servers themselves are distributed in different ways
(see, e.g., Sect.~\ref{sec:att-segment} for attitude), but all are
similar in that they accumulate normal equations by adding observation 
equations as outlined in Sects.~\ref{sec:att-normals}, \ref{sec:calupdate}, and
\ref{sec:globalupdate}, respectively.
When all \emph{DataTrain} jobs are finished, the \emph{RunManager}
signals to the servers that all observations have been sent.
This triggers that a Cholesky decomposition (Appendix~\ref{sec:cholesky}) 
is made of the accumulated normals matrices in every server, that the 
partial results on different servers are combined where necessary 
(e.g., the attitude segments are joined), and finally that the results are
persisted in the storage system. That marks the end of an iteration.
Subsequent iterations are then started in the same manner as the first
one, viz., through the creation of a set of processing jobs.

Iteration $k$ uses the computed updates from iteration $k-1$ and
generates updated parameters for use in the following iteration $k+1$.
This progress is monitored by the \emph{ConvergenceMonitor} through
the accumulation of a selected list of statistical quantities in the form of
graphical plots (e.g., histograms of the updates of all the astrometric parameters)
accessible in real time through a web interface. Naively, one may expect
that the system can be considered converged if the updates become
smaller than a pre-defined limit; however, finding an unambiguous
and automatically verifiable convergence criterion
has proven to be a surprisingly complex problem
\citep[see Sect.~4.4 in][]{bombrun+10}.
We believe now that human inspection is indispensable to assess the
convergence status of the system reliably and, ultimately, decide on
the termination of the iterative loop.

An important feature of the \emph{RunManager} is the ability to use different 
algebraic solution methods by selecting among different available iteration schemes. 
The description above explains the `kernel' computation of provisional updates 
to the unknowns employing
the Gauss-Seidel-type preconditioner approximation
(Eq.~\ref{eq:precondS(CAG)} in Sect.~\ref{sec:order})
to the full normal matrix of the system (Sect.~\ref{sec:iter}). How these 
provisional updates are actually combined at the end of an iteration to form 
the final updates to the unknowns depends on the chosen iteration scheme.
AGIS can use all four schemes outlined in Sect.~\ref{sec:CG}, viz., the  
simple iteration (SI), accelerated simple iteration (ASI), conjugate gradients (CG), 
and hybrid scheme (A/SI-CG). The previous description essentially refers to SI;
in the other schemes there are a few extra steps which however are immaterial 
for understanding the software system.

AGIS is controlled through a list of a few hundred key--value parameters
(`properties') which are configured before a run starts. Examples are:
the numbers of servers and threads, the size of \emph{DataTrain} jobs,
the starting values for the unknowns, and the employed solution method.
Also, which update blocks are active during a run is controlled
via properties. Any combination that involves at least a source update
is possible, e.g., SACG, SA, SCG.

The optimum number of data trains in a run is a complex trade-off between 
the available number of CPUs and memory, usable network bandwidth (more trains create
more inter-CPU traffic), and the given maximum storage system throughput.
By design of the system, the run time should scale inversely with the number of
data trains (assuming there are enough CPUs),
i.e., doubling the number of \emph{DataTrain}s should halve the run time.
In the tests done until now, the run times are very satisfactory;
however, more work remains for achieving the desired optimal scaling behaviour.
A first AGIS run using simulated data for a 5~yr mission with 50~million primary
sources was successfully completed in June 2011. This being only a factor 2 less than 
the baseline 100~million sources envisaged in the final AGIS run towards the
end of the mission, it marks an important
milestone in the development of the operational system.

\begin{figure}
\begin{center}
\includegraphics[width=0.98\columnwidth,height=0.46\columnwidth]{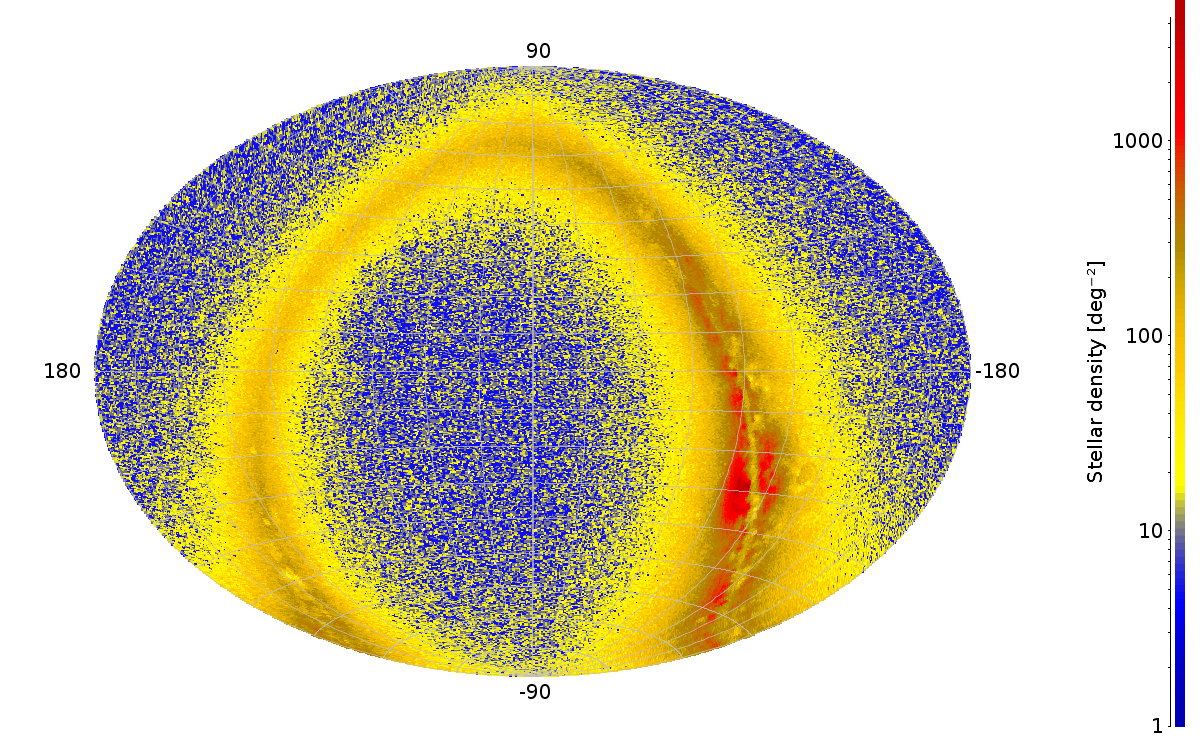}\\
\smallskip
\includegraphics[width=0.98\columnwidth,height=0.46\columnwidth]{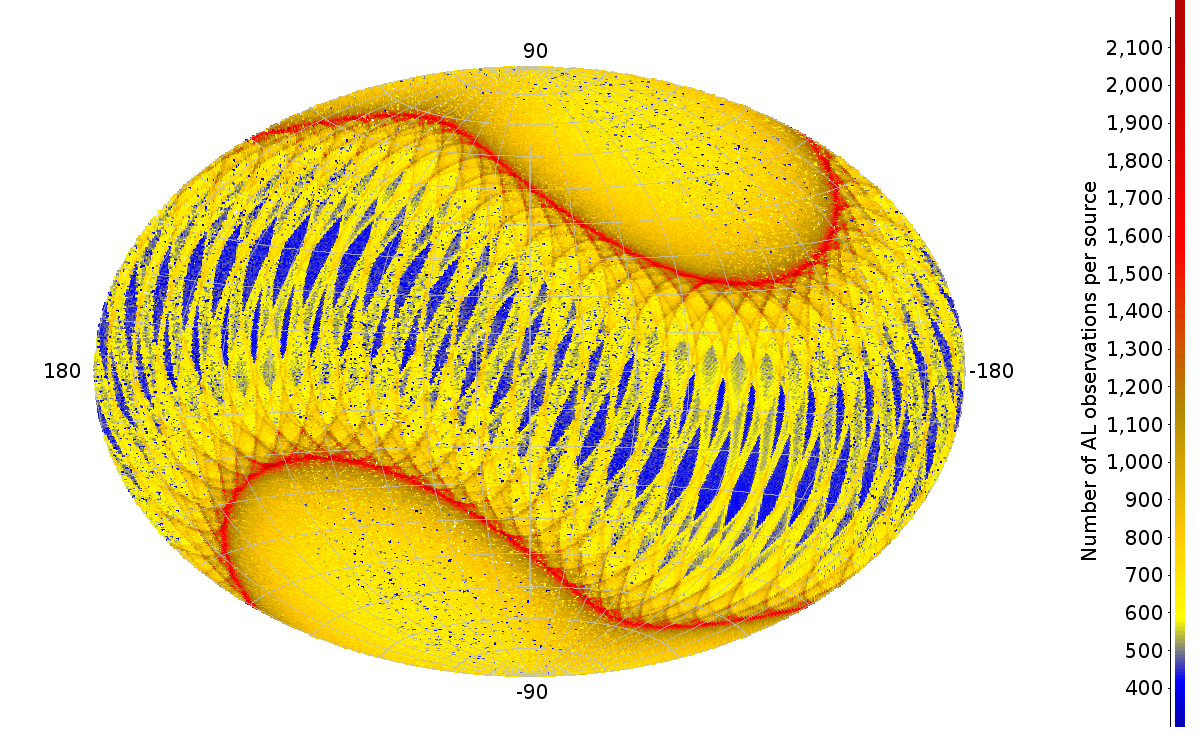}\\
\smallskip
\includegraphics[width=0.98\columnwidth,height=0.46\columnwidth]{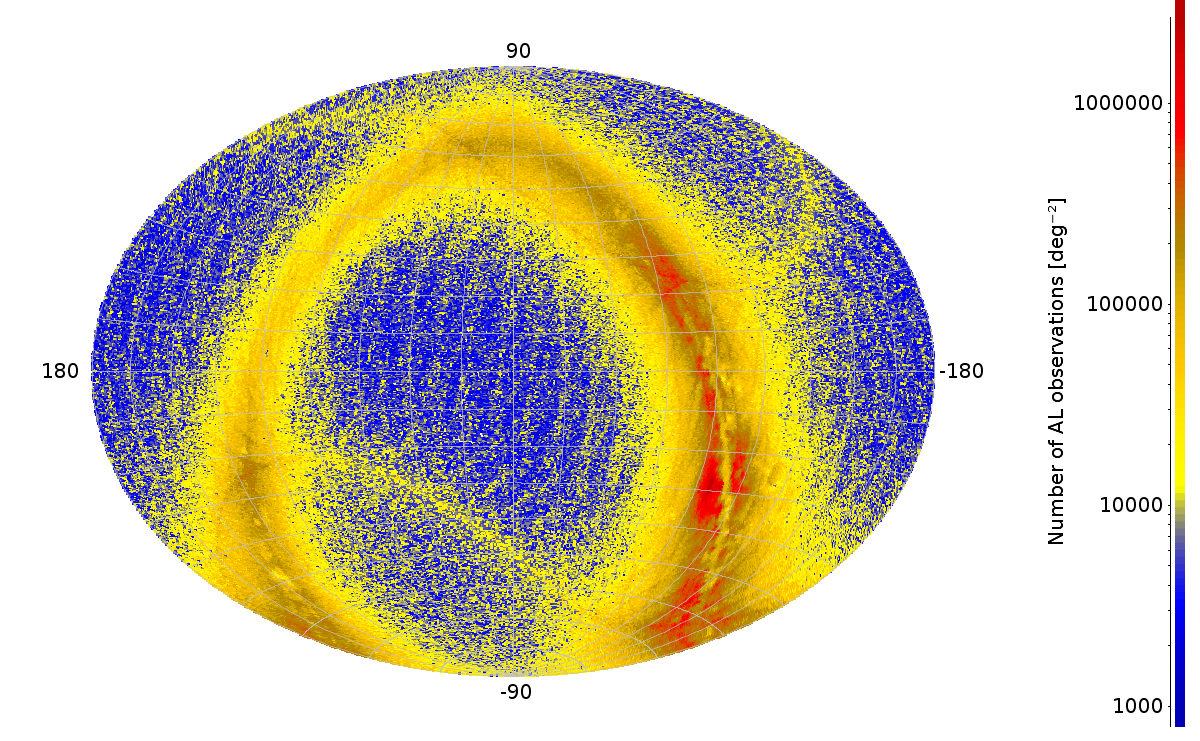}
\caption{All-sky projections of (from top to bottom) the total stellar density
in the input data to the demonstration solution, the number of AL observations per source, 
and the resulting spatial density of AL observations. 
These and all subsequent sky maps use the Hammer--Aitoff equal-area projection 
in equatorial coordinates, with $\alpha$ running from $-180^\circ$ to $+180^\circ$
right to left. {\bf Top:} The simulated sky contains some 2~million single stars 
covering the Gaia magnitude range 
$6\le G\le 20$. The density ranges from less than 1~deg$^{-2}$ around the 
galactic poles to a maximum of about 4800~deg$^{-2}$ near the galactic centre 
in the bottom-right quadrant of the map. {\bf Middle:} The number of along-scan
observations per source reflects the scanning law of Gaia, which is roughly symmetric
around the ecliptic plane and gives an over-abundance of observations at
ecliptic latitudes $\pm 45^\circ$. {\bf Bottom:} The combination of the source density
and the scanning law gives the displayed density of along-scan observations.}
\label{fig:res:dens-obs}
\end{center}
\end{figure}

\begin{figure}[t]
\begin{center}
\includegraphics[width=0.98\columnwidth,height=0.46\columnwidth]{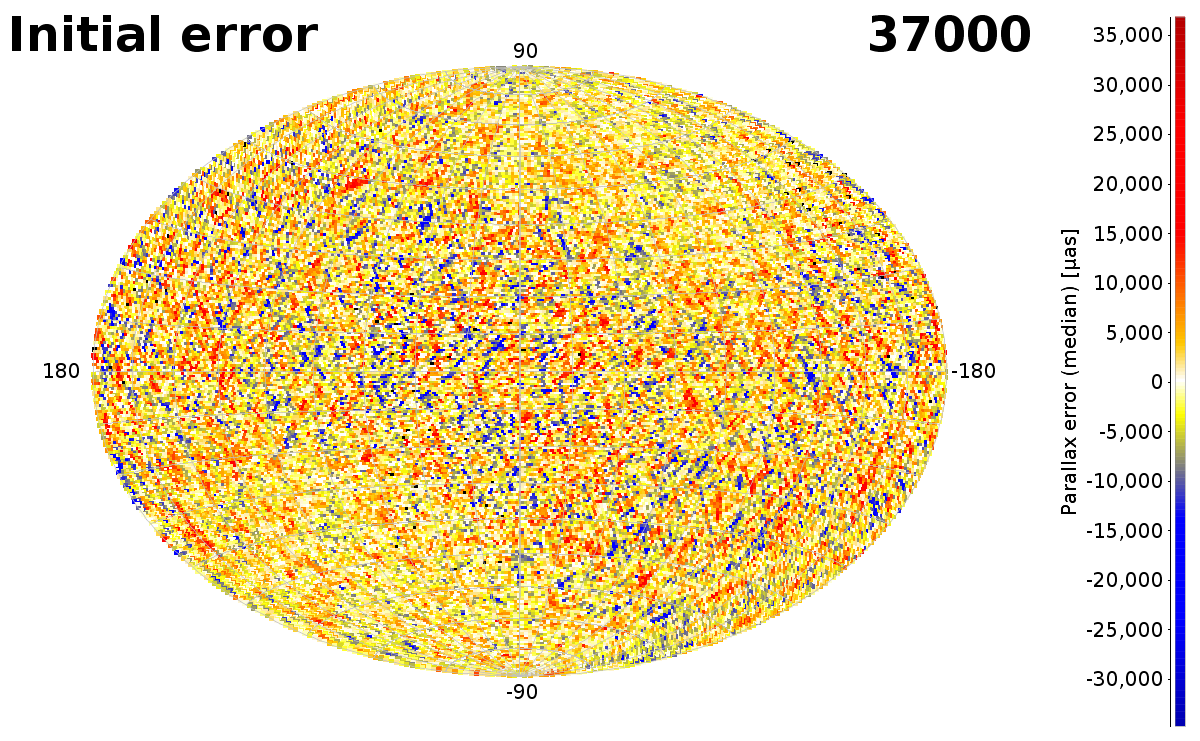}
\caption{Map of the parallax errors in iteration 1. The iterative astrometric solution
starts off with spatially correlated errors in the astrometric parameters, generated
as described in the text. These (initial) parallax errors have
amplitudes of a few tens of mas. The number at the top-right corner of this (and
subsequent) maps is the maximum value of the displayed range in $\mu$as.}
\label{fig:res:init}
\end{center}
\end{figure}

\begin{table}
\caption{Characteristics of the simulated input data and demonstration solution.}
\label{tab:sim}
\centering
\begin{tabular}{l r}
\hline\hline\noalign{\smallskip}
Quantity & Value \\
\noalign{\smallskip}\hline\noalign{\smallskip}
Duration of science mission & 5.0~yr \\
Number of sources & 2\,256\,222 \\
Number of along-scan (AL) observations & $1.625\times 10^9$ \\
Number of across-scan (AC) observations & $1.805\times 10^8$ \\
\multicolumn{2}{l}{Standard uncertainty per AL~/~AC observation (representative):}\\
\qquad $G\le 13$ &  92~$\mu$as~/~\phantom{00}520~$\mu$as \\
\qquad $G=15$ &   230~$\mu$as~/~\phantom{0}1350~$\mu$as \\
\qquad $G=17$ &   590~$\mu$as~/~\phantom{0}4000~$\mu$as \\
\qquad $G=18$ &   960~$\mu$as~/~\phantom{0}7600~$\mu$as \\
\qquad $G=19$ & 1600~$\mu$as~/~\phantom{}16000~$\mu$as \\
\qquad $G=20$ & 2900~$\mu$as~/~\phantom{}38000~$\mu$as \\
Number of astrometric parameters & $1.128\times 10^7$ \\
Number of attitude spline knots & $6.575\times 10^5$ \\
Number of attitude parameters & $2.630\times 10^6$ \\
Number of calibration parameters & 7\,812 \\
Number of global parameters & 1 \\
\noalign{\smallskip}\hline
\end{tabular}
\end{table}

\subsection{Demonstration solution\label{sec:agis-results}}

\subsubsection{Data simulation and model assumptions}\label{sec:simu}

Since the start of the development in early 2006, AGIS has been
tested continuously using simulated datasets of varying
complexity and size generated by the Gaia System Simulator
\citep{2005ESASP.576..457M} created by DPAC's dedicated coordination 
unit for Data Simulations \citep[CU2;][]{iaus248:FM,2011EAS....45...25L}. 
In the following we present the results of a test solution using 5~years of
simulated astrometric observations for about 2~million well-behaved (single) 
stars with a realistic distribution both in magnitude and coordinates,
based on the Besan\c{c}on galaxy model \citep{robin2003}.
Figure~\ref{fig:res:dens-obs} (top) shows the spatial source density of the
data set in equatorial coordinates. Of particular interest for the
AGIS run is the stark density contrast between $\sim\,$1 and 5000~deg$^{-2}$
mainly depending on galactic latitude, resulting in similarly high ratios in
the total astrometric weight of the sources in Gaia's two fields of view.
In \citet{bombrun+10} it was shown (using simulations on a smaller scale)
that a high weight contrast tends to reduce the convergence rate of the
astrometric solution compared to a situation where
the weights are more balanced; however, the solution always converges
to the correct solution provided that enough CG iterations are carried out.
We will show that this key result is confirmed in the demonstration solution.

The input data were generated using a fully-relativistic model of the observed
(proper) directions $\vec{u}_i(t)$ in Eq.~(\ref{eq:astro}), including gravitational
light deflection for PPN parameter $\gamma=1$, assuming the Nominal Scanning Law
(Sect.~\ref{sec:attmodel}), the nominal geometrical instrument model 
(Sect.~\ref{sec:instrumentmodel}) and nominal performance of the instrument 
(in particular the centroiding accuracies $\sigma_l^\text{AL}$, $\sigma_l^\text{AC}$ 
as functions of $G$; see Table~\ref{tab:sim}). However, in order to
test the capability to recover a varying basic angle, a step-wise perturbation 
was introduced corresponding to the sinusoidal variation of 
$\Delta\Gamma_j$ in Eq.~(\ref{eq:DeltaGamma}) with a period of 2.5~yr and
an (unrealistically large) amplitude of 500~$\mu$as. In the astrometric solution,
the large-scale AL calibration interval was set to 30~days, matching the step 
width of the perturbation signal. The solution used only one global parameter,
viz., $g_0=\gamma-1$ as described in Sect.~\ref{sec:globalupdate}.
Table~\ref{tab:sim} lists some statistics of the data
and solution, while the middle and bottom maps in Fig.~\ref{fig:res:dens-obs} 
show how the number of observations varies across the sky.

Not all elements of the numerical algorithms described in this paper have as yet 
been integrated into the running software system which otherwise implements
the basic model described in Sect.~\ref{sec:math}.
In particular, the estimation of source and attitude excess noise 
(Sects.~\ref{sec:source-inner} and \ref{sec:attNoise}) were not activated in 
the present solution. This was not a problem for the demonstration run, as the 
applied observation noise is well-behaved, purely Gaussian. Further
development of the AGIS software will to a large extent focus on making the 
solution robust against all kinds of unexpected input data.

\begin{figure}[t]
\begin{center}
\includegraphics[width=.9\columnwidth]{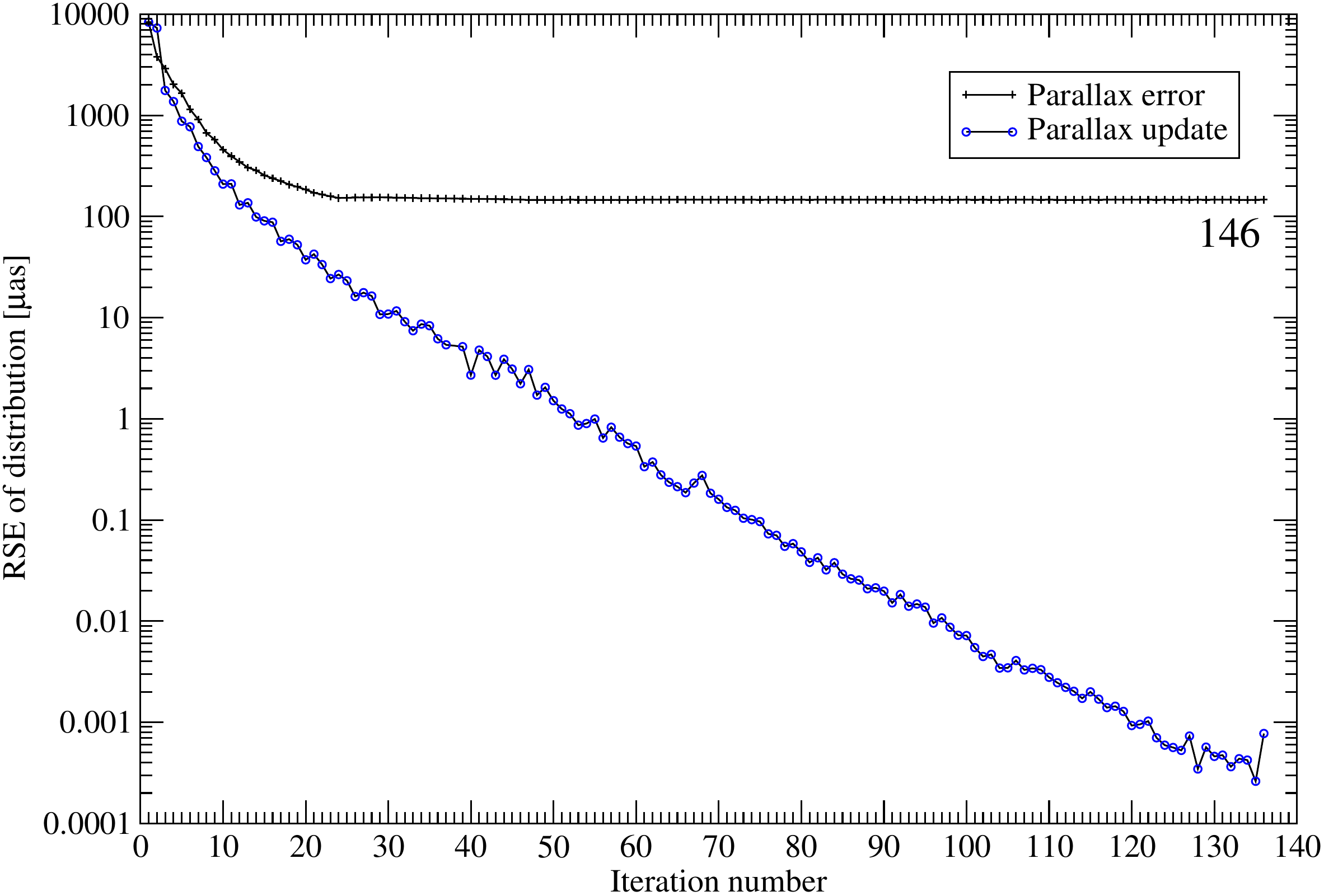}\\[6pt] 
\includegraphics[width=.9\columnwidth]{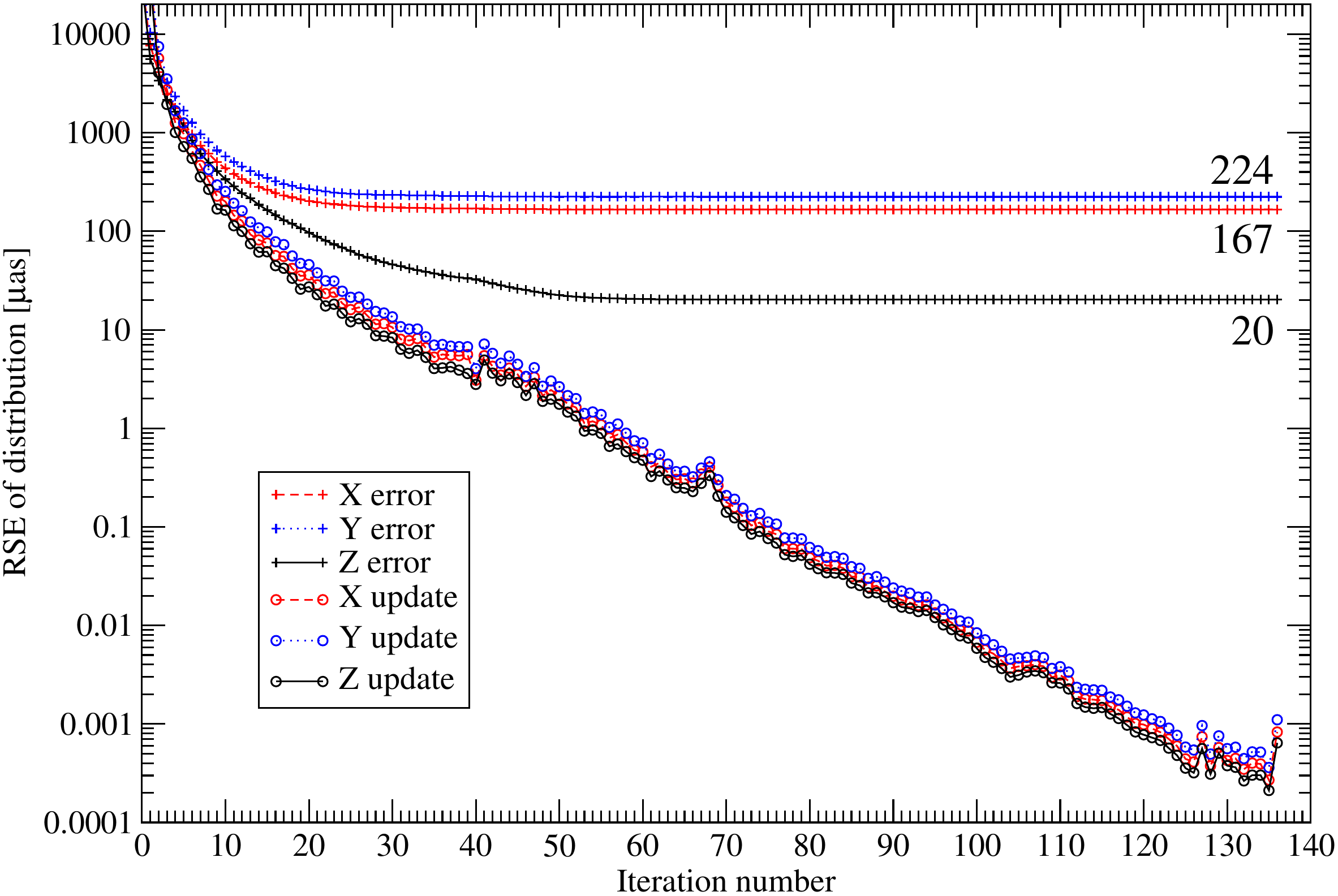}
\caption{{\bf Top:} Evolution of the typical parallax error (crosses) and 
parallax update (circles) as functions of the iteration number. The 
typical error settles at around 146~$\mu$as.
{\bf Bottom:} Evolution of the typical attitude error (crosses) and update 
(circles) as functions of the iteration number, for the three principal SRS axes
$x$ (red), $y$ (blue), and $z$ (black). The errors settle at around
167, 224, and 20~$\mu$as, respectively. In both these plots the typical
errors and updates are given by the Robust Scatter Estimate (RSE), similar
to an RMS value (see footnote~\ref{footnRSE}).
}
\label{fig:conv}
\end{center}
\end{figure}

The demonstration solution was run with all four update blocks (S, A, C, G) enabled,
using starting values for the attitude, calibration and global parameters that were
erroneous on the mas level. 
(As explained in Sect.~\ref{sec:iter} and footnote~\ref{footnote10},
the results of the subsequent iterations are independent of the initial values for the
source parameters, because the updating always starts with the source block.) These
initial values were created by adding Gaussian, uncorrelated errors to the
true attitude parameters, with a standard deviation of 50~mas, using the nominal 
calibration parameters (i.e., excluding the sinusoidal modulation of the basic angle), 
and a value of $g_0\equiv\gamma-1=0.1$ (Sect.~\ref{sec:globalupdate}).
An attitude knot interval of 240~s was used in order 
to have a sufficient number of observations per degree of freedom even at the galactic 
poles. This interval is short enough that the attitude splines are able to represent the
true attitude (i.e., the analytical nominal scanning law) with an RMS error 
of less than $9~\mu$as. Although this is larger than the modelling errors aimed 
at in the real data analysis, it is sufficiently small in comparison with the typical 
attitude estimation errors ($\ge 20~\mu$as; see Fig.~\ref{fig:conv}) to have a negligible 
impact on the overall astrometric accuracy of the present solution.

During the source update in the very first iteration, the initial errors in the 
attitude, calibration parameters and $\gamma$ propagate to the sources, creating 
astrometric errors of a few tens of mas (Fig.~\ref{fig:res:init}) that are spatially correlated
on a scale comparable to the attitude knot interval ($\sim\,$4$^\circ$). These
errors are quite hard to remove in subsequent iterations, but may be representative 
of the situation encountered by AGIS when processing the real mission data based on
a fairly uncertain initial attitude and instrument calibration.

With these starting conditions, 135 iterations were carried out using the conjugate
gradients (CG) scheme. A re-initialisation of the CG scheme was made after the
first 40 iterations to avoid the development of a much slower convergence phase observed
in some previous runs; after this, no further re-initialisation was made. 
At iteration 135 the 
typical updates of, for example, the parallaxes and the along-scan attitude were
at or below a level of $5\times10^{-4}~\mu$as (Fig.~\ref{fig:conv}). This is still
slightly above the numerical noise floor set by the double-precision arithmetic
(${\sim}\,10^{-16}~\text{rad}$ or ${\sim}\,10^{-5}~\mu$as).
As discussed by \citet{bombrun+10}, truncating the iterations 
before the numerical noise floor has been reached implies the presence of spatially 
correlated `truncation errors' having an amplitude of a few times the typical updates, 
or $\sim\! 10^{-3}~\mu$as in the present case.
We now proceed with a more detailed analysis of the results.

\subsubsection{Source results\label{sec:ressrc}}

\begin{figure*}[t!]
\begin{center}
\includegraphics[width=0.98\columnwidth,height=0.46\columnwidth]{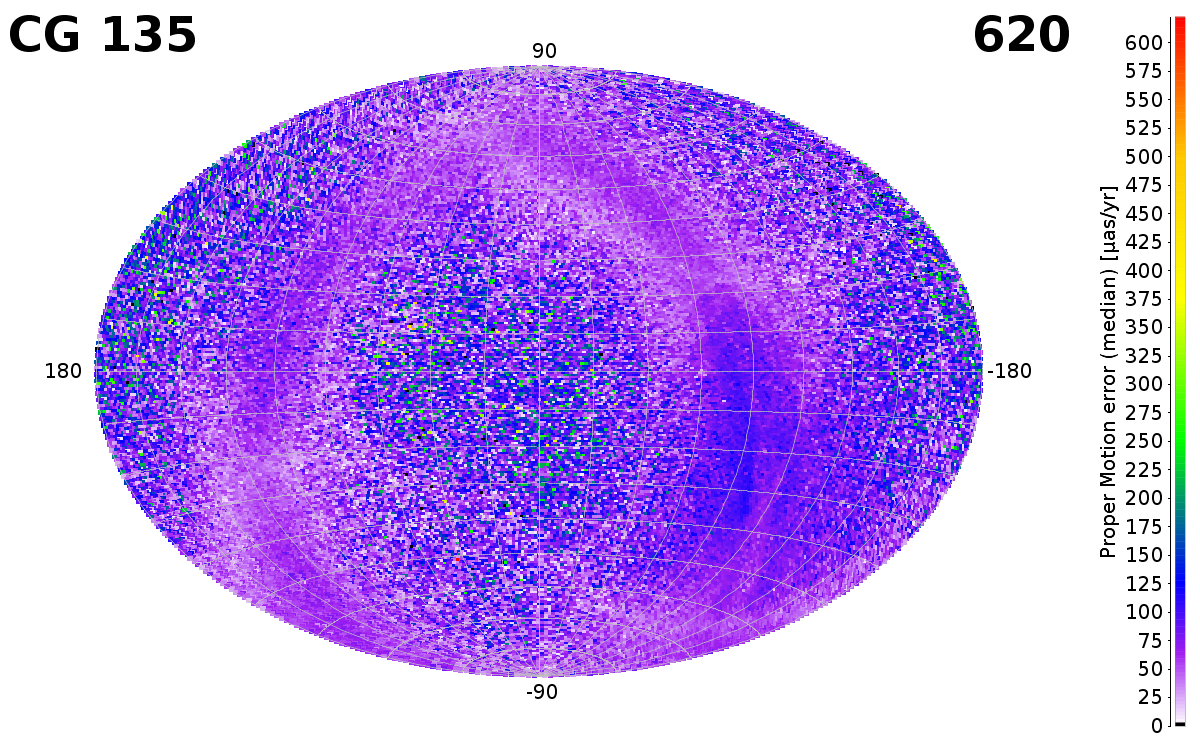}
\includegraphics[width=0.98\columnwidth,height=0.46\columnwidth]{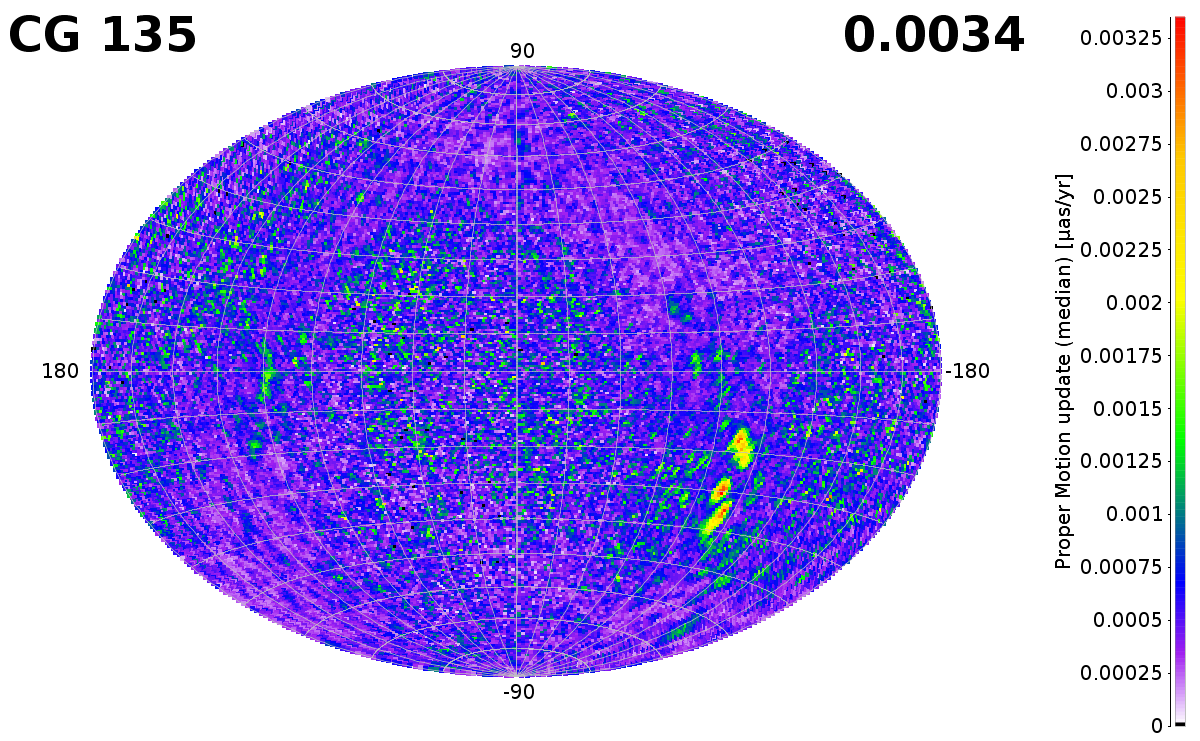}\\
\caption{Maps of the error (left) and update (right) in proper motion for iteration 135.
Each cell shows the colour-coded median error/update in units of $\mu$as~yr$^{-1}$,
where the individual error/update is computed in terms of the equatorial components
as $(\Delta\mu_{\alpha *}^2+\Delta\mu_\delta^2)^{1/2}$.}
\label{fig:res:pmmaps}
\end{center}
\end{figure*}

In this section we focus on the results for the parallax ($\varpi$), which is
arguably the most interesting parameter from an astrophysical viewpoint;
moreover, as a scalar quantity independent of epoch and reference frame, its 
statistics can be summarized compactly and without ambiguity. However, 
the behaviour of the position and proper motion parameters is qualitatively 
similar, and some results are given in Table~\ref{tab:res:varpistats} and 
Fig.~\ref{fig:res:pmmaps}.   

The top diagram in Fig.~\ref{fig:conv} shows the typical sizes of the
errors and updates in parallax versus iteration number, as measured by 
the Robust Scatter Estimate%
\footnote{The RSE is defined as 0.390152 times the difference between the
90th and 10th percentiles of the distribution of the variable. For a Gaussian
distribution it equals the standard deviation. The RSE is used as a standardized,
robust measure of dispersion in CU3.\label{footnRSE}} 
(RSE). The parallax errors settle relatively
quickly (around iteration 25) at a level of 146~$\mu$as and remain stable
till the end of the run with updates becoming successively smaller, reaching the 
level of $10^{-3}~\mu$as around iteration 120. 
The actual error distribution is symmetric
but strongly non-Gaussian (in fact more like a Laplace distribution) due to the
variation of star numbers and observational standard uncertainties with magnitude
(cf.\ Tables~\ref{tab:sim} and \ref{tab:res:varpistats}). The overall parallax error 
RSE of $\sim\,$150~$\mu$as is therefore representative for the median magnitude, 
$G\simeq 19$, in good agreement with current accuracy predictions 
\citep{iaus261:LL}.
The overall sizes of the errors and updates shown in Fig.~\ref{fig:conv} 
should however be complemented by more detailed statistics as functions of 
coordinate and magnitude.

\begin{figure*}[t!]
\begin{center}
\includegraphics[width=0.98\columnwidth,height=0.46\columnwidth]{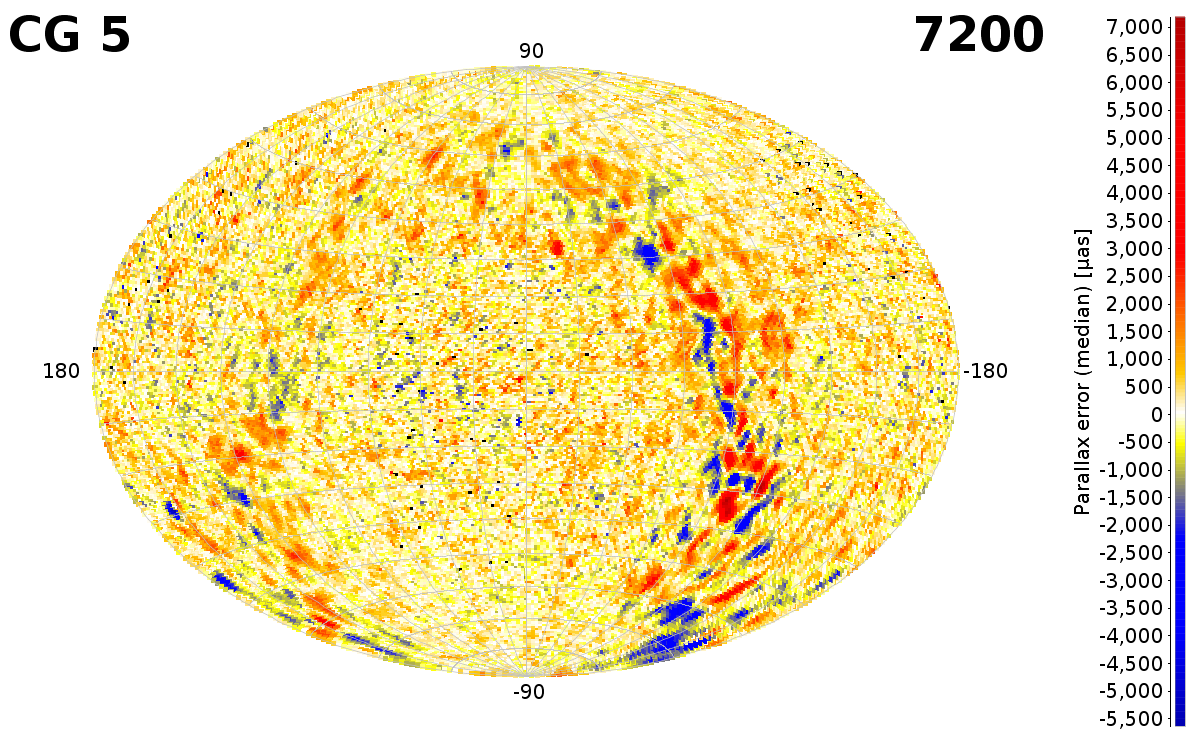}
\includegraphics[width=0.98\columnwidth,height=0.46\columnwidth]{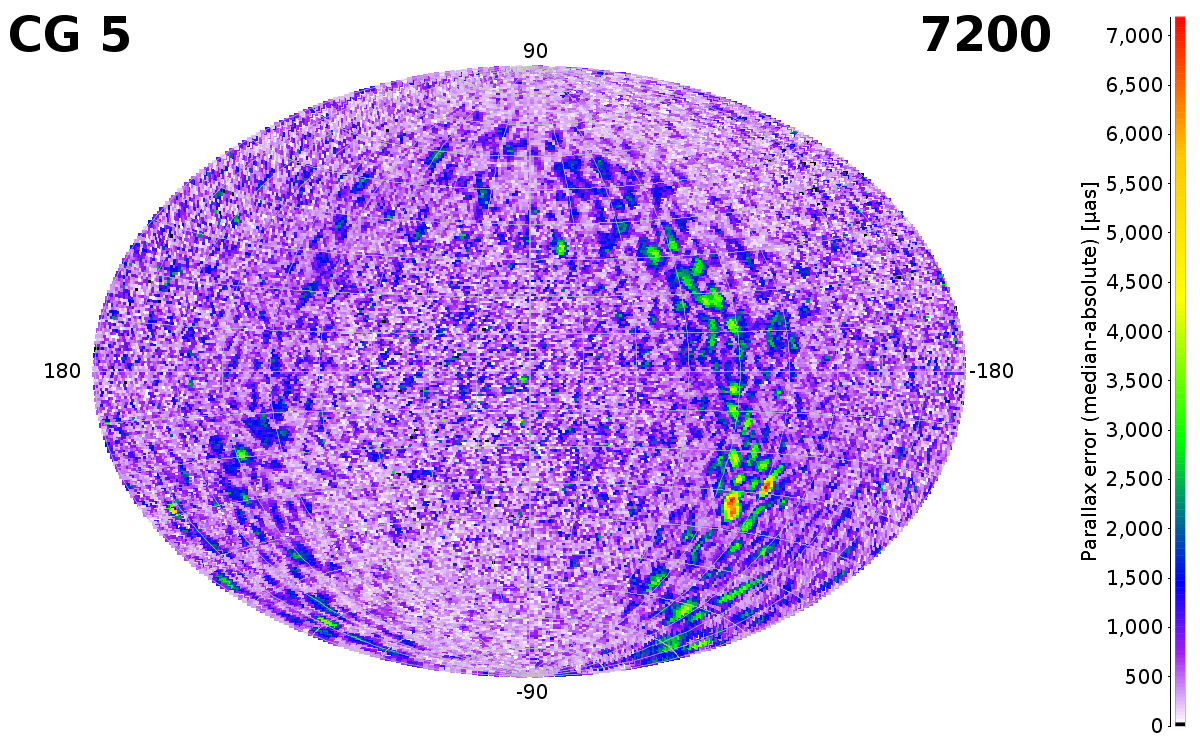}\\
\smallskip
\includegraphics[width=0.98\columnwidth,height=0.46\columnwidth]{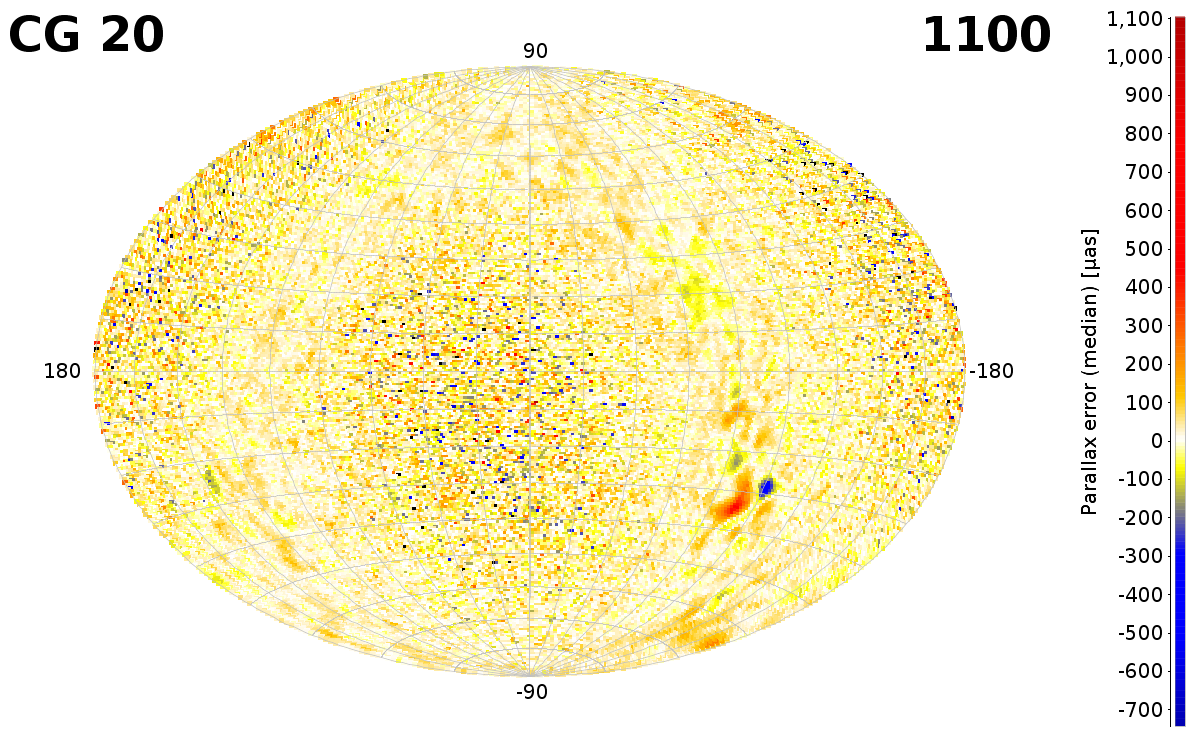}
\includegraphics[width=0.98\columnwidth,height=0.46\columnwidth]{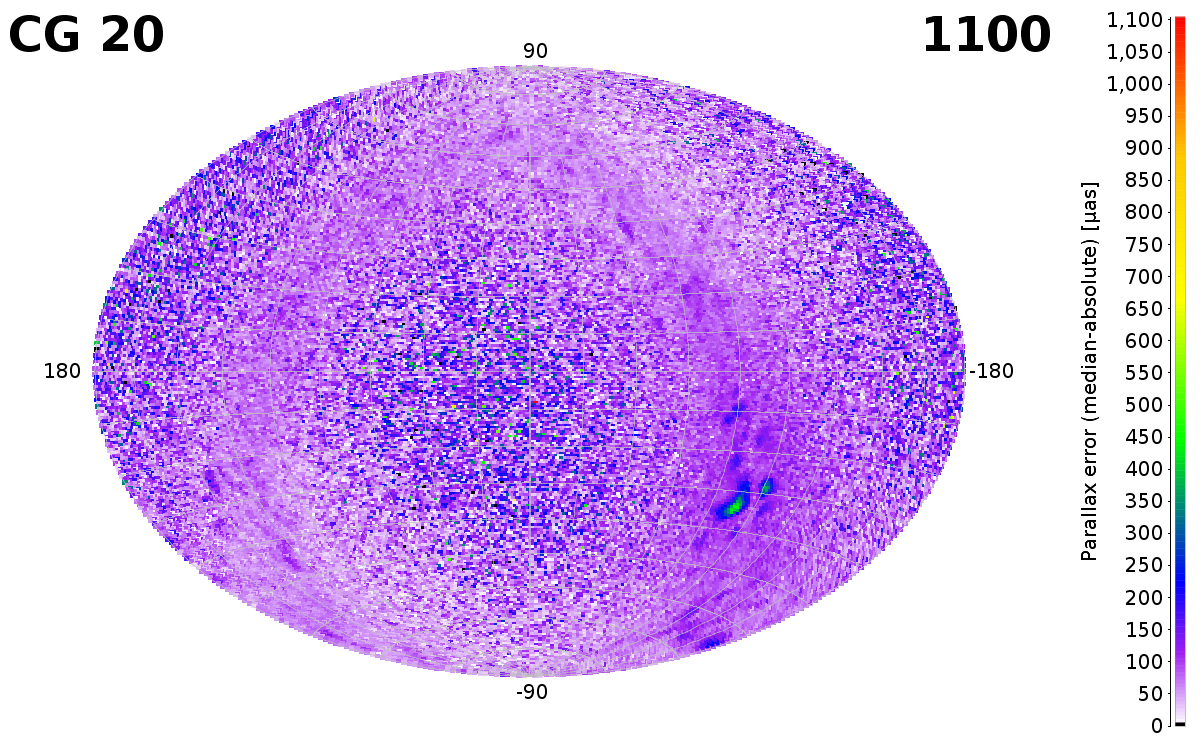}\\
\smallskip
\includegraphics[width=0.98\columnwidth,height=0.46\columnwidth]{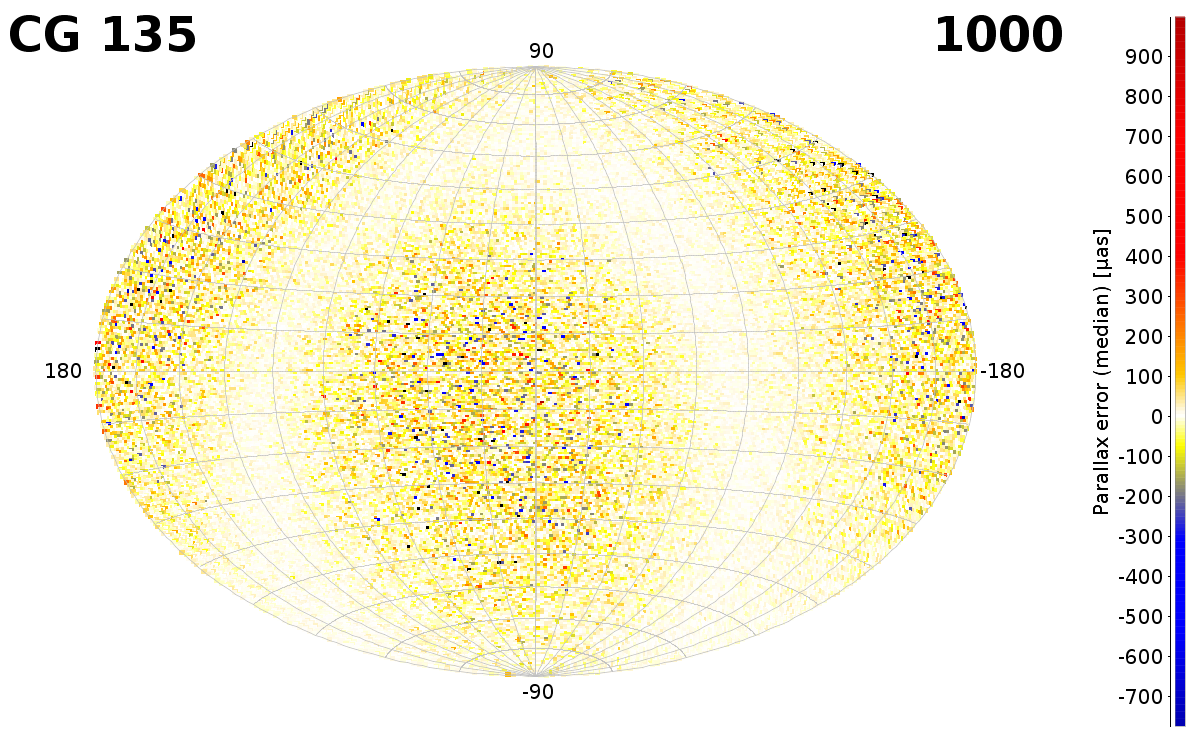}
\includegraphics[width=0.98\columnwidth,height=0.46\columnwidth]{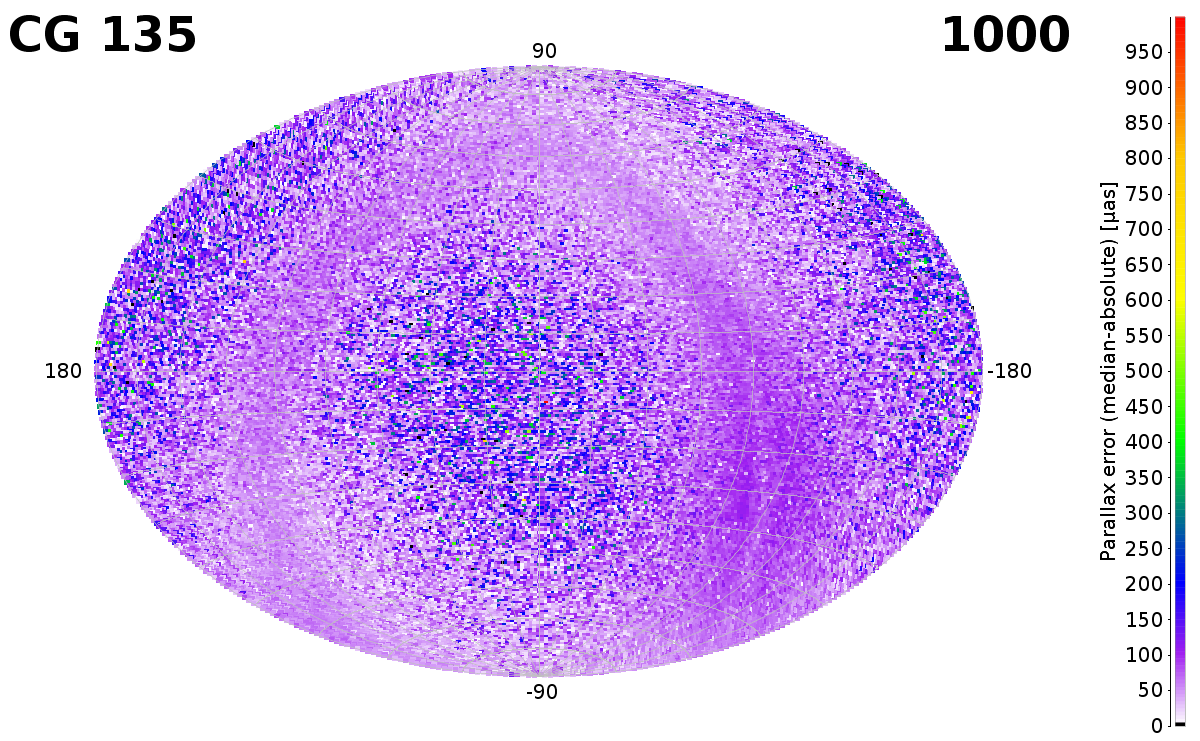}\\
\caption{Maps (in equatorial coordinates) of the parallax errors in the three selected iterations 
$k=5$, 20, 135 (top to bottom). The left column shows the median of the parallax error 
$\varpi^{(k)}-\varpi^\text{true}$, while the right column shows the median of the
absolute error $|\varpi^{(k)}-\varpi^\text{true}|$; each map cell (of about 0.84~deg$^2$)
shows the colour-coded value computed for the sources located in that cell. Empty cells
are shown in black.
On every map plot the top left label indicates the iteration number and the top right 
label is the maximum value of the displayed range in $\mu$as. See text for further details.
}
\label{fig:varpimaps-e}
\end{center}
\end{figure*}

\begin{figure*}[t!]
\begin{center}
\includegraphics[width=0.98\columnwidth,height=0.46\columnwidth]{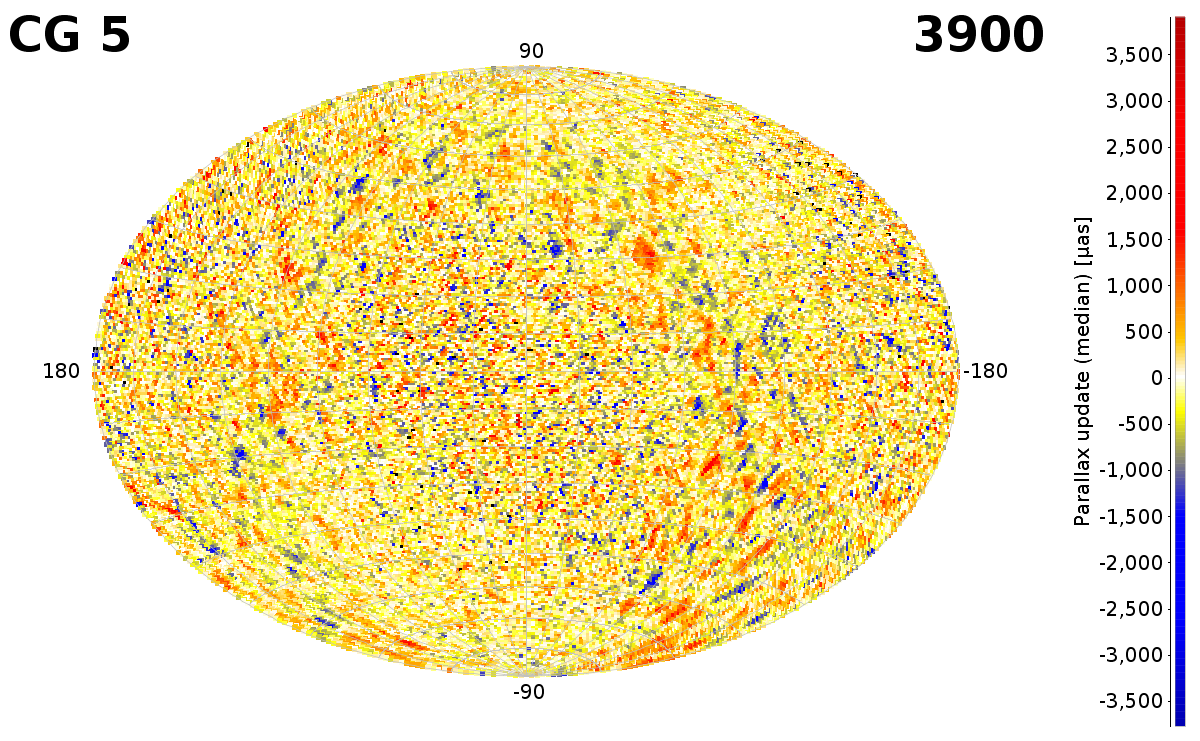}
\includegraphics[width=0.98\columnwidth,height=0.46\columnwidth]{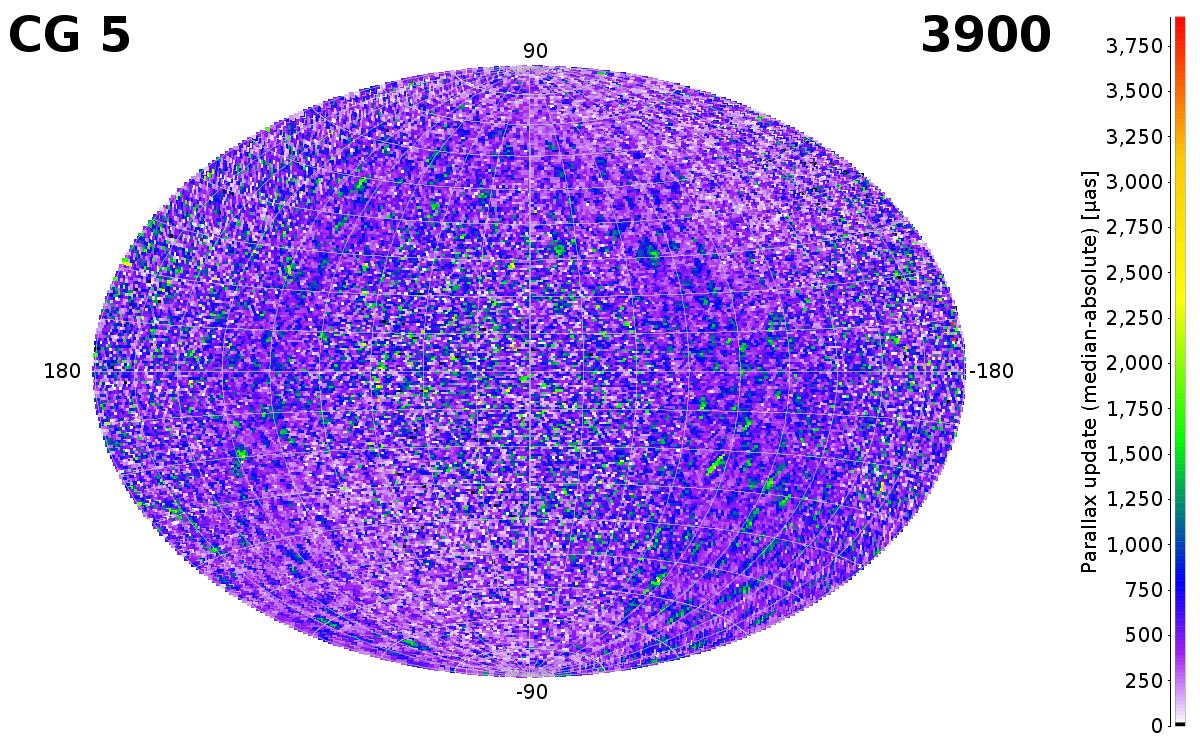}\\
\smallskip
\includegraphics[width=0.98\columnwidth,height=0.46\columnwidth]{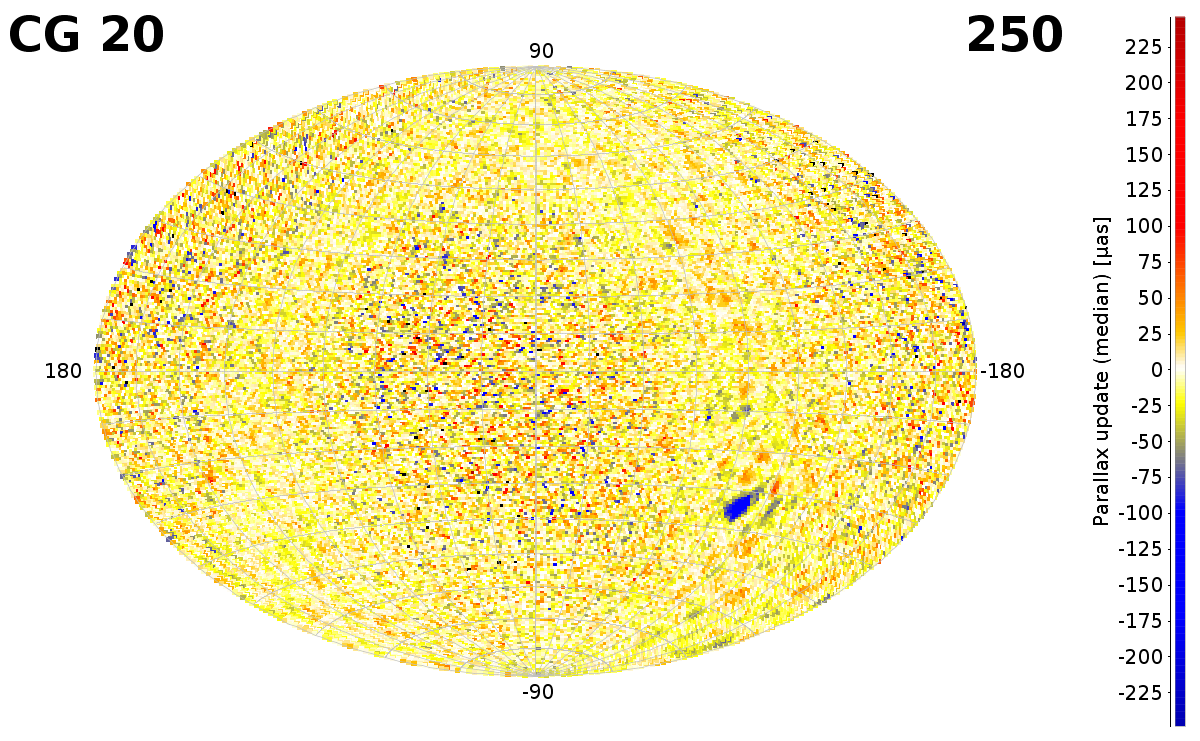}
\includegraphics[width=0.98\columnwidth,height=0.46\columnwidth]{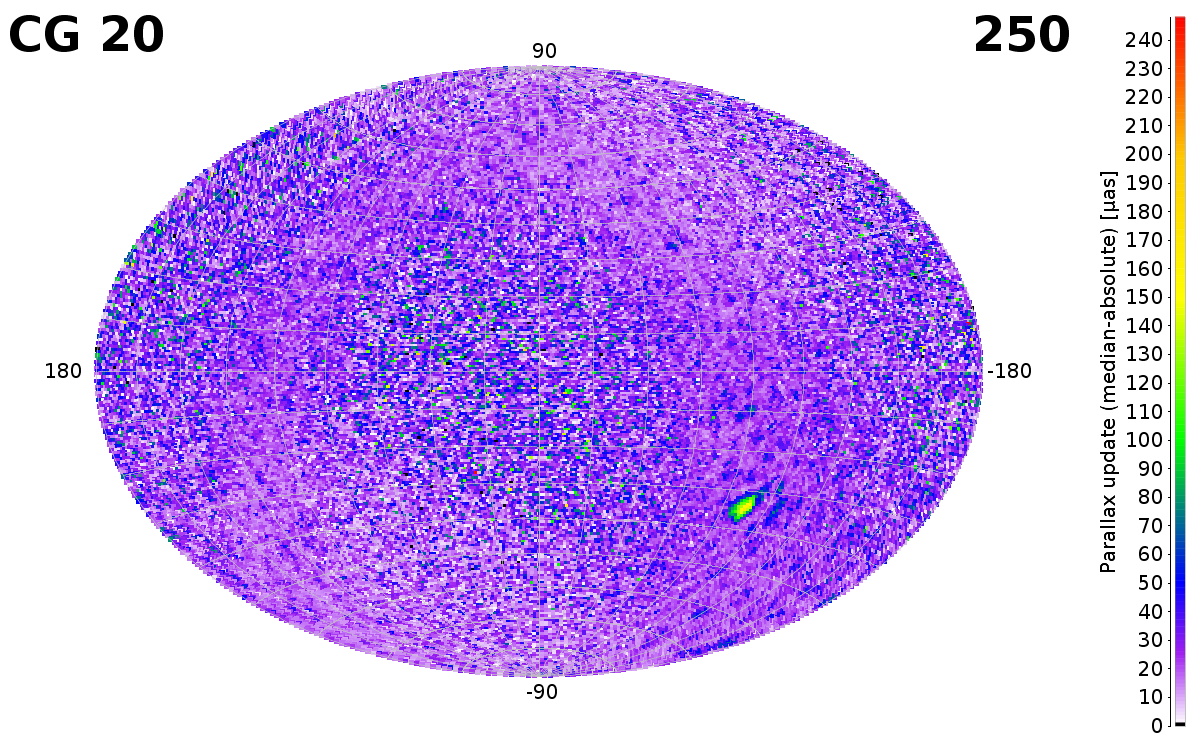}\\
\smallskip
\includegraphics[width=0.98\columnwidth,height=0.46\columnwidth]{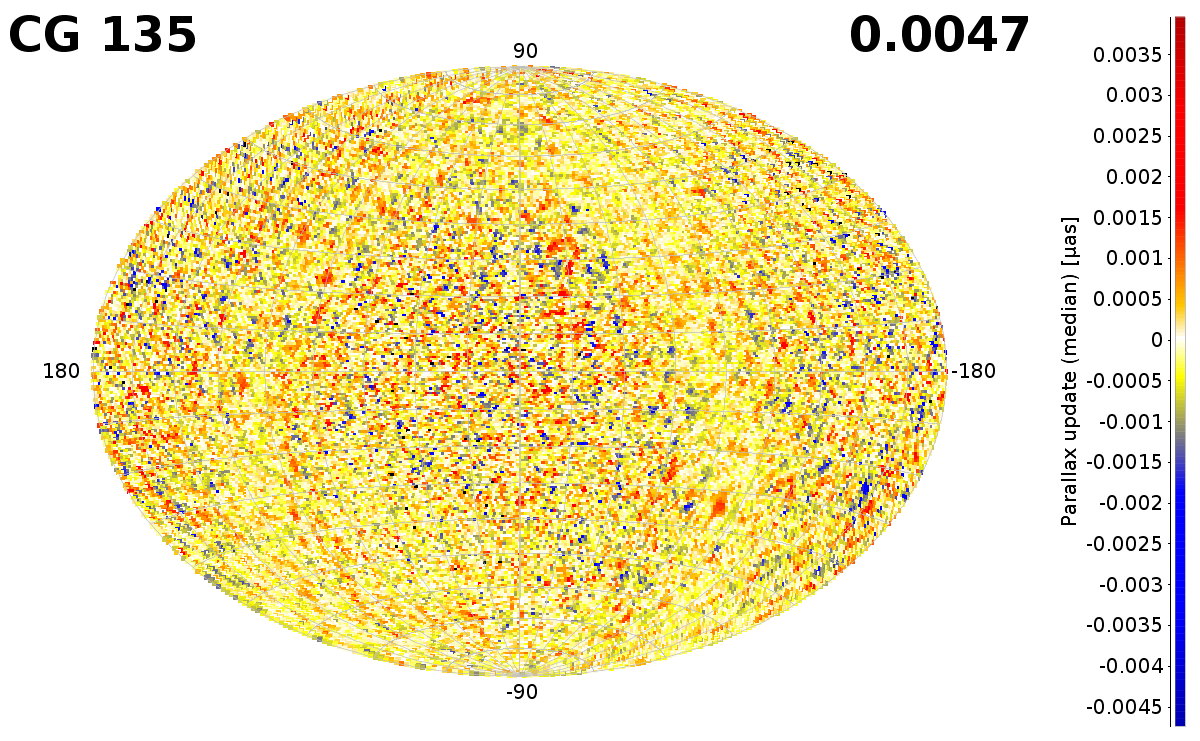}
\includegraphics[width=0.98\columnwidth,height=0.46\columnwidth]{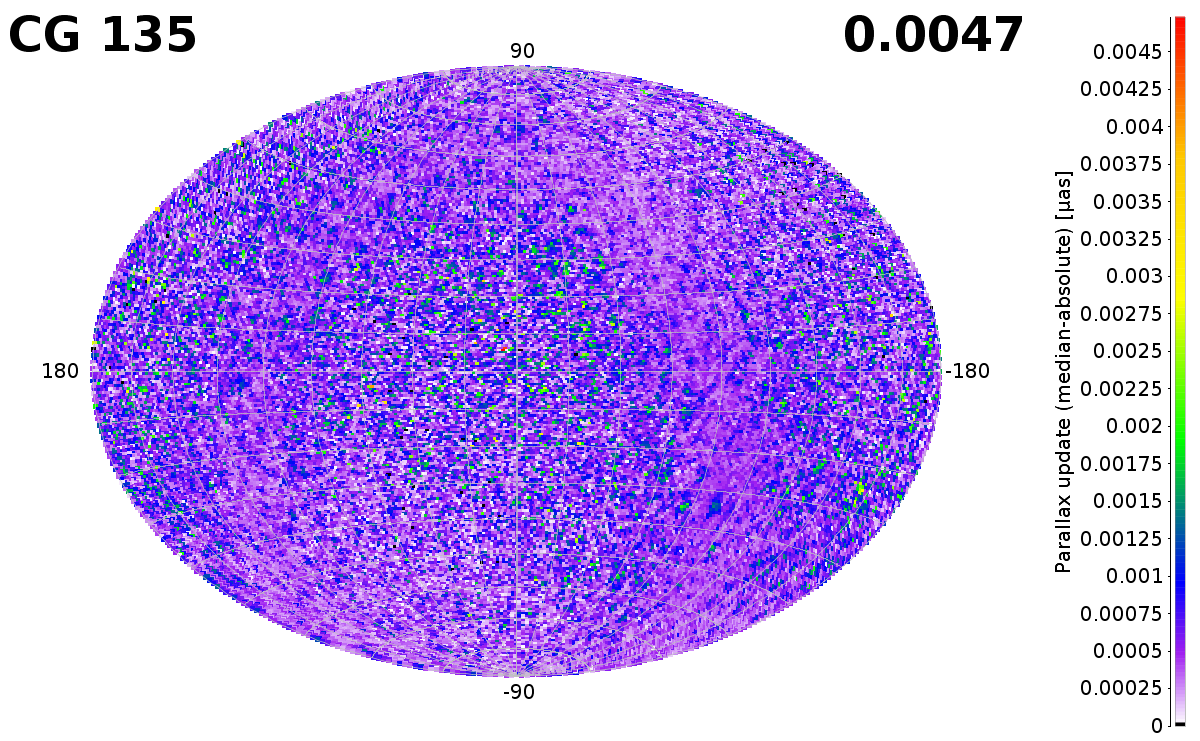}\\
\caption{Same as Fig.~\ref{fig:varpimaps-e} but showing the updates in parallax, i.e., 
the median values of $\varpi^{(k)}-\varpi^{(k-1)}$ (left column) and 
$|\varpi^{(k)}-\varpi^{(k-1)}|$ (right column). 
}
\label{fig:varpimaps-u}
\end{center}
\end{figure*}

Figure~\ref{fig:varpimaps-e} shows the spatial distribution of the 
parallax errors at a few selected iterations. The left column shows the
median error in each cell, while the right column shows the median
absolute value of the error. These quantities serve as robust proxies for
the mean and RMS values (the RSE is not used for the latter as many cells
have too few sources for this measure), and therefore may suggest the
levels of systematic and random errors as function of position.   
Figure~\ref{fig:varpimaps-u} shows the corresponding maps for the
parallax updates.
After a few iterations, when the overall parallax errors are already below
the 1~mas level, very significant systematic (i.e., spatially correlated)
errors of a few mas remain, especially in the high-density areas of the
galactic plane. These are damped in the subsequent iterations, but still
remain at a level of several hundred $\mu$as in the galactic centre 
region around iteration 20, when the overall parallax errors start to settle 
at their final value according to Fig.~\ref{fig:conv}. At iteration 135 the
regional errors have virtually disappeared, and the error map shows a 
characteristic pattern with the solution being seemingly better around 
the galactic equator than in the polar regions. This is purely an effect of 
number statistics: in the galactic pole areas the cells contain rather few 
sources (often just a single source, in which case the displayed
values are simply the individual parallax errors), while closer to the 
galactic plane the scatter from one cell to the next is reduced by the
median-averaging over many sources. 
 
The sequence of error maps for iterations 5, 20, and 135 corroborates
a key result of \citet{bombrun+10}, viz., that by iterating long enough 
the system can cope with large spatial imbalances in the astrometric weights,
provided that the minimum source density allows a good attitude 
determination at all points.

The median absolute values of the parallax errors shown in the
right column of Fig.~\ref{fig:varpimaps-e} quickly settle in a large-scale
pattern that mainly reflects the expected variation of parallax accuracy
with ecliptic latitude \citep[see, e.g., Table~3 in][]{iaus261:LL}. This,
in turn, depends on the scanning law, i.e., on a combination of the number
of observations per source (see Fig.~\ref{fig:res:dens-obs}, middle) 
and the geometric configuration of the scans -- for example, the over-density
of observations at $\pm 45^\circ$ ecliptic latitude does little to improve the 
parallaxes, which are then mainly producing shifts in the AC direction,
while Gaia is primarily sensitive to the AL displacement. Number statistics
reduce the between-cell scatter of the median absolute values as well,
which accounts for the smoother appearance along the galactic equator.

The update maps in Fig.~\ref{fig:varpimaps-u} conform with expectations
and the preceding discussion. Of particular interest from a diagnostic 
viewpoint is the observation that the amplitude and spatial distribution of 
the median updates in the non-converged solution give a fair indication of 
the (systematic) truncation errors. This is obviously useful for assessing the 
state of convergence, as the update maps can be constructed for the real 
mission data as well (whereas the error maps are of course unknown).
For example, based on the median updates in iteration 20 (middle left
map of Fig.~\ref{fig:varpimaps-u}) one might correctly conclude that 
truncation errors of a few hundred $\mu$as remain, especially in the 
galactic centre region, as shown in the middle left map of 
Fig.~\ref{fig:varpimaps-e}. By the same reasoning it appears that truncation
errors at iteration 135 should be well below $1~\mu$as.

In tests using other datasets with lower density contrasts we have seen a more
rapid convergence. A similar slowdown in convergence for a non-uniform weight
distribution was observed and discussed in \citet{bombrun+10} based on small-scale
simulations. A likely mechanism for this slowdown is related to the
weight contrast problem discussed by \citet{2005A&A...439..805V} in connection
with the new reduction of the Hipparcos data.  
Any of the fields of view scanning through a 
high-density area (or an area with many bright stars) creates a strong astrometric 
weight imbalance between the two viewing directions, as the other field usually 
points to an area of the sky with much lower source density (or fainter stars). 
Errors in the along-scan attitude create correlated 
errors in the parameters of all sources observed at that time, whether they are
in the preceding or following field of view. If these source parameters are then 
used to correct the attitude, with little counterbalancing effect from the 
(relatively few) sources in the other field, the attitude may get only marginally 
improved, with the net effect of slowing down the damping and de-correlation
of the errors in the high-density areas. Thus, more iterations are needed compared
to a more weight-balanced situation.%
\footnote{In the new reduction of the Hipparcos data by \citet{book:newhip} 
the weight ratio was artificially damped in order to improve the connectivity between 
the two fields of view in high-contrast situations. We have found that this is not 
required for Gaia provided that the solution is iterated to convergence; see 
\citet{bombrun+10}.}  

Figure~\ref{fig:res:pmmaps} shows the error and update maps of the proper
motions at iteration 135. Since proper motions are vectors, the displayed
quantities are the median lengths of the vectorial differences, which are 
non-negative by definition. The maps are in expected agreement with the 
corresponding parallax ones concerning visible patterns and structures. 
A prominent feature absent in the case of the parallaxes is the lighter 
bands of relatively smaller proper motion errors around ecliptic latitudes
$\pm 45^\circ$ caused by the oversampling of these parallels by the scanning
law (cf.\ Fig.~\ref{fig:res:dens-obs}). In contrast to the parallax case, these 
observations do contribute to the determination of the proper motion,
especially for the component in ecliptic longitude.

\begin{table*}
\caption{RSE of the errors in the astrometric parameters and of the normalized errors (i.e., after division by the formal standard deviations) for different magnitude ranges. The asterisk in $\alpha *$ indicates that the errors are true arcs on the sky; cf.\ footnote~\ref{footn:ast}. See text for further explanations.\label{tab:res:varpistats}}
\centering
\begin{tabular}{lrrrrrrrrrrrrr}
\hline\hline
\noalign{\smallskip}
\multirow{3}{*}{Magnitude range} & \multirow{3}{*}{No.~stars} && \multicolumn{5}{c}{RSE of error [$\mu$as and $\mu$as~yr$^{-1}$]} && \multicolumn{5}{c}{ $\varrho=$~RSE of normalized error [unitless]}\\
\noalign{\smallskip}
\cline{4-8}  \cline{10-14}
\noalign{\smallskip}
& && \multicolumn{1}{c}{$\alpha*$} & \multicolumn{1}{c}{$\delta$} & \multicolumn{1}{c}{$\varpi$} & \multicolumn{1}{c}{$\mu_{\alpha *}$} & \multicolumn{1}{c}{$\mu_{\delta}$} && \multicolumn{1}{c}{$\alpha *$} & \multicolumn{1}{c}{$\delta$} & \multicolumn{1}{c}{$\varpi$} & \multicolumn{1}{c}{$\mu_{\alpha*}$} & \multicolumn{1}{c}{$\mu_{\delta}$}\\
\noalign{\smallskip}
\hline
\noalign{\smallskip}
~~~\hspace{2.2em}$G<13$ & 18\,253 && 6.6 & 5.7 & 7.5 & 4.5 & 4.0 && 1.507 & 1.470 & 1.369 & 1.443 & 1.425 \\
~~~$13\le G<15$ & 70\,355 && 12.4 & 10.6 & 14.9 & 8.7 & 7.5 && 1.112 & 1.100 & 1.082 & 1.098 & 1.100 \\
~~~$15\le G<16$ & 88\,116 && 20.2 & 17.3 & 24.9 & 14.3 & 12.3 && 1.024 & 1.024 & 1.022 & 1.032 & 1.024 \\
~~~$16\le G<17$ & 151\,639 && 30.8 & 26.7 & 38.4 & 21.8 & 19.0 && 1.010 & 1.010 & 1.010 & 1.014 & 1.008 \\
~~~$17\le G<18$ & 272\,424 && 49.4 & 42.8 & 61.8 & 34.8 & 30.4 && 1.004 & 1.006 & 1.003 & 1.002 & 1.003 \\
~~~$18\le G<19$ & 489\,253 && 83.3 & 70.7 & 104.1 & 58.9 & 50.8 && 1.001 & 1.001 & 1.002 & 1.003 & 1.002 \\
~~~$19 \le G$ & 1\,166\,182 && 167.9 & 140.0 & 207.6 & 118.5 & 100.2 && 1.001 & 1.000 & 1.000 & 1.001 & 1.000 \\
\noalign{\smallskip}
\hline
\noalign{\smallskip}
~~~all $G$ & 2\,256\,222 && 116.8 & 98.7 & 145.6 & 82.4 & 70.6 && 1.009 & 1.009 & 1.007 & 1.009 & 1.008 \\
\noalign{\smallskip}
\hline
\end{tabular}
\end{table*}

All spatial maps in Figs.~\ref{fig:varpimaps-e}--\ref{fig:res:pmmaps}
show median values computed from distributions with stars of all
magnitudes. A lot more information is contained in the magnitude-resolved
versions of the maps, which are not presented here for brevity.
Instead, Table~\ref{tab:res:varpistats} shows the RSEs 
(see footnote \ref{footnRSE}) of the errors and normalized errors (see below)
of all the astrometric parameters in iteration 135, subdivided according to 
magnitude. The normalized error is defined as the error (in the case of
right ascension, $\Delta\alpha * \equiv\Delta\alpha\cos\delta$) divided 
by the corresponding formal standard uncertainty obtained from the inverse of
the source normal matrix, i.e., by using the approximation in 
Eq.~(\ref{eq:cov}). As discussed in Sect.~\ref{sec:cov} this approximation
underestimates the true standard uncertainties of the astrometric parameters
by neglecting the contribution from the attitude uncertainty, which may 
be particularly important for the bright stars where the photon noise 
is relatively less important.
  
The RSE of the errors in Table~\ref{tab:res:varpistats}
are in reasonable agreement with recent mission accuracy assessments.
For example, compared with Eqs.~(5.2)--(5.5) in \citet{iaus261:LL} the 
present values are a few per cent larger for $G<15$, and up to some 20\%
smaller for the fainter stars. This suggests a more conservative photon-statistical
error budget in \citet{iaus261:LL}, combined with a larger-than-nominal contribution
from the attitude uncertainty in the present solution. The latter effect, further 
discussed in Sect.~\ref{sec:resatt}, is indeed to be expected in the present 
demonstration solution using far fewer primary stars than planned for the real mission.

A more stringent test of the quality of the solution is obtained by 
considering the RSE of the normalized errors (i.e., after division by the formal
standard uncertainties as described above), which is here denoted
$\varrho$. Ideally we should have $\varrho\simeq 1$ for any parameter and
any magnitude. As shown in Table~\ref{tab:res:varpistats}, this is very nearly
the case in all parameters for $G>15$, meaning that the actual 
errors are roughly consistent with the standard uncertainties computed from
Eq.~(\ref{eq:cov}). For the brighter stars $\varrho$ becomes progressively larger,
supporting the interpretation that the attitude uncertainty has a significant
impact on the accuracy of the bright stars in this solution. For example, 
the value $\varrho=1.369$ obtained for the parallaxes of stars brighter than
$G=13$ suggests a quadratic attitude contribution to the parallax errors for 
these stars of $[7.5^2-(7.5/1.369)^2]^{1/2}=5.1~\mu$as, while a similar 
computation for the next three magnitude bins gives 5.7, 5.1 and 
5.4~$\mu$as (with a rapidly increasing uncertainty). Thus it appears that the
values $\varrho>1$ found for the parallaxes of the brighter stars can be 
accounted for by assuming a constant contribution, by about 5--6~$\mu$as RMS,
to the parallax errors from attitude and/or calibration errors. For the other
astrometric parameters we similarly find a constant RMS contribution to the 
positional errors of about 4--5~$\mu$as, and to the proper motion errors of about 
3--4~$\mu$as~yr$^{-1}$. It will be shown below
that these numbers are consistent with the actual attitude errors found in the solution. 

It should be noted that the reference epoch for the astrometric parameters
was set to exactly half-way into the mission, i.e., $t_\text{e}=2014.5$ for
the simulated mission interval 2012.0--2017.0. This is optimal in the sense 
that the positional uncertainties are minimized for approximately this epoch, 
and that the errors in position and proper motion are nearly uncorrelated. 
A reference epoch half-way through the mission was also assumed for the 
accuracy assessment in \citet{iaus261:LL}, with which the present results
have been compared.

In this solution, the median value of the parallax errors of all the 2.2~million 
sources was not significantly different from zero (the actual value was $+0.004~\mu$as).
This shows that, in the absence of systematic observational errors, the astrometric 
solution is able to determine the \emph{absolute} parallaxes of the sources, as
could be expected from the basic principles of the mission. It is especially worth
noting that this was achieved while simultaneously determining the PPN $\gamma$ 
parameter, known to be strongly correlated with the parallaxes 
(cf.\ Sect.~\ref{sec:globalupdate}).

\subsubsection{Attitude results\label{sec:resatt}}

The bottom diagram in Fig.~\ref{fig:conv} shows the RSE attitude errors and updates
as a functions of the iteration number, where the small differences of the attitude
quaternions have been transformed into small rotations along the SRS axes
($x$, $y$, $z$) according to Sect.~\ref{sec:diffrot} and expressed in 
$\mu$as. The error component around the $z$ axis corresponds to the AL
attitude error, while the $x$ and $y$ components are linear combinations of
the AC attitude errors in the PFoV and FFoV.%
\footnote{The attitude errors discussed here must not be confused with the 
excess attitude noise $\epsilon_a$ introduced in Sect.~\ref{sec:synthesismodel}.
The latter represents modelling errors, which are practically absent in the 
demonstration solution.}
The $z$ (AL) errors settle at an 
overall level of 20~$\mu$as around iteration 60, while the updates 
continue to decrease in a similar manner as for the parallaxes
(Fig.~\ref{fig:conv}, top).
The RSE values of the $x$ and $y$ errors converge to 167~$\mu$as and
224~$\mu$as, i.e., an order of magnitude larger than in $z$, reflecting the larger
observational errors in the AC direction (Table~\ref{tab:sim}) and the
smaller number of AC observations. The ratio of the errors about $y$ and
$x$,  $224/167\simeq 1.34$ is in perfect agreement with the value
expected from the geometry of the observations (Fig.~\ref{fig:srs}), viz.,
$\tan(\Gamma_{\rm c}/2)\simeq 1.34$ for a basic angle of
$\Gamma_{\rm c}=106.5^\circ$.

The converged AL attitude error of 20~$\mu$as is completely consistent with 
the previously inferred constant RMS contribution, by 5--6~$\mu$as, 
to the parallax errors (see Sect.~\ref{sec:ressrc}), as can
be seen from the following considerations. For most stars, the propagation of 
random observational errors from individual AL observations to the parallaxes 
(say) is largely governed by geometrical factors and the total number of 
observations per star, and can be statistically described by a `coefficient of
improvement' which can be estimated to $207.6/2300\simeq 0.09$
using the RSE of the parallax errors for the faintest bin in Table~\ref{tab:res:varpistats}
combined with the typical AL observational error at $G\simeq 19.6$ 
(cf.\ Table~\ref{tab:sim}). This factor assumes that the individual observational
errors (at each CCD) are uncorrelated, which is a very good approximation for
the photon-statistical centroiding errors, but not for the attitude errors, which 
have a correlation length determined by the knot interval of the attitude spline.
In the demonstration solution, the knot interval was 240~s, which is much longer
than the time it takes an image to cross the nine CCDs in the astrometric field
($\simeq 40$~s). Therefore it is a much better approximation to assume a 
\emph{constant} attitude error for the whole field crossing, corresponding to 
nine AL observations. As a result, the coefficient of improvement relevant for
the attitude error should be a factor three larger, or $\simeq 0.27$. The 
AL attitude uncertainty of 20~$\mu$as therefore corresponds to 
$20\times 0.27\simeq 5.4~\mu$as in the parallax, in very good agreement 
with the empirical result of 5--6~$\mu$as. For the other astrometric parameters a corresponding 
calculation yields a contribution of about 4~$\mu$as in position and 
3~$\mu$as~yr$^{-1}$ in proper motion. Although these numbers are 
somewhat smaller than the empirical estimates in Sect.~\ref{sec:ressrc} 
(possibly indicating an additional contribution from the calibration errors), 
the overall agreement is striking.

The attitude errors obtained in the solution ultimately come from the observational
errors of the individual observations, through the process of fitting the attitude
spline functions to these observations. If more observations (of the same quality)
are added, the attitude errors are expected to diminish inversely with the square
root of the number of observations (as long as the modelling errors are not a
limiting factor). The present AL attitude error of 20~$\mu$as is roughly what can 
be expected from the density and magnitudes of primary sources in the demonstration
solution, as can be seen from the following simple calculation. The AL attitude has 
essentially one degree of freedom per knot interval (240~s). The average number of 
AL observations per degree of freedom is therefore about 2500 (Table~\ref{tab:sim}). 
From the magnitude distribution in Table~\ref{tab:res:varpistats} and the AL 
observational uncertainties in Table~\ref{tab:sim} one can estimate that the 
average AL observation carries a statistical weight ($\sigma_\text{AL}^{-2}$) 
corresponding to an AL standard uncertainty of $\sigma_\text{AL}\simeq 650~\mu$as 
(taking into account the weight reduction by $\varrho^{-2}$ for the bright stars). 
The mean
flow of observations therefore should allow the AL attitude to be pinned down 
with an uncertainty of about $650\times 2500^{-1/2}\simeq 13~\mu$as. 
However, 
this rough calculation assumes uniform distribution of the stars across the sky,
with the same mean density (55~deg$^{-2}$) as in the demonstration run.
Considering that large parts of the sky have a much lower density (typically
5--10~deg$^{-2}$; see Fig.~\ref{fig:res:dens-obs}, top), which implies a less
precise attitude at the corresponding times, and that we have also 
neglected the attitude modelling errors, which here amount to at most $9~\mu$as 
RMS (Sect.~\ref{sec:simu}), it is not unreasonable that the
overall AL attitude uncertainty is about 50\% larger than according to 
the above calculation.

The demonstration run uses only 2\% of the stars envisaged
for the final AGIS solution, and the majority of them are faint, whereas the real
primary stars are preferentially selected among the brighter stars when
possible (cf.\ the discussion in Sect.~\ref{sec:relegate}). For a model distribution
of $10^8$ primary sources similar to the one described by 
\citet{hobbs+2010}, the AL observational uncertainty corresponding
to the average statistical weight is more like 200~$\mu$as, rather than the 650~$\mu$as
in the present data. On the other hand, the attitude knot interval will also be 
much shorter than the 240~s used in the present run, perhaps even as short 
as 5~s, which is about the shortest knot interval that can reasonably be used in 
view of the normal CCD integration time of 4.42~s. Combining these numbers
we estimate that the final AGIS run on the real Gaia data might obtain an 
AL attitude uncertainty, due to the photon noise, of about
$(20~\mu\text{as})\times(200/650)\times [0.02\times (240/15)]^{1/2}\simeq 6~\mu$as.
To this should be added the attitude modelling error (i.e., how well the spline can
represent the effective attitude), which is difficult to estimate without more reliable
information about attitude irregularities (Appendix~\ref{sec:attirr}). Generally speaking, the
optimum knot interval will roughly balance the estimation and modelling uncertainties, 
so that the total uncertainty is less than twice the estimation uncertainty. Assuming
a total AL attitude error of $12~\mu$as RMS, this would give less than $4~\mu$as RMS 
to be added quadratically to the parallax uncertainties. 
Thus, the attitude contribution to the final astrometric parameters appears to be
relatively small even for the bright stars. However, this does not take into account
the additional complications caused by the gated observations (Appendix~\ref{sec:TDI}) 
and residual calibration errors due to, for example, CTI effects 
(Appendix~\ref{sec:CTI}).

\begin{figure}
\begin{center}
\includegraphics[width=1.0\columnwidth]{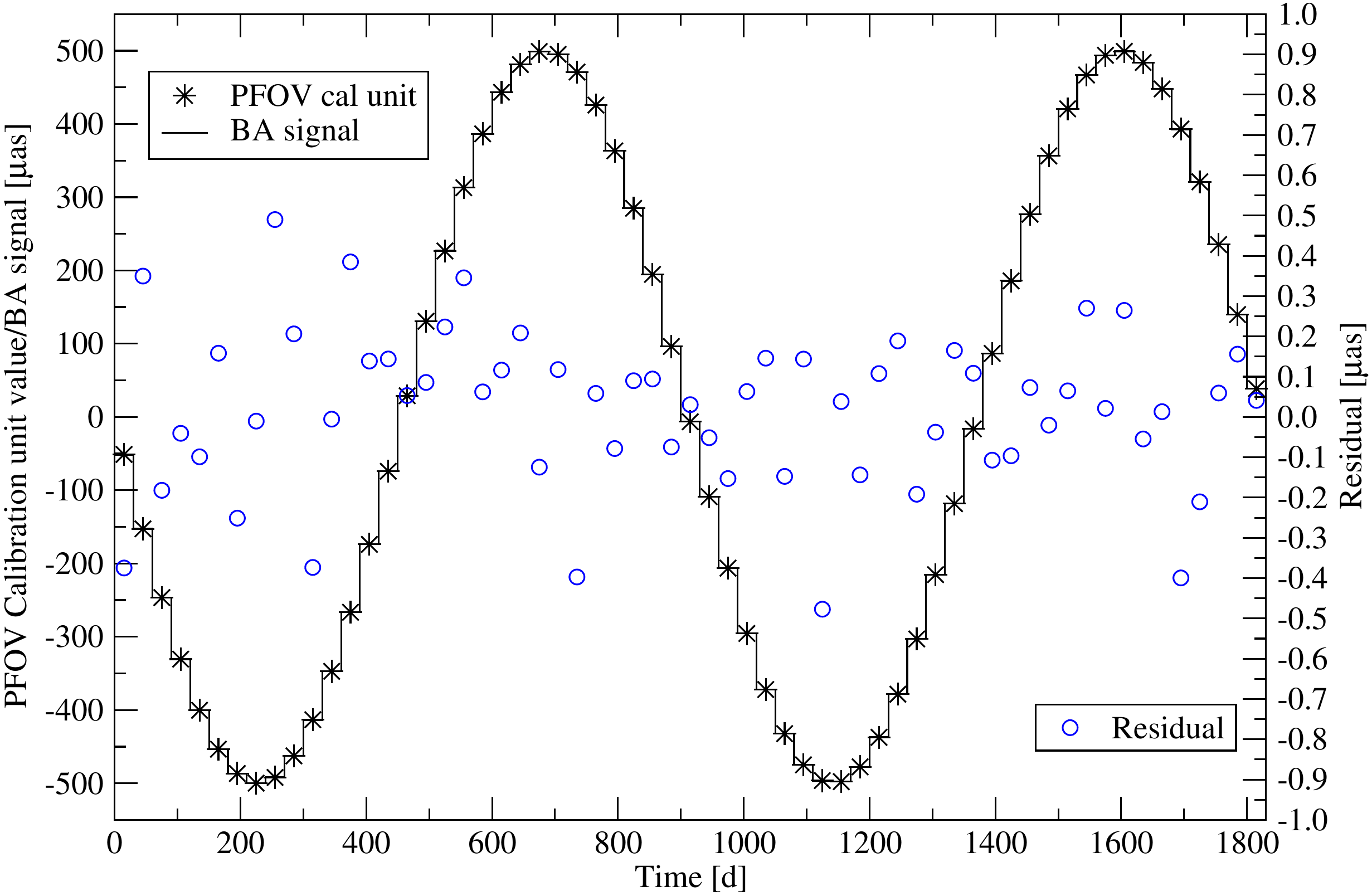}
\caption{Variation of the AL large-scale calibration parameters (averaged
over all CCDs) in the preceding field of view (PFoV), as a function of the time
since the beginning of the mission. The step-sinusoidal curve is the expected
variation due to the simulated basic-angle variation having a period of 2.5~yr
and an amplitude of 500~$\mu$as, but constant within each 30~day interval. 
The asterisks show the results of the solution (one value per 30~day interval), 
and the circles show the differences on the magnified scale to the right. 
Thanks to the constraint in Eq.~(\ref{eq:cc1}), the mean calibration parameters 
in the following field of view (FFoV) exactly mirror the displayed ones, and are
therefore not shown.}
\label{fig:res:callargescale}
\end{center}
\end{figure}

\subsubsection{Calibration results\label{sec:rescal}}

The calibration model used for the run merely contained one effect
($N_{\mbox{\scriptsize AL}}=1$ in Eq.~\ref{eq:generic1}),
viz., the large-scale calibration $\Delta\eta$ in Eq.~(\ref{eq:etazeta})
accounting for geometric distortions of the CCDs and optical effects
which are indistinguishable from geometric irregularities of the focal plane.
On this account, we expect the simulated basic angle signal (see
beginning of Sect.~\ref{sec:agis-results}) to manifest itself through
a corresponding time-dependence of the AL large-scale calibration 
parameters $\Delta\eta_{0fn0j}$.
The asterisks in Fig.~\ref{fig:res:callargescale} show, for $f=\mbox{PFoV}$ 
and for each time interval $j$, the calibration parameter values of the 
demonstration solution averaged 
over the 62 CCDs ($n$). The solid line depicts the step-sinusoidal input 
basic-angle signal applicable for PFoV. As anticipated, the estimated 
calibration parameters are in very good agreement with the input signal: 
the RMS value of the differences is $0.20~\mu$as, corresponding to an 
RMS error of $0.40~\mu$as for the basic angle offsets $\Delta\Gamma_j$ 
per calibration time interval (cf.\ Eq.~\ref{eq:DeltaGamma}). This is 
reasonably consistent with the expected precision of the large-scale
calibration based on the total weight of the observations, as shown
by the following calculation. The mean number of AL observations per
calibration time interval and field of view is $1.33\times 10^7$
(cf.\ Table~\ref{tab:sim}). Assuming, as we did in Sect.~\ref{sec:resatt}, 
that an AL observation of average weight corresponds to a standard
uncertainty of $\sigma_\text{AL}\simeq 650~\mu$as, the expected precision
of the basic angle determination is 
$2^{1/2}\times (650~\mu\text{as})\times [1.33\times 10^7]^{-1/2}\simeq 0.25~\mu$as.
The observed scatter, $0.40~\mu$as, is larger by roughly the same
factor as found for the AL attitude errors. 

The good agreement between the input calibration signal and the
recovered parameters demonstrates the correct functioning of the generic 
calibration model (see Sect.~\ref{sec:instrumentmodel}) in 
this simple case. We expect that many more validation runs will
be needed to confirm this result in more complex circumstances, i.e., with
more calibration effects of different functional compositions and
dependencies. A further important aspect is the practical study of possible
hidden correlations and degeneracies of calibration with source,
attitude, and global parameters which may not be fully obvious at the
mathematical level.

\begin{figure}
\begin{center}
\includegraphics[width=1.0\columnwidth]{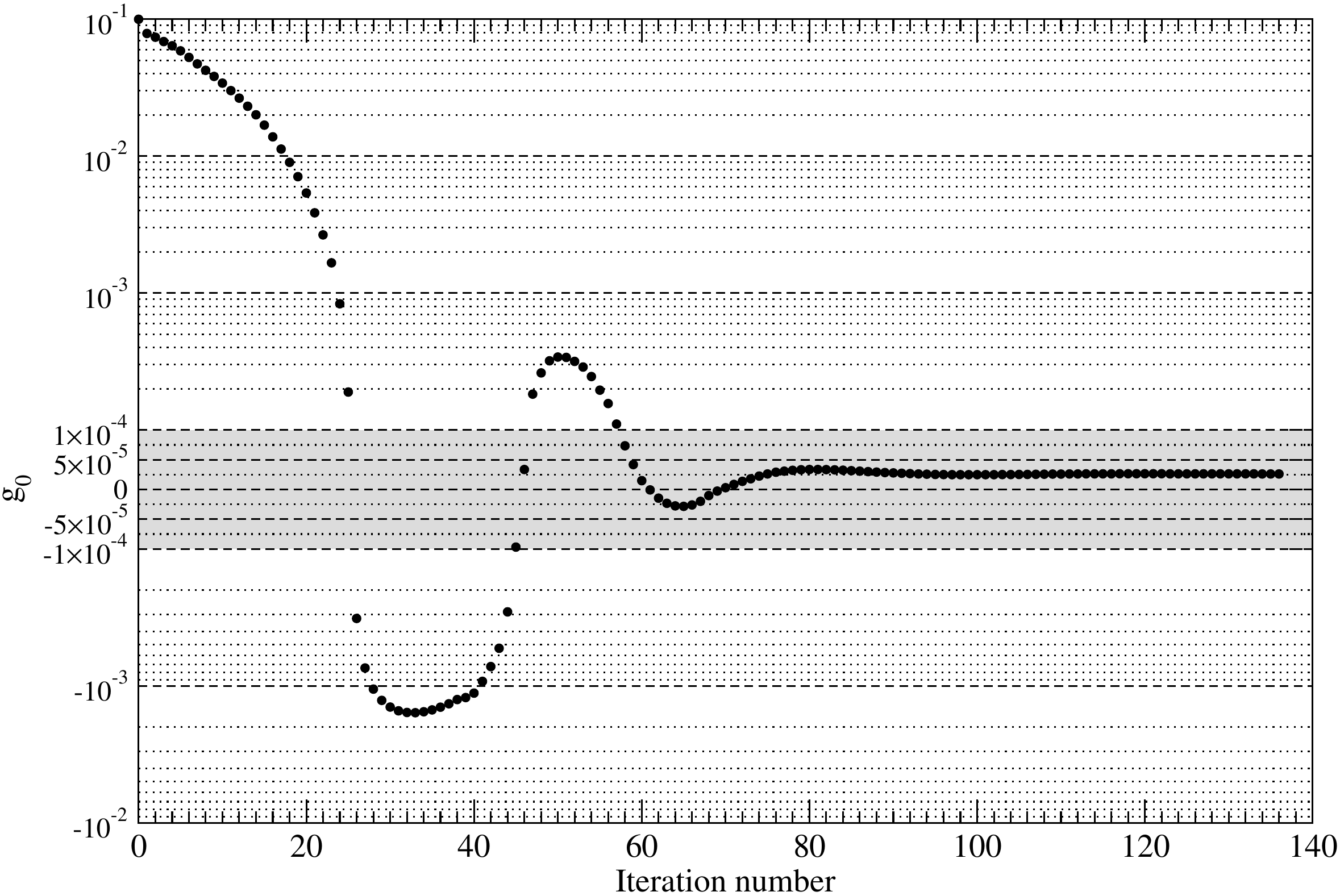}
\caption{Evolution of the estimated global parameter $g_0=\gamma-1$ 
(where $\gamma$ is the PPN parameter) as a function of the
iteration number. $g_0$ settles at a level of $2.6\times10^{-5}$ from a
starting value of $g_0=0.1$. The formal uncertainty of $g_0$ in this solution is
$2.15\times 10^{-5}$. Note that a linear scale is used for $|g_0|<10^{-4}$
(the grey area), while a logarithmic scale is used outside of this interval.}
\label{fig:res:global}
\end{center}
\end{figure}

\subsubsection{Global results\label{sec:resglb}}

Starting the iterations from a PPN $\gamma$ value of 1.1, the purpose of running
with the Global block was to see how such a grossly wrong initial value
would affect the (overall) convergence rate and to what level the correct
value could be recovered (recall that the input data were simulated using 
$\gamma=1$).

Figure~\ref{fig:res:global} presents the evolution of $g_0=\gamma-1$ 
during the run. In iteration 135 the parameter had settled on
the value $g_0=2.62\times 10^{-5}$, with a formal standard uncertainty%
\footnote{The formal uncertainty of $g_0$ was calculated as 
$\vec{N}_{gg}^{-1/2}$ (Eq.~\ref{eq:norm2}) times the factor 
$(1-\rho)^{-1/2}=\cot(\xi/2)=2.414$, where $\xi=45^\circ$ is the 
constant angle between Gaia's spin axis ($\vec{z}$) and the direction 
to the Sun; as explained by \citet{hobbs+2010}, this factor takes into 
account the statistical correlation ($\rho$) between PPN $\gamma$ and 
the parallaxes. The correction factor would not have been required, had 
the solution included the pseudo-parameter $g_1$ (Sect.~\ref{sec:globalupdate}).}
of $2.15\times 10^{-5}$. At $+1.2$~standard deviations, 
this value of $g_0$ is not significantly different from 0. The run
shows that the system is capable of recovering the `correct' value
with an error compatible with the statistical uncertainty, which in this 
case is set by the photon noise of the individual observations.

The total astrometric weight of the AL observations in the demonstration
run is $\simeq 4000~\mu\text{as}^{-2}$. According to \citet{hobbs+2010} 
this should yield an RMS uncertainty in $\gamma$ of about $3\times 10^{-5}$, 
in reasonable agreement with the formal standard uncertainty given above and the parameter 
value obtained in the solution.

\subsection{Processing times}

The demonstration solution was run on an IBM cluster at ESAC using 14 nodes 
(out of 32 available), each node having two processors%
\footnote{Intel Nehalem EP Xeon (quad-core, 2.93~GHz), with 32 GB RAM} 
with four cores each; thus in total 112~CPUs were engaged. 
This configuration of 14 nodes is estimated to have a total floating 
point performance of 0.65~Tflop~s$^{-1}$ ($0.65\times 10^{12}$ floating 
point operations per second). 
One iteration took about 1~hr (with high CPU occupancy, typically
90\%), so the total run time for 135 iterations was nearly 6~days, 
corresponding to about $3\times 10^{17}$ floating point operations (flop).

Scaled up to the projected $10^8$ primary sources of the final AGIS run
this would amount to $1.5\times 10^{19}$~flop. Using a more conservative 
estimate of $5\times 10^{19}$~flop to account for additional features not 
included in the demonstration run, this will require some 60~days on
a typically targeted 10~Tflop~s$^{-1}$ machine. 
On the other hand, there could also be a significant saving in computing 
time due to the fact that the final solution will start off from a previous solution, 
already very close to the final one; a smaller number of iterations might therefore 
suffice. In summary, the estimated processing time is clearly within the feasible range.

\section{Conclusions}\label{sec:concl}

A fundamental part of the scientific data processing for the Gaia mission is the
astrometric core solution, which will be run during and after the mission
(ca.\ 2013--2020) with successively larger datasets and eventually encompassing
at least some 100~million primary sources. This solution is central to the
performance of the mission as a whole, since it not only provides the astrometric
results for these primary sources, but also the reference frame (in the form of
the instrument attitude as a function of time) and the geometric calibration of
the instrument, for use by a large number of other processes in the overall
scientific reduction of the Gaia data.

In order to accomplish the astrometric core solution, a software 
system known as AGIS (Astrometric Global Iterative Solution) is being built 
within Coordination Unit 3 (CU3) of the Gaia Data Processing and Analysis
Consortium (DPAC). As detailed in this paper, the necessary mathematical 
models and numerical algorithms are well understood, and have been developed 
with sufficient rigour to allow the potential accuracy of Gaia to be fully
exploited. Most critical parts of this system have been implemented,
and numerous test runs have demonstrated the theoretical validity of 
the global iterative approach, as well as its practical feasibility in terms of 
data management and computations. While a number of additional 
features will have to be included in the software before it can be considered 
ready for the flight data, and many more complications will undoubtedly be 
discovered during the actual analysis of these data, all the fundamental parts
of AGIS are already in place.
   
In 2011, with roughly two years left until the launch of the satellite 
and time to mature
the concepts and software presented here into a robust operational system,
we have no reason to doubt that AGIS will be able to compute an
accurate astrometric solution, consistent with the ambitious goals of the 
Gaia mission.

\begin{acknowledgements}
The constant work of the Gaia Data Processing and Analysis Consortium (DPAC) has
played an important part in this work. We are particularly indebted to CU2 for
the production of independently simulated Gaia-like data for use in the system.
The data simulations have been done in the supercomputer Mare Nostrum at Barcelona
Supercomputing Center -- Centro Nacional de Supercomputaci{\'o}n
(The Spanish National Supercomputing Center). 
Research and development in Sweden is kindly supported by the Swedish National
Space Board (SNSB).

The successful development of concepts, algorithms and software incorporated in 
AGIS over a number of years, as well as the drafting of this manuscript, has benefited 
from the interaction with numerous colleagues, of which we especially wish to mention 
Sergei Klioner,
Alex Bombrun,
Alexey Butkevich,
Jos de Bruijne,
Berry Holl,
Floor van Leeuwen,
and Fran\c{c}ois Mignard.

We thank the referee, Dr.~F.~van Leeuwen, for numerous constructive comments 
which helped to improve the original version of the manuscript.

Our special thanks go to Gaia's former Project Scientist
Michael Perryman, whose vision, leadership, and enthusiasm in the early years
of the project laid the foundations for the excellent progress that is today seen
throughout DPAC and with AGIS in particular.
\end{acknowledgements}

\bibliographystyle{aa} 
\bibliography{aa17905-11} 

\begin{thebibliography}{71}
\expandafter\ifx\csname natexlab\endcsname\relax\def\natexlab#1{#1}\fi

\bibitem[{Axelsson(1996)}]{axelsson96}
Axelsson, O. 1996, Iterative Solution Methods (Cambridge University Press)

\bibitem[{{Bastian}(1995)}]{bastian95}
{Bastian}, U. 1995, in ESA Special Publication, Vol. 379, Future Possibilities
  for astrometry in Space, ed. {M.~A.~C.~Perryman \& F.~van Leeuwen}, 99

\bibitem[{{Bastian} \& {Biermann}(2005)}]{2005A&A...438..745B}
{Bastian}, U. \& {Biermann}, M. 2005, \aap, 438, 745

\bibitem[{Bj{\"o}rck(1996)}]{book:bjork-1996}
Bj{\"o}rck, {\AA}. 1996, Numerical Methods for Least Squares Problems (Society
  for Industrial and Applied Mathematics)

\bibitem[{{Bombrun} {et~al.}(2011){Bombrun}, {Lindegren}, Hobbs, {Holl},
  Lammers, \& Bastian}]{bombrun+10}
{Bombrun}, A., {Lindegren}, L., Hobbs, D., {et~al.} 2011, A\&A, submitted

\bibitem[{{Bombrun} {et~al.}(2010){Bombrun}, {Lindegren}, {Holl}, \&
  {Jordan}}]{bombrun+09}
{Bombrun}, A., {Lindegren}, L., {Holl}, B., \& {Jordan}, S. 2010, \aap, 516,
  A77

\bibitem[{Chambers {et~al.}(2006)Chambers, James, Lambert, \& {Vander
  Wiel}}]{chambers+06}
Chambers, J., James, D., Lambert, D., \& {Vander Wiel}, S. 2006, Statistical
  Science, 21, 463

\bibitem[{de~Boor(2001)}]{book:DB-01}
de~Boor, C. 2001, {A} {P}ractical {G}uide to {S}plines, Rev. ed., {A}pplied
  {M}athematical {S}ciences, Vol. 27 (Springer)

\bibitem[{{de Bruijne} {et~al.}(2010){de Bruijne}, {Siddiqui}, {Lammers},
  {Hoar}, {O'Mullane}, \& {Prusti}}]{2010IAUS..261..331D}
{de Bruijne}, J., {Siddiqui}, H., {Lammers}, U., {et~al.} 2010, in IAU
  Symposium, Vol. 261, IAU Symposium, ed. {S.~A.~Klioner, P.~K.~Seidelmann, \&
  M.~H.~Soffel}, 331--333

\bibitem[{Dongarra {et~al.}(1979)Dongarra, Bunch, Moler, \&
  Stewart}]{LINPACK-UG}
Dongarra, J., Bunch, J., Moler, C., \& Stewart, G. 1979, LINPACK Users' Guide
  (Philadelphia, PA, USA: Society for Industrial and Applied Mathematics), 367

\bibitem[{{Dravins} {et~al.}(1999){Dravins}, {Lindegren}, \&
  {Madsen}}]{drav+99}
{Dravins}, D., {Lindegren}, L., \& {Madsen}, S. 1999, \aap, 348, 1040

\bibitem[{ESA(1997)}]{hip:catalogue}
ESA. 1997, {T}he {H}ipparcos and {T}ycho {C}atalogues, {E}SA SP-1200

\bibitem[{{Feissel} \& {Mignard}(1998)}]{icrs1998}
{Feissel}, M. \& {Mignard}, F. 1998, \aap, 331, L33

\bibitem[{{Gilbert} {et~al.}(2002){Gilbert}, {Kotidis}, {Muthukrishnan}, \&
  {Strass}}]{gilbert+02}
{Gilbert}, A.~C., {Kotidis}, Y., {Muthukrishnan}, S., \& {Strass}, M.~J. 2002,
  in Proc. 28th International Conference on Very Large Data Bases, Hong Kong,
  China, 454

\bibitem[{{Golub} \& {van Loan}(1996)}]{golu+96}
{Golub}, G.~H. \& {van Loan}, C.~F. 1996, {Matrix computations, 3rd ed.} (The
  Johns Hopkins University Press, Baltimore)

\bibitem[{{G{\'o}rski} {et~al.}(2005){G{\'o}rski}, {Hivon}, {Banday},
  {Wandelt}, {Hansen}, {Reinecke}, \& {Bartelmann}}]{gor05}
{G{\'o}rski}, K.~M., {Hivon}, E., {Banday}, A.~J., {et~al.} 2005, \apj, 622,
  759

\bibitem[{Greenbaum(1997)}]{book:greenbaum-1997}
Greenbaum, A. 1997, Iterative Methods for Solving Linear Systems (Society for
  Industrial and Applied Mathematics)

\bibitem[{{Greenwald} \& {Khanna}(2001)}]{greenwald+01}
{Greenwald}, M. \& {Khanna}, S. 2001, in Proc. ACM SIGMOD International
  Conference on Management of Data, Santa Barbara, CA, 58

\bibitem[{{Gunn} {et~al.}(1998){Gunn}, {Carr}, {Rockosi}, {Sekiguchi}, {Berry},
  {Elms}, {de Haas}, {Ivezi{\'c}}, {Knapp}, {Lupton}, {Pauls}, {Simcoe},
  {Hirsch}, {Sanford}, {Wang}, {York}, {Harris}, {Annis}, {Bartozek},
  {Boroski}, {Bakken}, {Haldeman}, {Kent}, {Holm}, {Holmgren}, {Petravick},
  {Prosapio}, {Rechenmacher}, {Doi}, {Fukugita}, {Shimasaku}, {Okada}, {Hull},
  {Siegmund}, {Mannery}, {Blouke}, {Heidtman}, {Schneider}, {Lucinio}, \&
  {Brinkman}}]{gunn+98}
{Gunn}, J.~E., {Carr}, M., {Rockosi}, C., {et~al.} 1998, \aj, 116, 3040

\bibitem[{{Gwinn} {et~al.}(1997){Gwinn}, {Eubanks}, {Pyne}, {Birkinshaw}, \&
  {Matsakis}}]{gwinn+97}
{Gwinn}, C.~R., {Eubanks}, T.~M., {Pyne}, T., {Birkinshaw}, M., \& {Matsakis},
  D.~N. 1997, \apj, 485, 87

\bibitem[{Hamilton(1843)}]{hamilton1843}
Hamilton, W.~R. 1843, in Proceedings of the Royal Irish Academy, Vol.~2, 424

\bibitem[{Hansen(1998)}]{book:hansen1998}
Hansen, P.~C. 1998, Rank-Deficient and Discrete Ill-Posed Problems: Numerical
  Aspects of Linear Inversion, Monographs on Mathematical Modeling and
  Computation 4 (SIAM)

\bibitem[{{Higham}(1990)}]{higham1990}
{Higham}, N. 1990, in Reliable Numerical Computation, ed. {M.~G.~Cox \&
  S.~J.~Hammarling} (Oxford University Press), 161

\bibitem[{Higham(2002)}]{Higham:2002:ASN}
Higham, N.~J. 2002, Accuracy and Stability of Numerical Algorithms, 2nd edn.
  (Philadelphia, PA, USA: Society for Industrial and Applied Mathematics), 680

\bibitem[{{Hobbs} {et~al.}(2010){Hobbs}, {Holl}, {Lindegren}, {Raison},
  {Klioner}, \& {Butkevich}}]{hobbs+2010}
{Hobbs}, D., {Holl}, B., {Lindegren}, L., {et~al.} 2010, in IAU Symposium No.
  261, ed. {S.~A.~Klioner, P.~K.~Seidelmann, \& M.~H.~Soffel}, 315

\bibitem[{{H{\o}g}(2008)}]{iaus248:EH}
{H{\o}g}, E. 2008, in IAU Symposium No. 248, ed. {W.~J.~Jin, I.~Platais, \&
  M.~A.~C.~Perryman}, 300

\bibitem[{{H{\o}g} {et~al.}(2000){H{\o}g}, {Fabricius}, {Makarov}, {Urban},
  {Corbin}, {Wycoff}, {Bastian}, {Schwekendiek}, \& {Wicenec}}]{hog+00}
{H{\o}g}, E., {Fabricius}, C., {Makarov}, V.~V., {et~al.} 2000, \aap, 355, L27

\bibitem[{{Holl} {et~al.}(2010){Holl}, {Hobbs}, \& {Lindegren}}]{holl+2010}
{Holl}, B., {Hobbs}, D., \& {Lindegren}, L. 2010, in IAU Symposium No. 261, ed.
  {S.~A.~Klioner, P.~K.~Seidelmann, \& M.~H.~Soffel}, 320

\bibitem[{{Holl} {et~al.}(2012){Holl}, {Lindegren}, \& {Hobbs}}]{holl+2010a}
{Holl}, B., {Lindegren}, L., \& {Hobbs}, D. 2012, in preparation for \aap

\bibitem[{{Huber}(1981)}]{huber:1981}
{Huber}, P. 1981, Robust Statistics (Wiley)

\bibitem[{{Janesick}(2001)}]{2001sccd.book.....J}
{Janesick}, J.~R. 2001, {Scientific charge-coupled devices} (SPIE Optical
  Engineering Press)

\bibitem[{{Jordi} {et~al.}(2010){Jordi}, {Gebran}, {Carrasco}, {de Bruijne},
  {Voss}, {Fabricius}, {Knude}, {Vallenari}, {Kohley}, \&
  {Mora}}]{2010A&A...523A..48J}
{Jordi}, C., {Gebran}, M., {Carrasco}, J.~M., {et~al.} 2010, \aap, 523, A48+

\bibitem[{Kane {et~al.}(1983)Kane, Likins, \& Levinson}]{kane:1983}
Kane, T.~R., Likins, P.~W., \& Levinson, D.~A. 1983, {S}pacecraft dynamics
  (McGraw Hill Book Company)

\bibitem[{{Klioner}(2003)}]{klioner2003}
{Klioner}, S.~A. 2003, \aj, 125, 1580

\bibitem[{{Klioner}(2004)}]{klioner2004}
{Klioner}, S.~A. 2004, \prd, 69, 124001

\bibitem[{Klumpp(1976)}]{klumpp1976}
Klumpp, A.~R. 1976, J.~Spacecraft, 13, 754

\bibitem[{{Kopeikin} \& {Makarov}(2006)}]{kopeikin+06}
{Kopeikin}, S.~M. \& {Makarov}, V.~V. 2006, \aj, 131, 1471

\bibitem[{{Laborie} {et~al.}(2007){Laborie}, {Davancens}, {Pouny}, {V{\'e}tel},
  {Chassat}, {Charvet}, {Gar{\'e}}, \& {Sarri}}]{2007SPIE.6690E...8L}
{Laborie}, A., {Davancens}, R., {Pouny}, P., {et~al.} 2007, in Society of
  Photo-Optical Instrumentation Engineers (SPIE) Conference Series, Vol. 6690,
  Society of Photo-Optical Instrumentation Engineers (SPIE) Conference Series

\bibitem[{{Lammers} {et~al.}(2009){Lammers}, {Lindegren}, {O'Mullane}, \&
  {Hobbs}}]{Lammers+09}
{Lammers}, U., {Lindegren}, L., {O'Mullane}, W., \& {Hobbs}, D. 2009, in ASPC
  Series, Vol. 411, Astronomical Data Analysis Software and Systems XVIII, ed.
  {D.~A.~Bohlender, D.~Durand, \& P.~Dowler}, 55

\bibitem[{{Lawson} \& {Hanson}(1974)}]{laws+74}
{Lawson}, C.~L. \& {Hanson}, R.~J. 1974, {Solving Least Squares Problems}
  (Prentice-Hall, New Jersey)

\bibitem[{{Lindegren}(2005)}]{2005ESASP.576...29L}
{Lindegren}, L. 2005, in ESA Special Publication, Vol. 576, The
  Three-Dimensional Universe with Gaia, ed. {C.~Turon, K.~S.~O'Flaherty, \&
  M.~A.~C.~Perryman}, 29

\bibitem[{{Lindegren}(2010)}]{iaus261:LL}
{Lindegren}, L. 2010, in IAU Symposium No. 261, ed. {S.~A.~Klioner,
  P.~K.~Seidelmann, \& M.~H.~Soffel}, 296

\bibitem[{{Lindegren} {et~al.}(2008){Lindegren}, {Babusiaux}, {Bailer-Jones},
  {Bastian}, {Brown}, {Cropper}, {H{\o}g}, {Jordi}, {Katz}, {van Leeuwen},
  {Luri}, {Mignard}, {de Bruijne}, \& {Prusti}}]{iaus248:LL}
{Lindegren}, L., {Babusiaux}, C., {Bailer-Jones}, C., {et~al.} 2008, in IAU
  Symposium No. 248, ed. {W.~J.~Jin, I.~Platais, \& M.~A.~C.~Perryman}, 217

\bibitem[{{Lindegren} \& {Bastian}(2011)}]{2011EAS....45..109L}
{Lindegren}, L. \& {Bastian}, U. 2011, in EAS Publications Series, Vol.~45,
  109--114

\bibitem[{{Lindegren} \& {Dravins}(2003)}]{lindegren+2003}
{Lindegren}, L. \& {Dravins}, D. 2003, \aap, 401, 1185

\bibitem[{{Lindegren} \& {Kovalevsky}(1995)}]{ndac1995}
{Lindegren}, L. \& {Kovalevsky}, J. 1995, \aaps, 304, 189

\bibitem[{{Lindegren} \& {Perryman}(1996)}]{gaia1996}
{Lindegren}, L. \& {Perryman}, M.~A.~C. 1996, \aap, 116, 579

\bibitem[{{Luri} \& {Babusiaux}(2011)}]{2011EAS....45...25L}
{Luri}, X. \& {Babusiaux}, C. 2011, in EAS Publications Series, Vol.~45, EAS
  Publications Series, 25--30

\bibitem[{{Masana} {et~al.}(2005){Masana}, {Luri}, {Anglada-Escud{\'e}}, \&
  {Llimona}}]{2005ESASP.576..457M}
{Masana}, E., {Luri}, X., {Anglada-Escud{\'e}}, G., \& {Llimona}, P. 2005, in
  ESA Special Publication, Vol. 576, The Three-Dimensional Universe with Gaia,
  ed. {C.~Turon, K.~S.~O'Flaherty, \& M.~A.~C.~Perryman}, 457--+

\bibitem[{{Mignard} {et~al.}(2008){Mignard}, {Bailer-Jones}, {Bastian},
  {Drimmel}, {Eyer}, {Katz}, {van Leeuwen}, {Luri}, {O'Mullane}, {Passot},
  {Pourbaix}, \& {Prusti}}]{iaus248:FM}
{Mignard}, F., {Bailer-Jones}, C., {Bastian}, U., {et~al.} 2008, in IAU
  Symposium No. 248, ed. {W.~J.~Jin, I.~Platais, \& M.~A.~C.~Perryman}, 224

\bibitem[{{Murray}(1983)}]{murray1983}
{Murray}, C.~A. 1983, {Vectorial astrometry} (Bristol: Adam Hilger)

\bibitem[{{O'Mullane} {et~al.}(2001){O'Mullane}, {Banday}, {G{\'o}rski},
  {Kunszt}, \& {Szalay}}]{omu01}
{O'Mullane}, W., {Banday}, A.~J., {G{\'o}rski}, K.~M., {Kunszt}, P., \&
  {Szalay}, A.~S. 2001, in Mining the Sky, ed. A.~J. {Banday}, S.~{Zaroubi}, \&
  M.~{Bartelmann}, 638

\bibitem[{{O'Mullane} {et~al.}(2007){O'Mullane}, {Lammers}, {Bailer-Jones},
  {Bastian}, {Brown}, {Drimmel}, {Eyer}, {Huc}, {Katz}, {Lindegren},
  {Pourbaix}, {Luri}, {Torra}, {Mignard}, \& {van Leeuwen}}]{WOM+2007}
{O'Mullane}, W., {Lammers}, U., {Bailer-Jones}, C., {et~al.} 2007, in
  Astronomical Society of the Pacific Conference Series, Vol. 376, Astronomical
  Data Analysis Software and Systems XVI, ed. R.~A. {Shaw}, F.~{Hill}, \& D.~J.
  {Bell}, 99

\bibitem[{{O'Mullane} {et~al.}(2010){O'Mullane}, {Lammers}, {Hern{\'a}ndez},
  {Hoar}, {Parsons}, \& {Luri}}]{2010ASPC..434..135O}
{O'Mullane}, W., {Lammers}, U., {Hern{\'a}ndez}, J., {et~al.} 2010, in
  Astronomical Society of the Pacific Conference Series, Vol. 434, Astronomical
  Data Analysis Software and Systems XIX, ed. {Y.~Mizumoto, K.-I.~Morita, \&
  M.~Ohishi}, 135

\bibitem[{{O'Mullane} {et~al.}(2011){O'Mullane}, Luri, Parsons, Lammers, Hoar,
  \& Hern{\'a}ndez}]{WOM2011}
{O'Mullane}, W., Luri, X., Parsons, P., {et~al.} 2011, Experimental Astronomy,
  1, 10.1007/s10686-011-9241-6

\bibitem[{{Perryman} {et~al.}(2001){Perryman}, {de Boer}, {Gilmore}, {H{\o}g},
  {Lattanzi}, {Lindegren}, {Luri}, {Mignard}, {Pace}, \& {de Zeeuw}}]{gaia2001}
{Perryman}, M.~A.~C., {de Boer}, K.~S., {Gilmore}, G., {et~al.} 2001, \aap,
  369, 339

\bibitem[{Press {et~al.}(2007)Press, Teukolsky, Vetterling, \&
  Flannery}]{book:nr3}
Press, W., Teukolsky, S., Vetterling, W., \& Flannery, B. 2007, Numerical
  Recipes: The Art of Scientific Computing, 3rd edn. (Cambridge University
  Press)

\bibitem[{{Prod'homme} {et~al.}(2011){Prod'homme}, {Brown}, {Lindegren},
  {Short}, \& {Brown}}]{2011MNRAS.414.2215P}
{Prod'homme}, T., {Brown}, A.~G.~A., {Lindegren}, L., {Short}, A.~D.~T., \&
  {Brown}, S.~W. 2011, \mnras, 414, 2215

\bibitem[{{Prod'Homme} {et~al.}(2010){Prod'Homme}, {Weiler}, {Brown}, {Short},
  \& {Brown}}]{2010SPIE.7742E..29P}
{Prod'Homme}, T., {Weiler}, M., {Brown}, S.~W., {Short}, A.~D.~T., \& {Brown},
  A.~G.~A. 2010, in Society of Photo-Optical Instrumentation Engineers (SPIE)
  Conference Series, Vol. 7742, Society of Photo-Optical Instrumentation
  Engineers (SPIE) Conference Series

\bibitem[{{Proust} {et~al.}(2006){Proust}, {Quintana}, {Carrasco},
  {Reisenegger}, {Slezak}, {Muriel}, {D{\"u}nner}, {Sodr{\'e}}, {Drinkwater},
  {Parker}, \& {Ragone}}]{proust+06}
{Proust}, D., {Quintana}, H., {Carrasco}, E.~R., {et~al.} 2006, \aap, 447, 133

\bibitem[{{Robin} {et~al.}(2003){Robin}, {Reyl\'{e}}, {Derri\`{e}re}, \&
  Picaud}]{robin2003}
{Robin}, A.~C., {Reyl\'{e}}, C., {Derri\`{e}re}, S., \& Picaud, S. 2003,
  Astronomy and Astrophysics, 409, 523

\bibitem[{Schnabel \& Eskow(1999)}]{schnabel+99}
Schnabel, R.~B. \& Eskow, E. 1999, SIAM Journal on Optimization, 9, 1135

\bibitem[{{Seabroke} {et~al.}(2009){Seabroke}, {Holland}, {Burt}, \&
  {Robbins}}]{2009SPIE.7439E...3S}
{Seabroke}, G.~M., {Holland}, A.~D., {Burt}, D., \& {Robbins}, M.~S. 2009, in
  Society of Photo-Optical Instrumentation Engineers (SPIE) Conference Series,
  Vol. 7439, Society of Photo-Optical Instrumentation Engineers (SPIE)
  Conference Series

\bibitem[{{Short} {et~al.}(2010){Short}, {Prod'homme}, {Weiler}, {Brown}, \&
  {Brown}}]{short2010}
{Short}, A., {Prod'homme}, T., {Weiler}, M., {Brown}, S., \& {Brown}, A. 2010,
  in Presented at the Society of Photo-Optical Instrumentation Engineers (SPIE)
  Conference, Vol. 7742, Society of Photo-Optical Instrumentation Engineers
  (SPIE) Conference Series

\bibitem[{Soffel {et~al.}(2003)Soffel, Klioner, Petit, Wolf, Kopeikin,
  Bretagnon, Brumberg, Capitaine, Damour, Fukushima, Guinot, Huang, Lindegren,
  Ma, Nordtvedt, Ries, Seidelmann, Vokrouhlick{\'y}, Will, \& Xu}]{soffel+2003}
Soffel, M., Klioner, S.~A., Petit, G., {et~al.} 2003, \aj, 126, 2687

\bibitem[{{Stewart}(1998)}]{stewart:1998}
{Stewart}, G. 1998, Matrix Algorithms: Basic decompositions (Society for
  Industrial and Applied Mathematics)

\bibitem[{{Turon} {et~al.}(2005){Turon}, {O'Flaherty}, \&
  {Perryman}}]{2005ESASP.576.....T}
{Turon}, C., {O'Flaherty}, K.~S., \& {Perryman}, M.~A.~C., eds. 2005, ESA
  Special Publication, Vol. 576, {The Three-Dimensional Universe with Gaia}

\bibitem[{van~der Vorst(2003)}]{vanVorst:03}
van~der Vorst, H. 2003, {Iterative Krylov Methods for Large Linear Systems}
  (Cambridge University Press)

\bibitem[{{van Leeuwen}(2005)}]{2005A&A...439..805V}
{van Leeuwen}, F. 2005, \aap, 439, 805

\bibitem[{van Leeuwen(2007)}]{book:newhip}
van Leeuwen, F. 2007, {H}ipparcos, the {N}ew {R}eduction of the {R}aw {D}ata,
  Astrophysics and Space Science Laboratory. Vol. 350 (Springer)

\bibitem[{Wertz(1978)}]{wertz:1978}
Wertz, J.~R. 1978, {S}pacecraft {A}ttitude {D}etermination and {C}ontrol
  (Kluwer Academic Publishers)

\end{thebibliography}

\appendix

\section{Quaternions}\label{sec:quaternions}

Quaternions form a non-commutative algebra in $\mathbb{R}^4$. Invented by
W.~R.~Hamilton in 1843 as an extension of the complex numbers \citep{hamilton1843},
their most common usage today is for representing spatial rotations in a
particularly compact and convenient way, with applications for example in
computer graphics and spacecraft attitude control. Quaternions can be introduced
and understood in many different ways, with a correspondingly confusing multitude
of notations and conventions. Here we give just a brief introduction to the
subject, stating only the minimum results needed for our applications. For
more details, see for example \citet{wertz:1978} and \citet{kane:1983}.

\subsection{Quaternion algebra}\label{sec:quatalg}

A quaternion is a quadruple of real numbers for which the following algebraic
operations are defined. For any quaternions $\quat{a}=\bigl\{a_x,\,a_y,\,a_z,\,a_w\bigr\}$
and $\quat{b}=\bigl\{b_x,\,b_y,\,b_z,\,b_w\bigr\}$ we have addition
\begin{equation}\label{eq:quatadd}
\quat{a}+\quat{b}=\Bigl\{a_x+b_x,~a_y+b_y,~a_z+b_z,~a_w+b_w\Bigr\} \, ,
\end{equation}
multiplication
\begin{equation}\label{eq:quatmult}
\begin{split}
\quat{a}\quat{b}=\Bigl\{&\phantom{-}a_xb_w+a_yb_z-a_zb_y+a_wb_x,\\
&{-}a_xb_z+a_yb_w+a_zb_x+a_wb_y,\\
&\phantom{-}a_xb_y-a_yb_x+a_zb_w+a_wb_z,\\
&{-}a_xb_x-a_yb_y-a_zb_z+a_wb_w\Bigr\} \, ,
\end{split}
\end{equation}
and scalar multiplication
\begin{equation}\label{eq:quatscal}
s\quat{a}=\Bigl\{sa_x,~sa_y,~sa_z,~sa_w\Bigr\}
\end{equation}
for scalar $s$. Subtraction is analogous to addition, as
derived from $\quat{a}-\quat{b}=\quat{a}+(-1)\quat{b}$.
The conjugate of $\quat{a}$ is
\begin{equation}\label{eq:quatconj}
\quat{a}^*=\Bigl\{-a_x,~-a_y,~-a_z,~a_w\Bigr\} \, ,
\end{equation}
the norm (length) is
\begin{equation}\label{eq:quatnorm}
\lVert\quat{a}\rVert=(a_x^2+a_y^2+a_z^2+a_w^2)^{1/2} \, ,
\end{equation}
and the inverse (provided $\lVert\quat{a}\rVert>0$) is
\begin{equation}\label{eq:quatinv}
\quat{a}^{-1}=\lVert\quat{a}\rVert^{-2}\,\quat{a}^* \, .
\end{equation}
We have
\begin{equation}\label{eq:quatcom}
(\quat{a}\quat{b})^*=\quat{b}^*\quat{a}^* \, , \quad
\quad (\quat{a}\quat{b})^{-1}=\quat{b}^{-1}\quat{a}^{-1} \, .
\end{equation}
Any non-zero quaternion can be normalized to unit length. 
In analogy with the notation for vector normalization, we use 
angular brackets for this operation: 
\begin{equation}\label{eq:vecquatnorm}
\langle\quat{a}\rangle = \lVert\quat{a}\rVert^{-1}\quat{a} \, .
\end{equation}

The triplet of real numbers $(a_x,\,a_y,\,a_z)$ can be regarded 
as the coordinates of the vector $\vec{a}$ in some reference system 
$\tens{S}=[\vec{x}~\vec{y}~\vec{z}]$ (where $\vec{x}$, $\vec{y}$, 
$\vec{z}$ is a right-handed set of orthogonal unit vectors). 
Thus, we can write $\quat{a}=\bigl\{\tens{S}'\vec{a},\,a_w\bigr\}$, 
where $\tens{S}'\vec{a}=[a_x~a_y~a_z]'$ constitutes the so-called 
vector part of the quaternion, and $a_w$ the scalar part. Both scalars 
and vectors (in $\mathbb{R}^3$) can thus be seen as special
cases of quaternions, namely, when the vector or the scalar
part is zero. This allows us to write for example
$\lVert\quat{a}\rVert^2=\quat{a}\quat{a}^*=\quat{a}^*\quat{a}$.
In terms of the usual vector-scalar operations the quaternion
multiplication can also be written as
\begin{equation}\label{eq:quatmultvec}
\quat{a}\quat{b}=\Bigl\{\tens{S}'\left(\vec{a}\times\vec{b}+
\vec{a}b_w+\vec{b}a_w\right),\, a_wb_w-\vec{a}'\vec{b}\Bigr\} \, .
\end{equation}
Note that the vector part of a quaternion only makes sense when
expressed in some coordinate system ($\tens{S}$ in this example);
a physical vector cannot be part of a quaternion.

\subsection{Spatial rotations}\label{sec:quatrot}

Unit quaternions (of unit length)
are convenient for representing orientations and spins of
objects in three-dimensional space. This is compacter, numerically
more stable and requires fewer arithmetic operations than the use
of rotation matrices. Compared with the use of Euler angles,
much fewer or no trigonometric functions need to be evaluated,
and singularities are avoided.

According to Euler's rotation theorem any change in the orientation
of an object can be described as a rotation by a certain angle
around some fixed axis. Let this axis be represented by the unit
vector $\vec{u}$ and the rotation angle by $\phi$, reckoned
in the positive sense around the vector. In the given reference 
system $\tens{S}=[\vec{x}~\vec{y}~\vec{z}]$, the rotation is then
represented by the unit quaternion
\begin{equation}\label{eq:quatrot}
\quat{q}=\left\{\tens{S}'\vec{u}\sin\frac{\phi}{2},~\cos\frac{\phi}{2}
\right\} \, .
\end{equation}
The useful property here is that a sequence of rotations is 
represented by the product of the corresponding unit 
quaternions (see Sect.~\ref{sec:vecframe}).

From Eq.~(\ref{eq:quatinv}) it follows that the inverse of a unit
quaternion equals its conjugate, so
\begin{equation}\label{eq:quatrotinv}
\quat{q}^{-1}=\quat{q}^*
=\left\{-\tens{S}'\vec{u}\sin\frac{\phi}{2},~\cos\frac{\phi}{2}\right\} \, ,
\end{equation}
which represents a rotation by $-\phi$ around $\vec{u}$.

A rotation by the angle $2\pi$ around any axis is represented by the unit
quaternion $\bigl\{0,0,0,-1\bigr\}$. Since this operation is physically equivalent 
to no rotation at all, it implies a sign ambiguity in the quaternion representation
of any given rotation (modulo $2\pi$). This is potentially a problem only when
the quaternion is used to represent a continuously changing orientation as a
function of time, as is the case for the Gaia attitude (Sect.~\ref{sec:attmodel}).
It is then necessary to ensure that no sign flips occur, e.g., when converting
from some other representation of the orientations.

Equation~(\ref{eq:quatrot}) suggests an alternative, non-redundant, way 
of representing spatial rotation, namely by means of the components of
the vector $\vec{\phi}=\vec{u}\phi$. For any continuous 
rotation the three components could be continuous functions of time.
This formalism is mainly useful for small rotations
($\|\vec{\phi}\|\ll 1$);
when applied for example to a spinning satellite the length of the vector 
may grow indefinitely, causing unacceptable numerical errors in finite 
arithmetic. For the arbitrary vector $\vec{\phi}$ we nevertheless
introduce the special notation $\quat{Q}(\tens{S}'\vec{\phi})$ for the 
unit quaternion, in the reference system $\tens{S}$, representing a 
spatial rotation by the angle $\phi=\|\vec{\phi}\|$ about an axis parallel 
to $\vec{\phi}$:
\begin{equation}\label{eq:qrot}
\quat{Q}(\tens{S}'\vec{\phi}) = 
\begin{cases} \quad\displaystyle\left\{ \tens{S}'\langle\vec{\phi}\rangle
\sin\frac{\phi}{2},~\cos\frac{\phi}{2}\right\}
&\text{if $\phi>0$,}\\[6pt]
\quad\displaystyle\left\{0,0,0,1\right\} &\text{if $\phi=0$.}
\end{cases}
\end{equation} 
This notation is here only used when discussing the small rotational 
offset between the ICRS and AGIS frames in Sect.~\ref{sec:framerel}.

\subsection{Vector and frame rotations}\label{sec:vecframe}

Vector rotation and frame rotation are not standard notions in
vector or quaternion calculus, but we have found them helpful 
in order to clarify the relations between vectors and their 
representations in different reference systems.

\paragraph{Vector rotation.}
Let $\bigl\{\tens{S}'\vec{r}_0,\,0\bigr\}$ be the quaternion
representation in the reference system $\tens{S}$ of the arbitrary 
vector $\vec{r}_0$. Rotating the vector an angle $\phi$ around 
unit vector $\vec{u}$ results in a new vector $\vec{r}_1$, whose
coordinates in $\tens{S}$ can be calculated by two successive 
quaternion multiplications,
\begin{equation}\label{eq:quatvecrot}
\Bigl\{\tens{S}'\vec{r}_1,\,0\Bigr\}
=\quat{q}\Bigl\{\tens{S}'\vec{r}_0,\,0\Bigr\}\quat{q}^{-1} \, ,
\end{equation}
where $\quat{q}$ is given by Eq.~(\ref{eq:quatrot}) and $\quat{q}^{-1}=\quat{q}^*$.
We call the operation in Eq.~(\ref{eq:quatvecrot}) for the \emph{vector rotation} 
of $\vec{r}_0$ by the quaternion $\quat{q}$. This calculation requires the use 
of a particular reference system ($\tens{S}$ in this example) in which to
express the vectors and the quaternion; the resulting physical vector 
$\vec{r}_1$ is however independent of the reference system. 

Applying $n$ successive vector rotations, specified by the quaternions
$\quat{q}_1$, $\quat{q}_2$, $\dots,$ $\quat{q}_n$, gives the vector
$\vec{r}_n$ in
\begin{equation}\label{eq:quatvecrotn}
\Bigl\{\tens{S}'\vec{r}_n\,,0\Bigr\}
=\quat{q}_n\cdots\quat{q}_2\quat{q}_1
\,\Bigl\{\tens{S}'\vec{r}_0\,,0\Bigr\}\,
\quat{q}_1^{-1}\quat{q}_2^{-1}\cdots\quat{q}_n^{-1} \, .
\end{equation}
This is equivalent to a single vector rotation by 
$\quat{q}=\quat{q}_n\cdots\quat{q}_2\quat{q}_1$. All the
quaternions are here expressed in the fixed reference system
$\tens{S}$, and the order of multiplication is from right to left.

\paragraph{Frame rotation.}
A more common application in astrometry is where the reference
system itself is rotated, say from 
$\tens{S}_0=[\vec{x}_0~\vec{y}_0~\vec{z}_0]$ to
$\tens{S}_1=[\vec{x}_1~\vec{y}_1~\vec{z}_1]$, by the 
quaternion $\quat{q}$. Given the coordinates $\tens{S}_0'\vec{r}$
of the arbitrary vector $\vec{r}$ in reference system $\tens{S}_0$, we want
to compute the coordinates $\tens{S}_1'\vec{r}$ of the same physical
vector in the rotated reference system. By applying a vector rotation to 
each of the basis vectors we find
\begin{equation}\label{eq:quatframerot}
\Bigl\{\tens{S}_1'\vec{r},\,0\Bigr\}
=\quat{q}^{-1}\,\Bigl\{\tens{S}_0'\vec{r},\,0\Bigr\}\,\quat{q} \, ,
\end{equation}
which we refer to as the \emph{frame rotation} of $\vec{r}$ by the
quaternion $\quat{q}$. 

It is important to note that the numerical representation of the 
quaternion $\quat{q}$, representing a frame rotation from 
$\tens{S}_0$ to $\tens{S}_1$, is the same in the two frames. 
This follows because the components of the vector $\vec{u}$
are invariant under a frame rotation about the vector itself, i.e., 
$\tens{S}_1'\vec{u}=\tens{S}_0'\vec{u}$. In Eq.~(\ref{eq:quatrot}) 
the vector part of $\quat{q}$ can therefore 
be expressed in either of the two reference systems, i.e., we may
take $\tens{S}=\tens{S}_0$ or $\tens{S}=\tens{S}_1$. The scalar part 
$\cos(\phi/2)$ is of course independent of the reference system. 

Successive frame rotations can therefore be accomplished by referring
each rotation to the current set of axes, which is usually precisely
what is needed. Let $\quat{q}_1$, $\quat{q}_2$, $\dots,$ $\quat{q}_n$
be a sequence of frame rotations, from $\tens{S}_0$ to $\tens{S}_1$,
and then from $\tens{S}_1$ to $\tens{S}_2$, etc.; here $\quat{q}_1$ 
is expressed in $\tens{S}_0$ (or $\tens{S}_1$), $\quat{q}_2$ in
$\tens{S}_1$ (or $\tens{S}_2$), and so on. The corresponding 
transformation of the coordinates of the vector $\vec{r}$ is given by 
\begin{equation}\label{eq:quatframerot1}
\Bigl\{\tens{S}_n'\vec{r},\,0\Bigr\}
=\quat{q}_n^{-1}\cdots\quat{q}_2^{-1}\quat{q}_1^{-1}
\,\Bigl\{\tens{S}_0'\vec{r},\,0\Bigr\}\,
\quat{q}_1\quat{q}_2\cdots\quat{q}_n \, ,
\end{equation}
equivalent to a single frame rotation by 
$\quat{q}=\quat{q}_1\quat{q}_2\cdots\quat{q}_n$.
The quaternions are here expressed in the concurrent reference system,
and the order of multiplication is from left to right.

\subsection{Angular velocity}\label{sec:angvel}

Let $\quat{q}$ be a unit quaternion, expressed in the non-rotating 
reference system $\tens{C}$, and let us assume that $\quat{q}$ is a 
differentiable function of time. The angular velocity $\vec{\Omega}$ 
associated with the time-dependent vector rotation by $\quat{q}$ 
is the same for all vectors, and given by
\begin{equation}\label{eq:quatangvel}
\Bigl\{\tens{C}'\vec{\Omega},\,0\Bigr\}=2\quat{\dot{q}}\quat{q}^{-1} \, ,
\end{equation}
where the dot signifies the time derivative. This result can be derived
by taking the time derivative of the vector rotation formula for arbitrary
vector $\vec{r}$, using $\dot{\vec{r}}=\vec{\Omega}\times\vec{r}$
and Eq.~(\ref{eq:quatmultvec}).

Let $\tens{S}$ be the reference system obtained after rotation by 
$\quat{q}$. The coordinates of the angular velocity vector in the new 
system are found by performing a frame rotation of 
Eq.~(\ref{eq:quatangvel}) by $\quat{q}$; thus
\begin{equation}\label{eq:quatangvelrot}
\Bigl\{\tens{S}'\vec{\Omega},\,0\Bigr\}=
\quat{q}^{-1}\Bigl\{\tens{C}'\vec{\Omega},\,0\Bigr\}\quat{q}=
2\quat{q}^{-1}\quat{\dot{q}} \, .
\end{equation}
These expressions show, for example, how the instantaneous angular velocity
of Gaia can be calculated either in the celestial reference system (CoMRS),
using Eq.~(\ref{eq:quatangvel}), or in the instrument system (SRS),
using Eq.~(\ref{eq:quatangvelrot}), from a knowledge of the attitude
$\quat{q}$ and its time derivative $\quat{\dot{q}}$.

\subsection{The attitude matrix}\label{sec:attmat}

For completeness, we give here the full relations
between the attitude matrix defined in Sect.~\ref{sec:attmodel} and
the quaternion representation of the attitude. Let
$\tens{C}=[\vec{X}~\vec{Y}~\vec{Z}]$ be the celestial reference system 
(CoMRS; Sect.~\ref{sec:refsystems}) and $\tens{S}=[\vec{x}~\vec{y}~\vec{z}]$ 
the instrument system (SRS). The attitude matrix $\vec{A}$ is defined by
Eq.~(\ref{eq:attmat}). The attitude quaternion
$\quat{q}=\bigl\{q_x,\,q_y,\,q_z,\,q_w\bigr\}$ gives the 
rotation from the CoMRS to the SRS, i.e., 
$\bigl\{\tens{C}'\vec{x},0\bigr\}=
\quat{q}\,\bigl\{\tens{C}'\vec{X},0\bigr\}\,\quat{q}^{-1}$, etc. Then
\begin{equation}\label{eq:quat2mat}
\vec{A} =
\begin{bmatrix}
    1-2(q_y^2+q_z^2) & 2(q_xq_y+q_zq_w) &  2(q_xq_z-q_yq_w) \\
    2(q_xq_y-q_zq_w) & 1-2(q_x^2+q_z^2) &  2(q_yq_z+q_xq_w) \\
    2(q_xq_z+q_yq_w) & 2(q_yq_z-q_xq_w) &  1-2(q_x^2+q_y^2)
\end{bmatrix} .
\end{equation}
The conversion from $\vec{A}$ to $\quat{q}$ is less straightforward
if we want to avoid numerical problems for certain attitudes. A stable
algorithm was given by \citet{klumpp1976}. In our notation, using
pseudo-code, it is given by Algorithm~\ref{algo:A1}.
Note that $\bigl\{q_x,\,q_y,\,q_z,\,q_w\bigr\}$ and 
$\bigl\{-q_x,\,-q_y,\,-q_z,\,-q_w\bigr\}$
correspond to the same $\vec{A}$, while the algorithm always returns
a quaternion with $q_w\ge 0$; a reversal of the signs may therefore
sometimes be required to ensure the temporal continuity of the 
quaternion components.

\begin{algorithm}[t]
\caption{For given $\vec{A}=[ A_{xx}~A_{xy}~A_{xz};~A_{yx}~\cdots ]$,
this algorithm returns a unit quaternion $\quat{q}=\{q_x,\,q_y,\,q_z,\,q_w\}$
such that Eq.~(\ref{eq:quat2mat}) is satisfied.}
\label{algo:A1}
\begin{algorithmic}[1]
\STATE{$s \leftarrow A_{xx}+A_{yy}+A_{zz}$}
\FOR{$i=x,\,y,\,z$}
    \STATE{$|q_i| \leftarrow [(1-s)/4+A_{ii}/2]^{1/2}$}
\ENDFOR
\STATE{$q_w \leftarrow [(1+s)/4]^{1/2}$}
\IF{$|q_x|\ge\max(|q_y|,|q_z|)$}
    \STATE{$i \leftarrow x,~~j \leftarrow y,~~k \leftarrow z$}
\ELSIF{$|q_y|\ge\max(|q_x|,|q_z|)$}
    \STATE{$i \leftarrow y,~~j \leftarrow z,~~k \leftarrow x$}
\ELSE
    \STATE{$i \leftarrow z,~~j \leftarrow x,~~k \leftarrow y$}
\ENDIF
\STATE{$q_i \leftarrow |q_i|\,\mbox{sign}[A_{jk}-A_{kj}]$}
\STATE{$q_j \leftarrow |q_j|\,\mbox{sign}[q_i(A_{ij}+A_{ji})]$}
\STATE{$q_k \leftarrow |q_k|\,\mbox{sign}[q_i(A_{ik}+A_{ki})]$}
\end{algorithmic}
\end{algorithm}

\subsection{Differential rotation}\label{sec:diffrot}

Up until now, the quaternion formulae given in this Appendix hold rigorously 
for arbitrary rotations. We now derive a useful result, which however is only valid 
to first order in the (small) rotation angles. It can be used, for example, to compute 
the attitude error angles about the SRS axes, as was done in Sect.~\ref{sec:resatt}.

Let $\quat{q}_0$ and $\quat{q}_1$ be given unit quaternions, representing the two 
nearly co-aligned reference systems $\tens{S}_0=[\vec{x}_0~\vec{y}_0~\vec{z}_0]$ 
and $\tens{S}_1=[\vec{x}_1~\vec{y}_1~\vec{z}_1]$ with respect to some third 
(common) reference system such as the ICRS.  It is required to express the difference 
between $\tens{S}_1$ and $\tens{S}_0$ by means of three small angles 
$\phi_x$, $\phi_y$, $\phi_z$ representing rotations about the axes in either
system. More precisely, let $\vec{\phi}=[\phi_x~\phi_y~\phi_z]'$ be the spatial 
rotation that brings $\tens{S}_0$ into coincidence with $\tens{S}_1$. Assuming 
that $\|\vec{\phi}\|\ll 1$ and neglecting terms of order $\|\vec{\phi}\|^2$, we have 
$\tens{S}_1\simeq\tens{S}_0+\vec{\phi}\times\tens{S}_0$, or
\begin{equation}\label{eq:diffrot1}
\tens{S}_1'\tens{S}_0 \simeq \vec{I}+\left(\vec{\phi}\times\tens{S}_0\right)'\tens{S}_0
=\begin{bmatrix} 1 & \phi_z & -\phi_y \\ -\phi_z & 1 & \phi_x \\ \phi_y & -\phi_x & 1
\end{bmatrix} \, .
\end{equation}
According to Eq.~(\ref{eq:attmat}) this matrix describes the orientation of $\tens{S}_1$ 
with respect to $\tens{S}_0$. If $\quat{d}$ is the quaternion representing a frame rotation 
from $\tens{S}_0$ to $\tens{S}_1$, we have (Sect.~\ref{sec:vecframe}) 
$\quat{q}_1=\quat{q}_0\quat{d}$ and hence
\begin{equation}\label{eq:diffrot2}
\quat{d} \equiv \{d_x,\, d_y,\, d_z,\, d_w\} = \quat{q}_0^{-1}\quat{q}_1 \, .
\end{equation}
Given that $\|\vec{\phi}\|$ is small, $d_x$, $d_y$ and $d_z$ are also small quantities,
while $|d_w|\simeq 1$. Due to the sign ambiguity of the quaternion representation 
(Sect.~\ref{sec:quatrot}), $d_w$ could however have either sign. Comparing 
Eqs.~(\ref{eq:quat2mat}) and (\ref{eq:diffrot1}) we find, to first order in the small quantities,
\begin{equation}\label{eq:diffrot3}
\phi_x \simeq 2 d_x d_w\, , \quad
\phi_y \simeq 2 d_y d_w\, , \quad
\phi_z \simeq 2 d_z d_w\, ,
\end{equation}
where the factor $d_w$ (being close to $\pm 1$) guarantees that the angles 
obtain their correct signs. Equations~(\ref{eq:diffrot2})--(\ref{eq:diffrot3}) provide 
the required transformation from $(\quat{q}_0,~\quat{q}_1)$ to $(\phi_x,~\phi_y,~\phi_z)$.

\section{Splines}\label{sec:splines}

A spline is a piecewise polynomial function $S(t)$ defined on
some interval $[t_{\rm beg},t_{\rm end}]$. That is, if the interval
is divided into $K>0$ sub-intervals by means of the instants $t_k$
(called knots),
such that $t_{\rm beg} = t_0 < t_1 < \cdots < t_K = t_{\rm end}$,
then in the sub-interval $t_k \le t < t_{k+1}$ the spline
function $S(t)$ equals some polynomial $P_k(t)$, $k=0\dots K-1$.
The splines of interest here consist of polynomials of some fixed
order $M$ (or degree $M-1$); typically $M=4$, corresponding to cubic
splines. Moreover, the splines are usually maximally smooth, i.e.,
$S^{(m)}(t)\equiv{\rm d}^mS/{\rm d}t^m$ is continuous for
$t_{\rm beg}<t<t_{\rm end}$ and $m=0\dots M-2$.

The $K$ polynomials of order $M$ require $KM$ coefficients for their
specification. For a maximally smooth spline, there are $M-1$
continuity conditions for each interior knot, namely
$S^{(m)}(t_k-)=S^{(m)}(t_k+)$ for $m=0\dots M-2$ and $k=1\dots K-1$;
thus a total of $(K-1)(M-1)$ constraints. The spline consequently
has $KM-(K-1)(M-1)=K+M-1$ degrees of freedom.%
\footnote{When the spline is used for interpolation, it is
typically chosen to go exactly through the $K+1$ values $S(t_k)$
for $k=0\dots K$. This leaves $(K+M-1)-(K+1)=M-2$ degrees of freedom.
Thus, for a cubic spline ($M=4$), two more conditions must be imposed
for the spline to become unique. The most common choice is to make
the second derivative vanish at $t_{\rm beg}$ and $t_{\rm end}$.
This defines the `natural' interpolating spline. By contrast, when
splines are least-squares fitted to data, as in the attitude
determination problem, there are typically many data points per
sub-interval, and no need for special endpoint conditions.}
In the context of least-squares fitting, this equals the number of
unknowns (parameters to fit), which we denote by $N$. In the following
we take $N$ and $M$ to be the characteristic numbers of the spline,
rather than $K$ and $M$. For a maximally smooth spline we have
$K=N-M+1$.

\begin{figure}[t!]
\centerline{
\includegraphics[width=65mm,angle=-90]{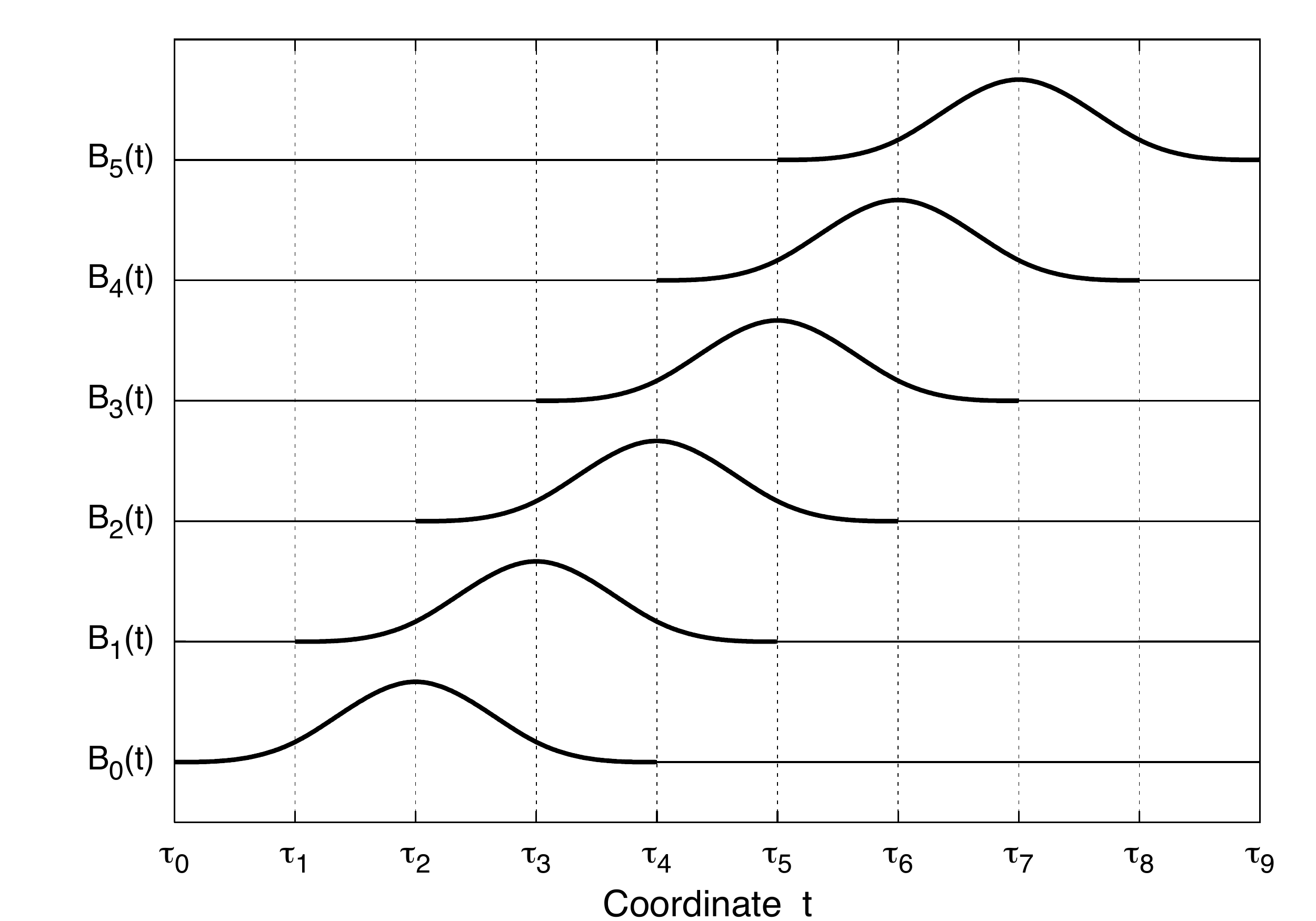}
}
\caption{The first six B-splines of order $M=4$ (cubic) defined on the
regular knot sequence $\tau_0,\,\tau_1,\,\dots$. For better visibility,
each B-spline is vertically displaced by one unit, and the non-zero parts
are drawn in thick lines.}
\label{GAIA-AppB0}
\end{figure}

\subsection{B-splines}\label{sec:bsplines}

There are many different ways in which a spline could be represented
(parametrized). The approach taken here is to consider $S(t)$ as a
linear combination of some basis functions $B_n(t)$,
\begin{equation}\label{eq:AppB0}
S(t) = \sum_{n=0}^{N-1} c_n\, B_n(t) \, ,
\end{equation}
where $c_n$ are coefficients to be determined. Choosing the basis
functions to have minimal support (see below) for the given order
and smoothness leads to the so-called B-splines \citep{book:DB-01}.

The B-splines are uniquely defined by the adopted knot sequence.
In the following we shall use $\tau_n$ (rather than $t_n$) to denote
the knots associated with the B-splines. The reason is that the knot
sequence for the B-splines has a slightly more general interpretation
than just the simple division of $[t_{\rm beg},t_{\rm end}]$ considered
above. Moreover, when fitting the spline to given data points, this
allows us to use $t_k$ (for example) to denote the time associated with
the $k$th data point, without ambiguity.

The knot sequence must be non-decreasing, $\tau_0 \le \tau_1 \le
\tau_2 \le \cdots$, and at least two knots must be different.
Very often we use a regular knot sequence in which
$\tau_{n+1}=\tau_n+\Delta\tau$ for some $\Delta\tau>0$ (the knot
interval). Figure~\ref{GAIA-AppB0} shows the first six cubic B-splines
defined on a regular knot sequence. Note that the non-zero part of
each B-spline, shown in heavy line, stretches over $M$ consecutive
knot intervals. We use the convention that the B-splines are
labelled with the same index as the left-most knot of its non-zero
part; therefore, the support of $B_n(t)$ is $(\tau_n,\tau_{n+M})$.

It is not possible to construct a maximally smooth spline function
of order $M$ that is non-zero over less than $M$ consecutive knot
intervals. This is what we mean by the statement that B-splines have
minimal support: they could not be shortened without loosing some
smoothness. This is an important property for the least-squares
fitting, as shown by the following consideration. Least-squares
fitting of (\ref{eq:AppB0}) to a given set of data points $(t_k,z_k)$
(where $t_k\in [t_{\rm beg},t_{\rm end}]$ for each $k$) will be
done by forming normal equations. The normal equations matrix
$\vec{N}$ is a symmetric, positive definite matrix of dimension%
\footnote{In the attitude determination problem, each of the four
components of the quaternion is represented by a spline, so the
number of parameters is actually $4N$ and the normal matrix is of
dimension $4N\times 4N$. Alternatively, we may take the elements
of $\vec{N}$ to be sub-matrices of dimension $4\times 4$. The
indices $i$ and $j$ used below refer to these sub-matrices.}
$N\times N$. Since $N$ may be very large, it is desirable that
the matrix is sparse, i.e., that most elements are zero. It is easy
to see that element $N_{ij}$ will be non-zero as soon as
$B_i(t_k)B_j(t_k)\ne 0$ for some data point $k$. To make
the matrix maximally sparse, we should therefore choose basis
functions with minimal support. B-splines have minimal support and
are therefore ideal for least-squares fitting using sparse matrix
algebra.

Since the support of a B-spline of order $M$ extends over at most $M$
consecutive knot intervals, we have $N_{ij}=0$ for $|i-j|>M-1$. This
shows that $\vec{N}$ is a symmetric banded matrix with (half-) bandwidth
equal to $M-1$. Cholesky decomposition preserves the band structure
of the matrix and is therefore ideal for solving the normals; it is
also numerically very efficient.

\subsection{Calculation of B-splines}\label{sec:calcBsplines}

At any given point $t$ there are at most $M$ non-zero B-splines,
namely $B_{\ell-M+1}(t)$, $B_{\ell-M+2}$, $\dots$ $B_{\ell}(t)$, where
$\ell$ is the `left index' of $t$, such that $\tau_\ell \le t < \tau_{\ell+1}$.
They can be computed simultaneously in a numerically stable way by
means of de~Boor's algorithm \citep{book:DB-01}, which is given as
pseudo-code in Algorithm~\ref{algo:B1}.
If needed, the derivatives of the B-splines with respect to $t$ can be
computed simultaneously with little additional effort.

\begin{algorithm}[t]
\caption{For given spline order $M$, knot sequence $\{\tau_n\}$, time $t$, 
and left index $\ell$ (such that $\tau_\ell \le t < \tau_{\ell+1}$), this algorithm 
returns $\{ b_0\dots b_{M-1}\}$ such that $b_0=B_{\ell-M+1}(t)$, $\dots$, $b_{M-1}=B_{\ell}(t)$.}
\label{algo:B1}
\begin{algorithmic}[1]
\STATE{$b_0 \leftarrow 1$}
\FOR{$i=0$ \TO $M-2$}
    \STATE{$R_i \leftarrow \tau_{\ell+i+1}-t$}
    \STATE{$L_i \leftarrow t-\tau_{\ell-i}$}
    \STATE{$s \leftarrow 0$}
    \FOR{$j=0$ \TO $i$}
    \STATE{$u \leftarrow b_j/(R_j + L_{i-j})$}
    \STATE{$b_j \leftarrow s + R_j \times u$}
    \STATE{$s \leftarrow L_{i-j} \times u$}
  \ENDFOR
  \STATE{$b_{i+1} \leftarrow s$}
\ENDFOR
\end{algorithmic}
\end{algorithm}

By inspection of the algorithm it is found that the knots used for computing
the B-spline values in the interval $[\tau_\ell,\,\tau_{\ell+1})$ are
$\tau_{\ell-M+2}$ through $\tau_{\ell+M-1}$. For example, with reference
to Fig.~\ref{GAIA-AppB0} (with $M=4$), the B-splines between $\tau_3$ and $\tau_4$
(i.e., for left index $\ell=3$) depend on
$\{\tau_1,\,\tau_2,\,\dots,\,\tau_6\}$, but not on $\tau_0$ or
$\tau_7$, even though $B_0(t)$ in general depends on $\tau_0$
and $B_3(t)$ depends on $\tau_7$.

\subsection{Use of multiple knots}\label{sec:multknot}

Algorithm~\ref{algo:B1} works also in the case when several consecutive
knots are placed at the same $t$ coordinate. Having a knot of multiplicity
$m$ (i.e., $m$ knots at the same $t$) removes the continuity for derivatives
of order $M-m$ and higher. The normal situation is that the knots have
multiplicity 1, which means that the spline is continuous at the knots up to
and including its $(M-2)$th derivative, but discontinuous in its $(M-1)$th
derivative. By inserting multiple knots at some
specific instant, one allows the spline to become less smooth at this point.
For example, in a cubic spline ($M=4$) a triple knot ($m=3$) allows the
first derivative to become discontinuous at that point, while leaving the
spline function itself continuous. In the Gaia
attitude processing, multiple knots will be used for modelling the effects
of micrometeoroid impacts, which cause (almost) instantaneous changes in
the angular velocity, corresponding to discontinuities in the first derivative
of the attitude spline.

Multiple knots may also be used at the endpoints of the spline interval
$[t_{\rm beg},t_{\rm end}]$. At any point in this interval, there must
be exactly $M$ non-zero B-splines in order that a linear combination of
them should to be able to represent an arbitrary spline of order $M$.
Again, with reference to Fig.~\ref{GAIA-AppB0} (for $M=4$), we see that this is
the case to the right of $\tau_3$ (or $\tau_{M-1}$ in general). Thus we
should put $\tau_{M-1}=t_{\rm beg}$. The first $M-1$ knots can in principle
be placed anywhere, as long as $\tau_0\le\tau_1\le\dots\le\tau_{M-1}$:
any such arrangement will produce $M$ non-zero B-splines in the next
sub-interval $(\tau_{M-1},\tau_M)$, and although the B-splines are
different depending on the arrangement of the knots, they are always
linearly independent and therefore can be used as a basis for the
spline. In particular, it is possible to put the first $M$ knots at
the same point, i.e., $\tau_0=\tau_1=\dots=\tau_{M-1}=t_{\rm beg}$.

\begin{figure}[t!]
\begin{center}
\includegraphics[width=\columnwidth,]{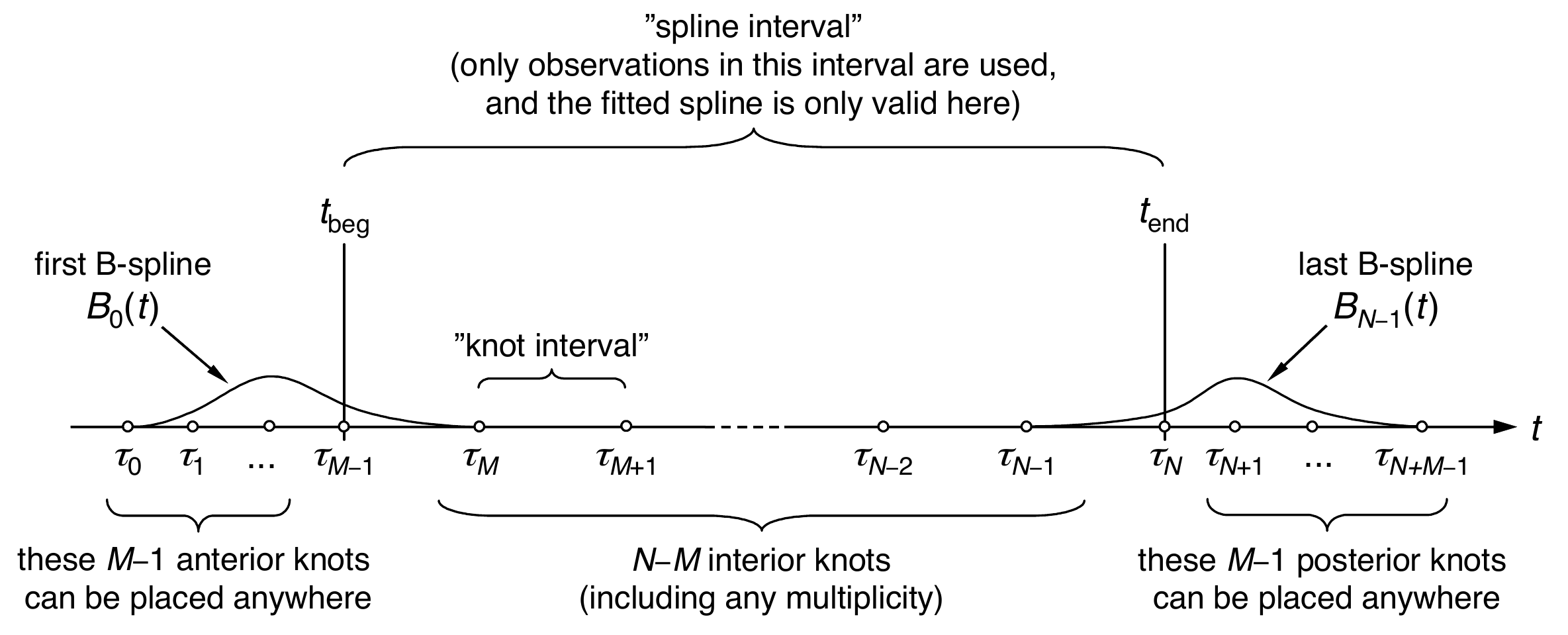}
\null\vspace*{3pt}
\caption{Illustration of the knot placement for a spline of order $M$
(e.g., $M=4$ for a cubic spline) with $N$ degrees of freedom.
$[t_{\rm beg},t_{\rm end}]$ is the spline interval over which the
spline is fitted to given data points. The end knots $\tau_{M-1}$ and
$\tau_N$ are at the endpoints of the spline interval.
The $N-M$ interior knots are chosen to give the spline the desired
flexibility, including multiple knots where required. The placement
of the anterior ($\tau\le t_{\rm beg}$) and posterior knots
($\tau\ge t_{\rm end}$) is in principle arbitrary: it does not change
the fitted spline in $[t_{\rm beg},t_{\rm end}]$, but may affect the
condition number of the least-squares fit. The parameters of the
fitted spline are the coefficients $c_0,\,\dots\,c_{N-1}$ of the
B-splines $B_0(t)$ through $B_{N-1}(t)$.}
\label{GAIA-AppB3}
\end{center}
\end{figure}

Corresponding considerations apply to the right limit of the spline
interval: with $N$ degrees of freedom, the support of the last
B-spline $B_{N-1}(t)$ extends from $\tau_{N-1}$ to $\tau_{N+M-1}$.
The spline interval must end at $\tau_N=t_{\rm end}$, and the
remaining $M-1$ knots can be placed anywhere provided that
$\tau_N\le\tau_{N+1}\le\dots\le\tau_{N+M-1}$. In particular, it
is possible to have an $M$-fold knot at the endpoint of the spline
interval, i.e., $t_{\rm end}=\tau_N=\tau_{N+1}=\dots=\tau_{N+M-1}$.
Figure~\ref{GAIA-AppB3} summarizes the placement of knots in relation to
a given spline interval for given order
$M$ and degrees of freedom $N$ (number of spline coefficients).

\begin{figure}[t!]
\begin{center}
\includegraphics[width=75mm,]{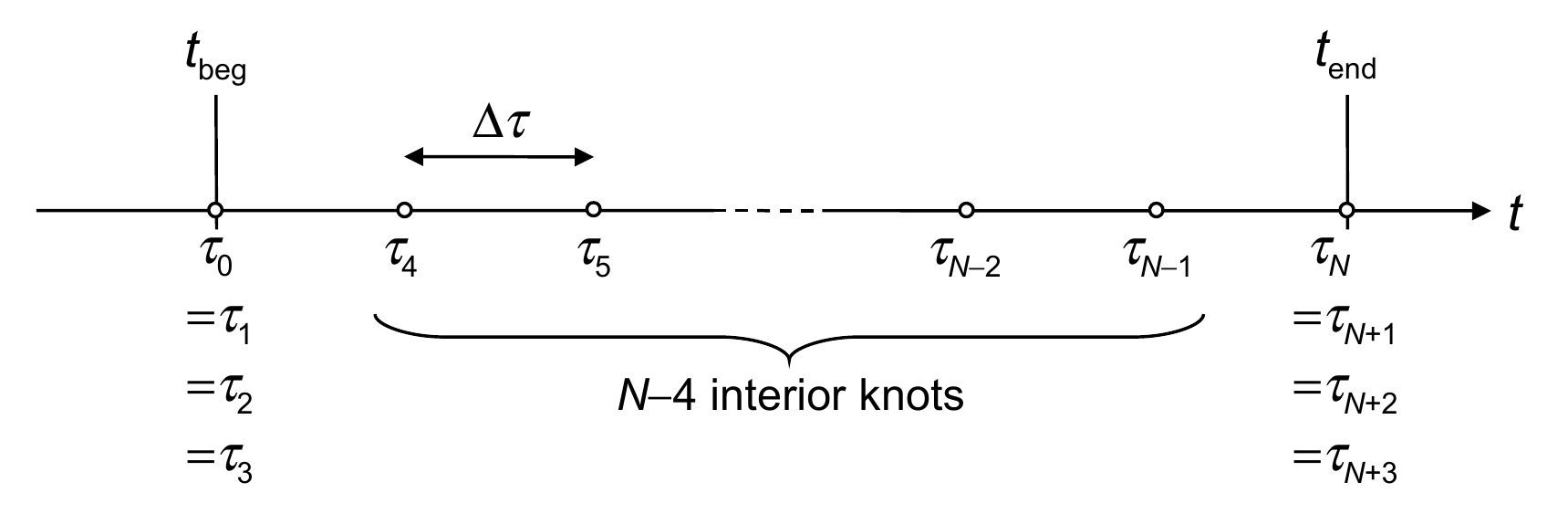}
\caption{Definition of a regular knot sequence for fitting a cubic
spline (order $M=4$) in the interval $[t_{\rm beg},t_{\rm end}]$.
The spline interval is divided into $N-3$ knot intervals of equal
length $\Delta\tau =(t_\text{end}-t_\text{beg})/(N-3)$.}
\label{GAIA-Model0}
\end{center}
\end{figure}

Although the placement of the first and last $M-1$ knots is arbitrary,
and does not change the resulting spline between $t_{\rm beg}$ and
$t_{\rm end}$, their placement does affect the numerical stability of
the resulting least-squares system. Butkevich \& Klioner
(unpublished technical note) 
has pointed out that collapsing the anterior and posterior knots into the
end knots, so that $\tau_0=\tau_1=\dots =\tau_{M-1}=t_{\rm beg}$
and $t_{\rm end}=\tau_N=\tau_{N+1}=\dots =\tau_{N+M-1}$) results in a
system with much smaller condition number than a regular sequence
extending beyond the endpoints.
For a cubic spline ($M=4$) the spline interval then begins and ends
with 4-fold knots as illustrated in Fig.~\ref{GAIA-Model0}. However,
the use of data segmentation, as described in Sect.~\ref{sec:att-segment},
may not permit this device.

\section{A robust Cholesky algorithm for positive semidefinite 
matrices without pivoting}\label{sec:cholesky}

\subsection{The use of normal equations}

The least-squares problems considered in this paper are solved 
by the method of normal equations (here denoted 
$\vec{N}\vec{x}=\vec{b}$), using the Cholesky algorithm
to decompose the symmetric normal matrix $\vec{N}$. This 
method is known to be computationally efficient and accurate 
for well-conditioned problems \citep[e.g.,][]{stewart:1998}, and 
is moreover well adapted to in-place 
manipulation of sparse matrices such as the band matrix 
obtained when fitting B-splines (Fig.~\ref{fig:Naa})

The method of normal equations for the general least-squares problem
is usually discouraged in the literature, due to the much superior 
stability and accuracy of alternative methods operating directly on 
the observation equations $\vec{A}\vec{x}\simeq\vec{h}$ (where 
$\vec{N}=\vec{A}'\vec{A}$ and $\vec{b}=\vec{A}'\vec{h}$),
e.g., using Householder orthogonal transformations 
\citep[e.g.,][]{book:bjork-1996}. However, when working with very
large problems that are inherently well-posed, thanks to a good 
design of experiment and reduction model, our experience is that 
the method of normal equations is nearly always adequate in terms 
of accuracy, and then has the edge over other methods in terms of
speed, storage and simplicity of the code. Moreover, in these
problems, iterative improvement of the solution is usually required
for other reasons (non-linearity, elimination of outliers), which 
partly compensates for the loss of precision when forming the
normal equations.

\subsection{The Cholesky algorithm}

The standard Cholesky algorithm 
\citep[e.g.,][]{book:bjork-1996,golu+96} requires that $\vec{N}$ is
positive definite, which is always the case for a well-conditioned 
least-squares problem. Given $\vec{N}$ and $\vec{b}$, the solution 
of the system $\vec{N}\vec{x}=\vec{b}$ proceeds in three steps:
\begin{description}
\item[C1.]
Use the Cholesky algorithm to find the unique upper-diagonal matrix 
$\vec{U}$, with positive diagonal entries, such that $\vec{N}=\vec{U}'\vec{U}$. 
\item[C2.]
Solve the lower-triangular system $\vec{U}'\vec{y}=\vec{b}$.
\item[C3.]
Solve the upper-triangular system $\vec{U}\vec{x}=\vec{y}$.
\end{description}
The matrix $\vec{U}$ (or its transpose) is known as the Cholesky factor 
or square root of $\vec{N}$. As all the computations can be made 
in-place, we have symbolically
\begin{equation}\label{eq:chol1}
\left[~\vec{N}~|~\vec{b}~\right]  \xrightarrow{\text{C1}}
\left[~\vec{U}~|~\vec{b}~\right]  \xrightarrow{\text{C2}}
\left[~\vec{U}~|~\vec{y}~\right]  \xrightarrow{\text{C3}}
\left[~\vec{U}~|~\vec{x}~\right] \, .
\end{equation}
The extension to an arbitrary number of right-hand sides (to the right of
the vertical line) is trivial. For example, the inverse $\vec{N}^{-1}$ can
be obtained by inserting the identity matrix for $\vec{b}$.
We also note that C1 and C2 are mathematically
equivalent to pre-multiplying the matrix and right-hand side, respectively, 
by $(\vec{U}')^{-1}$. Performing C2 on the unit vector 
$\vec{e}_i=[0,0,\dots,1,\dots,0]'$ (with 1 in position $i$) thus produces 
$\vec{\tilde{e}}_i=(\vec{U}')^{-1}\vec{e}_i$ such that 
$\vec{\tilde{e}}_i'\vec{\tilde{e}}_j=\vec{e}_i'\vec{U}^{-1}(\vec{U}')^{-1}\vec{e}_j
=\vec{e}_i'\vec{N}^{-1}\vec{e}_j=(\vec{N}^{-1})_{ij}$. Selected elements or
sub-matrices of $\vec{N}^{-1}$ can thus be obtained without computing the
full matrix.

The above three steps are accomplished by 
Algorithms~\ref{algo:C1}--\ref{algo:C3} for arbitrary symmetric
positive definite $\vec{N}$. (Actually, these algorithms include the
non-standard modification discussed below in order to handle 
semidefinite matrices gracefully.) A few remarks should be made concerning
its practical implementation. First, the matrices $\vec{U}$ and $\vec{N}$
in C1, $\vec{y}$ and $\vec{b}$ in C2, and $\vec{x}$ and $\vec{y}$ in C3
can share the same memory if it is acceptable that $\vec{U}$ overwrites 
$\vec{N}$, etc (in-place calculation). Second, since $\vec{N}$ is symmetric,
only the upper-diagonal part of it ($N_{ij}$ for $i\le j$) is used in C1, and 
similarly for $\vec{U}$. For large systems one can therefore save roughly 
half the memory by storing only the upper-diagonal parts of $\vec{N}$ 
and $\vec{U}$, e.g., as one-dimensional arrays. The code in 
lines~\ref{algoC1:19}--\ref{algoC1:21} of 
Algorithm~\ref{algo:C1} is then irrelevant.
Third, if $\vec{N}$ is a `skyline matrix' with envelope $E_j$ (i.e.,
$N_{ij}=0$ for $i<E_j$) then $\vec{U}$ has the same envelope: the
Cholesky decomposition gives no fill-in above the envelope. This
allows to store and decompose certain sparse matrices very efficiently,
such as the band matrix in Fig.~\ref{fig:Naa}.

\begin{algorithm}[t]
\caption{Returns upper-triangular $\vec{U}$ such that 
$\vec{U}'\vec{U}=\vec{N}$, where $\vec{N}\in\mathbb R^{n\times n}$ 
is symmetric and positive semidefinite. Also computes an estimate $d$
of the rank defect of $\vec{N}$.}
\label{algo:C1}
\begin{algorithmic}[1]
\STATE{$\vec{U} \leftarrow \vec{N}$, ~~~$d \leftarrow 0$}
\FOR{$j=0$ \TO $n-1$}
	\FOR{$i=0$ \TO $j$}
		\STATE{$s \leftarrow \sum_{k=0}^{i-1} U_{ki}U_{kj}$}
		\IF{$i<j$}
			\IF{$U_{ii}>0$}
				\STATE{$U_{ij} \leftarrow (U_{ij} - s) / U_{ii}$}
			\ELSE
				\STATE{$U_{ij} \leftarrow 0$}
			\ENDIF
		\ELSE
			\IF{$U_{jj} - s > 0$}
				\STATE{$U_{jj} \leftarrow \sqrt{U_{jj} - s}$}
			\ELSE
				\STATE{$U_{jj} \leftarrow 0$, ~~~$d \leftarrow d+1$}
			\ENDIF
		\ENDIF
	\ENDFOR
	\FOR{$i=j+1$ \TO $n-1$}\label{algoC1:19}
    	\STATE{$U_{ij} \leftarrow 0$}
    \ENDFOR\label{algoC1:21}
\ENDFOR
\end{algorithmic}
\end{algorithm}

\begin{algorithm}[t]
\caption{Returns $\vec{y}$ such that $\vec{U}'\vec{y}=\vec{b}$, where
$\vec{U}\in\mathbb R^{n\times n}$ is upper-triangular and $\vec{b}\in\mathbb R^{n}$.}
\label{algo:C2}
\begin{algorithmic}[1]
\STATE{$\vec{y} \leftarrow \vec{b}$}
\FOR{$i=0$ \TO $n-1$}
	\IF{$U_{ii}>0$}
	    \STATE{$y_i \leftarrow (y_i - \sum_{k=0}^{i-1} U_{ki}y_k) / U_{ii}$}
	\ELSE
	    \STATE{$y_i \leftarrow 0$}
	\ENDIF
\ENDFOR
\end{algorithmic}
\end{algorithm}

\begin{algorithm}[t]
\caption{Returns $\vec{x}$ such that $\vec{U}\vec{x}=\vec{y}$, where
$\vec{U}\in\mathbb R^{n\times n}$ is upper-triangular and $\vec{y}\in\mathbb R^{n}$.}
\label{algo:C3}
\begin{algorithmic}[1]
\STATE{$\vec{x} \leftarrow \vec{y}$}
\FOR{$i=n-1$ \TO $0$}
	\IF{$U_{ii}>0$}
	    \STATE{$x_i \leftarrow (x_i - \sum_{k=i+1}^{n-1} U_{ik}x_k) / U_{ii}$}
	\ELSE
	    \STATE{$x_i \leftarrow 0$}
	\ENDIF
\ENDFOR
\end{algorithmic}
\end{algorithm}

\subsection{Application to semidefinite systems}

In several of our applications the normal matrix is however not 
positive definite, either from a lack of observations (e.g., data gaps in the 
attitude spline representation) or by design (e.g., for the calibration 
parameters). Application of the standard Cholesky algorithm in such 
cases results in an exception which may be non-trivial to handle
(for example, by changing the knot sequence of the attitude spline) 
and which may leave the partially solved system in an undefined 
state. However, that the Cholesky algorithm fails does not mean that 
there is no solution to the normal equations: on the contrary, there 
is an infinitude of solutions and the problem is rather which one to 
pick. Indeed, in many situations we could in principle accept \emph{any} 
valid solution; for example, when the null space of the problem is 
known a~priori, and we are prepared to handle the associated 
non-uniqueness of the solution (cf.~Sect.~\ref{sec:framerotator}). 
For these and several other reasons it is advantageous if the 
computation can be continued in some sensible way, while of
course noting the detected rank deficiency. 

A number of methods are available to handle rank-deficient or
ill-conditioned least-squares problems. Singular Value Decomposition
\citep[SVD; e.g.,][]{book:bjork-1996,golu+96,book:nr3} is the method
often recommended; it provides the unique `pseudo-solution' with
the smallest Euclidean norm. However, SVD is computationally 
expensive and the pseudo-solution is not necessarily better than 
any other valid solution to the singular least-squares problem.

By construction, the normal matrix $\vec{N}=\vec{A}'\vec{A}$ is 
positive semidefinite: $\vec{x}'\vec{N}\vec{x}=\|\vec{A}\vec{x}\|^2\ge 0$ 
for any $\vec{x}$ (cf.\ footnote~\ref{footnote1}). A modification of the
standard Cholesky algorithm allows the decomposition in C1 to be made
also for such matrices, although $\vec{U}$ is no longer unique; similarly 
C2 and C3 can be modified to produce a valid (if non-unique) solution 
to the normal equations $\vec{N}\vec{x}=\vec{b}$. For example, the
LINPACK routine xCHDC \citep{LINPACK-UG} implements a robust
version of the Cholesky algorithm, using complete pivoting (i.e., a 
simultaneous permutation of the rows and columns of $\vec{N}$) 
to generate the unique square root with non-zero elements only in 
the first $r$ rows, if $r<n$ is the rank of the matrix 
\citep[for a detailed analysis, see][]{Higham:2002:ASN}.
Other modifications of the Cholesky algorithm 
\citep[e.g.,][]{schnabel+99} also uses pivoting.

Permuting the rows and/or columns of the matrix is highly 
undesirable in the present applications. For example, when applied 
to a band matrix (such as Fig.~\ref{fig:Naa}) it likely destroys the
band structure, and in general prevents the envelope-based storing
of the sparse matrices $\vec{N}$ and $\vec{U}$ outlined above.
While pivoting is never needed for the Cholesky factorization of a 
positive definite matrix, it is thought to be an essential ingredient 
in modified algorithms aimed at more general symmetric matrices. 
A simple modification of the Cholesky algorithm, which makes it 
applicable to the semidefinite case without pivoting was described
by \citet[][Eq.~19.5]{laws+74}; see also \citet[][Eq.~4.2.11]{golu+96}, 
who however warn that ``it may be preferable
to incorporate pivoting''. Nevertheless we have implemented the 
corresponding modifications in Algorithms~\ref{algo:C1}--\ref{algo:C3}
without pivoting, and find that they work quite well in our applications. 
Algorithm~\ref{algo:C1} includes a rough estimation of the rank defect 
$d=n-r$.

The numerical accuracy of the decomposition in Algorithm~\ref{algo:C1}
was tested in MATLAB for a range of rank-deficient random positive
semidefinite matrices $\vec{N}$ \citep[patterned after][]{higham1990} 
by computing the quantity
$\rho_N=\|\vec{N}-\vec{U}'\vec{U}\|_\text{F}/(u\|\vec{N}\|_\text{F})$
as a measure of the relative error in units of the floating point precision.
Here $\|\vec{N}\|_\text{F}=[\text{trace}(\vec{N}'\vec{N})]^{1/2}$ is
the Frobenius norm of $\vec{N}$, and $u=2^{-52}$ is the unit roundoff 
\citep{golu+96} of the double precision floating point arithmetic
used.  We find that the present algorithm, without 
pivoting, performs almost as well as LINPACK's xCHDC with complete 
pivoting, as judged from the statistics of our $\rho_N$ 
compared with the corresponding quantity $\rho_k$ reported by Higham.
However, C.1 is much less useful as a rank-revealing algorithm -- the 
estimated rank defect is often much too small.

Similarly, in order to check the numerical validity of the solution to
the rank-deficient normal equations computed by C.1--C.3, we made
the following experiments, using the same matrices as above. For 
random vectors $\vec{x}$ we first computed $\vec{b}=\vec{N}\vec{x}$, 
then used C.1--C.3 to recover a solution $\vec{\tilde{x}}$ (usually very
different from $\vec{x}$). Finally we computed
$\rho_b=\|\vec{b}-\vec{N}\vec{\tilde{x}}\|_\text{F}/(u\|\vec{b}\|_\text{F})$.
We find that our algorithm performs almost as well as MATLAB's 
basic solution {\tt xtilde = N$\backslash$b}, and much better than 
the minimum norm solution {\tt xtilde = pinv(N)*b} (with default tolerance). 
For example, the
99th percentile of $\rho_b$ was $\sim\!\!10^4$ for C.1--C.3, 
$\sim\!\!10^3$ for MATLAB's backslash ($\backslash$) operator, and
$\sim\!\!10^{11}$ (!) when using {\tt pinv}.

We conclude from these limited experiments
that the present version of the Cholesky algorithm is a useful,
simple and efficient substitute for much more sophisticated algorithms 
applicable in the semidefinite case. Since it does not use pivoting, it
preserves the envelope of the matrix and is therefore especially suited
for banded matrices and envelope-based sparse matrix methods. 
It provides an estimate of the rank of the matrix, which 
however is rather unreliable. In the positive definite case the algorithm
is equivalent to the standard Cholesky method.

\section{Complexities beyond the basic modelling} \label{sec:complexities}

Section~\ref{sec:math} describes a set of baseline models for the sources,
attitude, and geometrical instrument, which are believed to be realistic enough 
to serve as an acceptable first-order approximation of the actual data for primary 
sources. Due to the many complexities of the
real satellite and its operation, as well as the physical environment in space, there 
are however many additional effects that may affect the astrometric results at
the $\mu$as level, and which have to be considered in the final modelling. 
In this Appendix we discuss some of the known effects that will be 
addressed in future versions of the astrometric solution.

\subsection{Chromaticity}\label{sec:chrom}

Although the Gaia telescopes are all-reflective, with no refractive elements in the 
optical paths to the astrometric field, they are nevertheless not completely achromatic.
In the presence of odd wavefront errors, such as coma, the centroids of the optical images
do in fact depend on the wavelength, and hence on the spectral energy distributions
of the observed sources.%
\footnote{`Centroid' should here not be understood as the centre of gravity 
of the optical image; rather, it is a non-trivial function of the light distribution,
similar to the estimation of the image location $\kappa$ in Sect.~\ref{sec:signalmodel}.
For the sake of illustration, the centroid could for example be the point
obtained by fitting a Gaussian PSF to the image.}
For the typical wavefront errors expected in the astrometric
field of Gaia (about 50~nm RMS), the AL centroid shift from an early-type star to a late 
spectral type could amount to several mas. This systematic effect, known as chromaticity, 
can therefore be many times larger than the photon-statistical uncertainty of the estimated
image location (cf.\ Table~\ref{tab:sim}). It is thus essential to have a very good calibration
of the chromaticity, for which it is necessary to know the spectral energy distribution of every 
observed source. This is obtained from the photometric observations in the BP and RP
fields (see Fig.~\ref{fig:fpa}).

Chromaticity is eliminated in the CCD signal analysis by using a 
Line Spread Function (LSF), $L(x)$ in Eq.~(\ref{eq:signal01}), which is correspondingly shifted
depending on the (known) spectrum of the source. The resulting AL pixel coordinate
$\kappa$ is therefore in principle achromatic, and the effect need not be further considered 
in the astrometric solution. However, as mentioned in Sect.~\ref{sec:instrumentmodel}, 
it is envisaged to have diagnostic colour-dependent terms in the geometric instrument 
model of the astrometric solution. These calibration parameters should obtain negligible values in 
the solution if the chromaticity has been correctly accounted for in the calibration of $L(x)$.
Conversely, non-zero values can be used to improve the LSF calibration in the next 
processing cycle.

\subsection{Charge Transfer Inefficiency of the CCDs}\label{sec:CTI}

The CCD signal model in Eq.~(\ref{eq:signal01}) assumes a perfectly linear
detector, which is not exactly the case for the real detectors and especially not 
in the presence of radiation damage on the CCDs. Traps in the silicon substrate,
produced by particle radiation in the space environment, will capture charges
during the TDI operation of the CCDs, and release them with delays ranging
from a fraction of the TDI period to minutes. The charge capture and release
processes introduce a number of phenomena, collectively referred to as
Charge Transfer Inefficiency \citep[$\text{CTI}=1-\text{CTE}$, where CTE is
the Charge Transfer Efficiency;][]{2001sccd.book.....J}. CTI will affect all kinds of
observations in Gaia (astrometric, photometric, spectroscopic). The most 
important phenomena for the astrometric observations are the
apparent charge loss (because part of the charges are released outside the
observed window) and centroid shift (because some charges are released with
a delay of one or a few TDI periods). These and more general effects of the
radiation damage are the subject of extensive theoretical and experimental
studies within the Gaia community 
\citep[e.g.,][]{2009SPIE.7439E...3S,2010SPIE.7742E..29P,2011MNRAS.414.2215P}. 
The adopted method to
handle these effects in the Gaia data processing is by means of forward modelling   
using a so-called Charge Distortion Model \citep[CDM;][]{short2010}. In the context
of the CCD signal model of Sect.~\ref{sec:signalmodel}, the CDM may be represented 
by the (non-linear) operator $D$:
\begin{equation}
    \lambda_k \equiv \mbox{E}\left( N_k \right) = D\left[
    \bigl\{\lambda_{k'}^0\bigr\}_{k'\le k}\, \middle\vert \, \Psi\right] \, ,
    \label{eq:signal02}
\end{equation}
where $\lambda_{k}^0$ is the signal model at sample $k$ in the absence of 
radiation-damage effects, e.g., according to Eq.~(\ref{eq:signal01}). 
The variable $\Psi$ represents the state of the CCD, e.g., in terms of how
much radiation damage it has suffered. Note that the expected value of sample
$k$ depends on the CCD illumination history up to and including sample $k$,
which is expressed by the CDM taking as input the (undamaged) value
not only for sample $k$ but also for the preceding samples ($k'\le k$).%
\footnote{A further complication is that most AF observations are binned in the
AC direction before read-out, while the CTI effects operate independently on each
pixel column. Thus the CDM should ideally be applied to the two-dimensional
charge image, and the distorted charge image then binned for comparison with
the one-dimensional data. This requires the use of a PSF replacing the LSF
in Eq.~(\ref{eq:signal01}).} 
One of the methods employed for mitigating CTI effects in Gaia 
is through a periodic (e.g., once per second) electronic injection of charges 
in a few consecutive TDI lines. As the lines of charges travel across the CCD, 
most of the harmful traps are temporarily filled, thus reducing the CTI of
subsequent charge transfers \citep{2007SPIE.6690E...8L}.
The method has the additional benefit of periodically resetting the illumination 
history of the pixels, so that in Eq.~(\ref{eq:signal02}) only the samples since the 
previous charge injection need to be considered \citep{short2010}.

The CDM depends on a moderate number of parameters that will be
estimated in parallel with the LSF (or PSF) calibration prior to the astrometric 
solution (the `Instrument response parameters' in Fig.~\ref{fig:blockdiagram}).
\emph{In principle}, the subsequently estimated image location $\kappa$ should 
then not only be achromatic, as discussed in Appendix~\ref{sec:chrom}, but also free 
of CTI effects, so that the astrometric solution can use a purely geometric 
instrument model, as required by the primary source model. Although this is 
obviously a highly idealised condition, it it nevertheless what the final data 
analysis must aim to achieve.

For the simple image of a primary source, the
centroid shift due to the CTI depends mainly on the magnitude of the source,
the time since the previous charge injection, and the accumulated radiation
dose experienced by the CCD \citep{2011MNRAS.414.2215P}. It is expected 
that imperfections in the CDM calibration will likewise show up in systematic 
shifts depending primarily on these (known) quantities, and can be represented 
by a set of diagnostic calibration functions in the generic calibration model of 
Sect.~\ref{sec:instrumentmodel}. Non-negligible values of the diagnostic
parameters in the astrometric solution indicate that the CDM is not doing 
its job properly, and they can then be used to improve the model.
The reader is referred to the cited papers for quantitative information on 
the expected level of CTI effects in Gaia data, the effectiveness of different 
mitigation strategies, and the associated performance degradations.

\subsection{Effects of the finite CCD integration time} \label{sec:TDI}

Up until now we have regarded the astrometric observations of Gaia as 
instantaneous measurements of the crossings of the source images 
over the fiducial `observation line' (Fig.~\ref{fig:instrcal}) 
at the centre of the CCD (or of the gated portion of the CCD) in the AL 
direction. In reality, due to the finite integration time ($T$)  of the 
CCD observations, any measurement clocked into the
CCD readout register at time $t$ actually depends on the average attitude
and source position over the preceding integration time interval, $[t-T,t]$. The time
delay is, to first order, taken into account by associating the measurement with the
observation time $t-T/2$ (Sect.~\ref{sec:signalmodel}). 
Following \citet{2005A&A...438..745B} we should, more precisely, 
assume that the observed location $\kappa$ of the image centre in the
pixel stream is a weighted mean of the \emph{instantaneous} location 
$\kappa_*(t)$ of the optical image relative to the charge image during the
preceding integration interval:
\begin{equation}\label{eq:kappa}
\kappa = \frac{\int_0^T e(\tau)\kappa_*(t-\tau)\,\text{d}\tau}{\int_0^T e(\tau)\,\text{d}\tau}\, ,
\end{equation}
where $e(\tau)$ is the (nominally flat) `exposure function' for look-back time 
$\tau$, i.e., the rate at which electrons are produced, and transported to the
read-out register, for constant illumination. The instantaneous location is 
given by
\begin{equation}\label{eq:kappa-ast}
\kappa_*(t) =  s\left[ \eta_*(t) - \eta_{fng} \right] + k(t)  \, ,
\end{equation}
where $s$ is the local scale factor (pixels per radian), $\eta_*(t)$ the 
instantaneous AL field angle of the optical image centre, and $\eta_{fgn}$ 
the AL coordinate of the fiducial observation line for the appropriate AC 
coordinate, field of view, etc. The function $k(t)$ is the inverse of $t_k$, 
relating the time coordinate to the TDI period index $k$. For the present 
discussion $k(t)$ is regarded as a continuous function, ignoring the 
step-wise transportation of the charge image in TDI mode.%
\footnote{The corresponding expression in \citet{2005A&A...438..745B}  is
their Eq.~(6), in which $k(t)$ is the (integer) index of the last TDI clock 
stroke before time $t$. Thus their $\kappa_*(t)$ oscillates with an amplitude 
of $\pm 0.5$~unit for every TDI period. The continuous approximation 
adopted in our Eq.~(\ref{eq:kappa-ast}) is acceptable since we are always 
considering integrals covering a whole number of oscillations.}

Recalling that $\eta_*(t)$ is decreasing with time (cf.\ Fig.~\ref{fig:fpa}),
while $k(t)$ is increasing, it is seen that $\kappa_*(t)$ remains 
approximately constant throughout the integration. Let us denote by
$t_\text{c}$ the exact time when the centre of the optical image crosses the
fiducial observation line, so that $\eta_*(t_\text{c})=\eta_{fng}$, and let
$\kappa_\text{c}=k(t_\text{c})$ be the corresponding pixel coordinate. If the speed 
of the optical image exactly matches the speed of the charge image, or 
$s\dot{\eta}_*+\spdot{k}=0$, it is seen that $\kappa_*(t)$ is 
indeed constant and equal to $\kappa_\text{c}$. Let us now consider what
happens if there is a mismatch between the speeds. This could be caused
by (i) a deviation in the local scale value $s$ due to optical distortion; 
(ii) a non-nominal local scan rate; and (iii) that the object itself has 
significant motion (e.g., an asteroid). Assuming that $\spdot{k}$ is constant 
and adopting a Taylor series expansion for the AL field angle over the 
exposure time, we have
\begin{align}\label{eq:kappa-model}
k(t) &= \kappa_\text{c} + (t-t_\text{c})\spdot{k} \, , \\
\eta_*(t) &= \eta_{fng} + (t-t_\text{c})\dot{\eta}_* + \frac{1}{2}(t-t_\text{c})^2\ddot{\eta}_* + \cdots\, .
\end{align} 
Inserting in Eq.~(\ref{eq:kappa-ast}), and assuming that $s$ is constant across 
the section of the CCD in question, we obtain by means of Eq.~(\ref{eq:kappa})
\begin{equation}\label{eq:kapp-eff}
\kappa = \kappa_\text{c} + (s\dot{\eta}_*+\spdot{k})e_1 + \frac{1}{2}s\ddot{\eta}_*e_2 + \cdots\, ,
\end{equation}
where
\begin{equation}\label{eq:e-moments}
e_m = \frac{\int_0^T e(\tau)(T/2-\tau)^m\,\text{d}\tau}{\int_0^T e(\tau)\,\text{d}\tau}\, , \quad m = 1,\, 2,\, \dots
\end{equation}
are the normalized moments of $e(\tau)$ relative to
the exposure mid-time at $\tau=T/2$. For a constant and matching image speed
we recover $\kappa=\kappa_\text{c}$ as expected. In the general case of imperfect
speed matching and non-constant scan rate there is a difference which should 
be taken into account in the astrometric solution. If we assume that $s$ is known, 
the speed mismatch $s\dot{\eta}_*+\spdot{k}$ can be computed for every observation, 
and the factor $e_1$ can then be estimated as an instrument calibration parameter 
using the generic calibration model in Sect.~\ref{sec:instrumentmodel}. $e_1$ will
depend (at least) on the CCD and gate used; but due to the accumulating radiation
damage it is likely to evolve with time and could possibly have a magnitude-dependent
component as well. In the next (quadratic) term we may know $\ddot{\eta}_*$ (from the
attitude) and $e_2 \simeq T^2/12$ (for constant exposure function) to sufficient 
accuracy that it might be applied as a correction to the observed $\kappa$. However,
since most observations are ungated (giving maximum $T$), it may be better to
adopt the uncorrected $\kappa$ for this maximum $T$ as defining the 
\emph{effective} attitude,%
\footnote{The effective attitude is then the physical attitude convolved 
with the (average) exposure function for maximum $T$. It corresponds closely 
to the `astrometric attitude' introduced by \citet{2005A&A...438..745B}.}
and only correct the gated observations for the difference in $e_2$; hence they, 
too, will refer to the effective attitude. A complication is that $\ddot{\eta}_*$
in Eq.~(\ref{eq:kapp-eff}) should be computed from (unknown) 
\emph{physical} attitude, but to first order it can be obtained from the effective 
attitude. Attitude irregularities on time scales shorter than $T$ add further 
complications (see Appendix~\ref{sec:attirr}), but it is unlikely that higher-order 
terms in Eq.~(\ref{eq:kapp-eff}) can profitably be accounted for.

\subsection{Attitude irregularities}\label{sec:attirr} 

The basic attitude model described in Sect.~\ref{sec:attmodel} uses a spline
representation which is (normally) continuous in the angles specifying the
instantaneous orientation of the instrument, as well as in the first $M-2$
derivatives of the angles, where $M$ is the order of the spline (typically
cubic splines are used, for which $M=4$). The actual (physical) attitude
is much more complex and in particular there may be discontinuities and
irregularities on time scales that are too short to be adequately represented
by a spline with the knot separations considered in the basic model.
Low-frequency perturbations ($\lesssim 0.01$~Hz) are of no concern here,
as they can be perfectly represented by splines. The most important contributors 
to perturbations at higher frequencies are thruster noise in the micro-propulsion
system used for the attitude control; the discrete and partially stochastic nature 
of the control system (for example that the commanded thrusts are updated once
per second); micrometeoroid impacts on the satellite; and various dynamical 
effects such as fuel sloshing and structural vibrations.

The high-frequency perturbations due to the thruster noise and control system
generate angular jitter of the physical attitude that has a significant amplitude
relative to the astrometric accuracies ultimately aimed at, but still small in
comparison with the AL pixel size ($\simeq 59$~mas). Thanks to the
TDI integration this high-frequency jitter is largely removed from the effective 
attitude by the averaging over the exposure time $T$. As a result, the shortest
knot interval needed to accurately represent the effective attitude is also of 
the order of the exposure time, or about 5~s. The optimum knot interval may 
be longer, depending on the number and magnitudes of the primary sources 
available for the attitude determination, and on the actual level of perturbations.

The expected frequency of micrometeoroid impacts of various sizes can be 
predicted from the known velocity and mass spectrum of interplanetary particles. 
Each impact produces a quasi-instantaneous change in the angular velocity
of the satellite, 
while the attitude angles are continuous across the impact time. It is
expected that a few hundred impacts will occur every year producing a change
in the AL angular velocity exceeding 1~mas~s$^{-1}$ (which should be easily
detectable), with the frequency roughly inversely proportional to the minimum 
change considered. Discontinuities in the \emph{physical} attitude rate can 
be represented in the spline model by inserting multiple knots at the estimated 
times of impact, $t_\text{i}$ (Sect.~\ref{sec:attInit}). However, the \emph{effective} 
attitude will instead see a linear change in the attitude rates over an interval 
equal to the exposure time $T$, centred on $t_\text{i}$, which requires that 
multiple knots are inserted both at $t_\text{i}-T/2$ and $t_\text{i}+T/2$. 
(This treatment becomes more problematic in connection with gated observations, 
when $T$ is non-nominal.)  
Impacts will be detected by inspecting the observation residuals in connection
with the attitude updating (Sect.~\ref{sec:attNoise}).  

The detailed re-examination of the Hipparcos attitude by
\citet{2005A&A...439..805V,book:newhip} revealed numerous discontinuities
of the along-scan attitude angle (scan phase) of several tens of mas. A large
fraction of them could be linked to the beginning or end of eclipses experienced
by Hipparcos in its highly elliptic orbit around the Earth. A likely cause is thermal 
re-adjustment of the solar-panel hinges, following a sudden change in temperature. 
As there is no change in the net angular momentum, but only a re-distribution of
inertia, the attitude rates are the same before and after a discontinuity.
For Gaia it is estimated that such `clanks' will be negligible, but the attitude
processing should nevertheless be capable of identifying instances, should they
occur, and to take appropriate measures. Due to the finite CCD integration time,
the apparent effects of a clank will be two discontinuities in the attitude rates,
equal but of opposite sign, and separated by the integration time $T$.
Again, this can be handled by suitable modification of the knot sequence of
the attitude spline. Like the micrometeoroid impacts, clanks will be detected
during the attitude updating by means of the characteristic patterns that they 
generate in the observation residuals versus time.

\section{Tables of acronyms and variables}\label{sec:vars}

Table~\ref{tab:acr} is a list of acronyms used in the paper. 
Table~\ref{tab:vars} lists the most important variables, with a short 
description and a reference to where it is introduced or explained.

\begin{table}
\caption{\label{tab:acr} List of acronyms}
\centering
\begin{tabular}{ll}
\hline\hline
\noalign{\smallskip} 
Acronym & Description \\
\noalign{\smallskip} 
\hline 
\noalign{\smallskip} 
AC&ACross scan direction (Fig.~\ref{fig:fpa}) \\
ACF&ACross scan coordinate in the Following FoV \\
ACP&ACross scan coordinate in the Preceding FoV \\
AF&Astrometric Field \\
AGIS&Astrometric Global Iterative Solution \\
AGN&Active Galactic Nucleus \\
AL&ALong scan direction (Fig.~\ref{fig:fpa}) \\
ASI&Accelerated Simple Iteration \\
BAM&Basic-Angle Monitor (Fig.~\ref{fig:fpa}) \\
BCRS&Barycentric Celestial Reference System \\
BP&Blue Photometer (Fig.~\ref{fig:fpa}) \\
CCD&Charge-Coupled Device \\
CDM&Charge Distortion Model \\
CFS&Calibration Faint Star \\
CG&Conjugate Gradient \\
CPU&Central Processing Unit \\
CoMRS&Centre of Mass Reference System \\
CTE & Charge Transfer Efficiency (of a CCD) \\
CTI & Charge Transfer Inefficiency (of a CCD) \\
CU2 & DPAC Coordination Unit 2, `Data Simulations' \\
CU3 & DPAC Coordination Unit 3, `Core Processing' \\
DPAC&Data Processing and Analysis Consortium \\
EADS&European Aeronautic Defence and Space company  \\
ESA&European Space Agency \\
ESAC&European Space Astronomy Centre \\
FFoV&Following Field of View (Fig.~\ref{fig:srs}) \\
FPA&Focal Plane Assembly  (Fig.~\ref{fig:fpa})\\
FoV&Field of View \\
GCRS&Geocentric Celestial Reference System \\
HEALPix&Hierarchical Equal-Area iso-Latitude Pixelisation \\
IAU&International Astronomical Union \\
ICRS&International Celestial Reference System \\
LSF&Line Spread Function \\
NSL&Nominal Scanning Law \\
PFoV&Preceding Field of View (Fig.~\ref{fig:srs}) \\
PPN&Parametrised Post-Newtonian (relativity formalism) \\
PSF&Point Spread Function \\
RMS&Root-Mean-Square \\
RP&Red Photometer \\
RSE&Robust Scatter Estimate (footnote \ref{footnRSE}) \\
RVS&Radial Velocity Spectrometer (Fig.~\ref{fig:fpa}) \\
SI&Simple Iteration \\
SM&Sky Mapper \\
SRS&Scanning Reference System (Fig.~\ref{fig:srs}) \\
SVD&Singular Value Decomposition \\
TB&TeraByte \\
TCB&Barycentric Coordinate Time \\
TDI&Time-Delayed Integration (CCD operation mode)\\
VLBI&Very Long Baseline Interferometry \\
WFS&WaveFront Sensor (Fig.~\ref{fig:fpa}) \\
XML&eXtensible Markup Language \\
\noalign{\smallskip} 
\hline 
\end{tabular} 
\end{table}

\begin{longtable}{lll}
\caption{\label{tab:vars} List of mathematical variables}\\ 
\hline\hline
\noalign{\smallskip} 
Var. & Description& Ref. \\
\noalign{\smallskip} 
\hline 
\noalign{\smallskip} 
\endfirsthead 
\caption{List of mathematical variables (continued)}\\ 
\hline\hline
\noalign{\smallskip} 
Var. & Description& Ref. \\
\noalign{\smallskip} 
\hline
\noalign{\smallskip} 
\endhead
\noalign{\smallskip} 
\hline
\endfoot
$A_\text{u}$ & the astronomical unit & Sect.~\ref{sec:astromodel}\\
$\vec{A}$ & attitude matrix & Eq.~(\ref{eq:attmat})\\
$\vec{a}$ & galactocentric acceleration vector & Sect.~\ref{sec:framedet}\\
$\vec{a}$ & attitude parameters & Sect.~\ref{sec:attmodel}\\
$\quat{a}_n$ & quaternion coefficient for $B_n(t)$ & Eq.~(\ref{eq:qspline})\\
$B_n(t)$ & B-spline function & Sect.~\ref{sec:bsplines} \\
$\vec{B}$ & iteration matrix & Sect.~\ref{sec:iter} \\
$\vec{b}$ & right-hand side of normal equations & Eq.~(\ref{eq:norm1}) \\
$\vec{b}_G(t)$ & barycentric coordinate of Gaia at time $t$ & Eq.~(\ref{eq:astro-standard}) \\
$\tens{C}$ & celestial reference system & Sect.~\ref{sec:refsystems} \\
$\vec{C}$ & constraint matrix for the calibration parameters & Eq.~(\ref{eq:genericConstraint}) \\
$\vec{c}$ & calibration parameters & Sect.~\ref{sec:instrumentmodel}\\
$D$ & non-linear Charge Distortion Model  & Eq.~(\ref{eq:signal02}) \\
$\vec{d}$ & update vector & Sect.~\ref{sec:iter}\\
$\mbox{E}$ & expectation operator   & Eq.~(\ref{eq:signal01})\\
$e(t)$ & CCD exposure function & Appendix~\ref{sec:TDI}\\
$\vec{e}$ & error vector & Sect.~\ref{sec:iter}\\
$f$ & field index ($\pm 1$ for preceding/following) & Eq.~(\ref{eq:srspropeta}) \\
$\vec{f}$ & detector coordinates  & Eq.~(\ref{eq:generalobs})\\
$\vec{f}_\text{P}$, $\vec{f}_\text{F}$ & preceding, following viewing direction & Fig.~\ref{fig:srs} \\
$G$ & Gaia broadband magnitude & Sect.~\ref{sec:instrumentmodel} \\
$g$ & gate index & Eq.~(\ref{eq:srspropeta}) \\
$\vec{g}$ & global parameters & Sect.~\ref{sec:instrumentmodel} \\
$\vec{h}$ & auxiliary data, e.g., ephemerides &Eq.~(\ref{eq:astro})\\
$\vec{I}$ & the identity matrix  & Sect.~\ref{sec:iter} \\
$i$ & subscript denoting some source  & Sect.~\ref{sec:astromodel} \\
$j$ & `short' calibration time interval index  & Eq.~(\ref{eq:etazeta}) \\
$K$ & no.\ of TDI periods for CCD integration  & Sect.~\ref{sec:signalmodel} \\
$\vec{K}$ & preconditioner matrix & Sect.~\ref{sec:iter} \\
$k$ & TDI index in CCD pixel stream  & Sect.~\ref{sec:signalmodel} \\
$k$ & `long' calibration time interval index  & Eq.~(\ref{eq:etazeta}) \\
$k$ & iteration index  & Eq.~(\ref{eq:iter}) \\
$L$& Line Spread Function  & Eq.~(\ref{eq:signal01}) \\
$L_r^\ast$ & shifted Legendre polynomial of degree $r$ & Eq.~(\ref{eq:etazeta}) \\
$l$ & subscript denoting some observation  & Sect.~\ref{sec:synthesismodel} \\
$\ell$ & left index in a knot sequence  & Sect.~\ref{sec:calcBsplines} \\
$M$ & spline order -- $M=4$ for cubic spline & App.~\ref{sec:splines} \\
$N$ & number of degrees of freedom for a spline & App.~\ref{sec:splines} \\
$\vec{N}$ & normal equations matrix & Eq.~(\ref{eq:norm1}) \\
$n$ & attitude parameter index & Eq.~(\ref{eq:qspline}) \\
$n$ & CCD index & Eq.~(\ref{eq:etazeta0}) \\
$n$ & dimension of the normal matrix $\vec{N}$ & Eq.~(\ref{eq:null}) \\
$\vec{n}$ & nuisance parameters & Eq.~(\ref{eq:generalobs}) \\
$\vec{p}_i$ &  normal-triad component  & Eq.~(\ref{eq:normal-triad}) \\
$Q$ & objective function to be minimized  & Eq.~(\ref{eq:synth}) \\
$\quat{q}$ & attitude quaternion & Eq.~(\ref{eq:qspline}) \\
$\vec{q}_i$ &  normal-triad component  & Eq.~(\ref{eq:normal-triad}) \\
$r$ & degree of the small-scale calibration polynomial & Eq.~(\ref{eq:etazeta}) \\
$\vec{r}_i$ &  normal-triad component  & Eq.~(\ref{eq:normal-triad}) \\
$R_l$ & residual of observation $l$  &  Sect.~\ref{sec:synthesismodel} \\
$\tens{S}$ & Scanning Reference System & Sect.~\ref{sec:refsystems} \\
$s$ & local scale factor & Eq.~(\ref{eq:kappa-ast}) \\
$\vec{s}$ & astrometric (`source') parameters & Sect.~\ref{sec:astromodel} \\
$\vec{s}_i$ & the astrometric parameters for source $i$ & Sect.~\ref{sec:astromodel} \\
$T$ & light integration (exposure) time on CCD & Sect.~\ref{sec:instrumentmodel}, \ref{sec:TDI}\\
$t_k$ & time for sample $k$  & Sect.~\ref{sec:signalmodel} \\
$t_l$ & time for observation $l$ & Sect.~\ref{sec:signalmodel} \\
$t_\text{ep}$ & reference epoch for astrometric parameters &Sect.~\ref{sec:astromodel} \\
$t_\text{fr}$ & reference epoch for frame rotator & Sect.~\ref{sec:framerotator} \\
$t_\text{P}$ & reference epoch for non-rotating ICRS sources & Sect.~\ref{sec:framedet}\\
$t_\text{PM}$ & reference epoch for moving ICRS sources  & Sect.~\ref{sec:framedet}\\
$U$ & relegation factor for primary-star selection & Sect.~\ref{sec:relegate} \\
$\vec{u}_i$ & proper direction to source $i$  &  Sect.~\ref{sec:astromodel} \\
$\vec{\bar{u}}_i$ & geometric direction to source $i$  &  Sect.~\ref{sec:astromodel} \\
$\vec{V}$ & nullspace of the normal matrix & Sect.~\ref{sec:iter} \\
$\vec{W}_i$ & weight matrix for source $i$ & Eq.~(\ref{eq:normSmat})\\
$w_l$ & downweighting factor for observation $l$ & Sect.~\ref{sec:synthesismodel}, \ref{sec:source-inner}\\
$W_l$ & statistical weight of observation $l$ & Sect.~\ref{sec:synthesismodel}\\
$\vec{x}$ & differential parameter vector& Sect.~\ref{sec:norm}\\
$[\vec{X~Y~Z}]$ & celestial reference system (ICRS or CoMRS) & Sect.~\ref{sec:refsystems} \\
$[\vec{x~y~z}]$ & Scanning Reference System (SRS) & Fig.~\ref{fig:srs} \\
$\alpha$ &  flux (image parameter)  &Eq.~(\ref{eq:signal01})\\
$\alpha_i$ & right ascension of source $i$ at $t_\text{ep}$  &Sect.~\ref{sec:astromodel} \\
$\beta$ &  background level (image parameter)  &Eq.~(\ref{eq:signal01})\\
$\Gamma_{\rm c}$ &  (conventional) basic angle  & Eq.~(\ref{eq:srspropeta})\\
$\gamma$ &  PPN curvature parameter  & Eq.~(\ref{eq:gammaPPN})\\
$\delta_i$ &  declination of source $i$ at $t_\text{ep}$ &Sect.~\ref{sec:astromodel} \\
$\epsilon_a$ &  excess attitude noise & Eq.~(\ref{eq:excess}) \\
$\epsilon_i$ &  excess source noise (for source $i$) & Eq.~(\ref{eq:excess}) \\
$\epsilon_l$ &  excess noise in observation $l$ & Eq.~(\ref{eq:excess}) \\
$\vec{\varepsilon}$ & orientation correction (frame rotator) &Sect.~\ref{sec:framerotator} \\
$\zeta$ & across-scan (AC) field angle & Eq.~(\ref{eq:srsprop1}) \\
$\eta$ & along-scan (AL) field angle & Eq.~(\ref{eq:srspropeta}) \\
$\kappa$ &  along-scan (AL) pixel coordinate (image location)  &Eq.~(\ref{eq:signal01})\\
$\lambda$ & regularization parameter for the attitude update &Eq.~(\ref{eq:JfuncMod})\\
$\lambda_k$ & intensity for CCD sample values  & Eq.~(\ref{eq:signal01})\\
$\mu$ & across-scan (AC) pixel coordinate (image location) & Sect.~\ref{sec:instrumentmodel}\\
$\mu_{\alpha* i}$ & proper motion in $\alpha$ ($\times\cos\delta_i$)
for source $i$ &Sect.~\ref{sec:astromodel} \\
$\mu_{\delta i}$ & proper motion in $\delta$ for source $i$ & Sect.~\ref{sec:astromodel} \\
$\mu_{ri}$ & radial proper motion for source $i$ &Sect.~\ref{sec:astromodel} \\
$\vec{\mu}_0$ & proper motion due to galactocentric acceleration & Sect.~\ref{sec:framedet}\\
$\nu$ &  number of degrees of freedom & Sect.~\ref{sec:source-inner}\\
$\varpi_i$ & parallax for source $i$  &Sect.~\ref{sec:astromodel} \\
$\varrho$ & normalized RSE error of astrometric parameters & Sect.~\ref{sec:ressrc}\\
$\sigma_l$ & formal standard uncertainty for observation $l$ & Sect.~\ref{sec:synthesismodel} \\
$\tau_n$ & attitude spline knot $n$ & Sect.~\ref{sec:bsplines} \\
$\varphi$ & along-scan instrument angle & Eq.~(\ref{eq:srsprop1}) \\
$\vec{\omega}$ & spin correction (frame rotator)  &Sect.~\ref{sec:framerotator} \\
\end{longtable}
 
\end{document}